\documentclass[12pt,a4paper]{article}

\usepackage[UKenglish]{babel}
\usepackage{geometry}
\usepackage[toc, page]{appendix}
\usepackage{cite}
\usepackage[bf]{caption}
\usepackage{subcaption}
\usepackage[bookmarksopen, bookmarksnumbered, bookmarksopenlevel=2, hidelinks]{hyperref}

\hypersetup{
	pdftitle={Non-minimal Elliptic Threefolds at Infinite Distance II: Asymptotic Physics},
	pdfauthor={Rafael \'Alvarez-Garc\'ia, Seung-Joo Lee and Timo Weigand}
}

\def\fnote#1#2{\begingroup\def\thefootnote{#1}\footnote{#2}\addtocounter{footnote}{-1}\endgroup}

\usepackage{amsmath}
\usepackage{amssymb}
\usepackage{amsthm}
\usepackage{mathtools}
\usepackage{thmtools}
\usepackage{mathrsfs}

\numberwithin{equation}{section}
\numberwithin{table}{section}

\newtheorem{theorem}{Theorem}[section]

\theoremstyle{definition}

\theoremstyle{definition}

\theoremstyle{remark}

\makeatletter
\def\th@plain{%
  \thm@notefont{}
  \itshape 
}
\def\th@definition{%
  \thm@notefont{}
  \normalfont 
}
\makeatother

\usepackage[capitalise,noabbrev]{cleveref}
\usepackage{graphicx}
\usepackage{xcolor}
\usepackage{tabularray}
\UseTblrLibrary{diagbox}
\usepackage{tcolorbox}
\tcbset{title filled, arc=0mm}
\usepackage{enumitem}
\usepackage{accents}
\usepackage{trimclip}
\usepackage{rotating}
\usepackage[all]{nowidow}
\usepackage{microtype}

\definecolor{diagDarkPurple}{RGB}{159,64,116}
\definecolor{diagLightPurple}{RGB}{201,127,232}
\definecolor{diagDarkGreen}{RGB}{80,82,34}
\definecolor{diagLightGreen}{RGB}{110,139,32}
\definecolor{diagPurpleBlue}{RGB}{69,26,121}
\definecolor{diagDarkBlue}{RGB}{45,35,133}
\definecolor{diagLightBlue}{RGB}{32,107,190}
\definecolor{diagCyan}{RGB}{64,215,169}
\definecolor{diagDarkBrown}{RGB}{91,71,56}
\definecolor{diagLightBrown}{RGB}{119,107,84}
\definecolor{diagDarkRed}{RGB}{138,52,31}
\definecolor{diagMediumRed}{RGB}{214,69,32}
\definecolor{diagLightRed}{RGB}{214,127,127}
\definecolor{diagDarkYellow}{RGB}{116,75,26}
\definecolor{diagMediumYellow}{RGB}{181,124,16}
\definecolor{diagLightYellow}{RGB}{181,153,63}
\definecolor{mixPurpleGreen}{RGB}{156,133,132}
\definecolor{mixYellowGreen}{RGB}{146,146,48}

\usepackage{tikz}
\usetikzlibrary{cd, decorations.pathmorphing, shapes.misc}

\tikzstyle{black line}=[-, draw=black, line width=1.5pt, line cap=rect]
\tikzstyle{black fine-line}=[-, draw=black, line width=0.75pt, line cap=rect]
\tikzstyle{dashed black line}=[dashed, draw=black, line width=1.5pt, line cap=rect]
\tikzstyle{dotted black line}=[loosely dotted, draw=black, line width=1.5pt, line cap=rect]
\tikzstyle{light-yellow thick-line}=[-, draw=diagLightYellow, line width=1.5pt, line cap=rect]
\tikzstyle{medium-red thick-line}=[-, draw=diagMediumRed, line width=1.5pt, line cap=rect]

\tikzstyle{dark-purple line}=[-, draw=diagDarkPurple, line width=1.25pt]
\tikzstyle{light-purple line}=[-, draw=diagLightPurple, line width=1.25pt]
\tikzstyle{dark-green line}=[-, draw=diagDarkGreen, line width=1.25pt]
\tikzstyle{light-green line}=[-, draw=diagLightGreen, line width=1.25pt]
\tikzstyle{purple-blue line}=[-, draw=diagPurpleBlue, line width=1.25pt]
\tikzstyle{dark-blue line}=[-, draw=diagDarkBlue, line width=1.25pt]
\tikzstyle{light-blue line}=[-, draw=diagLightBlue, line width=1.25pt]
\tikzstyle{cyan line}=[-, draw=diagCyan, line width=1.25pt]
\tikzstyle{dark-brown line}=[-, draw=diagDarkBrown, line width=1.25pt]
\tikzstyle{light-brown line}=[-, draw=diagLightBrown, line width=1.25pt]
\tikzstyle{dark-red line}=[-, draw=diagDarkRed, line width=1.25pt]
\tikzstyle{medium-red line}=[-, draw=diagMediumRed, line width=1.25pt]
\tikzstyle{light-red line}=[-, draw=diagLightRed, line width=1.25pt]
\tikzstyle{dark-yellow line}=[-, draw=diagDarkYellow, line width=1.25pt]
\tikzstyle{medium-yellow line}=[-, draw=diagMediumYellow, line width=1.25pt, line cap=round]
\tikzstyle{light-yellow line}=[-, draw=diagLightYellow, line width=1.25pt]
\tikzstyle{medium-yellow snake}=[-, draw=diagMediumYellow, line width=1.25pt, decorate, decoration=snake]
\tikzstyle{light-yellow snake}=[-, draw=diagLightYellow, line width=1.25pt, decorate, decoration=snake]
\tikzstyle{light-yellow snake dotted}=[densely dotted, draw=diagLightYellow, line width=1.25pt, decorate, decoration=snake]
\tikzstyle{light-yellow dotted}=[densely dotted, draw=diagLightYellow, line width=1.25pt]
\tikzstyle{medium-red zigzag}=[-, draw=diagMediumRed, line width=1.25pt, decorate, decoration=zigzag]
\tikzstyle{medium-red zigzag thick line}=[-, draw=diagMediumRed, line width=1.5pt, decorate, decoration=zigzag]

\tikzset{cross-purple/.style={cross out, draw=diagLightPurple, line width=1.5pt, minimum size=2*(#1-\pgflinewidth), inner sep=0pt, outer sep=0pt},cross/.default={1pt}}
\tikzset{cross-green/.style={cross out, draw=diagLightGreen, line width=1.5pt, minimum size=2*(#1-\pgflinewidth), inner sep=0pt, outer sep=0pt},cross/.default={1pt}}
\tikzset{cross-yellow/.style={cross out, draw=diagLightYellow, line width=1.5pt, minimum size=2*(#1-\pgflinewidth), inner sep=0pt, outer sep=0pt},cross/.default={1pt}}
\tikzset{cross-blue/.style={cross out, draw=diagLightBlue, line width=1.5pt, minimum size=2*(#1-\pgflinewidth), inner sep=0pt, outer sep=0pt},cross/.default={1pt}}
\tikzset{cross-red/.style={cross out, draw=diagMediumRed, line width=1.5pt, minimum size=2*(#1-\pgflinewidth), inner sep=0pt, outer sep=0pt},cross/.default={1pt}}
\tikzset{cross-cyan/.style={cross out, draw=diagCyan, line width=1.5pt, minimum size=2*(#1-\pgflinewidth), inner sep=0pt, outer sep=0pt},cross/.default={1pt}}
\tikzset{cross-purple-green/.style={cross out, draw=mixPurpleGreen, line width=1.5pt, minimum size=2*(#1-\pgflinewidth), inner sep=0pt, outer sep=0pt},cross/.default={1pt}}
\tikzset{cross-yellow-green/.style={cross out, draw=mixYellowGreen, line width=1.5pt, minimum size=2*(#1-\pgflinewidth), inner sep=0pt, outer sep=0pt},cross/.default={1pt}}
\tikzset{cross-node/.style={path picture={\draw[diagLightYellow] (path picture bounding box.south east) -- (path picture bounding box.north west) (path picture bounding box.south west) -- (path picture bounding box.north east);}}}

\newcommand{\ord}[1]{\mathrm{ord}_{#1}}
\newcommand{\fphys}{f_{\mathrm{phys}}}
\newcommand{\gphys}{g_{\mathrm{phys}}}

\newcommand{\Dphys}{\Delta_{\mathrm{phys}}}

\newcommand{\bigslant}[2]{{\left.\raisebox{.2em}{$#1$}\middle/\raisebox{-.2em}{$#2$}\right.}}

\makeatletter
\newcommand{\smallbullet}{} 
\DeclareRobustCommand\smallbullet{%
  \mathord{\mathpalette\smallbullet@{0.5}}%
}
\newcommand{\smallbullet@}[2]{%
  \vcenter{\hbox{\scalebox{#2}{$\m@th#1\bullet$}}}%
}
\makeatother
\DeclareMathOperator{\rank}{rank}
\DeclareMathOperator{\tr}{tr}

\begin{document}

\begin{flushright}
{\tt\normalsize CTPU-PTC-23-54}\\
{\tt\normalsize ZMP-HH/23-22}
\end{flushright}

\vskip 40 pt
\begin{center}
{\large \bf%
Non-minimal Elliptic Threefolds at Infinite Distance II:\\ \vspace{2mm} Asymptotic Physics
} 

\vskip 11 mm

Rafael \'Alvarez-Garc\'ia,${}^{1}$
Seung-Joo Lee,${}^{2}$
and Timo Weigand${}^{1,3}$

\vskip 11 mm
\small ${}^{1}${\textit{II. Institut f\"ur Theoretische Physik, Universit\"at Hamburg,\\  Luruper Chaussee 149, 22607 Hamburg, Germany}}\\[3 mm]
\small ${}^{2}${\textit{%
Particle Theory and Cosmology Group, Center for Theoretical Physics of the Universe,\\ 
Institute for Basic Science (IBS), Daejeon 34126, Korea
}}\\[3 mm]
\small ${}^{3}${\textit{Zentrum f\"ur Mathematische Physik, Universit\"at Hamburg,\\ Bundesstrasse 55, 20146 Hamburg, Germany}}\\[3 mm]

\fnote{}{Email: \href{mailto:rafael.alvarez.garcia@desy.de}{rafael.alvarez.garcia{\fontfamily{ptm}\selectfont @}desy.de}, \href{mailto:seungjoolee@ibs.re.kr}{seungjoolee{\fontfamily{ptm}\selectfont @}ibs.re.kr}, \href{mailto:timo.weigand@desy.de}{timo.weigand{\fontfamily{ptm}\selectfont @}desy.de}}
\end{center}

\vskip 7mm
\begin{abstract}
We interpret infinite-distance limits in the complex structure moduli space of F-theory compactifications to six dimensions in the light of general ideas in quantum gravity. The limits we focus on arise from non-minimal singularities in the elliptic fiber over curves in a Hirzebruch surface base, which do not admit a crepant resolution. Such degenerations take place along infinite directions in the non-perturbative brane moduli space in F-theory. A blow-up procedure, detailed generally in Part I of this project \cite{Alvarez-Garcia:2023gdd}, gives rise to an internal space consisting of a union of log Calabi-Yau threefolds glued together along their boundaries. We geometrically classify the resulting configurations for genus-zero single infinite-distance limits. Special emphasis is put on the structure of singular fibers in codimension zero and one. As our main result, we interpret the central fiber of these degenerations as endpoints of a decompactification limit with six-dimensional defects. The conclusions rely on an adiabatic limit to gain information on the asymptotically massless states from the structure of vanishing cycles. We also compare our analysis to the heterotic dual description where available. Our findings are in agreement with general expectations from quantum gravity and provide further evidence for the Emergent String Conjecture.
\end{abstract}

\vfill
\thispagestyle{empty}
\pagenumbering{roman}
\setcounter{page}{0}
\newpage

\tableofcontents
\newpage

\pagenumbering{arabic}
\setcounter{page}{1}


\section{Introduction and summary}

In this article, we continue the systematic analysis of a large class of infinite-distance degenerations of elliptic Calabi-Yau threefolds and their F-theory interpretation initiated in \cite{Alvarez-Garcia:2023gdd}. The degenerations studied are those in which the threefold develops non-minimal, or non-Kodaira, singularities in the elliptic fiber over loci of complex codimension one in the base of the elliptic fibration.
Our interest in these degenerations is largely motivated by the goal to understand asymptotic regions in the moduli spaces of string compactifications which lie at infinite distance. According to general ideas in the Swampland Program of quantum gravity \cite{Vafa:2005ui, Brennan:2017rbf, Palti:2019pca, vanBeest:2021lhn, Grana:2021zvf, Agmon:2022thq}, near such asymptotic branches infinite towers of states should become light, and the theory should asymptote to a dual description \cite{Ooguri:2006in}. The towers should furthermore admit an interpretation as Kaluza-Klein towers, possibly in some dual frame, or as excitations of a unique critical and asymptotically weakly coupled string \cite{Lee:2019wij}. If true, this Emergent String Conjecture would greatly constrain the moduli space dynamics, at least in all asymptotic regions.

These and related ideas have so far withstood numerous quantitative tests, mostly in the closed string moduli space of Calabi-Yau compactifications. This includes the complex structure and K\"ahler moduli space probed by string and M-theory, see for instance  \cite{Grimm:2018ohb,Blumenhagen:2018nts,Lee:2018urn,Grimm:2018cpv,Joshi:2019nzi,Font:2019cxq,Lee:2019xtm,Erkinger:2019umg,Grimm:2019ixq,Baume:2019sry,Lee:2019wij, Klawer:2021ltm, Alvarez-Garcia:2021mzv, Alvarez-Garcia:2021pxo}. In the present work, we aim to advance our understanding of infinite-distance limits in what to first approximation can be viewed as the open string moduli space. As argued in \cite{Lee:2021qkx,Lee:2021usk} in the context of F-theory compactifications to eight dimensions, non-compact directions in the open string moduli space occur at the non-perturbative level. Along such non-compact directions, suitable mutually non-local $[p,q]$ 7-branes coalesce. In F-theory on elliptic K3 surfaces, the resulting brane configurations are encoded geometrically in non-Kodaira singularities over points on the rational base. Their study is interesting by itself, which furnishes an independent motivation for the present work.

More precisely, in the complex structure moduli space of elliptic K3 surfaces, any fibral degenerations at infinite distance involve such non-Kodaira fibers in codimension-one and/or minimal Kodaira singularities in codimension-zero. As such, they are classified as Type II and Type III Kulikov models \cite{Kulikov1977, Persson1977,Kulikov1981, PerssonPinkham1981}, with a more recent refinement as Type~II.a, Type~II.b, Type~III.a and Type~III.b \cite{Lee:2021qkx} (see  \cite{Brunyantethesis, alexeev2021compactifications, ascher2021compact,odaka2020pl, odaka2021collapsing} for independent treatments). Furthermore, the physical interpretation of this refinement indeed confirmed the ideas of the Emergent String Conjecture in a non-trivial manner \cite{Lee:2021qkx, Lee:2021usk}, as illustrated in \cref{fig:Kulikov-models-summary}.

In the first part of the current analysis \cite{Alvarez-Garcia:2023gdd} we have provided a classification of the analo\-gous codimension-one non-Kodaira degenerations of elliptic Calabi-Yau threefolds, underlying \mbox{F-theory} compactifications to six dimensions. This geometric analysis is complementary to the Hodge theoretic approach to studying complex structure degenerations of Calabi-Yau varieties, explored in  \cite{Grimm:2018ohb,Grimm:2018cpv,Grimm:2019ixq} in the context of string theory. A succinct summary of \cite{Alvarez-Garcia:2023gdd}, focusing on so-called single infinite-distance limits, will be provided in \cref{sec:review-part-I}. The goal of the present article is to interpret these geometric results from the physics point of view. As we will see, the situation is considerably richer compared with the degenerations of elliptic K3 surfaces. Our strategy will therefore be to focus on a favourable corner of the moduli space, where a particularly clear picture emerges. 

Specifically, the main part of our analysis is devoted to what we call horizontal models. The base of the degenerating elliptic fibration is taken to be a Hirzebruch surface $\mathbb{F}_{n}$, which is, topologically, a fibration of a rational curve $\mathbb{P}^{1}_{f}$ over a rational base $\mathbb{P}^{1}_{b}$. Consequently, the threefold admits a fibration by K3 surfaces over $\mathbb{P}^{1}_{b}$. This allows us to leverage, in favourable cases, the insights gained in \cite{Lee:2021qkx,Lee:2021usk} on the degenerations of K3 surfaces. Such favourable cases include the horizontal models, where the non-minimal fibers are engineered to appear over certain section divisors of the base $\mathbb{F}_{n}$. After performing a suitable chain of blow-ups and line bundle shifts in order to resolve the degeneration, as described generally in \cite{Alvarez-Garcia:2023gdd}, the endpoint of the limit is described by a Calabi-Yau threefold free of non-minimal singularities in codimension one. Its base is an open chain of surfaces $B^{p}$, each of which has the topology of an $\mathbb{F}_{n}$ surface. It is furthermore guaranteed that the elliptic fiber can degenerate at worst to a Kodaira type~$\mathrm{I}_{n_{p} \geq 0}$ fiber over the generic points of each base component $B^{p}$. Those components~$B^{p}$ for which $n_{p} > 0$ represent a local region of asymptotic weak coupling in F-theory, while in those for which $n_{p} = 0$ the physics remains strongly coupled. The infinite-distance limits that we take lead to consistent patchings of these so-called log Calabi-Yau spaces. The resulting asymptotic physics is encoded both in the nature of the codimension-zero fibers of the different \mbox{log Calabi-Yau} spaces, and in the way how the 7-branes extend through the intersection loci between them. The degeneration furthermore turns out to be responsible for the appearance of asymptotically massless particle~towers.

Our first task is to classify the patterns of generic $\mathrm{I}_{n_{p}}$ fibers that can appear at the endpoints of the limits. The resulting models are found to fall into four classes, examples of which can also be obtained by fibering the four types of refined Kulikov models for elliptic K3 surfaces over the base of the original $\mathbb{F}_{n}$ surface. They are the following:
\begin{itemize}
    \item \textbf{Type II.a model:} There are at least two components and all $n_{p}=0$, such that the physics is globally at strong coupling.
    
    \item \textbf{Type II.b model:} There is a single component with $n_{0} > 0$; the models of this type are a bit of an outlier and correspond to global Type IIB orientifold limits.
    
    \item \textbf{Type III.a model:} There are at least two components, with $n_{p} = 0$ for one or both end-components and $n_{p} > 0$ for the intermediate components, so that weak and strong coupling regions are mixed at the global level.
    
    \item \textbf{Type III.b model:} There are at least two components and all $n_{p} > 0$, such that the physics is globally at weak coupling; as we explain, such limits are only possible for models constructed over $\mathbb{F}_{n}$ with $0 \leq n \leq 4$.
\end{itemize}

Apart from the structure of codimension-zero fibers over the base components, we must also account for the types of divisors wrapped by 7-branes in the open-chain formed by them. According to their position with respect to the rational fibration of the base, we distinguish between horizontal, vertical and mixed divisors. Horizontal branes are the direct analogue of the 7-branes in the eight-dimensional degenerations of \cite{Lee:2021qkx,Lee:2021usk}, while the most interesting new effects of the degenerations of threefolds are associated with the vertical branes. Before coming to their interpretation in the infinite-distance limits, we first find general constraints on the patterns of enhancements that can occur, and in particular establish bounds on the rank of the gauge algebra supported on the vertical divisors.

With this geometric understanding at our disposal, we approach the interpretation of the infinite-distance limits, beginning with the horizontal Type II.a models. To retain control, we first take an adiabatic limit, in which the base curve $\mathbb{P}^{1}_{b}$ is taken to be asymptotically large. This regime in the moduli space affords us a clear geometric picture of the degenerating cycles in the infinite-distance limit, and also allows us to compare our conclusions to the heterotic dual side. Over a generic point on $\mathbb P^{1}_{b}$, one can locally define two vanishing 2-cycles of the topology of a torus, as in the eight-dimensional analysis of \cite{Lee:2021qkx,Lee:2021usk}. In the M-theory picture, two towers of asymptotically massless states arise from M2-branes wrapped multiple times around these local 2-cycles. By comparison with the heterotic dual, we can identify these as \mbox{Kaluza-Klein} towers of a decompactification limit. More precisely, the heterotic dual is defined on an elliptic K3 surface which undergoes an adiabatic decompactification limit of large generic fiber with $\mathcal{V}_{\mathbb{P}^{1}_{b}} \gg \mathcal{V}_{T^{2}_{\mathrm{het}}} \rightarrow \infty$. The two towers associated with the locally defined vanishing 2-cycles on the F-theory side map to the two Kaluza-Klein towers along the decompactifying heterotic torus. However, unlike for the simpler eight-dimensional models, these towers are not globally well-defined because the construction of the two-cycles breaks down over special points on $\mathbb{P}^{1}_{b}$. These special points include, in particular, the location of vertical 7-branes. Our interpretation is that these vertical branes do not participate in the decompactification process, but remain as six-dimensional defects. In the adiabatic limit of large $\mathbb{P}^{1}_{b}$, these defects can be thought of as being pushed to infinity so that we end up with a decompactification to ten dimensions with six-dimensional defects. As noted above, the total rank of the defect algebras is bounded, unlike in generic theories with lower-dimensional defects. This reflects the special status of the asymptotic theories as decompactification limits of lower-dimensional ones, which have to obey more stringent quantum gravity bounds on the ranks of the allowed gauge algebras.

On the dual heterotic side, the vertical branes map to non-perturbative gauge sectors which are localised at the geometric singularities of the heterotic K3 surface probed by point-like instantons with discrete holonomy. We explain this phenomenon, which is interesting by itself, in a separate appendix.

Away from the adiabatic regime, the two (dual) Kaluza-Klein towers associated with the locally defined 2-cycles reorganise in a complicated manner, which reflects the fact that the globally defined objects degenerating in the infinite-distance limits are 3-cycles. We argue that, on the heterotic side, departure from adiabaticity still leads to a decompactification limit to ten dimensions. While this constitutes no proof, we take it nonetheless as pointing towards a ten-dimensional decompactification also on the F-theory side. An independent identification of the towers away from the adiabatic regime remains as a challenge, but our analysis confirms that, at least modulo the assumption of adiabaticity, the towers indeed admit an interpretation as Kaluza-Klein towers, as postulated by the Emergent String Conjecture.

The analysis can be carried out in a very similar fashion for horizontal Type III.a models, which contain both weak and strong coupling regions. In the adiabatic regime, we propose a decompactification to nine dimensions, together with lower-dimensional defects. The remaining horizontal limits are found to be all weak coupling limits, possibly superimposed with a decompactification limit.

Finally, in addition to the horizontal models, there exist three more types of genus-zero single infinite-distance limit degenerations of Hirzebruch models \cite{Alvarez-Garcia:2023gdd}. Two of them admit a straightforward generalisation of the adiabatic analysis, with a corresponding physical interpretation.

The article is organised as follows. We begin, in  \cref{sec:Kulikov-models-review}, with a review of the four types of Kulikov models for elliptic K3 surfaces. We also briefly summarise the main results of the geometric analysis of the analogous \mbox{non-minimal} degenerations of elliptic Calabi-Yau threefolds carried out in \cite{Alvarez-Garcia:2023gdd}. In \cref{sec:horizontal-models-general-properties}, we then elaborate on the geometry of what we call horizontal models over Hirzebruch base spaces. This includes a derivation of rank bounds for the gauge algebras supported on vertical branes. Various technical details are relegated to \cref{sec:discriminant-weakly-coupled-components,sec:bounds-codimension-zero-pattern,sec:sec:bounds-vertical-gauge-rank-examples}. The interpretation of horizontal Type II.a models as decompactification limits and their dual heterotic interpretation is the subject of \cref{sec:horizontal-models-heterotic-duals}. As a spin-off of our analysis of the defects, we elaborate on the heterotic dual interpretation of vertical gauge algebras as non-perturbative heterotic gauge theories at the location of ADE singularities probed by point-like instantons with discrete holonomy. We include this analysis in \cref{sec:defect-algebras-heterotic-dual}. In \cref{sec:horizontal-models-systematics} we generalise our interpretation of the infinite-distance limits to horizontal Type III.a models, as well as the globally weakly coupled Type II.b models and Type III.b models. Their relation to perturbative Type IIB orientifold limits is analysed in \cref{sec:horizontal-IIIb-models-orientifold-geometry}. Generalisations to genus-zero single infinite-distance limits that are not of horizontal type are the subject of \cref{sec:cases-B-C-D-models}. \cref{sec:conclusions} contains our conclusions and points out avenues for future studies.

\section{Review: Complex structure degenerations in F-theory}
\label{sec:Kulikov-models-review}

Our study of open-moduli infinite-distance limits in six-dimensional F-theory will make frequent use of the results obtained in \cite{Lee:2021qkx,Lee:2021usk} for the analogous problem in eight dimensions. We therefore briefly recall, in \cref{sec:K3-degenerations}, the main conclusions of \cite{Lee:2021qkx,Lee:2021usk}. This is followed, in \cref{sec:review-part-I}, by a succinct summary of key results on the degenerations of elliptic Calabi-Yau threefolds in the language of \cite{Alvarez-Garcia:2023gdd}.

To set notation, let us recall that an elliptic fibration $\pi: Y \rightarrow B$ can be described in terms of a Weierstrass model
\begin{equation}
    y^2 = x^3 + f x z^4 + g z^6\,.
\label{sec:Weierstrass-geb}
\end{equation}
Here $[x : y : z]$ are homogenous coordinates in the weighted projective space $\mathbb{P}_{231}$, which is the ambient space of the elliptic fiber cut out by the hypersurface \eqref{sec:Weierstrass-geb}.
Furthermore, the defining polynomials $f$ and $g$ depend on the coordinates on the base $B$ of the fibration. 
When the vanishing orders of $f$, $g$ and of the discriminant 
\begin{equation}
    \Delta = 4 f^3 + 27 g^2
\end{equation}
at a locus $D$ exceed the Kodaira bound,
\begin{equation}
    \ord{Y}(f,g,\Delta)_{D} \geq (4,6,12)\,,
\label{eq:non-Kodaira-bound}
\end{equation}
the resulting singularity in the elliptic fibers over $D$ is called non-minimal. Below this bound, the singularities give rise to degenerate elliptic fibers classified, for surfaces, by Kodaira and N\'eron. In F-theory, the vanishing locus of the discriminant $\Delta$ on the base is interpreted as the location of 7-branes. The non-abelian gauge algebra can be read off from the Kodaira type of the degenerate fibers. For more details on elliptic fibrations and their appearance in F-theory we refer to the review \cite{Weigand:2018rez} and references therein.

\subsection{Degenerations of elliptic K3 surfaces}
\label{sec:K3-degenerations}

For F-theory compactified to eight dimensions, the elliptically fibered internal space is an elliptic K3 surface with base $B = \mathbb{P}^{1}$. The infinite-distance limits in the open string moduli space\footnote{Throughout this text we use this term despite the fact that in non-perturbative string theory the open and closed moduli spaces are, of course, not clearly distinguishable.} of 7-branes are encoded in the complex structure deformations of the elliptic K3 surface at infinite distance. For simplicity, we focus on degenerations described by a single complex parameter $u \in D$, with $D \subset \mathbb C$ an open disk. Our starting point is a family of Weierstrass models $\hat{\mathcal{Y}}$, whose elements $\hat{Y}_u$ are Weierstrass models over the base $\mathbb{P}^1$. The central element $\hat{Y}_{0}$ of the family, located at $u=0$, is a Weierstrass model with certain degenerations. These include, in particular, the non-minimal fibral singularities \eqref{eq:non-Kodaira-bound} over points on the base. Possibly after \mbox{performing a base change $u \mapsto u^k$}, with $k \in \mathbb{Z}_{\geq 1}$, one can find a birational transformation to a related\footnote{More precisely, the original degeneration $\hat{\rho}: \hat{\mathcal{Y}} \rightarrow D$ and the resolved degeneration $\rho: \mathcal{Y} \rightarrow D$ are equivalent, representing the same limit of algebraic varieties but with different geometrical representatives for its endpoint \cite{Alvarez-Garcia:2023gdd}.} family of Weierstrass models, that we denote $\mathcal{Y}$, whose central element $Y_{0}$ is free of infinite-distance non-minimal singularities. After resolving the remaining Kodaira fibral singularities, one arrives at a family $\mathcal{X}$ of smooth K3 surfaces $X_{u}$ whose central element is reduced with local normal crossings,
\begin{equation}
	X_{0} = \bigcup_{p=0}^{P} X^{p}\,,
\end{equation}
with a very complete description of the possible types of central fiber available \cite{Kulikov1977,Persson1977,FriedmanMorrison1983}. The resulting so-called Kulikov models are classified into Type I, Type II and Type III. Models of Type I correspond to finite-distance degenerations, and their central fiber $X_{0}$ is a smooth, irreducible variety. Type~II and III models constitute the complex structure infinite-distance limits studied in \cite{Lee:2021qkx,Lee:2021usk}. 

We are mostly interested in the family $\mathcal{Y}$ of Kulikov Weierstrass models $Y_{u}$, whose geometry and associated physics we describe below. By slight abuse of nomenclature, we refer to these as Kulikov models as well.

Before doing so, let us denote the two independent 1-cycles of an elliptic curve $\mathcal{E}$ by \mbox{$\sigma_{i} \in H_{1}(\mathcal{E}, \mathbb{Z})$}, for $i = 1, 2$. If $\mathcal{E}$ is the elliptic fiber of the central element $Y_{0}$ of a Kulikov Weierstrass model, $\sigma_{1}$ and $\sigma_{2}$ will be trivial in $H_{1}(Y_{0},\mathbb{Z}) = 0$. However, in $Y_0$ the base of the elliptic K3 surface degenerates into an open chain\footnote{This picture applies to Type II.a and Type III models, with Type II.b models described below.}
\begin{equation}
	B_{0} = \bigcup_{p=0}^{P} B^{p}
\end{equation}
of $\mathbb{P}^{1}$ curves intersecting at points. The presence of 7-branes in this chain allows us to define a 1-chain $\Sigma$ on $B_0$ that cannot be slipped off into triviality. Fibering $\sigma_{i}$ over $\Sigma$ we obtain a non-trivial 2-cycle $\gamma_{i} \in H_{2}(Y_{0},\mathbb{Z})$ of the topology of a torus that we will often reference in the remainder of the text.

\begin{figure}[t!]
    \centering
    \begin{tikzpicture}[every text node part/.style={align=right}]
        \node[inner sep=0pt] at (0,0) {\includegraphics[width=0.5\textwidth]{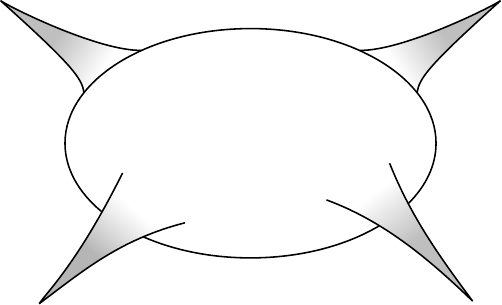}};
        \node[anchor=south east] at (-1.5,3) {II.a: $8\mathrm{D} \rightarrow 10\mathrm{D}$\\ $T_{\mathrm{het}} \rightarrow i\infty$\\ $G_{\infty} = \left( \hat{\mathrm{E}}_{9} \oplus \hat{\mathrm{E}}_{9} \right)/\sim$};
        \node[anchor=north east] at (-0.8,-3) {III.a: $8\mathrm{D} \rightarrow 9\mathrm{D}$\\ $T_{\mathrm{het}}, U_{\mathrm{het}} \rightarrow i\infty$\\ $G_{\infty} = \left( \hat{\mathrm{E}}_{9+n_{0}-n_{1}} \oplus \hat{\mathrm{E}}_{9 + n_{P} - n_{P-1}} \oplus \hat{H} \right)/\sim$};
        \node[anchor=south west] at (3.5,3) {II.b: $8\mathrm{D}$ emergent string\\ $g_{\mathrm{IIB}} \rightarrow 0$};
        \node[anchor=north west] at (3.5,-3) {III.b: $8\mathrm{D} \rightarrow 10\mathrm{D}$\\ $g_{\mathrm{IIB}} \rightarrow 0$\\ $U_{\mathrm{IIB}} \rightarrow 0$};
        \node[anchor=south] at (0,-0.1) {F-theory on K3};
    \end{tikzpicture}
    \caption{Infinite-distance complex structure degenerations for F-theory on an elliptic K3 surface and their associated physics. Figure adapted from \cite{Lee:2021usk}.}
    \label{fig:Kulikov-models-summary}
\end{figure}

The geometry and physics of Type II and Type III Kulikov models is reviewed below, and succinctly summarised in \cref{fig:Kulikov-models-summary}.

\paragraph{Type II.a Kulikov models:} These degenerations can be brought into a canonical form in which the central fiber $Y_{0}$ has two components intersecting over an elliptic curve $\mathcal{E}$ over the intersection point $B^{0} \cap B^{1}$ of the two base components, i.e.\ $Y_{0} = Y^{0} \cup_{\mathcal{E}} Y^{1}$. The components $Y^{0}$ and $Y^{1}$ have $\mathrm{I}_{0}$ type fibers in codimension-zero. Each component has 12 7-branes of arbi\-trary ADE type located at points in the base components $B^{p}$. The 2-cycles $\gamma_{1}$ and $\gamma_{2}$ can be globally defined in $Y_{0}$, and their calibrated volume vanishes in the limit. Since the $\gamma_{i}$ have the topology of a torus, M2-branes can wrap arbitrarily often around them, leading to two towers of asymptotically massless BPS particles. These are, in a dual frame, understood as Kaluza-Klein towers signalling a decompactification from eight to ten dimensions. One can argue using string junctions that the towers lead to a double loop enhancement of the gauge algebras associated to the 7-branes of each $Y^{p}$ component taken together (since their separation is an artefact of the resolution process). In the infinite-distance limit, this leads to the gauge algebra
\begin{equation}
    G_{\infty} = \left( \hat{\mathrm{E}}_{9} \oplus \hat{\mathrm{E}}_{9} \right)/\sim\,,
\end{equation}
where the quotient indicates that the imaginary roots are to be identified. The decompactified theory has a gauge algebra
\begin{equation}
    G_{\mathrm{10D}} = \mathrm{E}_{8} \oplus \mathrm{E}_{8}\,.
\end{equation}
These models can be understood in terms of a dual heterotic picture in which the $\mathrm{E}_{8} \times \mathrm{E}_{8}$ heterotic string compactified on a torus $T^{2}_{\mathrm{het}}$ undergoes a large volume limit in which the complexified K\"ahler modulus diverges, i.e.\ $T_\mathrm{het} \rightarrow i\infty$.

\paragraph{Type II.b Kulikov models:} In their canonical form, the central fiber $Y_{0}$ of these models also consists of two components $Y_{0} = Y^{0} \cup_{E} Y^{1}$  meeting over an elliptic curve $E$ whose two 1-cycles we denote $\sigma_{i}^{E} \in H_{1}(E,\mathbb{Z})$, for $i = 1,2$. However, the base $B_{0}$ consists of a single rational curve that acts as the base of both components. The elliptic curve $E = Y^{0} \cap Y^{1}$ is a bisection of the fibration and gives a double cover of $B_{0}$ branched at four points. $Y_{0}$ has $\mathrm{I}_{2}$ type fibers in codimension-zero, signalling a global weak coupling limit. Due to the codimension-zero $\mathrm{I}_{2}$~type singularities, only one of the $\{ \sigma_{i} \}_{i=1,2}$ 1-cycles of the elliptic fiber $\mathcal{E}$ is monodromy invariant, namely the collapsed one, say $\sigma_{1}$. Combining the $\sigma_{i}^{E}$ with $\sigma_{1}$ we still obtain two 2-cycles $\gamma_{i} \in H_{2}(Y_{0},\mathbb{Z})$ with torus topology and vanishing calibrated volume in the limit. M2-branes multiply wrapping the $\gamma_{i}$ lead again to two towers of asymptotically massless BPS particles that are dually interpreted as Kaluza-Klein towers. However, we can also wrap an M2-brane around $\sigma_{1}$, leading to an asymptotically tensionless fundamental Type II string that is non-BPS due to the triviality of $\sigma_{1} \in H_{1}(Y_{0},\mathbb{Z}) = 0$. The excitations of this string are at the same parametric scale as the Kaluza-Klein towers, leading to the interpretation of these models as equidimensional weak coupling emergent string limits.

\paragraph{Type III.a Kulikov models:} The central fiber $Y_{0}$ is a chain of surfaces $Y_{0} = \bigcup_{p=0}^{P} Y^p$, with $P \geq 1$, in which the middle components and up to one of the end-components have $\mathrm{I}_{n>0}$ type fibers in codimension-zero. The end-component(s) with $\mathrm{I}_{0}$ type fibers in codimension-zero are rational elliptic surfaces. The middle components can only have $\mathrm{I}_{n}$ type fibers in codimen\-sion-one, while an end-component with codimension-zero singular fibers, if present, always has two $\mathrm{I}_{n}^{*}$ type, and possibly also additional $\mathrm{I}_{n}$ type, fibers in codimension-one. The presence of the codimension-zero singularities means that only one of the 1-cycles of the elliptic fiber $\mathcal{E}$ is monodromy invariant, say $\sigma_{1}$, and by fibering it over a base 1-chain $\Sigma$ we obtain a single two-cycle $\gamma_{1} \in H_{2}(Y_{0},\mathbb{Z})$ with torus topology. This leads to a single tower of asymptotically massless BPS particles with a dual interpretation as a Kaluza-Klein tower. The theory partially decompactifies from eight to nine dimensions, with the gauge algebra undergoing a single loop enhancement \cite{Lee:2021usk,Collazuol:2022oey,Collazuol:2022jiy}. If $n_0=n_P=0$, the enhanced algebra takes the form,  
\begin{equation}
    G_{\infty} = \left( \hat{\mathrm{E}}_{9-n_{1}} \oplus \hat{H} \oplus \hat{\mathrm{E}}_{9 - n_{P-1}} \right)/\sim\,,
\end{equation}
where the $H$ factor accounts for the A-type gauge algebras associated to the 7-branes located in the middle components; in this case, from the point of view of the decompactified theory in nine dimensions we have
\begin{equation}
    G_{\mathrm{9D}} = \mathrm{E}_{9-n_{1}} \oplus H \oplus \mathrm{E}_{9 - n_{P-1}}\,.
\end{equation}
If, instead, $n_{p} = 0$ and $n_{q} > 0$, for $p, q \in \{ 0,P \}$ and $p \neq q$, the enhanced algebra is
\begin{equation}
    G_{\infty} = \left( \hat{E}_{9-n_{p}} \oplus \hat{H} \right)/\sim\,,
\end{equation}
leading in the decompactified theory to
\begin{equation}
    G_{\mathrm{9D}} = E_{9-n_{p}} \oplus H\,.
\end{equation}
These models correspond in the dual $\mathrm{E}_{8} \times \mathrm{E}_{8}$ heterotic string compactified on $T^{2}_{\mathrm{het}}$ to a large volume limit $T_{\mathrm{het}} \rightarrow i\infty$ in which, simultaneously, the complex structure $U_{\mathrm{het}}$ scales such that $T_{\mathrm{het}}/U_{\mathrm{het}}$ remains finite.

\paragraph{Type III.b Kulikov models:} The geometry and 7-brane content of the central fiber $Y_{0}$ is the one described above for Type III.a models, but with all components having $\mathrm{I}_{n > 0}$ type fibers in codimension-zero, including both end-components. This signals a global weak coupling limit, allowing us to reinterpret the endpoint of the limit as a perturbative Type IIB orientifold compactification on a torus $T^{2}_{\mathrm{IIB}}$. This torus undergoes a large complex structure limit at constant volume. There are two towers of asymptotically massless particles, one coming from the winding modes of the $\mathrm{F1}$-string around the shrinking 1-cycle of $T^{2}_\mathrm{IIB}$, and an additional supergravity Kaluza-Klein tower from the dual 1-cycle of $T^{2}_\mathrm{IIB}$; the volume of the latter is inversely proportional to that of the shrinking 1-cycle. Only the former tower can be seen in the F-theory picture, from M2-branes wrapping the single shrinking 2-cycle $\gamma_{1}$ obtained analogously as for Type III.a models. Altogether, this indicates a full decompactification to ten-dimensional weakly coupled Type IIB string theory.

\subsection{Degenerations of elliptic Calabi-Yau threefolds}
\label{sec:review-part-I}

In analogy to the degenerations of elliptic K3 surfaces, one can consider degenerations $\hat{\rho}: \hat{\mathcal{Y}} \rightarrow D$ of Weierstrass models whose elements $\hat{Y}_{u}$ are Calabi-Yau threefolds elliptically fibered over a base $\hat{B}_{u}$. The family degenerates to its central element $\hat{Y}_{0}$, which, in particular, may present infinite-distance non-minimal singularities. The compositions of base changes and birational transformations required to turn this into an equivalent degeneration $\rho: \mathcal{Y} \rightarrow D$ whose central fiber $Y_{0}$, again an elliptic fibration $\pi_{0}: Y_{0} \rightarrow B_{0}$, is free of infinite-distance non-minimal singularities is now considerably more involved, and was studied systematically in \cite{Alvarez-Garcia:2023gdd}. Notice that the base spaces of these elliptic fibrations form themselves families $\hat{\mathcal{B}}$ and $\mathcal{B}$; we will denote the divisor classes corresponding to $\hat{B}_{0}$ and $B_{0}$ in said family varieties by $\mathcal{U}$ and $\tilde{\mathcal{U}}$, respectively.

Focusing on non-minimal fibers supported over a curve $C \subset \hat{B}_{0}$, an important technicality is the distinction between so-called Class 5 models, in which the vanishing orders of the defining polynomials $f$ and $g$ over $C$ lie strictly above the minimality bound \eqref{eq:non-Kodaira-bound}, i.e.\
\begin{equation}
    \ord{\hat{\mathcal{Y}}}(f)_{C} > 5\qquad \text{and}\qquad \ord{\hat{\mathcal{Y}}}(g)_{C} > 7\,,
\end{equation}
and those where at least one (Class 2 or 3) or both of them (Class 1 or 4) saturate the bound. Class 5 models can be turned, after a sequence of base changes and modifications, into an equivalent Class 1--4 model, or into one presenting only minimal degenerations \cite{Alvarez-Garcia:2023gdd,ALWClass5}. Focusing therefore on Class 1--4 degenerations, it was shown in \cite{Alvarez-Garcia:2023gdd} that, for so-called single infinite-distance limits, the central fiber $Y_0$ is again an open chain of intersecting components after suitable birational transformations. Single infinite-limits distance arise when, morally speaking, \mbox{non-minimal} singular elliptic fibers in $\hat{Y}_{0}$ are supported over non-intersecting curves. For the precise definitions and proofs, we refer to \cite{Alvarez-Garcia:2023gdd}.

In this article, we will focus on single infinite-distance limit degenerations of Hirzebruch models, i.e.\ those in which the degenerating Calabi-Yau threefolds $\hat{Y}_{u}$ are elliptic fibrations over Hirzebruch surfaces $\hat{B}_{u} = \mathbb{F}_{n}$
\begin{equation}
\begin{tikzcd}
    \mathcal{E} \arrow[r] & \hat{Y}_{u} \arrow[d, "\pi_{\mathrm{ell}}"]\\
     & \mathbb{F}_{n}\mathrlap{\,.}
\end{tikzcd}
\end{equation}
Recall that a Hirzebruch surface $\mathbb{F}_{n}$ is a rational fibration over the complex projective plane, whose fiber and base we denote by $\mathbb P^1_f$ and $\mathbb P^1_b$, respectively. It has two distinguished sections: The $(-n)$-curve, in the divisor class $h$, and the $(+n)$-curve, in the divisor class $h + nf$, where $f$ is the class of the fiber. Depending on the set of curves $\hat{\mathscr{C}}_{r}$ in $\hat{B}_{0} = \mathbb{F}_{n}$ over which the non-minimal elliptic fibers are supported, the resulting models fall into four classes\footnote{In the present work, we focus on genus-zero single infinite-distance limit degenerations of Hirzebruch models. In addition, non-minimal singularities can arise over an anti-canonical divisor, which is a genus-one curve, see \cite{Alvarez-Garcia:2023gdd} for further comments.} of degenerations, which are listed in \cref{tab:genus-zero-Hirzebruch-summary} alongside their main properties.

After a suitable composition of base changes and blow-ups, the central fiber $Y_{0}$ of the resolved degeneration $\rho: \mathcal{Y} \rightarrow D$ is a union
\begin{equation}
	Y_{0} = \bigcup_{p=0}^{P} Y^{p}\,.
\end{equation}
The components $Y^{p}$ are themselves the total spaces of elliptic fibrations $\pi^{p}: Y^{p} \rightarrow B^{p}$, and the bases $B^{p}$ are again Hirzebruch surfaces. Correspondingly, the base $B_{0}$ of the central fiber can be expressed as the union
\begin{equation}
	B_{0} = \bigcup_{p=0}^{P} B^{p}\,.
\end{equation}
Each component $B^{p}$ of the base corresponds to the divisor $E_{p}$ associated to the vanishing locus of the exceptional coordinate $e_{p}$, introduced in the $p$-th blow-up during the resolutions process necessary to remove the non-minimal singularities. Renaming $e_{0} := u$, the (strict transform of) the original base component $\hat{B}_{0}$ corresponds to $B^{0}$, in the divisor class $E_{0}$.

In Class 1--4 models, the generic elliptic fiber over each of the base components $B^{p}$ can only be of Kodaira type $\mathrm{I}_{n_{p}}$, for some value $n_{p} \geq 0$. Each component $Y^{p}$ is not a Calabi-Yau variety by itself, but rather a log Calabi-Yau space; their union along specific divisors, however, makes $Y_{0}$ a reducible Calabi-Yau variety. To the Weierstrass model of each component $\pi^{p}: Y^{p} \rightarrow B^{p}$, there is an associated holomorphic line bundle $\mathcal{L}_{p}$ over $B^{p}$, in the sense that its defining polynomials $f_{p}$ and $g_{P}$ are sections of $\mathcal{L}_{p}^{\otimes 4}$ and $\mathcal{L}_{p}^{\otimes 6}$, respectively. Furthermore, the discriminant $\Delta'_{p}$ is obtained, after factoring out all overall powers of the exceptional coordinates, by restriction of the discriminant of $\mathcal{Y}$. Since the base components $B^{p}$ are all Hirzebruch surfaces, we distinguish the $(-n)$-curve, $(+n)$-curve and fiber divisor class of each of $B^{p}$ denoting them by $S_{p}$, $T_{p}$ and $V_{p} = W_{p}$, respectively. This is the notation employed in \cref{tab:genus-zero-Hirzebruch-summary} while summarising the relevant line bundles for each class of models. The divisors of the base family $\mathcal{B}$ that are the natural extension of the aforementioned classes will be denoted calligraphically, e.g.\ $\mathcal{V}$ is the divisor of $\mathcal{B}$ restricting to the fiber class in the elements $B_{\tilde{u}}$ of the family.

\begin{sidewaystable}[p!]
    \centering
    \resizebox{\linewidth}{!}{
    \begin{tblr}{columns={c,m},
				row{1}={abovesep=3pt,belowsep=3pt},
    				row{2-5}={abovesep=7.5pt,belowsep=7.5pt},
				hline{1}={2-4}{solid},
				hline{2-6}={solid},
				vline{1}={2-5}{solid},
				vline{2-5}={solid},
				}
         & Non-minimal curves & Central component structure & Component line bundles and discriminants\\
        \begin{tabular}{c}Horizontal\\ (Case A)\end{tabular} &
        $\begin{aligned}
            \hat{\mathscr{C}}_{1} &= \{h\}\\
            \hat{\mathscr{C}}_{1} &= \{h + nf\}\\
            \hat{\mathscr{C}}_{2} &= \{h, h + nf\}
        \end{aligned}$ &
        $\begin{tikzcd}[column sep=1em, ampersand replacement=\&]
            \mathrm{I}_{n_{0}} \arrow[r, dash] \arrow[d, dash] \& \cdots \arrow[r, dash] \& \mathrm{I}_{n_{p}} \arrow[r, dash] \arrow[d, dash] \& \cdots \arrow[r, dash] \& \mathrm{I}_{n_{P}} \arrow[d, dash]\\
            \mathbb{F}_{n} \arrow[r, dash] \& \cdots \arrow[r, dash] \& \mathbb{F}_{n} \arrow[r, dash] \& \cdots \arrow[r, dash] \& \mathbb{F}_{n}
        \end{tikzcd}$ &
        {$\begin{aligned}
            \mathcal{L}_{0} &= S_{0} + (2+n)V_{0}\\
            \mathcal{L}_{p} &= 2V_{p}\\
            \mathcal{L}_{P} &= S_{P} + 2V_{P}
        \end{aligned}$\\[0.5cm]
        $\begin{aligned}
            \Delta'_{0} &= (12 + n_{0} - n_{1})S_{0} + (24 + 12n)V_{0}\\
            \Delta'_{p} &= (2n_{p} - n_{p-1} - n_{p+1})S_{p} + (24 + n(n_{p} - n_{p-1}))V_{p}\\
            \Delta'_{P} &= (12 + n_{P} - n_{P-1})S_{P} + (24 + n(n_{P} - n_{P-1}))V_{P}
        \end{aligned}$}\\
         \begin{tabular}{c}Vertical\\ (Case B)\end{tabular} &
        $\hat{\mathscr{C}}_{1} = \{f\}$ &
        $\begin{tikzcd}[column sep=1em, ampersand replacement=\&]
            \mathrm{I}_{n_{0}} \arrow[r, dash] \arrow[d, dash] \& \cdots \arrow[r, dash] \& \mathrm{I}_{n_{p}} \arrow[r, dash] \arrow[d, dash] \& \cdots \arrow[r, dash] \& \mathrm{I}_{n_{P}} \arrow[d, dash]\\
            \mathbb{F}_{n} \arrow[r, dash] \& \cdots \arrow[r, dash] \& \mathbb{F}_{0} \arrow[r, dash] \& \cdots \arrow[r, dash] \& \mathbb{F}_{0}
        \end{tikzcd}$ &
        {$\begin{aligned}
            \mathcal{L}_{0} &= 2S_{0} + (1+n)W_{0}\\
            \mathcal{L}_{p} &= 2S_{p}\\
            \mathcal{L}_{P} &= 2S_{P} + W_{P}
        \end{aligned}$\\[0.5cm]
        $\begin{aligned}
            \Delta'_{0} &= 24S_{0} + (12 + 12n + n_{0} - n_{1})W_{0}\\
            \Delta'_{p} &= 24S_{p} + (2n_{p} - n_{p-1} - n_{p+1})W_{p}\\
            \Delta'_{P} &= 24S_{P} + (12 + n_{P} - n_{P-1})W_{P}
        \end{aligned}$}\\
        \begin{tabular}{c}Mixed sectional\\ (Case C)\end{tabular} &
        \begin{tabular}{c}
            $\hat{\mathscr{C}}_{1} = \{h+(n+\alpha)f\}$\\[0.25cm]
            $\alpha = 1\quad \textrm{with}\quad n \leq 6$\\
            $\alpha = 2\quad \textrm{with}\quad n = 0$
        \end{tabular} &
        $\begin{tikzcd}[column sep=1em, ampersand replacement=\&]
            \mathrm{I}_{n_{0}} \arrow[r, dash] \arrow[d, dash] \& \cdots \arrow[r, dash] \& \mathrm{I}_{n_{p}} \arrow[r, dash] \arrow[d, dash] \& \cdots \arrow[r, dash] \& \mathrm{I}_{n_{P}} \arrow[d, dash]\\
            \mathbb{F}_{n+2\alpha} \arrow[r, dash] \& \cdots \arrow[r, dash] \& \mathbb{F}_{n+2\alpha} \arrow[r, dash] \& \cdots \arrow[r, dash] \& \mathbb{F}_{n}
        \end{tikzcd}$ &
        {$\begin{aligned}
            \mathcal{L}_{0} &= S_{0} + (2+(n+2\alpha))V_{0}\\
            \mathcal{L}_{p} &= 2V_{p}\\
            \mathcal{L}_{P} &= S_{P} + (2-\alpha)V_{P}
        \end{aligned}$\\[0.5cm]
        $\begin{aligned}
            \Delta'_{0} &= (12 + n_{0} - n_{1})S_{0} + (24 + 12(n+2\alpha))V_{0}\\
            \Delta'_{p} &= (2n_{p} - n_{p-1} - n_{p+1})S_{p} + (24 + (n+2\alpha)(n_{p} - n_{p-1}))V_{p}\\
            \Delta'_{P} &= (12 + n_{P} - n_{P-1})S_{P} + ((24-12\alpha) + (n+\alpha)(n_{P} - n_{P-1}))V_{P}
        \end{aligned}$}\\
        \begin{tabular}{c}Mixed bisectional\\ (Case D)\end{tabular}
         &
        \begin{tabular}{c}
            $\hat{\mathscr{C}}_{1} = \{ 2h + bf \}$\\[0.25cm]
            $(n,b) = (0,1)$\\
            $(n,b) = (1,2)$
        \end{tabular} &
        \begin{tikzcd}[column sep=1em, ampersand replacement=\&]
            \mathrm{I}_{n_{0}} \arrow[r, dash] \arrow[d, dash] \& \cdots \arrow[r, dash] \& \mathrm{I}_{n_{p}} \arrow[r, dash] \arrow[d, dash] \& \cdots \arrow[r, dash] \& \mathrm{I}_{n_{P}} \arrow[d, dash]\\
            \mathbb{F}_{4} \arrow[r, dash] \& \cdots \arrow[r, dash] \& \mathbb{F}_{4} \arrow[r, dash] \& \cdots \arrow[r, dash] \& \mathbb{F}_{n}
        \end{tikzcd} &
        {$\begin{aligned}
            \mathcal{L}_{0} &= S_{0} + (2+4)V_{0}\\
            \mathcal{L}_{p} &= 2V_{p}\\
            \mathcal{L}_{P} &= V_{P}
        \end{aligned}$\\[0.5cm]
        $\begin{aligned}
            \Delta'_{0} &= (12 + n_{0} - n_{1})S_{0} + (24 + 12 \cdot 4)V_{0}\\
            \Delta'_{p} &= (2n_{p} - n_{p-1} - n_{p+1})S_{p} + (24 + 4(n_{p} - n_{p-1}))V_{p}\\
            \Delta'_{P} &= 2(n_{P} - n_{P-1})S_{P} + (12 + (n+1)(n_{P} - n_{P-1}))V_{P}
        \end{aligned}$}
    \end{tblr}
    }
    \caption{Genus-zero single infinite-distance limit degenerations of Hirzebruch models \cite{Alvarez-Garcia:2023gdd}. The horizontal models are analysed in detail in the bulk of the present article. The remaining three classes are studied in \cref{sec:cases-B-C-D-models}.  }
    \label{tab:genus-zero-Hirzebruch-summary}
\end{sidewaystable}

\section{General properties of horizontal models}
\label{sec:horizontal-models-general-properties}

Horizontal models, as defined in \cite{Alvarez-Garcia:2023gdd}, are single infinite-distance limit degenerations of Hirzebruch models $\hat{\rho}: \hat{\mathcal{Y}} \rightarrow D$ in which the non-minimal elliptic fibers are supported over either the \mbox{$(-n)$-curve} in the base central fiber $\hat{B}_{0} = \mathbb{F}_{n}$, a $(+n)$-curve, or both. This class of degenerations is of particular interest because it has a well-controlled heterotic dual, which we explore in \cref{sec:horizontal-models-heterotic-duals}. Moreover, it is possible to interpret the degenerations fiberwise as the generalization of the Kulikov models reviewed in \cref{sec:K3-degenerations}, a point of view that we exploit in \cref{sec:horizontal-models-systematics}. This makes horizontal models a natural starting point for the analysis of infinite-distance limits in the complex structure moduli space of six-dimensional F-theory.

Before we delve into the aforementioned aspects, we study some general properties of horizontal models in this section. 
The busy reader interested mostly in the physics interpretation of the infinite-distance limits is invited to skip these slightly more technical details in a first read and jump directly to \cref{sec:horizontal-models-heterotic-duals} after \cref{sec:classification-horizontal-models}.

The fine-grained classification of Kulikov models of \cite{Lee:2021qkx,Lee:2021usk} attending to their physical interpre\-tation is mirrored for horizontal models in \cref{sec:classification-horizontal-models}. The Kodaira type of singular fibers in codimension-zero in the components $\{Y^{p}\}_{0 \leq p \leq P}$ of the central fiber $Y_{0}$  encodes impor\-tant information about the background value of the axio-dilaton; in \cref{sec:horizontal-restrictions-codimension-zero} we restrict the possible patterns of codimension-zero singular fibers that horizontal models can present. This results in constraints on the existence of horizontal models representing global weak-coupling limits. As emphasized in \cite{Alvarez-Garcia:2023gdd}, the assignment of non-abelian gauge algebra factors to divisors (taking a six-dimensional standpoint prior to considering any enhancements to higher-dimensional algebras) should be carried out taking into account the global components of the physical discriminant, whose the possible types we list in \cref{sec:horizontal-global-divisors}. We then analyse in \cref{sec:horizontal-local-global-brane-content} the restrictions on the local and global 7-brane content. Out of the possible non-abelian gauge algebra factors that a horizontal model can present, the so-called vertical ones have a distinguished interpretation in the analyses of \cref{sec:horizontal-models-heterotic-duals,sec:horizontal-models-systematics}; in \cref{sec:bounds-vertical-gauge-rank} we derive some bounds on their gauge rank for later reference. A number of technical details are relegated to \cref{sec:discriminant-weakly-coupled-components,sec:bounds-codimension-zero-pattern,sec:sec:bounds-vertical-gauge-rank-examples}. Many aspects of this analysis carry over, \textit{mutatis mutandis}, to the remaining three types of degenerations listed in \cref{tab:genus-zero-Hirzebruch-summary}, as we explain in \cref{sec:cases-B-C-D-models}.

\subsection{Classification of horizontal models}
\label{sec:classification-horizontal-models}

An elliptic fibration over a Hirzebruch surface can be equivalently seen as a K3-fibration over the complex projective line. Both fibrations naturally extend to the family variety $\hat{\mathcal{Y}}$ of a degeneration of Hirzebruch models $\hat{\rho}: \hat{\mathcal{Y}} \rightarrow D$. If we restrict to a point on the base $\mathbb{P}^{1}_{b}$ of the Hirzebruch surface, we arrive at a degeneration of elliptic K3 surfaces. Concretely, given a point $p_{b} := [v_{0}:w_{0}] \in \mathbb{P}^{1}_{b}$, the degeneration of elliptic K3 surfaces is obtained by the restriction
\begin{equation}
    \hat{\sigma}_{p_{b}} := \left. \hat{\rho} \right|_{\hat{\mathcal{Z}}_{p_{b}}}: \hat{\mathcal{Z}}_{p_{b}} \longrightarrow D\,,\qquad \hat{\mathcal{Z}}_{p_{b}} := \Pi_{\text{K3}}^{-1}(p_{b} \times D) \,.
\end{equation}
Here $\Pi_{\mathrm{K3}}: \hat{\mathcal{Y}} \rightarrow \mathbb{P}^{1}_{b} \times D$ is the K3-fibration of the family variety. 
For the reasons reviewed in \cref{sec:review-part-I}, it suffices to only consider Class~1--4 horizontal models. Their induced degeneration of elliptic K3 surfaces $\hat{\sigma}_{p_{b}}: \hat{\mathcal{Z}}_{p_{b}} \rightarrow D$ obtained for a generic $p_{b} \in \mathbb{P}^{1}_{b}$ will then be a Class~1--4 Kulikov Weierstrass model, in the language of \cite{Lee:2021qkx}. In what follows, we drop the $p_{b}$ subscript, understanding that we always refer to the generic restriction unless stated otherwise.

We recall from \cref{tab:genus-zero-Hirzebruch-summary} that the central fiber $Y_{0}$ of the open-chain resolution $\rho: \mathcal{Y} \rightarrow D$ of the horizontal model has the structure
\begin{equation}
    \begin{tikzcd}[column sep=1em]
        \mathrm{I}_{n_{0}} \arrow[r, dash] \arrow[d, dash] & \cdots \arrow[r, dash] & \mathrm{I}_{n_{p}} \arrow[r, dash] \arrow[d, dash] & \cdots \arrow[r, dash] & \mathrm{I}_{n_{P}} \arrow[d, dash]\\
        \mathbb{F}_{n} \arrow[r, dash] & \cdots \arrow[r, dash] & \mathbb{F}_{n} \arrow[r, dash] & \cdots \arrow[r, dash] & \mathbb{F}_{n}\mathrlap{\,.}
    \end{tikzcd}
\end{equation}
Due to the way in which the resolution process works for horizontal models, the generic vertical slice $\sigma: \mathcal{Z} \rightarrow D$ of $\rho: \mathcal{Y} \rightarrow D$ corresponds to the resolution of the Kulikov Weierstrass model $\hat{\sigma}: \hat{\mathcal{Z}} \rightarrow D$ that we would have obtained following the steps described in \cite{Lee:2021qkx}. Its central fiber $Z_{0}$ presents then the pattern of codimension-zero singular fibers
\begin{equation}
    \begin{tikzcd}[column sep=1em]
        \mathrm{I}_{n_{0}} \arrow[r, dash] \arrow[d, dash] & \cdots \arrow[r, dash] & \mathrm{I}_{n_{p}} \arrow[r, dash] \arrow[d, dash] & \cdots \arrow[r, dash] & \mathrm{I}_{n_{P}} \arrow[d, dash]\\
        \mathbb{P}^{1} \arrow[r, dash] & \cdots \arrow[r, dash] & \mathbb{P}^{1} \arrow[r, dash] & \cdots \arrow[r, dash] & \mathbb{P}^{1}\mathrlap{\,,}
    \end{tikzcd}
\end{equation}
which can be readily classified into Type II.a, II.b, III.a or III.b, see \cref{sec:K3-degenerations}. The same would be true for finite-distance degenerations of Hirzebruch surfaces, whose generic vertical restriction would be a Type I Kulikov model.

The classification of Kulikov models can therefore be inherited by the horizontal degenerations of Hirzebruch surfaces. That is, we say that a horizontal model $\hat{\rho}: \hat{\mathcal{Y}} \rightarrow D$ is of Type~I, II.a, II.b, III.a or III.b if its generic vertical restriction $\sigma: \mathcal{Z} \rightarrow D$ is of the corresponding Kulikov type. Since the pattern of codimension-zero singular elliptic fibers is the same for both $Y_{0}$ and $Z_{0}$, the criterion is equal to the one reviewed in \cref{sec:K3-degenerations}, which we now write in a condensed form.\pagebreak

Let $\hat{\rho}: \hat{\mathcal{Y}} \rightarrow D$ be a Class~1--4 horizontal model and $\rho: \mathcal{Y} \rightarrow D$ its open-chain resolution, with codimension-zero $\mathrm{I}_{n_{0}} - \cdots - \mathrm{I}_{n_{P}}$ fibers in the components $\{ Y^{p} \}_{0 \leq p \leq P}$ of its central fiber $Y_{0}$. It can be classified into one of the following types:
\begin{itemize}
    \item \textbf{Horizontal Type II.a model:} If $P \geq 1$ and $n_{p} = 0$ for all $p \in \{0, \dotsc, P\}$.

    \item \textbf{Horizontal Type II.b model:} If $P = 0$ and $n_{0} > 0$.

    \item \textbf{Horizontal Type III.a model:} If $P \geq 1$, $n_{p} > 0$ for all $p \in \{1, \dotsc, P-1\}$, and $n_{p} = 0$ for $p = 0$ and/or $ p = P$.\footnote{A horizontal Type III.a model with $P=1$ is not allowed to have both $n_{0}$=0 and $n_{P}$=0.}

    \item \textbf{Horizontal Type III.b model:} If $P \geq 1$ and $n_{p} > 0$ for all $p \in \{0, \dotsc, P\}$.
\end{itemize}
Note that the horizontal Type II.a model with $P = 1$ and $n_{0} = n_{1} = 0$ is the stable degeneration limit that is usually taken in the F-theory literature \cite{Morrison:1996na,Morrison:1996pp,Bershadsky:1996nh,Friedman:1997yq,Aspinwall:1997ye,Clingher:2003ui} when considering the duality to the heterotic string.

As we will see in the subsequent sections, this classification of the geometry of the central fiber of the degeneration reflects the properties of the asymptotic physics, just like in the eight-dimensional scenario.

\subsection{Restrictions on components at strong and weak coupling}
\label{sec:horizontal-restrictions-codimension-zero}

The Kodaira type of codimension-zero elliptic fibers in the components $\{ Y^{p} \}_{0 \leq p \leq P}$ of the central fiber $Y_{0}$ of a resolved horizontal model encodes important physical information. In F-theory, the base of the elliptic fibration is identified with the internal Type IIB spacetime. The complex structure $\tau$ of the elliptic curve over a given point corresponds to the value of the axio-dilaton and, hence, the string coupling $g_{s}$. The codimension-zero elliptic fibers therefore inform us about the generic (i.e.\ background) value of the axio-dilaton in a given spacetime component. Since we are considering Class~1--4 (horizontal) models, said fibers can only be of Kodaira type~$\mathrm{I}_{m}$~\cite{Alvarez-Garcia:2023gdd}. We need to distinguish the following cases:
\begin{itemize}
    \item \textbf{$Y^{p}$ with codimension-zero $\mathrm{I}_{m>0}$ fibers:} In these components the complex structure $\tau$ of the generic elliptic fiber attains a value for which $j(\tau) \rightarrow \infty$, implying that $\tau \rightarrow i\infty$ and therefore $g_{s} \rightarrow 0$. This means that these components are at weak string coupling.

    \item \textbf{$Y^{p}$ with codimension-zero $\mathrm{I}_{0}$ fibers:} In these components the $j(\tau)$ associated to the generic elliptic fiber is finite, and hence so are $\tau$ and the string coupling $g_{s}$. These components present non-perturbative string coupling.
\end{itemize}
A global weak coupling limit is therefore one in which all components $\{ Y^{p} \}_{0 \leq p \leq P}$ of $Y_{0}$ have codimension-zero $\mathrm{I}_{n_{p}>0}$ fibers. These are either horizontal Type II.b or Type III.b models.

The background value of the axio-dilaton also affects the local types of 7-brane stacks that can exist in a given component, since their associated type of singular elliptic fiber must be compatible with the generic one; we explore this further in \cref{sec:horizontal-local-global-brane-content}.

Given the importance of the pattern $\mathrm{I}_{n_{0}} - \cdots - \mathrm{I}_{n_{P}}$ of codimension-zero fibers of $Y_{0}$ for the asymptotic physics, we use the remainder of this section to constrain it in different ways. It will be useful to keep in mind from \cref{sec:review-part-I} and \cref{tab:genus-zero-Hirzebruch-summary} that the holomorphic line bundles over the components $\{ B^{p} \}_{0 \leq p \leq P}$ of $B_{0}$ are
\begin{subequations}
\begin{alignat}{2}
    \mathcal{L}_{0} &= S_{0} + (2+n)V_{0}\,, & \label{eq:line-bundle-hor-0}\\
    \mathcal{L}_{p} &= 2V_{p}\,, &\qquad p = 1, \dotsc, P-1\,, \label{eq:line-bundle-hor-1}\\
    \mathcal{L}_{P} &= S_{P} + 2V_{P}\,, & \label{eq:line-bundle-hor-2}
\end{alignat}
\label{eq:horizontal-line-bundles}%
\end{subequations}
with the modified discriminant in each component lying in the divisor class
\begin{subequations}
\begin{alignat}{2}
    \Delta'_{0} &= (12 + n_{0} - n_{1})S_{0} + (24 + 12n)V_{0}\,, & \label{eq:Delta-hor-1}\\
    \Delta'_{p} &= (2n_{p} - n_{p-1} - n_{p+1})S_{p} + (24 + n(n_{p} - n_{p-1}))V_{p}\,, &\qquad p = 1, \dotsc, P-1\,, \label{eq:Delta-hor-2}\\
    \Delta'_{P} &= (12 + n_{P} - n_{P-1})S_{P} + (24 + n(n_{P} - n_{P-1}))V_{P}\,. & \label{eq:Delta-hor-3}
\end{alignat}
\label{eq:Delta-hor}%
\end{subequations}
The polynomial form of the restrictions $\{\Delta'_{p}\}_{0 \leq p \leq P}$ of the modified discriminant $\Delta'$ to those components at weak string coupling has a particular structure, as we detail in \cref{sec:discriminant-weakly-coupled-components}.

\subsubsection{Effectiveness bounds}
\label{sec:effectiveness-bounds-horizontal}

An immediate constraint for the $\mathrm{I}_{n_{0}} - \cdots - \mathrm{I}_{n_{P}}$ pattern of codimension-zero fibers of $Y_{0}$ can be obtained by taking into account that the divisor classes \eqref{eq:Delta-hor} of the restrictions $\{ \Delta'_{p} \}_{0 \leq p \leq P}$ of the modified discriminant $\Delta'$ must be effective.

From the horizontal part of \eqref{eq:Delta-hor} we then obtain the bounds
\begin{subequations}
\begin{align}
    n_{0} &\geq n_{1} - 12\,,\label{eq:horizontal-model-effectiveness-constraint-horizontal-1}\\
    n_{p} &\geq \frac{n_{p-1} + n_{p+1}}{2}\,,\qquad p = 1, \dotsc, P-1\,,\label{eq:horizontal-model-effectiveness-constraint-horizontal-2}\\
    n_{P} &\geq n_{P-1} - 12\,,\label{eq:horizontal-model-effectiveness-constraint-horizontal-3}
\end{align}
\label{eq:horizontal-model-effectiveness-constraint-horizontal}%
\end{subequations}
where $n_{p} \in \mathbb{Z}_{\geq 0}$ for all $p \in \{ 0, \dotsc, P\}$. These inequalities also apply to the eight-dimensional models studied in \cite{Lee:2021qkx,Lee:2021usk}. One consequence of these constraints is that, if one component $Y^{p}$ has codimension-zero $\mathrm{I}_{n_{p}>0}$ fibers, all the intermediate components $Y_{p}$ with $p \in \{1, \dotsc, P-1 \}$ must be at local weak coupling as well. Moreover, tuning a component to have higher codimension-zero singularities may force an enhancement in the adjacent components . For example, it is not possible to further tune the pattern $\mathrm{I}_{0} - \mathrm{I}_{1} - \mathrm{I}_{1} - \mathrm{I}_{1} - \mathrm{I}_{0}$ to achieve the pattern $\mathrm{I}_{0} - \mathrm{I}_{2} - \mathrm{I}_{1} - \mathrm{I}_{1} - \mathrm{I}_{0}$, since the latter does not satisfy the bounds; instead, the tuning will actually result in the pattern $\mathrm{I}_{0} - \mathrm{I}_{2} - \mathrm{I}_{2} - \mathrm{I}_{1} - \mathrm{I}_{0}$. These constraints also imply that $n(p) := n_{p}$ is a concave function, as also discussed in \cite{ALWClass5}.

A bound new to the six-dimensional models can be obtained considering the vertical part of \eqref{eq:Delta-hor}, from which we infer that
\begin{equation}
    n_{p-1} - n_{p} \leq \frac{24}{n}\,,\qquad p = 1, \dotsc, P\,.
\label{eq:horizontal-model-effectiveness-constraint-vertical}
\end{equation}
For example, a pattern of codimension-zero fibers $\mathrm{I}_{0} - \mathrm{I}_{3} - \mathrm{I}_{0}$ cannot occur for a model constructed over $\hat{B} = \mathbb{F}_{9}$, even if it satisfies the constraints \eqref{eq:horizontal-model-effectiveness-constraint-horizontal}.

Satisfying the effectiveness constraints is a necessary condition to obtain a consistent model, but not a sufficient one. To illustrate this, consider a two-component model constructed over $\hat{B} = \mathbb{F}_{12}$, with codimension-zero fibers $\mathrm{I}_{1} - \mathrm{I}_{0}$. Such a model fulfils the above inequalities. We could try to tune it further to obtain a model with the pattern $\mathrm{I}_{1} - \mathrm{I}_{1}$, which also satisfies the effectiveness constraints. The result of this tuning is, however, a model with three components and codimension-zero fibers $\mathrm{I}_{1} - \mathrm{I}_{1} - \mathrm{I}_{0}$.

This last example shows that when we try to construct a horizontal Type III.b model over $\hat{B} = \mathbb{F}_{12}$, the geometry has forced a new component with codimension-zero fibers of Kodaira type $\mathrm{I}_{0}$, preventing us from doing so. Indeed, models in which the asymptotic physics is at global weak coupling, like horizontal Type III.b models, cannot be constructed over arbitrary Hirzebruch surfaces. We revisit this aspect in \cref{sec:horizontal-restrictions-global-weak-coupling}.

\subsubsection{Tighter bounds on \texorpdfstring{$\left| n_{p} - n_{p+1} \right|$}{|np - np+1|}}
\label{sec:tighter-np-npplusone-horizontal}

While the effectiveness bounds \eqref{eq:horizontal-model-effectiveness-constraint-horizontal} and \eqref{eq:horizontal-model-effectiveness-constraint-vertical} can be obtained rather directly from \eqref{eq:Delta-hor}, they can be improved by exploiting our knowledge of the resolution process for horizontal models. Said class of models is amenable to toric methods, as was studied in \cite{Alvarez-Garcia:2023gdd}. This leads to a global description in terms of homogeneous coordinates, both for the initial degeneration $\hat{\rho}: \hat{\mathcal{Y}} \rightarrow D$ and for its open-chain resolution $\rho: \mathcal{Y} \rightarrow D$. In this description, the exceptional coordinates $\{ e_{p} \}_{0 \leq p \leq P}$ do not appear arbitrarily, but their powers follow specific patterns, a fact that threads the discussion in \cite{ALWClass5}. Studying this structure we can obtain tighter bounds for $\left| n_{p} - n_{p+1} \right|$, with $p \in \{0, \dotsc, P-1\}$.

Postponing the technical details to \cref{sec:bounds-codimension-zero-pattern-horizontal}, the inequalities that we find from the resolution structure are
\begin{equation}
	n_{p} - n_{p+1} \leq
	\begin{cases}
		8\,, & 0 \leq n \leq 2\,,\\
		4\,, & 3 \leq n \leq 4\,,\\
		2\,, & 5 \leq n \leq 8\,,\\
		1\,, & 9 \leq n \leq 12\,,
	\end{cases}
	\qquad p = 0, \dotsc, P-1\,,
\label{eq:npminusnpplusone-bound-horizontal}
\end{equation}
and
\begin{equation}
	n_{p} - n_{p-1} \leq 8\,,\qquad p = 1, \dotsc, P\,.
\label{eq:npminusnpminusone-bound-horizontal}
\end{equation}
These bounds apply when the codimension-zero $\mathrm{I}_{n_{p}>0}$ fibers in the $Y^{p}$ component arise from a single or double accidental cancellation structure, see \cref{sec:discriminant-weakly-coupled-components}. It would be interesting to know if allowing for higher accidental cancellations would relax the inequalities to be fulfilled, equating them to the effectiveness bounds \eqref{eq:horizontal-model-effectiveness-constraint-horizontal} and \eqref{eq:horizontal-model-effectiveness-constraint-vertical}.

\subsubsection{Restrictions on global weak coupling limits}
\label{sec:horizontal-restrictions-global-weak-coupling}

As mentioned earlier, in a global weak coupling limit all components $\{Y^{p}\}_{0 \leq p \leq P}$ of $Y_{0}$ have codimension-zero $\mathrm{I}_{n_{p}>0}$ fibers. Within Class~1--4 horizontal models, this property corresponds to either horizontal Type II.b or Type III.b models. However, such models cannot be constructed over arbitrary Hirzebruch surfaces $\hat{B} = \mathbb{F}_{n}$. As we will now show, they cannot be engineered over Hirzebruch surfaces $\hat{B} = \mathbb{F}_{n}$ with $n \geq 5$, while for  $n=3$ and $4$ they necessarily enforce a horizontal line of $\mathrm{D}$ type enhancements of rank at least $3$ and $4$, respectively. For $n=0$, $1$, $2$, no restrictions arise.

In a Hirzebruch surface $\mathbb{F}_{n}$ there is a single irreducible curve with negative self-intersection, the $(-n)$-curve that is the unique representative of the $h$ class. Given another curve $C$, the intersection product $C \cdot h$ is negative if and only if $C$ contains $h$ as a component.\footnote{We review this and other useful facts about algebraic surfaces in Appendix~A of \cite{Alvarez-Garcia:2023gdd}.} Considering the divisors $F$, $G$ and $\Delta$ associated with a Weierstrass model over $\mathbb{F}_{n}$, one can see that
\begin{equation}
    F \cdot h < 0\,,\quad G \cdot h < 0\,,\quad \Delta \cdot h < 0\,,\qquad n \geq 3\,.
\end{equation}
This inevitably leads to the presence of gauge algebra factors supported on $h$, the so-called non-Higgsable clusters \cite{Morrison:2012np}. We list the generic vanishing orders\footnote{To be more precise, we are printing the family vanishing orders for a degeneration $\rho: \hat{\mathcal{Y}} \rightarrow D$ of Hirzebruch models. In the variety $\hat{\mathcal{Y}}$, the base of the elliptic fibration is the threefold $\hat{B} = \mathbb{F}_{n} \times D$. Each Weierstrass model $\hat{Y}_{u}$ presents a non-Higgsable cluster over $\{s=0\}_{\hat{B}_{u}}$, with the component vanishing orders printed in \cref{tab:non-Higgsable-clusters}; since this applies to all fibers of the degeneration, it results in the corresponding family vanishing orders being realised over $\{s=0\}_{\mathcal{B}}$ too. See \cite{Alvarez-Garcia:2023gdd} for the differences between the various notions of vanishing orders.} associated with the non-Higgsable clusters supported on $h$ for the different Hirzebruch surfaces $\hat{B} = \mathbb{F}_{n}$ in \cref{tab:non-Higgsable-clusters}.
\begin{table}[t!]
    \centering
    \begin{tblr}{columns={c}, row{1,2,4,5} = {5mm,c}, row{3} = {1mm}, hline{1,2,3,4,5,6} = {solid}, vlines = {1,2,4,5}{solid}, stretch = 0}
        $n$ & $3$ & $4$ & $5$ & $6$ & $7$\\
        $\ord{\hat{\mathcal{Y}}}(f,g,\Delta)_{s=0}$ & $(2,2,4)$ & $(2,3,6)$ & $(3,4,8)$ & $(3,4,8)$ & $(3,5,9)$\\
        \\
        $n$ & $8$ & $9$ & $10$ & $11$ & $12$\\
        $\ord{\hat{\mathcal{Y}}}(f,g,\Delta)_{s=0}$ & $(3,5,9)$ & $(4,5,10)$ & $(4,5,10)$ & $(4,5,10)$ & $(4,5,10)$
    \end{tblr}
    \caption{Generic vanishing orders associated with the non-Higgsable clusters over the \mbox{$(-n)$-curve} $h$ of $\mathbb{F}_{n}$.}
    \label{tab:non-Higgsable-clusters}
\end{table}

Consider now how this applies to models with components at global weak coupling. First, take a horizontal Type III.b model $\hat{\rho}: \hat{\mathcal{Y}} \rightarrow D$, which we are considering, without loss of generality, to be one in which the non-minimal elliptic fibers are supported over $\hat{\mathscr{C}}_{1} = \{ h \}$. In the central fiber $Y_{0}$ of its open-chain resolution $\rho: \mathcal{Y} \rightarrow D$, the $h$ class in the end-component $B^{P}$ is the only one that can support non-abelian gauge enhancements, since the remaining $(-n)$-curves in the $\{B^{p}\}_{0 \leq p \leq P-1}$ correspond to the interface curves between components.\footnote{The interface curves can be arranged to have trivial vanishing orders by a combination of base changes and modifications of the degeneration, see \cite{Alvarez-Garcia:2023gdd}.} To achieve a global weak coupling limit, all components of $Y_{0}$, and $Y^{P}$ in particular, must support codimension-zero $\mathrm{I}_{n_{P}>0}$ fibers. This means that, at the very least, the single accidental cancellation structure
\begin{equation}
    f_{p} = -3h_{p}^{2}\,,\qquad g_{p} = 2h_{p}^{3}\,,\qquad h_{p} \in H^{0}\left(B^{p},\mathcal{L}_{p}^{\otimes 2}\right)\,,
\end{equation}
must be enforced. This  restricts the modified discriminant to be of the form
\begin{equation}
	\Delta_{p}^{\prime} = h_{p}^{k} \Delta_{p}^{\prime\prime}\,,\qquad k \geq 2\,.
\end{equation}
For details see \cref{sec:discriminant-weakly-coupled-components}. Denoting the divisor class associated to $h_{p}$ by $H_{p}$, this means, in view of \eqref{eq:horizontal-line-bundles}, that the relevant divisors are
\begin{equation}
    F_{p} = 2H_{p}\,,\qquad G_{p} = 3H_{p}\,,\qquad H_{p} = 2\mathcal{L}_{p} =
    \begin{cases}
        2S_{0} + (4+2n)V_{0}\,, &p=0\,,\\
        4V_{p}\,, &p=1,\dotsc,P-1\,,\\
        2S_{P} + 4V_{P}\,, &p=P\,.
    \end{cases}
\label{eq:hp-classes-horizontal}
\end{equation}
Note that tuning the components $\{Y^{p}\}_{0 \leq p \leq P-1}$ to be at local weak coupling does not lead to enhancements of type II, III, IV, $\mathrm{I}^{*}_{m}$, $\mathrm{IV}^{*}$, $\mathrm{III}^{*}$ or $\mathrm{II}^{*}$ at the interfaces $\{ B^{p} \cap B^{p+1} \}_{0 \leq p \leq P-1}$ between components, which would turn the model into a (possibly obscured) Class~5 model. This is a consequence of
\begin{equation}
    H_{p} \cdot h \geq 0\,,\qquad p=0,\dotsc,P-1\,,
\end{equation}
which implies that no forced factorizations occur. The discriminant $\Delta'_{p}$, which is not simply a multiple of $H_{p}$, does not lead to special fibers at the intersections with the adjacent components either: Indeed, \eqref{eq:Delta-hor-1} and \eqref{eq:Delta-hor-2} imply
\begin{equation}
    \Delta'_{p} \cdot S_{p} = -n(n_{p} - n_{p+1}) + 24 \geq 0 \Leftrightarrow n_{p} - n_{p+1} \leq \frac{24}{n}\,,\qquad p = 0, \dotsc, P-1\,,
\end{equation}
and the last inequality is guaranteed by the effectiveness bounds, namely by \eqref{eq:horizontal-model-effectiveness-constraint-vertical}. For the $Y^{P}$ component, instead, 
\begin{align}
    H_{P} \cdot S_{P} < 0 &\Leftrightarrow n \geq 3\,,\\
    \left( H_{P} - S_{P} \right) \cdot S_{P} < 0 &\Leftrightarrow n \geq 5\,,
\end{align}
leading to
\begin{equation}
    h_{P} \propto
    \begin{cases}
        s\,, &n = 3,4\,,\\
        s^{2}\,, &n \geq 5\,.
    \end{cases}
\label{eq:hP-s-powers}
\end{equation}
Given the structure of $f_{P}$, $g_{P}$ and $\Delta'_{P}$, we find that the minimal component vanishing orders over $S_{P}$ are
\begin{equation}
    \ord{Y^{P}}(f_{P},g_{P},\Delta'_{P})_{s=0} \geq
    \begin{cases}
        (2,3,k+\alpha)\,, &n=3,4\,,\\
        (4,6,2k+\alpha)\,, &n \geq 5\,,
    \end{cases}
\label{eq:end-component-vanishing-orders}
\end{equation}
where $\alpha$ accounts for additional factorisations forced by the reducibility of
\begin{equation}
    \Delta_{P}^{\prime\prime} := \Delta'_{P} - kH_{P}\,.
\end{equation}
Its value can be explicitly computed by once again analysing the intersection numbers of the classes $\Delta_{P}^{\prime\prime}$ and $S_{P}$. The class $\Delta_{P}^{\prime\prime} - \alpha S_{P}$ will continue to contain $S_{P}$ components as long as
\begin{equation}
    \left( \Delta_{P}^{\prime\prime} - \alpha S_{P} \right) \cdot S_{P} < 0 \Leftrightarrow \alpha < \frac{4k - 24}{n} + 12 - 2k\,.
\label{eq:SP-factorization-criterion}
\end{equation}
Therefore, the final value of $\alpha$, taking all the mandatory factorizations into account, is
\begin{equation}
    \alpha = \max \left\{ \left\lceil \frac{4k - 24}{n} + 12 - 2k \right\rceil, 0 \right\}\,.
\label{eq:end-component-alpha}
\end{equation}

A similar analysis can be performed for horizontal Type II.b models. Their central fiber $Y_{0}$ consists of a single component, whose Weierstrass model is associated to the line bundle
\begin{equation}
    \mathcal{L}_{B_{0}} = h + 2f + (h+nf) = \mathcal{L}_{P} + (h+nf)\,.
\end{equation}
Since $(h+nf) \cdot h = 0$, the same conclusions follow. Hence, in the following paragraphs we write $Y^{P}$ using the notation pertaining to the horizontal Type III.b case, but it is interchangeable with the $Y_{0}$ central fiber of the horizontal Type II.b model.

Let us summarize what the above considerations imply for the horizontal models constructed over the Hirzebruch surfaces $\hat{B} = \mathbb{F}_{n}$, with $0 \leq n \leq 12$.

\paragraph{Models with $\mathbf{5 \leq n \leq 12}$}

The horizontal models constructed over these Hirzebruch surfaces have component vanishing orders in the $Y^{P}$ component
\begin{equation}
    \ord{Y^{P}}(f_{P},g_{P})_{s=0} = (4,6)\,.
\end{equation}
This means that after tuning the $Y^{P}$ component to be at local weak coupling, one must  perform at least one further base blow-up\footnote{The need for a further base blow-up, accompanied by an appropriate line bundle shift, may only be apparent after a base change, see the comments on obscured infinite-distance limits in \cite{Alvarez-Garcia:2023gdd}.} in order to arrive at the open-chain resolution of the degeneration.

\begin{figure}[p!]
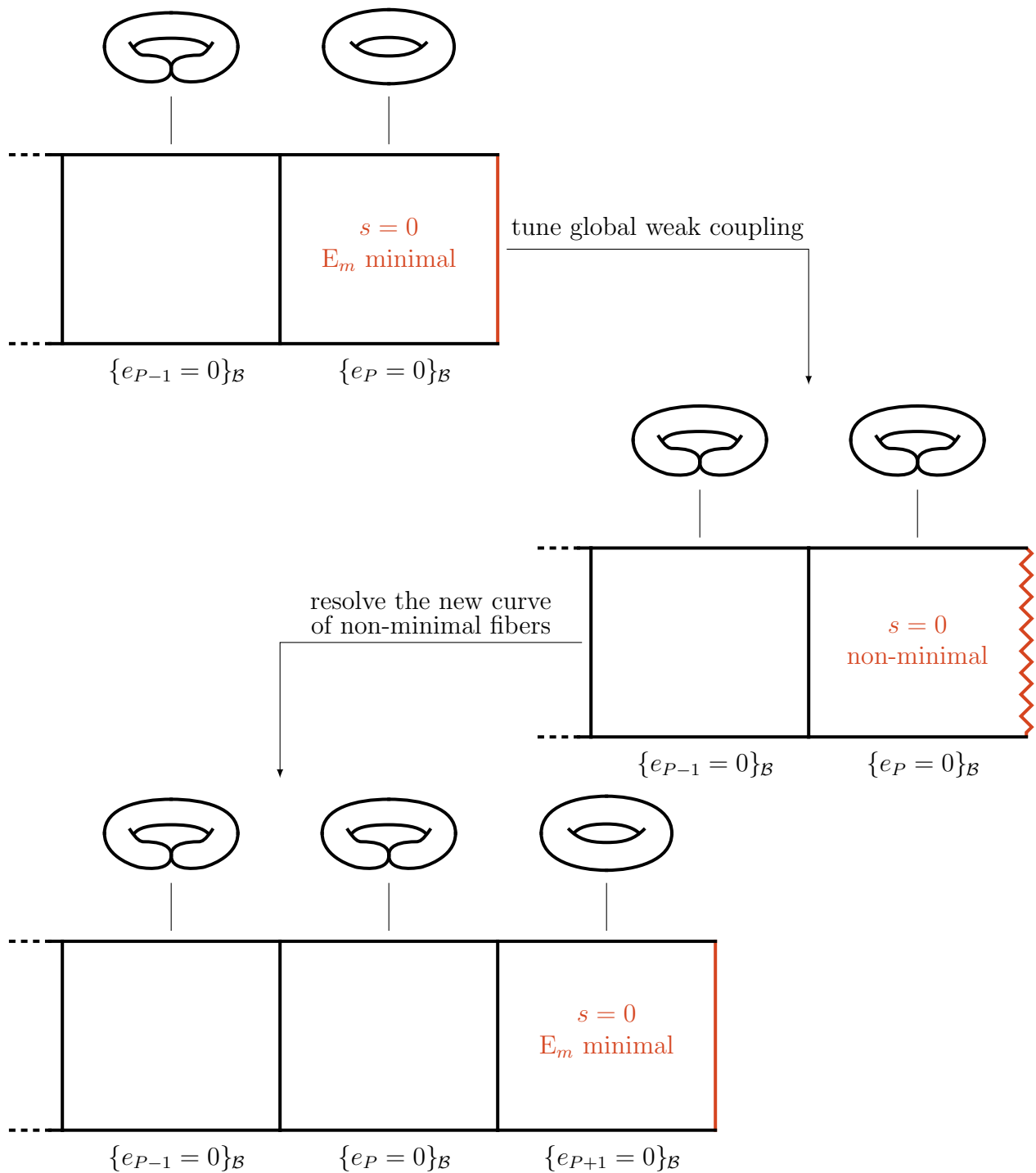

    \centering

    \caption{The open-chain resolutions of horizontal models constructed over the Hirzebruch surfaces $\hat{B} = \mathbb{F}_{n}$, with $5 \leq n \leq 12$, present a non-Higgsable cluster over the $(-n)$-curve of the $Y^{P}$ component of the central fiber $Y^{0}$. It corresponds to an exceptional algebra $\mathfrak{e}_{m}$, with $m=6$, $7$ or $8$ depending on the value of $n$, see \cref{tab:non-Higgsable-clusters}. Forcing said component to be at weak coupling enhances the non-Higgsable cluster to be non-minimal. The resolution process demands then, possibly after a base change, at least a further base blow-up. In other words, the geometry prevents global weak coupling limits by shedding a new component at strong coupling.}
    \label{fig:restrictions-global-weak-coupling}
\end{figure}

As a consequence, global weak coupling limits cannot be realised in horizontal models constructed over $\hat{B} = \mathbb{F}_{n}$, with $5 \leq n \leq 12$. If we try to forcefully tune one, the geometry prevents this by shedding a new component for $Y_{0}$ that is at strong coupling, preventing us from constructing a horizontal Type II.b or Type III.b model. We schematically summarise the discussion in \cref{fig:restrictions-global-weak-coupling}.

From \eqref{eq:end-component-vanishing-orders} and \eqref{eq:end-component-alpha} we observe that $\{s=0\}_{B^{P}}$ can present component vanishing orders
\begin{equation}
    \ord{Y^{P}}(f_{P},g_{P},\Delta_{P}^{\prime})_{s=0} = (4,6,<12)\,,
\end{equation}model and some generic representatives of a subset of the global divisor classes discussed in
which seem pathological. These hold no physical significance, since the component vanishing orders should be used to read off the physics once the degeneration has been fully resolved as explained in \cite{Alvarez-Garcia:2023gdd}, at which point these vanishing orders cannot appear. Nonetheless, one can arrange via an appropriate base change for the component vanishing orders to not present this pathological behaviour even for the intermediate steps of the resolution process.

\paragraph{Models with $\mathbf{3 \leq n \leq 4}$}

These Hirzebruch surfaces lead to horizontal models whose $Y^{P}$ component exhibits component vanishing orders
\begin{equation}
    \ord{Y^{P}}(f_{P},g_{P})_{s=0} = (2,3)\,.
\end{equation}
When $n=4$,  \eqref{eq:end-component-vanishing-orders} and \eqref{eq:end-component-alpha} imply that
\begin{equation}
    \ord{Y^{P}}(f_{P},g_{P},\Delta_{P}^{\prime})_{s=0} = (2,3,\geq 6)\,,
\end{equation}
meaning that we either find the non-Higgsable cluster present at finite distance or an enhancement of it. For $n=3$, we infer instead that
\begin{equation}
    \ord{Y^{P}}(f_{P},g_{P},\Delta_{P}^{\prime})_{s=0} = (2,3,\geq 5)\,.
\end{equation}
The non-Higgsable cluster associated to Kodaira type IV singularities present at finite distance is always enhanced to be of Kodaira type $\mathrm{I}_{m}^{*}$, a fact that we comment on again in \cref{sec:horizontal-local-global-brane-content}. While the $n=4$ case starts off with $\mathrm{D}_{4}$ singularities that may worsen in particular models, the $n=3$ case allows for $\mathrm{D}_{3}$ singularities over $\{s=0\}_{B^{P}}$. Tuning horizontal Type~II.b or Type~III.b global weak coupling limits is therefore possible over these Hirzebruch surfaces.

\paragraph{Models with $\mathbf{0 \leq n \leq 2}$}

For these Hirzebruch surfaces no forced factorisation of $S_{P}$ classes occurs, and we generically have
\begin{equation}
    \ord{Y^{P}}(f_{P},g_{P},\Delta_{P}^{\prime})_{s=0} = (0,0,0)\,.
\end{equation}
Hence, constructing horizontal Type II.b or Type III.b global weak coupling limits is possible.

\vspace{\baselineskip}
The same conclusions can be reached directly for horizontal Type II.a models by using the Sen limit \cite{Sen:1996vd,Sen:1997gv} in its formulation for Tate models \cite{Donagi:2009ra,Esole:2012tf} and performing a similar analysis. However, the above treatment is more appropriate for horizontal Type III.b models, since their Sen limit presentation does not properly distinguish them from horizontal Type III.a models, as we discuss in \cref{sec:horizontal-IIIb-models-orientifold}. Moreover, the analysis carried out above generalises for the non-horizontal models listed in \cref{tab:genus-zero-Hirzebruch-summary}, as we exploit in \cref{sec:cases-B-C-D-models}, for which the constraints can actually be stricter. The absence of horizontal Type II.a models constructed over $\hat{B} = \mathbb{F}_{n}$ with $5 \leq n \leq 12$ was also commented on in \cite{Braun:2009bh} from the consideration of Nikulin involutions \cite{nikulin1979factor,nikulin1979quotient,nikulin1986discrete}.

\subsection{Types of global divisors}
\label{sec:horizontal-global-divisors}

To fix the notation that we will employ in subsequent sections, let us review the types of global divisors that can occur in the multi-component base $B_{0}$ of the central fiber $Y_{0}$ of the resolved degeneration $\rho: \mathcal{Y} \rightarrow D$ of a horizontal model.

Given a fibration $\pi: V \rightarrow W$, horizontal divisors $D_{\mathrm{hor}}$ of $V$ are those fulfilling $\pi_{*}(D_{\mathrm{hor}}) = W$, while vertical divisors $D_{\mathrm{ver}}$ of $V$ map to a proper subvariety $\pi_{*}(D_{\mathrm{ver}}) \neq W$ of $W$. Any effective divisor $D$ in $V$ can be decomposed as a sum $D = D_{\mathrm{hor}} + D_{\mathrm{ver}}$. We can classify the global divisors of $B_{0}$ into horizontal, vertical and mixed divisors using a similar nomenclature.

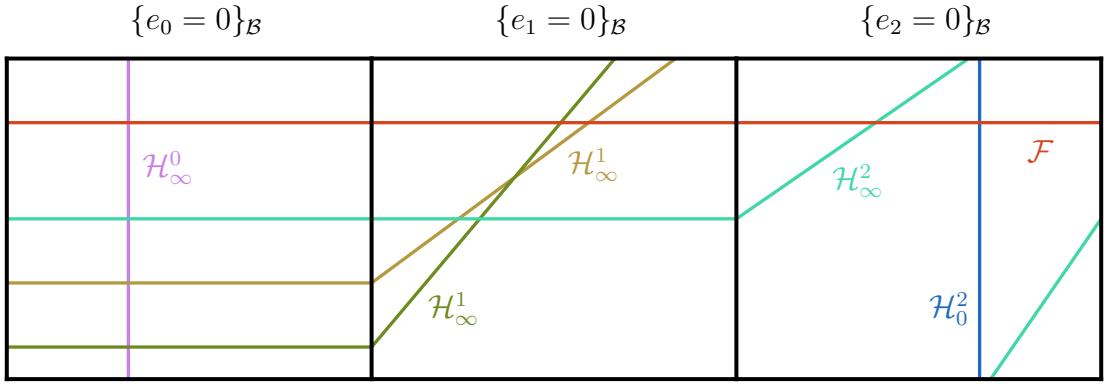
\begin{figure}[t!]
    \centering
	\begin{tikzpicture}
		\def\scalereliiiahor{0.8}
		\def\scalereliiiaver{0.85}
		
		\node (26) at (7*\scalereliiiahor, 2.5*\scalereliiiaver) {};
		\node (27) at (7*\scalereliiiahor, -2.5*\scalereliiiaver) {};
		\node (28) at (-7*\scalereliiiahor, 2.5*\scalereliiiaver) {};
		\node (29) at (-7*\scalereliiiahor, -2.5*\scalereliiiaver) {};
		\node (30) at (2*\scalereliiiahor, 2.5*\scalereliiiaver) {};
		\node (31) at (-3*\scalereliiiahor, -1*\scalereliiiaver) {};
		\node (32) at (-9*\scalereliiiahor, -1*\scalereliiiaver) {};
		\node (33) at (1*\scalereliiiahor, 2.5*\scalereliiiaver) {};
		\node (34) at (-3*\scalereliiiahor, -2*\scalereliiiaver) {};
		\node (35) at (-9*\scalereliiiahor, -2*\scalereliiiaver) {};
		\node (36) at (7*\scalereliiiahor, 2.5*\scalereliiiaver) {};
		\node (37) at (3*\scalereliiiahor, 0*\scalereliiiaver) {};
		\node (38) at (-9*\scalereliiiahor, 0*\scalereliiiaver) {};
		\node (39) at (-9*\scalereliiiahor, 1.5*\scalereliiiaver) {};
		\node (40) at (9*\scalereliiiahor, 1.5*\scalereliiiaver) {};
		\node (48) at (9*\scalereliiiahor, 0*\scalereliiiaver) {};
		
		\draw [style=light-blue line] (26.center) to (27.center);
		\draw [style=light-purple line] (28.center) to (29.center);
		\draw [style=light-yellow line] (30.center) to (31.center);
		\draw [style=light-yellow line] (31.center) to (32.center);
		\draw [style=light-green line] (33.center) to (34.center);
		\draw [style=light-green line] (34.center) to (35.center);
		\draw [style=cyan line] (36.west) to (37.center);
		\draw [style=cyan line] (37.center) to (38.center);
		\draw [style=cyan line] (27.east) to (48.center);
		\draw [style=medium-red line] (39.center) to (40.center);

        \node [label={[yshift=0cm]\textcolor{diagLightBlue}{$\mathcal{H}_{0}^{2}$}}] (49) at (6.5*\scalereliiiahor, -2*\scalereliiiaver) {};
        \node [label={[yshift=0cm]\textcolor{diagLightPurple}{$\mathcal{H}_{\infty}^{0}$}}] (50) at (-6.35*\scalereliiiahor, 0.2*\scalereliiiaver) {};
        \node [label={[yshift=0cm]\textcolor{diagLightGreen}{$\mathcal{H}_{\infty}^{1}$}}] (51) at (-1.65*\scalereliiiahor, -2*\scalereliiiaver) {};
        \node [label={[yshift=0cm]\textcolor{diagLightYellow}{$\mathcal{H}_{\infty}^{1}$}}] (52) at (0.65*\scalereliiiahor, 0.25*\scalereliiiaver) {};
        \node [label={[yshift=0cm]\textcolor{diagCyan}{$\mathcal{H}_{\infty}^{2}$}}] (53) at (5*\scalereliiiahor, 0*\scalereliiiaver) {};
        \node [label={[yshift=0cm]\textcolor{diagMediumRed}{$\mathcal{F}$}}] (54) at (8*\scalereliiiahor, 0.55*\scalereliiiaver) {};

		\node (0) at (-9*\scalereliiiahor, 2.5*\scalereliiiaver) {};
		\node (1) at (-3*\scalereliiiahor, 2.5*\scalereliiiaver) {};
		\node (2) at (3*\scalereliiiahor, 2.5*\scalereliiiaver) {};
		\node (3) at (9*\scalereliiiahor, 2.5*\scalereliiiaver) {};
		\node (4) at (9*\scalereliiiahor, -2.5*\scalereliiiaver) {};
		\node (5) at (3*\scalereliiiahor, -2.5*\scalereliiiaver) {};
		\node (6) at (-3*\scalereliiiahor, -2.5*\scalereliiiaver) {};
		\node (7) at (-9*\scalereliiiahor, -2.5*\scalereliiiaver) {};

		\draw [style=black line] (0.center) to (1.center);
		\draw [style=black line] (1.center) to (2.center);
		\draw [style=black line] (2.center) to (3.center);
		\draw [style=black line] (7.center) to (6.center);
		\draw [style=black line] (6.center) to (5.center);
		\draw [style=black line] (5.center) to (4.center);
		\draw [style=black line] (0.center) to (7.center);
		\draw [style=black line] (1.center) to (6.center);
		\draw [style=black line] (2.center) to (5.center);
		\draw [style=black line] (3.center) to (4.center);
		
		\node [label={[yshift=0cm]$\{e_{0} = 0\}_{\makebox[0pt]{$\scriptstyle \;\;\mathcal{B}$}}$}] (41) at (-6*\scalereliiiahor, 2.5*\scalereliiiaver) {};
        \node [label={[yshift=0cm]$\{e_{1} = 0\}_{\makebox[0pt]{$\scriptstyle \;\;\mathcal{B}$}}$}] (42) at (0*\scalereliiiahor, 2.5*\scalereliiiaver) {};
        \node [label={[yshift=0cm]$\{e_{2} = 0\}_{\makebox[0pt]{$\scriptstyle \;\;\mathcal{B}$}}$}] (43) at (6*\scalereliiiahor, 2.5*\scalereliiiaver) {};
	\end{tikzpicture}
    \caption{We schematically represent the base of the central fiber of a resolved horizontal model and some generic representatives of a subset of the global divisor classes discussed in \cref{sec:horizontal-global-divisors}. The depiction is based on models constructed over $\hat{B} = \mathbb{F}_{1}$.}
    \label{fig:horizontal-global-divisors}
\end{figure}

The discussion is illustrated in \cref{fig:horizontal-global-divisors}, in which we depict generic representatives of a subset of the global divisor classes listed below.

\subsubsection*{Vertical divisors}

These divisors project to points $p_{b} \in \mathbb{P}^{1}_{b}$, and correspond to the divisor class
\begin{equation}
    \mathcal{F} := \sum_{p=0}^{P} V_{p} = \left. \mathcal{V} \right|_{\tilde{\mathcal{U}}}\,.
\label{eq:horizontal-vertical-divisor}
\end{equation}
Here and below we use the notation of \cite{Alvarez-Garcia:2023gdd}, reviewed in \cref{sec:review-part-I}.

They extend through the whole base $B_{0}$ of the multi-component central fiber $Y_{0}$,
\begin{equation}
    \mathcal{F} \cdot E_{p} \neq 0\,,\qquad p = 0, \dotsc, P\,.
\end{equation}
Vertical divisors are $P+1$ copies of $\mathbb{P}^{1}_{f}$ intersecting in an open-chain, i.e.\ a collection of genus-zero curves; they conform the base of the central fiber $Z_{0}$ of the associated degeneration of K3 surfaces $\sigma_{p_{b}}: \mathcal{Z}_{p_{b}} \rightarrow D$, see \cref{sec:classification-horizontal-models}. We will refer to the gauge enhancements supported on this class of divisors as the vertical gauge algebras.

\subsubsection*{Horizontal divisors}

We call horizontal divisors those that restrict to a horizontal divisor in a single component $B^{p}$ of $B_{0}$, not extending to the adjacent ones. The interface curves given by the classes $\{S_{p}\}_{0 \leq p \leq P-1}$ are an edge case, but they are not relevant to the discussion, since they do not support gauge algebras in an open-chain resolution of the horizontal model \cite{Alvarez-Garcia:2023gdd}. The possibilities are:
\begin{itemize}
    \item The genus-zero curves in the class
    \begin{equation}
        \mathcal{H}_{\infty}^{0} := T_{0} = \left. \mathcal{T} \right|_{\tilde{\mathcal{U}}}\,,
    \end{equation}
    which have the intersections
    \begin{equation}
        \mathcal{H}_{\infty}^{0} \cdot E_{0} \neq 0\,, \qquad \mathcal{H}_{\infty}^{0} \cdot E_{p} = 0\,,\quad p = 1, \dotsc, P\,,\qquad \mathcal{H}_{\infty}^{0} \cdot \mathcal{F} = 1\,.
    \end{equation}

    \item The genus-zero curve in the class
    \begin{equation}
        \mathcal{H}_{0}^{P} := S_{P} = \left. \mathcal{S} \right|_{\tilde{\mathcal{U}}}\,,
    \end{equation}
    with intersections
    \begin{equation}
        \mathcal{H}_{0}^{P} \cdot E_{P} \neq 0\,,\qquad \mathcal{H}_{0}^{p} \cdot E_{p} = 0\,,\quad p = 0, \dotsc, P-1\,,\qquad \mathcal{H}_{0}^{P} \cdot \mathcal{F} = 1\,.
    \end{equation}

    \item When the model has been constructed over $\hat{B}_{0} = \mathbb{F}_{0}$, the definition also encompasses the genus-zero curves
    \begin{equation}
        \mathcal{H}_{\infty}^{p}:= T_{p} = S_{p} = \left.\left( T + \sum_{i=0}^{p-1} (p-i) E_{i} \right)\right|_{\tilde{\mathcal{U}}}\,,\qquad p = 1, \dotsc, P\,,
    \end{equation}
    which intersect
    \begin{equation}
        \mathcal{H}_{\infty}^{p} \cdot E_{p} \neq 0\,,\qquad \mathcal{H}_{\infty}^{p} \cdot E_{q} = 0\,,\quad p \neq q = 0, \dotsc, P\,,\qquad \mathcal{H}_{\infty}^{p} \cdot \mathcal{F} = 1\,.
    \end{equation}

    \item Multiples $\alpha \mathcal{H}_{\infty}^{p}$ of the above curve classes, whose genus can be computed via adjunction resulting in
    \begin{equation}
        g\left( \alpha \mathcal{H}_{\infty}^{p} \right) = \frac{1}{2} (\alpha -1) (\alpha  n-2)\,.
    \end{equation}
\end{itemize}

\subsubsection*{Mixed divisors}

The remaining divisors we call mixed divisors. Among these, we can highlight two classes:
\begin{itemize}
    \item When the model has been constructed over $\hat{B}_{0} = \mathbb{F}_{n}$ with $n \geq 1$, we have the curve classes
    \begin{equation}
        \mathcal{H}_{\infty}^{p}:= \sum_{i=0}^{p-1}n V_{i} + T_{p} = \left.\left( T + \sum_{i=0}^{p-1} (p-i) E_{i} \right)\right|_{\tilde{\mathcal{U}}}\,,\qquad p = 1, \dotsc, P\,,
    \label{eq:horizontal-Hpinf-n-geq-1}
    \end{equation}
    with intersections
    \begin{equation}
        \mathcal{H}_{\infty}^{p} \cdot E_{q} \neq 0\,,\quad q = 0, \dotsc, p\,,\qquad \mathcal{H}_{\infty}^{p} \cdot E_{q} = 0\,,\quad q = p+1, \dotsc, P\,,\qquad \mathcal{H}_{\infty}^{p} \cdot \mathcal{F} = 1\,.
    \end{equation}
    They restrict to genus-zero curves in all the components that they intersect non-trivially.

    \item Divisors arising from a combination of all the previously listed classes. For example, the divisors in the classes $\mathcal{H}_{\infty}^{p} + \alpha \mathcal{F}$, that restrict in $B^{p}$ to the sections $h+(n+\alpha)f$ and extend vertically to the left.
\end{itemize}
Mixed divisors can be split via a finite-distance complex structure deformation into horizontal and vertical divisors, and we will therefore interpret the gauge factors supported on the former as a Higgsing of the latter.

\subsubsection*{Exceptional divisors}

In models presenting codimension-two finite-distance non-minimal points, we may also have exceptional curves arising from their resolution. These curves are completely contained within a component $\{ B^{p} \}_{0 \leq p \leq P}$ and are therefore global divisors. The gauge algebras supported over them are not subject to any subtleties relating to the multi-component nature of $Y_{0}$, and we mostly ignore them in what follows by keeping the exceptional curves blown-down, i.e.\ by maintaining $\hat{B} = \mathbb{F}_{n}$. In \cref{sec:defect-algebras-heterotic-dual} we briefly comment on them.

\subsection{Restrictions on the local and global 7-brane content}
\label{sec:horizontal-local-global-brane-content}

The gauge algebra content at the endpoints of the infinite-distance limits that we study is analysed in two steps. First, we take a six-dimensional standpoint and determine the gauge algebras as read off from the central fiber $Y_{0}$ of the open-chain resolution $\rho: \mathcal{Y} \rightarrow D$ of the degeneration of Hirzebruch surfaces. Second, we consider any partial decompactifications that may occur along the limit, leading to gauge enhancements from the point of view of the asymptotic, higher-dimensional gauge theory.

The first of these points was addressed in \cite{Alvarez-Garcia:2023gdd}. In the reducible variety $Y_{0}$, individual gauge factors are associated to the global divisors that appear as components of the physical discriminant $\Dphys$, and that in a horizontal model will be (multiples of) the classes just listed in \cref{sec:horizontal-global-divisors}. These divisors are obtained by consistently gluing the irreducible components of the restrictions $\{\Delta_{p}^{\prime}\}_{0 \leq p \leq P}$ of the modified discriminant $\Delta'$ to the different base components $\{B^{p}\}_{0 \leq p \leq P}$. While the analysis in terms of the global divisors clarifies the complications arising in a componentwise study, the restrictions on the 7-brane content occurring in single compo\-nents are still useful to consider, as they provide valuable information on the possible global 7-branes of the model. In this section, we comment on how this works for horizontal models.

The results can be succinctly summarised as follows: Those components at local strong coupling are not subject to any restrictions on their local 7-brane content. The components at local weak coupling can only present local 7-branes associated with Kodaira type $\mathrm{I}_{m}$ and $\mathrm{I}_{m}^{*}$ fibers. If a global 7-brane passes through various components it has to obey the most stringent of the individual local constraints found in them.

The effectiveness bounds \eqref{eq:horizontal-model-effectiveness-constraint-horizontal-2} imply that, as soon as a single component is at weak coupling, all intermediate components of the horizontal model must be as well. This leaves us with the cases with no codimension-zero singular elliptic fibers, those in which only the intermediate components are at weak coupling, and finally those in which the end-components also are, \mbox{cf.\ the classification in \cref{sec:classification-horizontal-models}}.

\subsubsection{Models with no components at weak coupling}

When the generic elliptic fiber in all components $\{Y^{p}\}_{0 \leq p \leq P}$ is of Kodaira type $\mathrm{I}_{0}$, we see from \eqref{eq:line-bundle-hor-1} and \eqref{eq:Delta-hor-2} that the local 7-brane content in the intermediate components $\{Y^{p}\}_{1 \leq p \leq P-1}$ can only consist of vertical classes. This means that no global gauge enhancements can occur over the divisors classes $\{\mathcal{H}_{\infty}^{p}\}_{1 \leq p \leq P-1}$ or combinations containing them. The local 7-branes over the intermediate components can extend to global 7-branes in the divisor classes $\mathcal{F}$, $\mathcal{H}_{\infty}^{P}$, or combinations of them. The global divisors $\mathcal{H}_{\infty}^{0}$, $\mathcal{H}_{0}^{P}$, and combinations involving them also can support gauge enhancements. Since no components are at weak coupling, all listed divisors can support any of the Kodaira type elliptic fibers, leading to the associated simply-laced ADE Lie algebras or, possibly, the non-simply-laced ones resulting from folding the corresponding Dynkin diagrams. Horizontal Type II.a models fall within this category.

\subsubsection{Models with the intermediate components at weak coupling}

For the intermediate components $\{Y^{p}\}_{1 \leq p \leq P-1}$ to be at weak coupling we need, at the very least, the single accidental cancellation structure \eqref{eq:codimension-zero-accidental-cancellation} to hold, meaning that $f_{p} = -3h_{p}^{2}$, $g_{p} = 2h_{p}^{3}$ and $\Delta_{p}^{\prime} = h_{p}^{k} \Delta_{p}^{\prime\prime}$, with $k \geq 2$ and $p \in \{ 1, \dotsc, P-1\}$. From \eqref{eq:hp-classes-horizontal} we read that
\begin{equation}
    H_{p} = 4V_{p} \Rightarrow h_{p} = p_{4}([v:w])\,,\qquad p \in \{1, \dotsc, P-1\}\,,
\end{equation}
which is always reducible. The polynomials $\{h_{p}\}_{1 \leq p \leq P-1}$ must all be identical, since the local 7-branes need to consistently extend between components. None of the four roots of these polynomials may coincide in a horizontal model; this would lead to non-minimal component vanishing orders over a representative of the fiber class of the intermediate components and, as a consequence, the model would not be a single infinite-distance limit, see \cite{Alvarez-Garcia:2023gdd}. The structure \eqref{eq:codimension-zero-accidental-cancellation} hence implies that we have four local Kodaira type $\mathrm{D}_{m}$ enhancements in the intermediate components, where $m$ can also attain the values $m = 0$, $1$, $2$ and $3$. Additionally, we observe from \eqref{eq:Delta-hor-2} that the intermediate components may also present Kodaira type $\mathrm{I}_{m}$ enhancements depending on the values of $\{n_{p}\}_{0 \leq p \leq P}$.

The end-components $Y^{0}$ and $Y^{P}$ do not fulfil the structure \eqref{eq:codimension-zero-accidental-cancellation}, and their Weierstrass model is generic enough that the aforementioned local enhancements in the intermediate components generically extend without forcing a global enhancement.

Let us now list the types of singular elliptic fibers that can occur over global divisors in such a model. We only list a few divisor classes, with the understanding that divisor classes built as combinations of them are subject to the most stringent of the restrictions applying to their individual pieces. The global divisors $\mathcal{H}_{\infty}^{0}$ and $\mathcal{H}_{0}^{P}$ are fully contained in the $Y^{0}$ and $Y^{P}$ component, respectively; since these components are not at weak coupling, no constraints apply, and they can support any of the Kodaira type elliptic fibers. The divisor classes $\mathcal{F}$ and $\{\mathcal{H}_{\infty}^{p}\}_{1 \leq p \leq P}$ extend through the intermediate components, meaning that their associated global gauge factor must be compatible with the local weak coupling found in them (an exception occurs for horizontal models constructed over $\hat{B} = \mathbb{F}_{0}$, for which the $\mathcal{H}_{\infty}^{P}$ class is completely contained in the $Y^{P}$ component and subject to the same considerations as $\mathcal{H}_{0}^{P}$). Hence, $\mathcal{F}$ and $\mathcal{H}_{\infty}^{P}$ can only support A and D type singular elliptic fibers, while $\{\mathcal{H}_{\infty}^{p}\}_{1 \leq p \leq P-1}$ only allow for those of A type. The A type fibers of Kodaira type III and IV are also possible, as long as they suffer a local enhancement to Kodaira type $\mathrm{I}_{m}^{*}$ in the intermediate components. Horizontal Type III.a models fall within this category.

\subsubsection{Models with the end-components at weak coupling}

If, in addition to the intermediate components $\{Y^{p}\}_{1 \leq p \leq P-1}$, one of the two end-components $\{ Y^{p} \}_{p=0,P}$ is also at weak coupling, the situation does not vary much with respect to the previous case. The degenerations still correspond to horizontal Type III.a models, and the gauge enhancements possible over the global divisors are essentially the ones previously listed. The differences arise for enhancements over the divisor classes $\mathcal{H}_{\infty}^{0}$ or $\mathcal{H}_{0}^{P}$ (also $\mathcal{H}_{\infty}^{P}$ if $\hat{B} = \mathbb{F}_{0}$), depending on which of the end-components is at weak coupling; these classes can only support A and D type singular elliptic fibers. The component that is not at weak coupling still prevents the forced local enhancements of the intermediate components from generically extending to a global enhancement.\footnote{This is different from the behaviour of the end-components at weak coupling in eight-dimensional models, which inevitably carry two Kodaira type $\mathrm{I}_{m}^{*}$ singularities \cite{Lee:2021qkx,Lee:2021usk} that manifest from the global perspective. This is because 7-branes are points in the internal space in eight-dimensions, and therefore the relevant $h_{p}$ polynomial always factorizes, similarly to how it occurs for the vertical classes in the intermediate components at weak coupling, with the difference that the points are fully contained in a component.}

The situation changes once all components $\{Y^{p}\}_{0 \leq p \leq P}$ are at weak coupling, i.e.\ once we are dealing with a horizontal Type III.b model, which we saw in \cref{sec:horizontal-restrictions-global-weak-coupling} can only occur when $\hat{B} = \mathbb{F}_{n}$ with $0 \leq n \leq 4$. Then, all components satisfy at least the single accidental cancellation structure \eqref{eq:codimension-zero-accidental-cancellation}, and the global divisor
\begin{equation}
    \mathcal{H} := \sum_{p=0}^{P} H_{p} -
    \begin{cases}
        0\,, & 0 \leq n \leq 2\,,\\
        S_{p}\,, & 3 \leq n \leq 4\,,
    \end{cases}
\end{equation}
that we have defined as the gluing of the $\{H_{p}\}_{0 \leq p \leq P}$ with the non-Higgsable clusters subtracted, appears with physical vanishing orders
\begin{equation}
    \ord{Y_{0}}(\fphys,\gphys)_{\mathcal{H}} = (2,3)\,.
\end{equation}
Taking into account \eqref{eq:hp-classes-horizontal}, we see that
\begin{equation}
    \mathcal{H} = \left(2T_{0} + 4V_{0}\right) + \left( \sum_{p=1}^{P-1} 4V_{p} \right) + 
    \begin{cases}
        2T_{P} + 4V_{P}\,, & n=0\,,\\
        2T_{P} + 2V_{P}\,, & n=1\,,\\
        2T_{P}\,, & n=2\,,\\
        T_{P} + V_{P}\,, & n=3\,,\\
        T_{P}\,, & n=4\,,
    \end{cases}
\end{equation}
or, in terms of global divisors,
\begin{equation}
    \mathcal{H} = 2\mathcal{H}_{\infty}^{0} + 
    \begin{cases}
        2\mathcal{H}_{\infty}^{P} + 4\mathcal{F}\,, & n=0\,,\\
        2\mathcal{H}_{\infty}^{P} + 2\mathcal{F}\,, & n=1\,,\\
        2\mathcal{H}_{\infty}^{P}\,, & n=2\,,\\
        \mathcal{H}_{\infty}^{P} + \mathcal{F}\,, & n=3\,,\\
        \mathcal{H}_{\infty}^{P}\,, & n=4\,,
    \end{cases}
\end{equation}
meaning that we generically have a D type enhancement (or its folding) over a global mixed divisor. In models for which only single accidental cancellations occur
\begin{equation}
    \ord{Y_{0}}(\fphys,\gphys,\Dphys)_{\mathcal{H}} = (2,3,3)
\end{equation}
are the generic physical vanishing orders, corresponding to $\mathrm{D}_{1}$ type singular elliptic fibers. When higher accidental cancellations take place, this can be lowered to
\begin{equation}
    \ord{Y_{0}}(\fphys,\gphys,\Dphys)_{\mathcal{H}} = (2,3,2)\,,
\end{equation}
corresponding to $\mathrm{D}_{0}$ type fibers, see the discussion in \cref{sec:discriminant-weakly-coupled-components}. All gauge enhancements appearing in a horizontal Type III.b model are associated with Kodaira type $\mathrm{I}_{m}$ or $\mathrm{I}_{m}^{*}$ singular elliptic fibers.

In a horizontal Type II.b model we have a single component $Y_{0}$ for the central fiber of the resolved degeneration, which is at weak coupling. The generic elliptic fiber is of Kodaira type~$\mathrm{I}_{m}$, with $\mathrm{I}_{m'}^{*}$ fibers over a divisor in the class
\begin{equation}
    H = 2h+(2+n)f -
    \begin{cases}
        0\,, & 0 \leq n \leq 2\,,\\
        h\,, & 3 \leq n \leq 4\,,
    \end{cases}
\end{equation}
a $\mathrm{I}_{m''}^{*}$ non-Higgsable cluster over $h$ when $\hat{B} = \mathbb{F}_{n}$ with $3 \leq n \leq 4$, and some additional $\mathrm{I}_{m'''}$~type fibers over curves.

\subsection{Physical interpretation of the constraints}
\label{sec:physical-interpretation-of-the-constraints}

As we have just seen, the most notable effect of having codimension-zero $\mathrm{I}_{n_{p}>0}$ singular elliptic fibers in a component is to restrict the local gauge enhancements in it to be associated with Kodaira type $\mathrm{I}_{m}$ or $\mathrm{I}_{m}^{*}$ singular elliptic fibers; this constraint affects in some form all global divisors traversing a component at weak coupling, and applies to all possible gauge enhancements in those models at global weak coupling.

In terms of the geometry of the elliptic fibration of the component at weak coupling, the restriction arises from the compatibility of the $j$-invariant of the generic elliptic fibers with the one found over the 7-brane loci. At weak coupling the generic elliptic fiber is such that $j(\tau) \rightarrow \infty$, a property that is shared by Kodaira type $\mathrm{I}_{m}$ and $\mathrm{I}_{m}^{*}$ fibers. From the point of view of the physics, the constraint stems from the fact that the exceptional gauge groups $\mathrm{G}_{2}$, $\mathrm{F}_{4}$, $\mathrm{E}_{6}$, $\mathrm{E}_{7}$ and $\mathrm{E}_{8}$ are non-perturbative in Type IIB, and therefore cannot appear in a component in which locally $g_{s} \rightarrow 0$.

While it is clear that $\mathrm{F}_{4}$, $\mathrm{E}_{6}$, $\mathrm{E}_{7}$ and $\mathrm{E}_{8}$ cannot appear if only Kodaira type $\mathrm{I}_{m}$ and $\mathrm{I}_{m}^{\ast}$ singular fibers are allowed, the same is not true for $\mathrm{G}_{2}$; it can be produced by folding the $\mathrm{D}_{4}$ Dynkin diagram associated to a $\mathrm{I}_{0}^{*}$ fiber, see \cref{tab:AD-monodromy-cover}.
\begin{table}[t!]
    \centering
	\begin{tblr}{columns = {c}, hlines, vlines}
        Type & Split & Semi-split & Non-split\\
        $\mathrm{I}_{m}$ & $\mathfrak{su}$-algebra (A) & --- & $\mathfrak{sp}$-algebra (C)\\
        $\mathrm{I}_{0}^{*}$ & $\mathfrak{so}$-algebra (D) & $\mathfrak{so}$-algebra (B) & $\mathfrak{g}_{2}$-algebra (G)\\
        $\mathrm{I}_{m>0}^{*}$ & $\mathfrak{so}$-algebra (D) & --- & $\mathfrak{so}$-algebra (B)
    \end{tblr}
    \caption{Gauge algebras associated to $\mathrm{I}_{m}$ and $\mathrm{I}_{m}^{*}$ fibers depending on the number of irreducible components of the monodromy cover.}
    \label{tab:AD-monodromy-cover}
\end{table}
However, one can check that the monodromy cover in a component $Y^{p}$ at weak coupling never allows for $\mathrm{I}_{0}^{*\, \mathrm{ns}}$ fibers. If we have $\mathrm{I}_{0}^{*}$ fibers over a divisor $\mathcal{D} = \{p_{\mathcal{D}} = 0\}_{B^{p}}$, the single accidental cancellation structure \eqref{eq:codimension-zero-accidental-cancellation} means that the defining polynomials will take the form
\begin{equation}
	f_{p} = -3 \left( p_{\mathcal{D}} h'_{p} \right)^{2}\,,\qquad g_{p} = 2 \left( p_{\mathcal{D}} h'_{p} \right)^{3}\,,\qquad h'_{p} \in H^{0}\left(B^{p},\mathcal{L}_{p}^{\otimes 2} - \mathcal{D}\right)\,.
\end{equation}
Then, the monodromy cover for the $\mathrm{I}_{0}^{*}$ fiber is
\begin{equation}
	\psi^{3} + \psi \left. \frac{f_{p}}{p_{\mathcal{D}}^{2}} \right|_{p_{\mathcal{D} = 0}} + \left. \frac{g_{p}}{p_{\mathcal{D}}^{3}} \right|_{p_{\mathcal{D} = 0}} = \left( \psi + 2 \left. h'_{p} \right|_{p_{\mathcal{D} = 0}} \right) \left( \psi - \left. h'_{p} \right|_{p_{\mathcal{D} = 0}} \right)^{2}\,,
\end{equation}
and hence always at least semi-split. This ensures that a $\mathrm{I}_{0}^{*\, \mathrm{ns}}$ fibers will never be supported on $\mathcal{D}$, in alignment with the considerations above.

The fact that horizontal global weak coupling limits can only be constructed over $\hat{B} = \mathbb{F}_{n}$ with $0 \leq n \leq 4$, as discussed in \cref{sec:horizontal-restrictions-global-weak-coupling}, can also be understood from a physical standpoint. Global weak coupling limits must have an interpretation as perturbative Type IIB orientifold compactifications; the models constructed over $\hat{B} = \mathbb{F}_{n}$ with $5 \leq n \leq 12$ present non-Higgsable clusters with gauge algebra $\mathfrak{f}_{4}$, $\mathfrak{e}_{6}$, $\mathfrak{e}_{7}$ or $\mathfrak{e}_{8}$, see \cref{tab:non-Higgsable-clusters}, and should therefore be incompatible with the global weak coupling limit.

While in a generic F-theory model with internal space $\pi: Y \rightarrow B$ we have that
\begin{equation}
    \ord{Y}(f,g)_{D} = (2,3) \Rightarrow \ord{Y}(\Delta)_{D} \geq 6
\end{equation}
and the $\mathrm{D}_{m}$ type singularities start therefore at $\mathrm{D}_{4}$, we have observed that in a component $Y^{p}$ at weak coupling we have instead
\begin{equation}
    \ord{Y^{p}}(f,g)_{D} = (2,3) \Rightarrow \ord{Y^{p}}(\Delta)_{D} \geq 2\,,
\end{equation}
meaning that $\mathrm{D}_{0}$, $\mathrm{D}_{1}$, $\mathrm{D}_{2}$ and $\mathrm{D}_{3}$ type singularities are also possible. This was observed already in \cite{Lee:2021qkx,Lee:2021usk} for eight-dimensional models, and the explanation is the same: The $\mathrm{D}_{m}$ type singularities are interpreted as O7-planes with $m$ mutually local 7-branes on top. While this is allowed in the strict weak coupling limit $g_{s} \rightarrow 0$ for any value of $m$, such configurations split up into mutually non-local branes if $0 \leq m \leq 3$ when $g_{s}$ takes a non-zero value \cite{Sen:1996vd,Sen:1997gv}.

\subsection{Bounds on the vertical gauge rank}
\label{sec:bounds-vertical-gauge-rank}

With the classification of \cref{sec:horizontal-global-divisors} in mind, gauge algebra factors can be classified into horizontal and vertical if they are supported on the homonymous divisor classes, with gauge enhancements over mixed divisors arising from the Higgsing of these contributions. From the point of view of the heterotic dual theory, a correspondence that we review in \cref{sec:F-theory-heteoric-duality}, this distinction is also pertinent: Horizontal and vertical gauge factors on the F-theory side correspond to perturbative and non-perturbative gauge contributions, respectively, on the dual heterotic side.\footnote{The non-perturbative heterotic gauge sector is completely accounted for once the gauge algebra factors supported over the exceptional curves resulting from the resolution of codimension-two finite-distance non-minimal points are also considered.}

Obtaining a rough bound for the rank of the horizontal gauge algebras is therefore direct from heterotic considerations, from where we see that
\begin{equation}
    \rank(\mathfrak{g}_{\mathrm{hor}}) \leq 18\,.
\label{eq:horizontal-gauge-rank-bound}
\end{equation}
This bound can be saturated in models at finite distance by tuning two lines of $\mathrm{II}^{*}$ fibers over representatives of the curves classes $C_{0}$ and $C_{\infty}$ in $\hat{B} = \mathbb{F}_{n}$, and a line of $\mathrm{IV}$ fibers over a representative of $C_{\infty}$. In models at infinite distance the bound becomes more stringent, as we discuss later for the six-dimensional models under study, and was already observed in the analysis of eight-dimensional limits in \cite{Lee:2021qkx,Lee:2021usk}, and from the point of view of heterotic toroidal compactifications in \cite{Collazuol:2022oey}.

Due to the non-perturbative nature (from a heterotic standpoint) of the vertical gauge algebras, obtaining bounds for them is less immediate. They can, however, be extracted from the geometry on the F-theory side and lead to bounds displayed in \cref{tab:vertical-rank-bounds}. Attempting to surpass said bounds sends us to infinite-distance in the moduli space, so they can be regarded as a measure of how much tuning of the vertical sector is possible at finite-distance.

To understand the origin of these bounds, we recall that in a Hirzebruch surface $\hat{B} = \mathbb{F}_{n}$ a curve $C$ can have negative intersection product $C \cdot h$ with the $(-n)$-curve $h$ if and only if $C$ contains $h$ as a component, a fact that was already exploited in \cref{sec:horizontal-restrictions-global-weak-coupling}. For generic models this leads to the appearance of the non-Higgsable clusters of \cref{tab:non-Higgsable-clusters}, or their appropriate enhancements if the component is at weak coupling. Suppose now that a vertical gauge factor
\begin{equation}
    \ord{Y_{0}}(\fphys,\gphys,\Dphys)_{\mathcal{F}} = (\alpha,\beta,\gamma)
\end{equation}
has been tuned. The global divisor $\mathcal{F}$ traverses all components, and in particular goes through the $Y^{P}$ component, where we will have component vanishing orders
\begin{equation}
    \ord{Y^{P}}(f_{P},g_{P},\Delta_{P}^{\prime})_{V_{P}} \geq (\alpha,\beta,\gamma)\,.
\end{equation}
We are focusing on the component $Y^{P}$ because it is the one in which forced factorizations of $S_{P}$ in $\Delta_{P}^{\prime}$ will occur. Indeed, the vertical tuning favours this even further, since
\begin{equation}
    V_{P} \cdot S_{P} = 1 \Rightarrow \left( \Delta_{P}^{\prime} - \gamma V_{P} \right) \cdot S_{P} < \Delta_{P}^{\prime} \cdot S_{P}\,.
\label{eq:vertical-bound-intersection}
\end{equation}
As a consequence, a high enough vertical tuning will lead to a non-minimal enhancement over the curve $S_{P}$, limiting the vertical gauge rank that can be realized in the class of models.

\begin{figure}[p!]
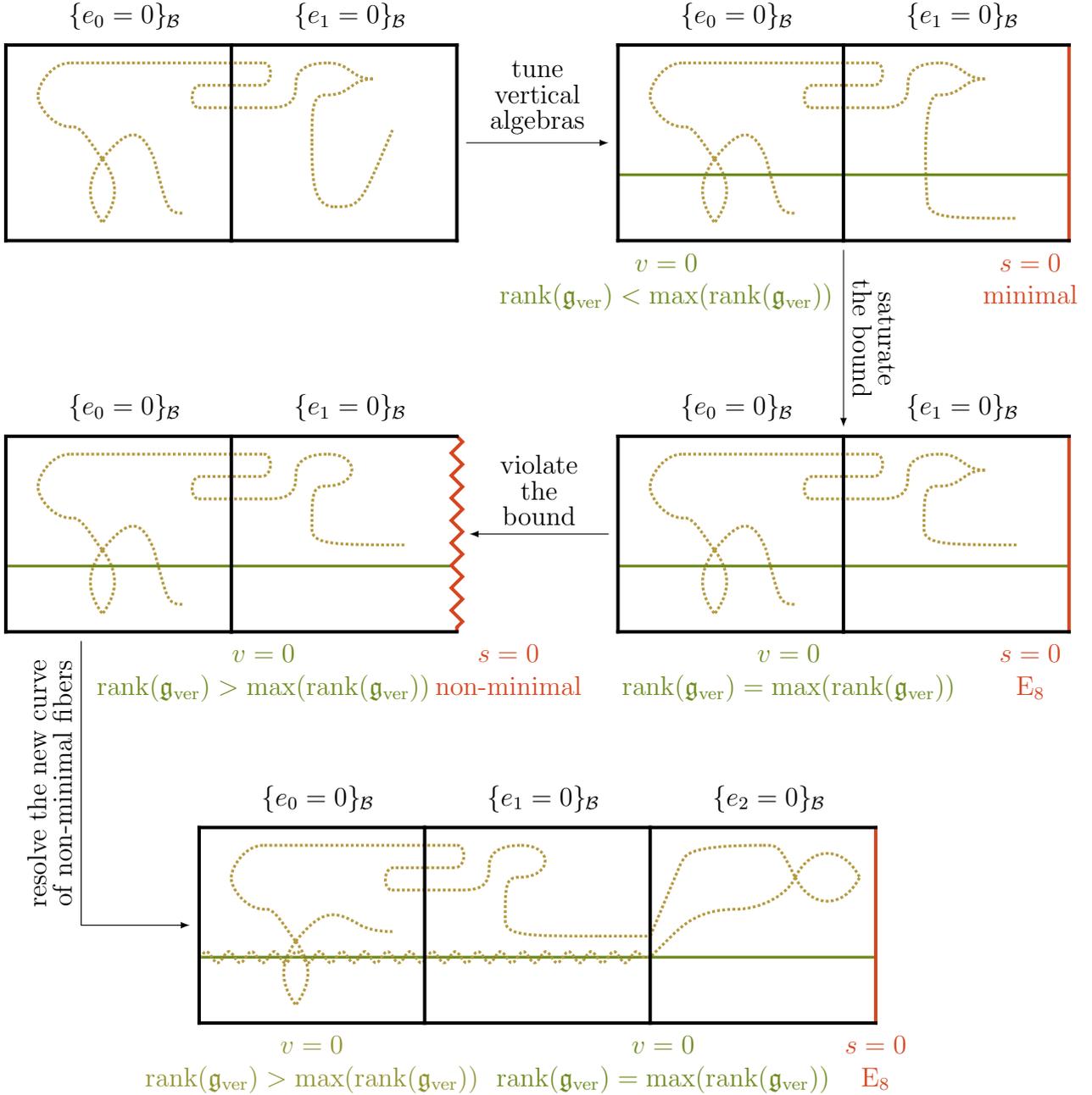

    \centering

    \caption{Tuning vertical algebras over representatives of the global divisor $\mathcal{F}$ leads to forced enhancements over the $S_{P}$ curve in the end-component. Eventually, the gauge factor supported on $S_{P}$ corresponds to $\mathrm{E}_{8}$, meaning that further forced factorizations of $S_{P}$ in $\Delta_{P}^{\prime}$ will make it non-minimal. If we try to exceed the vertical gauge rank by a higher tuning over the representatives of $\mathcal{F}$, the model sheds a new component in which the bound is still satisfied, rendering the tuning a local enhancement. As a consequence, the bound is still respected from the global point of view that is used to assign the gauge algebras.}
    \label{fig:vertical-gauge-rank-bounds}
\end{figure}

Given a bound $\max(\rank(\mathfrak{g}_{\mathrm{ver}}))$ for the vertical gauge rank computed by an analysis along these lines, the geometry enforces the bound in a similar fashion to how it prevents certain global weak coupling limits. Consider a horizontal model whose open-chain resolution leads to $\{Y^{p}\}_{0 \leq p \leq P}$ components for the central fiber $Y_{0}$ of the degeneration. Tuning vertical gauge factors over $\mathcal{F} = \sum_{p=0}^{P} V_{p}$ leads, as we have just discussed, to forced gauge enhancements over $S_{P}$ in the $Y^{P}$ component. As long the vertical gauge factors are such that
\begin{equation}
    \rank(\mathfrak{g}_{\mathrm{ver}}) \leq \max(\rank(\mathfrak{g}_{\mathrm{ver}}))\,,
\end{equation}
the component vanishing orders over $S_{P}$ can be minimal.\footnote{Some vertical enhancement patterns can saturate the bound $\max(\rank(\mathfrak{g}_{\mathrm{ver}}))$, while others are less efficient in their expenditure of the available vertical classes, and lead to a non-minimal enhancement over $S_{P}$ before the bound on the vertical gauge rank is reached.} Tuning a higher vertical enhancement such that
\begin{equation}
    \rank(\mathfrak{g}_{\mathrm{ver}}) > \max(\rank(\mathfrak{g}_{\mathrm{ver}}))
\end{equation}
is possible, but the component vanishing orders over $S_{P}$ will then be non-minimal,\footnote{Given \eqref{eq:Delta-hor-3}, we see that some patterns of codimension-zero singular elliptic fibers subtract $S_{P}$ divisor classes from $\Delta_{P}^{\prime}$. When this occurs, a vertical enhancement surpassing the bound leads to pathological vanishing orders over $S_{P}$; as discussed in \cref{sec:horizontal-restrictions-global-weak-coupling}, these are removed via an appropriate base change, after which the need for additional base blow-ups in order to resolve the degeneration becomes evident.} meaning that (possibly after a base change) a new base blow-up must be performed in the resolution process of the horizontal model. The central fiber $Y_{0}$ then has $\{Y^{q}\}_{0 \leq q \leq Q}$ components, with $Q > P$. In the subset $\{Y^{p}\}_{0 \leq p \leq P} \subset \{Y^{q}\}_{0 \leq q \leq Q}$ of these (intuitively) corresponding to the ``original" components, the local vertical gauge enhancements over representatives of $\sum_{p=0}^{P} V_{p}$ can exceed the vertical gauge rank bound due to the tuning. However, $\sum_{p=0}^{P} V_{p}$ no longer is the global vertical class, which after the new resolution process is given by $\mathcal{F} = \sum_{q=0}^{Q} V_{q}$. In the new end-component, the vertical gauge rank bound must be respected by the local vertical enhancements. Otherwise, the arguments above would apply, and we would have a non-minimal enhancement over $S_{P}$, at the very least, at the level of the component vanishing orders. This would mean that we would not have an open-chain resolution free of obscured infinite-distance limits. Hence, the attempted global vertical enhancement is rendered a local enhancement by the model shedding a new component in which the bound on the vertical gauge rank is respected. This discussion is summarised below in \cref{fig:vertical-gauge-rank-bounds}.

While we have framed the discussion in terms of the horizontal models with a multi-component central fiber, similar bounds also apply to the six-dimensional F-theory models over the Hirzebruch surfaces at finite-distance. In fact, since for such models we have that
\begin{equation}
    \mathcal{L}_{B_{0}} = h + 2f + (h+nf) = \mathcal{L}_{P} + (h+nf)
\end{equation}
and $(h+nf) \cdot h = 0$, the resulting bounds are the same that hold for multi-component horizontal models in which no components of the central fiber are at weak coupling. This means that tuning too many vertical gauge factors in a Hirzebruch surface F-theory model leads to non-minimal elliptic curves over $h$, and therefore drives us to infinite distance in the moduli space.

Given that the mechanism enforcing the vertical gauge rank bounds is \eqref{eq:vertical-bound-intersection}, it is clear that the bounds for models constructed over $\hat{B} = \mathbb{F}_{n}$ must become more stringent as we increase the value of $n$. In fact, for models constructed over $\hat{B} = \mathbb{F}_{0}$ the argument does not apply, since $h \cdot h = 0$. For these models, the distinction between horizontal and vertical divisors is arbitrary, since $\mathbb{F}_{0} = \mathbb{P}^{1}_{f} \times \mathbb{P}^{1}_{b}$; from the heterotic dual side, this corresponds to heterotic/heterotic duality \cite{Morrison:1996na,Morrison:1996pp,Duff:1996rs}. Hence, the bound \eqref{eq:horizontal-gauge-rank-bound} also applies to the vertical gauge rank, and we have
\begin{equation}
    \rank(\mathfrak{g}_{\mathrm{ver}}) \leq 18\,,
\label{eq:vertical-gauge-rank-bound}
\end{equation}
which applies to all horizontal models constructed over $\hat{B} = \mathbb{F}_{n}$ and is saturated for $\hat{B} = \mathbb{F}_{0}$.\footnote{In a model constructed over $\hat{B} = \mathbb{F}_{0}$ and choosing to call the direction along which the splitting of the components occurs horizontal, the bound \eqref{eq:horizontal-gauge-rank-bound} becomes stricter in models at infinite distance, as we commented on above. The analogous vertical bound \eqref{eq:vertical-gauge-rank-bound} remains, however, unaltered, since the vertical directions are ``orthogonal" to the splitting of the components, and hence not really affected by it.}

This bound can be refined by analysing the geometry in more detail. For example, for horizontal models in which no components are at weak coupling the bounds given in \cref{tab:vertical-rank-bounds} apply, and we have checked in specific examples that they can be saturated. They are derived via the arguments given above, and we give an illustrative example of such a discussion in \cref{sec:B-F8-no-weak-coupling}.
\begin{table}[t!]
    \centering
    \begin{tblr}{columns = {c}, column{2-13} = {4mm}, hlines, vlines}
        $n$ & $0$ & $1$ & $2$ & $3$ & $4$ & $5$ & $6$ & $7$ & $8$ & $9$ & $10$ & $11$ & $12$ \\
        $\max(\rank(\mathfrak{g}_{\mathrm{ver}}))$ & $18$ & $17$ & $16$ & $16$ & $10$ & $9$ & $8$ & $8$ & $4$ & $2$ & $1$ & $0$ & $0$
    \end{tblr}
    \caption{Bounds on the rank of the vertical gauge contribution in horizontal models, where $n$ indicates the type of Hirzebruch surface $\hat{B} = \mathbb{F}_{n}$ over which the model is constructed.}
    \label{tab:vertical-rank-bounds}
\end{table}
Once some components are at weak coupling, not all types of global gauge factors can be realised, as we elaborated on in \cref{sec:horizontal-local-global-brane-content}. This makes the bounds on the vertical gauge rank become more stringent. For instance, for models constructed over $\hat{B} = \mathbb{F}_{7}$ the bound $\rank(\mathfrak{g}_{\mathrm{ver}}) \leq 8$ becomes $\rank(\mathfrak{g}_{\mathrm{ver}}) \leq 5$ as soon as one component is at weak coupling, see \cref{sec:B-F7-weak-coupling}. In addition to this effect, which depends solely on the existence of some components at weak coupling, the concrete pattern of codimension-zero singular elliptic fibers can reduce the maximal vertical gauge rank even further. For instance, consider again the models constructed over $\hat{B} = \mathbb{F}_{7}$. The bound $\rank(\mathfrak{g}_{\mathrm{ver}}) \leq 5$ can be saturated by tuning a vertical line of Kodaira type $\mathrm{I}_{2}^{*\, \mathrm{ns}}$ fibers in a model with a $\mathrm{I}_{0} - \mathrm{I}_{1} - \mathrm{I}_{0}$ pattern of codimension-zero singular elliptic fibers. The tuning of such a vertical gauge algebra forces an $\mathrm{E}_{8}$ enhancement over the curve $S_{P}$. If the codimension-zero singular elliptic fibers appear instead in the pattern $\mathrm{I}_{0} - \mathrm{I}_{2} - \mathrm{I}_{0}$, which can be achieved via a double cancellation structure according to the discussion in \cref{sec:bounds-codimension-zero-pattern-horizontal}, the number of local horizontal classes in $\Delta'_{P}$ is insufficient to realise an $\mathrm{E}_{8}$ enhancement of the $\mathrm{E}_{7}$ non-Higgsable cluster supported over $S_{P}$, meaning that any enhancement of it is non-minimal. This makes the vertical line of fibers saturating the bound $\rank(\mathfrak{g}_{\mathrm{ver}}) \leq 5$ not tunable in such a configuration.

\section{Type II.a models as decompactifications with defects}
\label{sec:horizontal-models-heterotic-duals}

Our interest in the degenerations $\hat{\rho}: \hat{\mathcal{Y}} \rightarrow D$ of elliptically fibered Calabi-Yau threefolds stems from the fact that they represent infinite-distance limits in the complex structure moduli space of six-dimensional F-theory. The central fiber $Y_{0}$ of the resolved degeneration $\rho: \mathcal{Y} \rightarrow D$ allows us to extract information about the asymptotic physics that these limits lead to. So far, we have mainly focused on two aspects: First, the possibility of realising global weak coupling limits, and second the gauge algebra content of the asymptotic models. Both facets of the problem were addressed in \cref{sec:horizontal-models-general-properties}, with the discussion supported on the general analysis of the degenerations carried out in \cite{Alvarez-Garcia:2023gdd}.

In determining the gauge algebra, we have thus far taken a six-dimensional standpoint. As we traverse an infinite distance in the moduli space, however, the theory may undergo a decompactification process that can lead to gauge enhancements as viewed from the perspective of the higher-dimensional theory. In fact, according to the Emergent String Conjecture \cite{Lee:2019wij}, we expect the limits to have an interpretation either as decompactification or emergent string limits. In this section, we study how the information about the asymptotic physics extracted while insisting on the six-dimensional point of view reorganises itself once a more accurate picture for the endpoint of the limits is considered, and how the results fit with the Emergent String Conjecture. This was also the ultimate goal of the analogous systematic analysis in the moduli space of eight-dimensional F-theory carried out in \cite{Lee:2021qkx,Lee:2021usk}.

To this end, the single infinite-distance limit degenerations of Hirzebruch models that we focus on in the present work constitute an advantageous starting point for our survey for two reasons: F-theory compactified on an elliptically fibered Calabi-Yau threefold whose base is a Hirzebruch surface is dual to heterotic string theory compactified on a K3 surface. Additionally, horizontal degenerations of Hirzebruch models are a fiberwise generalisation of the degenerations of K3 surfaces studied in \cite{Lee:2021qkx,Lee:2021usk}, as we discussed in \cref{sec:classification-horizontal-models}. This offers two complementary perspectives on the limits from which to extract the asymptotic physics. As we will explain, horizontal Type II.a limits are the perfect subclass to exploit this approach, and we focus on them in the remainder of this section.

We start in \cref{sec:F-theory-heteoric-duality} by reviewing F-theory/heterotic duality and the conditions that need to be fulfilled in order to have explicit control over the corresponding duality map. In \cref{sec:horizontal-IIa-generic-vertical-slices} we analyse the generic vertical slices of horizontal Type II.a models, which encode information about the bulk asymptotic physics. We focus first on studying the endpoints of horizontal Type~II.a models constructed over $\hat{B} = \mathbb{F}_{0}$, providing a complete analysis in the so-called adiabatic regime and pointing out the challenges arising in a naive extension of these results away from it; the models constructed over $\hat{B} = \mathbb{F}_{n}$, whose asymptotic physics differs only slightly, are considered in \cref{sec:horizontal-IIa-Fn}. Our analysis also offers interesting new insights into the heterotic theory by itself. First, as we note in \cref{sec:non-minimal-points-heterotic-K3-surface}, non-minimal singularities of the heterotic K3 surface correspond to codimension-one infinite-distance degenerations on the \mbox{F-theory} side, after an appropriate base change has been performed. Furthermore, we comment on the role of vertical gauge algebras on the F-theory side as part of the non-perturbative heterotic gauge sector. This discussion is relegated to \cref{sec:defect-algebras-heterotic-dual}.

\subsection{F-theory/heterotic duality}
\label{sec:F-theory-heteoric-duality}

F-theory/heterotic duality in six dimensions is a valuable tool to gain intuition about the asymptotic physics associated with the limits under study. In this section, we review how this duality works.\footnote{A clear and succinct summary of the duality in 8D, 6D and 4D can also be found in \cite{Anderson:2014gla}.} Our focus is on the $\mathrm{E}_{8} \times \mathrm{E}_{8}$ heterotic string, although the duality can also be taken to the $\mathrm{Spin}(32)/\mathbb{Z}_{2}$ theory.

In eight dimensions, F-theory compactified on a K3 surface is dual to  heterotic string theory compactified on $T^{2}_{\mathrm{het}}$ \cite{Vafa:1996xn}. The F-theory side is defined by pure geometry, which is mapped  to geometric, B-field and gauge bundle data on the heterotic side. If we demand the heterotic gauge group to be unbroken, i.e.\ we freeze the 16 Wilson line moduli at an appropriate point in the moduli space, the remaining complex structure and complexified K\"ahler modulus of $T^{2}_{\mathrm{het}}$ can be very explicitly matched to the F-theory side, with the map between the two eight-dimensional theories constructed in \cite{LopesCardoso:1996hq,Lerche:1998nx,Shioda2006}.

The duality map becomes more subtle once we allow the heterotic gauge group to be broken, and hence need to also take into account the heterotic gauge bundle moduli. The data on both sides can be matched in a particular limit, which in the F-theory side is the stable degeneration limit of the K3 surface. On the heterotic side, this limit corresponds to the decompactification limit obtained by sending the area of $T^{2}_{\mathrm{het}}$ to infinity. These are the Kulikov Type II.a decompactification limits studied in \cite{Lee:2021qkx,Lee:2021usk} and reviewed in \cref{sec:K3-degenerations}. The K3 surface $Y_{\tilde{u} \neq 0}$ on the F-theory side degenerates into two log Calabi-Yau components $Y_{0} = Y^{0} \cup_{\mathcal{E}} Y^{1}$ glued along their boundaries, in this case an elliptic curve $\mathcal{E}$ which is identified with the compactification space on the heterotic side. Each of the components $\{Y^{p}\}_{0 \leq p \leq 1}$ is a $\mathrm{dP}_{9}$ surface. The data of the two $\mathrm{E}_{8}$ heterotic bundles $V_{0}$ and $V_{1}$ is encoded in $\mathrm{Def}(Y^{0})$ and $\mathrm{Def}(Y^{1})$, respectively. One can intuitively think of this limit on the F-theory side as an elongation of the base $\mathbb{P}^{1}$ of the internal K3 surface separating the two poles of the sphere; the elongated $\mathbb{P}^{1}$ is then identified with the Ho\v{r}ava-Witten interval, with each of the poles associated with the information of one of the $\mathrm{E}_{8}$ heterotic factors \cite{Morrison:1996na,Morrison:1996pp}. Hence, the ten-dimensional heterotic dilaton is mapped to the volume of the base of the F-theory K3 surface measured in Type IIB string units
\begin{equation}
    \mathcal{V}_{\mathbb{P}^{1}}^{\mathrm{IIB}} \sim g_{\mathrm{het}}\,.
\end{equation}

The stable degeneration limit separates the Wilson line moduli information and ensures that the duals of the F-theory backgrounds are given by geometric compactifications of the heterotic string. Away from this boundary of the moduli space, the $\mathrm{O}(\Lambda^{2,18})$ symmetry of the heterotic compactification on $T^{2}_{\mathrm{het}}$ mixes the complex structure and complexified K\"ahler modulus of $T^{2}_{\mathrm{het}}$ with the Wilson line moduli, making the heterotic compactification non-geometric.  The connection between the two sides of the duality can still be made explicit when only a single Wilson line modulus is allowed to be non-zero, leading to an $\mathrm{E}_{7} \times E_{8}$ gauge symmetry \cite{Malmendier:2014uka,Gu:2014ova,Garcia-Etxebarria:2016ibz}.

This duality is the starting point to obtain a lower-dimensional duality as follows. Consider a theory A in $D_{\mathrm{A}}$ dimensions and a theory B in $D_{\mathrm{B}}$ dimensions that are dual to each other when compactified on the varieties $X$ and $Y$, respectively, with dimensions $\dim(X) = D_{\mathrm{A}} - d$ and $\dim(Y) = D_{\mathrm{B}} - d$. Fibering the compactifications spaces over a common base $B$, the theories compactified on the total spaces
\begin{equation}
    \begin{tikzcd}
        X \arrow[r] & \mathcal{X} \arrow[d, "\pi_{\mathrm{A}}"]\\
         & B
    \end{tikzcd}
    \qquad \text{and} \qquad
    \begin{tikzcd}
        Y \arrow[r] & \mathcal{Y} \arrow[d, "\pi_{\mathrm{B}}"]\\
         & B
    \end{tikzcd}
\end{equation}
live in $d - \dim(B)$ dimensions. In a local patch, the fibrations can be trivialised to look like the product spaces $X \times \mathbb{R}^{\dim(B)}$ and $Y \times \mathbb{R}^{\dim(B)}$. If the parameters of the $d$-dimensional theories are varying slowly over $B$ and the volumes exhibit the hierarchy \mbox{$\mathcal{V}_{B} \gg \mathcal{V}_{X}$} and \mbox{$\mathcal{V}_{B} \gg \mathcal{V}_{Y}$}, a low-energy observer will not be able to locally distinguish between the theories A and B on the spaces \mbox{$\mathcal{X} \times \mathbb{R}^{1,d-\dim(B)-1}$} and \mbox{$\mathcal{Y} \times \mathbb{R}^{1,d-\dim(B)-1}$}, respectively, and the theories on the spaces \mbox{$X \times \mathbb{R}^{\dim(B)} \times \mathbb{R}^{1,d-\dim(B)-1}$} and \mbox{$Y \times \mathbb{R}^{\dim(B)} \times \mathbb{R}^{1,d-\dim(B)-1}$}. Since for the latter two the \mbox{$d$-dimensional} duality applies, the low-energy observer can use it to translate from one description to the other. Given the slow variation of the $d$-dimensional parameters, it is reasonable to assume that the equivalence will remain valid globally as long as the low-energy observer suitably redefines the physical quantities while looping around $B$. This is a fiberwise application of the $d$-dimensional duality enabled by an adiabaticity assumption.\footnote{Such a type of argument was used early on in the study of string dualities to establish connections between Type II and heterotic string theories \cite{Vafa:1995gm}.} Note that if there are loci in $B$ over which the fiber degenerates, the compactification space will not look like a product variety in a neighbourhood around them, and the adiabaticity assumption hence fails. It is therefore reasonable to expect the physics new to $d-\dim(B)$ dimensions to be concentrated at these loci, as we will see in \cref{sec:horizontal-IIa-F0} is indeed the case.

In our context, this reasoning leads to the well-known duality between the heterotic string compactified on an elliptically fibered manifold $X_{d}$,
\begin{equation}
    \begin{tikzcd}
        \mathcal{E} \arrow[r] & X_{d} \arrow[d, "\pi_{\mathrm{het}}"]\\
         & B_{d-1}\mathrlap{\,,}
    \end{tikzcd}
\end{equation}
and F-theory compactified on a manifold $Y_{d+1}$ fibered by elliptic K3 surfaces. The two fibrations 
\begin{equation}
    \begin{tikzcd}
        \mathcal{E} \arrow[r] & Y_{d+1} \arrow[d, "\pi_{\mathrm{ell}}"]\\
         & B_{d}
    \end{tikzcd}
    \qquad
    \textrm{and}
    \qquad
    \begin{tikzcd}
        \mathrm{K3} \arrow[r] & Y_{d+1} \arrow[d, "\pi_{\mathrm{K3}}"]\\
         & B_{d-1}
    \end{tikzcd}
\end{equation}
are compatible in the sense that they fit into the diagram
\begin{equation}
    \begin{tikzcd}
    Y_{d+1} \arrow[rr, "\pi_{\mathrm{ell}}"] \arrow[dd, "\pi_{\mathrm{K3}}"'] &  & B_{d} \arrow[dd, "\pi_{\mathbb{P}^{1}}"] \\
     &  & \\
    B_{d-1} \arrow[rr, "=", <->] & & B_{d-1}\mathrlap{\,.}
    \end{tikzcd}
\end{equation}
In particular, the base manifold $B_{d}$ is, therefore, $\mathbb{P}^{1}$-fibered itself.

When $d=2$, the resulting lower-dimensional effective theories are six-dimensional, with the internal space on the F-theory side being an elliptically fibered Calabi-Yau threefold with base $B_{2} = \mathbb{F}_{n}$, and the internal space on the heterotic side being a K3 surface that is an elliptic fibration over $B_{1} = \mathbb{P}_{b}^{1}$. On the F-theory side, these are the types of compactifications whose complex structure infinite-distance limits we study, which therefore correspond to infinite-distance limits of heterotic string theory compactified on K3 upon taking the duality.

This six-dimensional duality was considered for the first time in \cite{Morrison:1996na,Morrison:1996pp}, with a comparison of the moduli spaces carried out in \cite{Bershadsky:1996nh} by counting parameters on the two sides. The matching of the moduli spaces was made fully precise in the stable degeneration limit in \cite{Friedman:1997yq,Friedman:1997ih}. In this limit, the K3 fibers on the F-theory side undergo the stable degeneration limit discussed for the eight-dimensional duality above, i.e.\ they become Kulikov Type II.a models. The F-theory Calabi-Yau threefold splits into two log Calabi-Yau components $Y_{0} = Y^{0} \cup_{\mathrm{K3}} Y^{1}$ glued along their boundaries, which are the K3 surface that is identified with the heterotic compactification space. This geometry corresponds to that of horizontal Type II.a models, in the language of \cref{sec:classification-horizontal-models}, which are the relative version of the Kulikov Type II.a models found for the fiber, in accordance with the fiberwise construction of the duality. The data of the two $\mathrm{E}_{8}$ heterotic bundles $V_{0}$ and $V_{1}$ is once again encoded in $\mathrm{Def}(Y^{0})$ and $\mathrm{Def}(Y^{1})$, respectively. Furthermore, we have the relation
\begin{equation}
	\frac{\mathcal{V}_{\mathbb{P}^{1}_{f}}}{\mathcal{V}_{\mathbb{P}^{1}_{b}}} \sim \left( g_{\mathrm{het}}^{\mathrm{6D}} \right)^{2}\,.
\label{eq:6D-F-het-coupling-matching}
\end{equation}
Note that when $B_{2} = \mathbb{F}_{0}$, the two $\mathbb{P}^{1}$ factors present in the internal geometry on the F-theory side of the duality are on equal grounds. Exchanging them corresponds to heterotic/heterotic duality, which inverts the heterotic string coupling, as can be read from \eqref{eq:6D-F-het-coupling-matching}. Although the duality between F-theory and heterotic string theory is expected to hold in general, the precise map between the two is available only when we have the hierarchy of volumes
\begin{align}
	\mathcal{V}_{\mathbb{P}^{1}_{b}} &\gg \mathcal{V}_{\mathrm{K3}} \quad \textrm{(F-theory side)}\,,\\
	\mathcal{V}_{\mathbb{P}^{1}_{b}} 	&\gg \mathcal{V}_{T^{2}_{\mathrm{het}}} \quad \textrm{(heterotic side)}\,,
\end{align}
which implements the adiabaticity of the fibrations\footnote{In the precise formulation of the duality using vector bundles obtained via the spectral cover construction \cite{Friedman:1997yq,Friedman:1997ih}, this guarantees the stability of said bundles.} on both sides, as is necessary for the fiberwise application of the eight-dimensional duality. Additionally, the volume of the heterotic elliptic curve $\mathcal{V}_{T^{2}_{\mathrm{het}}} \rightarrow \infty$ as a consequence of the stable degeneration limit, and $\mathcal{V}_{\mathbb{P}^{1}_{b}} \gg \mathcal{V}_{\mathbb{P}^{1}_{f}}$ in order to be at weak heterotic coupling.

On the heterotic side, the $\mathrm{E}_{8}$ bundles $V_{0}$ and $V_{1}$ are poly-stable, with
\begin{equation}
    c_{1}(V_{i}) = 0 \mod 2\,,\qquad i = 0,1\,.
\end{equation}
On the F-theory side, this maps to the usual condition that the total compactification space $Y_{0}$ must be a (possibly singular) Calabi-Yau manifold, even if the individual components after the stable degeneration are not. The integrated Bianchi identity on the heterotic side reads
\begin{equation}
	c_{2}(V_{0}) + c_{2}(V_{1}) = c_{2}(\mathrm{K3}) = 24\,,
\end{equation}
which is fulfilled by a distribution of instanton numbers $c_{2}(V_{0,1}) = 12 \pm n$. On the F-theory side, the choice of how the heterotic instanton numbers are distributed corresponds to the choice of the Hirzebruch surface $\hat{B} = \mathbb{F}_{n}$ over which the horizontal Type II.a model is constructed. In our notation, the log Calabi-Yau component $Y^{1}$ corresponds to the gauge bundle $V_{1}$ with the smaller instanton number. As a result, it is this component that harbours the non-Higgsable cluster present when $n \geq 3$, i.e.\ when $c_{2}(V_{1}) < 10$.

The following point will be particularly important for our analysis: The heterotic gauge bundles are allowed to become singular by concentrating their curvature (or at least part of it) at a point, a situation in which the vector bundle would be more precisely described as a sheaf, with the aforementioned points of concentrated curvature being skyscraper sheaves. The subset of these that are point-like instantons with trivial holonomy contribute 1 unit each to $c_{2}(V_{\mathrm{het}})$. Through a small instanton transition, a singular bundle contribution corresponding to a point-like instanton with trivial holonomy can be regarded as an NS5-brane or, in the language of heterotic M-theory, an M5-brane that can be separated from the Ho\v{r}ava-Witten walls and moved into the $S^{1}/\mathbb{Z}_{2}$ interval. This is allowed because a point-like instanton with trivial holonomy has 1 hypermultiplet parametrizing its position on the heterotic K3 surface, and 29 hypermultiplets parametrizing the deformations that would give finite volume to the curvature support. In the gravitational anomaly cancellation formula, we can trade the latter 29 hypermultiplets for 1 tensor, signalling that the configuration can be seen as an M5-brane with 1 hypermultiplet and 1 tensor multiplet in its worldvolume. Their scalars parametrize its position on the K3 surface and the Ho\v{r}ava-Witten interval, respectively \cite{Duff:1996rs,Ganor:1996mu}. Having $m$ such M5-branes modifies the integrated Bianchi identity to
\begin{equation}
	c_{2}(V_{1}) + c_{2}(V_{2}) + m = c_{2}(\mathrm{K3}) = 24\,.
\end{equation}
We see that each brane contributes like 1 unit of instanton charge in this equation. 
Importantly, there can be additional types of singular gauge bundle contributions beyond point-like instantons with trivial holonomy. Apart from being interesting by themselves, these play a central role in our understanding of the gauge algebras on vertical divisors on the F-theory side. We describe them in more detail in \cref{sec:defect-algebras-heterotic-dual-broken-E8xE8}.

Trading a point-like instanton with trivial holonomy associated to one of the Ho\v{r}ava-Witten walls for an M5-brane, moving it to the opposite Ho\v{r}ava-Witten wall, and dissolving it back into gauge bundle changes the instanton number distribution between $V_{0}$ and $V_{1}$. On the F-theory side, the point-like instantons with trivial holonomy that can be traded for the \mbox{M5-branes} correspond to codimension-two finite-distance non-minimal singularities. The possibility of blowing up the base in order to resolve them corresponds to moving the associated M5-branes through the Ho\v{r}ava-Witten interval; the distance to the Ho\v{r}ava-Witten walls is parametrized by a scalar in a tensor multiplet on the heterotic side, that on the F-theory side corresponds to the volume of the exceptional $\mathbb{P}^{1}$ curves introduced in the base blow-ups. Taking an M5-brane to the opposite Ho\v{r}ava-Witten wall and dissolving it into gauge bundle is identified with blowing up the base to produce an exceptional $\mathbb{P}^{1}$ and shrinking some other curve in order to recover a different Hirzebruch surface than the starting one. This coincides with the heterotic picture of redistributing the instanton numbers between $V_{0}$ and $V_{1}$. The points in the moduli space in which the exceptional curves are kept contracted correspond to the origin of the tensor branch of the SCFTs associated with the codimension-two finite-distance non-minimal singularities, see \cite{Heckman:2018jxk} for a review.

\subsection{Horizontal Type II.a models: generic vertical slices}
\label{sec:horizontal-IIa-generic-vertical-slices}

The preceding discussion shows how useful it can be, when dealing with a relative model, to exploit our knowledge of the physics associated with the fibers as much as possible. As advanced in the introduction, this is the reason we focus on horizontal Type II.a models as the starting point of our analysis: We have just seen them naturally appear in the fiberwise extension of the eight-dimensional F-theory/heterotic duality, and they also are a relative version of the degenerations of K3 surfaces that were studied in detail in \cite{Lee:2021qkx,Lee:2021usk}. In this spirit, we commence by analysing the physics associated with their vertical slices.

Horizontal Type II.a models with an open-chain resolution $\rho: \mathcal{Y} \rightarrow D$ were defined in \cref{sec:classification-horizontal-models} to be those whose generic vertical restriction $\sigma: \mathcal{Z} \rightarrow D$ is a Kulikov Type II.a model. In terms of the pattern of codimension-zero singular elliptic fibers, this is equivalent to demanding that no component is at weak coupling. Taking into consideration the expressions \eqref{eq:Delta-hor} for the restrictions $\{ \Delta_{p}^{\prime} \}_{0 \leq p \leq P}$ of the modified discriminant $\Delta'$ and the discussion on the local and global 7-brane content in \cref{sec:horizontal-local-global-brane-content}, the intermediate components offer redundant information, and can be blown down to present the central fiber $Y_{0}$ as a two-component model. We assume in the remainder of the section that this has been done.

\begin{figure}[t!]
    \centering
	\begin{tikzpicture}
		\def\scalereliiahor{0.7}
		\def\scalereliiaver{0.8}

		\node (0) at (-6*\scalereliiahor, 2.5*\scalereliiaver) {};
		\node (1) at (-6*\scalereliiahor, -2.5*\scalereliiaver) {};
		\node (2) at (0*\scalereliiahor, 2.5*\scalereliiaver) {};
		\node (3) at (0*\scalereliiahor, -2.5*\scalereliiaver) {};
		\node (4) at (6*\scalereliiahor, 2.5*\scalereliiaver) {};
		\node (5) at (6*\scalereliiahor, -2.5*\scalereliiaver) {};
		\node (6) at (-2*\scalereliiahor, 2.5*\scalereliiaver) {};
		\node (7) at (-0.75*\scalereliiahor, 2.5*\scalereliiaver) {};
		\node (8) at (-6*\scalereliiahor, -1.5*\scalereliiaver) {};
		\node (9) at (-6*\scalereliiahor, -2*\scalereliiaver) {};
		\node (10) at (4.5*\scalereliiahor, 2.5*\scalereliiaver) {};
		\node (11) at (4.5*\scalereliiahor, -2.5*\scalereliiaver) {};
		\node (13) at (0*\scalereliiahor, 0.5*\scalereliiaver) {};
		\node (14) at (-6*\scalereliiahor, 0.5*\scalereliiaver) {};
		\node (15) at (-6*\scalereliiahor, 1.17*\scalereliiaver) {};
		\node (16) at (6*\scalereliiahor, 1.17*\scalereliiaver) {};
		
		\node (17) at (3*\scalereliiahor, 6.875*\scalereliiaver) {};
		\node (18) at (6.5*\scalereliiahor, 5.625*\scalereliiaver) {};
		\node (19) at (5.5*\scalereliiahor, 5.625*\scalereliiaver) {};
		\node (20) at (9*\scalereliiahor, 6.875*\scalereliiaver) {};
		
		\node (21) at (9*\scalereliiahor, 3.75*\scalereliiaver) {};
		\node (22) at (12.5*\scalereliiahor, 2.5*\scalereliiaver) {};
		\node (23) at (11.5*\scalereliiahor, 2.5*\scalereliiaver) {};
		\node (24) at (15*\scalereliiahor, 3.75*\scalereliiaver) {};
		
		\node (25) at (9*\scalereliiahor, 0.625*\scalereliiaver) {};
		\node (26) at (12.5*\scalereliiahor, -0.625*\scalereliiaver) {};
		\node (27) at (11.5*\scalereliiahor, -0.625*\scalereliiaver) {};
		\node (28) at (15*\scalereliiahor, 0.625*\scalereliiaver) {};
		
		\node (29) at (9*\scalereliiahor, -2.5*\scalereliiaver) {};
		\node (30) at (12.5*\scalereliiahor, -3.75*\scalereliiaver) {};
		\node (31) at (11.5*\scalereliiahor, -3.75*\scalereliiaver) {};
		\node (32) at (15*\scalereliiahor, -2.5*\scalereliiaver) {};
		
		\node (33) at (3*\scalereliiahor, -4.875*\scalereliiaver) {};
		\node (34) at (6.5*\scalereliiahor, -6.125*\scalereliiaver) {};
		\node (35) at (5.5*\scalereliiahor, -6.125*\scalereliiaver) {};
		\node (36) at (9*\scalereliiahor, -4.875*\scalereliiaver) {};
		
		\node (37) at (-6*\scalereliiahor, 1.84*\scalereliiaver) {};
		\node (38) at (6*\scalereliiahor, 1.84*\scalereliiaver) {};
		\node (39) at (6*\scalereliiahor, 0.5*\scalereliiaver) {};
		\node (40) at (-6*\scalereliiahor, -0.45082*\scalereliiaver) {};
		\node (41) at (6*\scalereliiahor, -0.45082*\scalereliiaver) {};
		\node (42) at (-6*\scalereliiahor, -1.2*\scalereliiaver) {};
		\node (43) at (6*\scalereliiahor, -1.2*\scalereliiaver) {};
		
		\node [label={[yshift=0cm]$\{e_{0} = 0\}_{\makebox[0pt]{$\scriptstyle \;\;\mathcal{B}$}}$}] (44) at (-3*\scalereliiahor, 2.5*\scalereliiaver) {};
        \node [label={[yshift=0cm]$\{e_{1} = 0\}_{\makebox[0pt]{$\scriptstyle \;\;\mathcal{B}$}}$}] (45) at (3*\scalereliiahor, 2.5*\scalereliiaver) {};
		
		\draw [style=dashed] (37.center) to (38.center);
		\draw [style=dashed] (14.center) to (39.center);
		\draw [style=dashed] (40.center) to (41.center);
		\draw [style=dashed] (42.center) to (43.center);

		\draw [style=light-green line]  (8.center) to (7.center);
		\draw [style=light-purple line]  (6.center) to (9.center);
		\draw [style=light-blue line] (10.center) to (11.center);
		\draw [style=light-yellow line] (10.west) to (13.center);
		\draw [style=light-yellow line] (13.center) to (14.center);
		\draw [style=light-yellow line] (11.east) to (39.center);
		\draw [style=medium-red line] (15.center) to (16.center);
		
		\draw [style=black line] (19.center) to (20.center);
		\draw [style=black line] (18.center) to (17.center);
		\draw [style=medium-red thick-line] (23.center) to (24.center);
		\draw [style=medium-red thick-line] (22.center) to (21.center);
		\draw [style=black line] (27.center) to (28.center);
		\draw [style=light-yellow thick-line] (26.center) to (25.center);
		\draw [style=black line] (31.center) to (32.center);
		\draw [style=black line] (30.center) to (29.center);
		\draw [style=black line] (35.center) to (36.center);
		\draw [style=black line] (34.center) to (33.center);
		
		\draw [style=black line] (0.center) to (2.center);
		\draw [style=black line] (2.center) to (4.center);
		\draw [style=black line] (4.center) to (5.center);
		\draw [style=black line] (5.center) to (3.center);
		\draw [style=black line] (3.center) to (1.center);
		\draw [style=black line] (1.center) to (0.center);
		\draw [style=black line] (2.center) to (3.center);
		
		\draw [-latex] (38.east) to [out=19.6538,in=-19.6538] ([xshift=0.15cm, yshift=-0.0535714cm]18.center);
		\draw [-latex] (16.east) to ([xshift=0cm, yshift=-0.05cm]23.west);
		\draw [-latex] (39.east) to (25.west);
		\draw [-latex] (41.east) to [out=0,in=90] (29.north);
		\draw [-latex] (43.east) to [out=-19.6538,in=90] (36.north);
		
		\node [cross-purple=5pt, rotate=-19.6538, anchor=center] (46) at (4.4*\scalereliiahor,6.375*\scalereliiaver) {};
		\node [cross-green=5pt, rotate=-19.6538, anchor=center] (47) at (5.45*\scalereliiahor,6*\scalereliiaver) {};
		\node [cross-blue=5pt, rotate=19.6538, anchor=center] (48) at (7.6*\scalereliiahor,6.375*\scalereliiaver) {};
		\node [cross-yellow=5pt, rotate=19.6538, anchor=center] (49) at (7.075*\scalereliiahor,6.1875*\scalereliiaver) {};
		
		\node [cross-purple=5pt, rotate=-19.6538, anchor=center] (50) at (10.225*\scalereliiahor,3.3125*\scalereliiaver) {};
		\node [cross-green=5pt, rotate=-19.6538, anchor=center] (51) at (11.1*\scalereliiahor,3*\scalereliiaver) {};
		\node [cross-blue=5pt, rotate=19.6538, anchor=center] (52) at (13.6*\scalereliiahor,3.25*\scalereliiaver) {};
		\node [cross-yellow=5pt, rotate=19.6538, anchor=center] (53) at (12.55*\scalereliiahor,2.875*\scalereliiaver) {};
		
		\node [cross-purple=5pt, rotate=-19.6538, anchor=center] (54) at (10.12*\scalereliiahor,0.225*\scalereliiaver) {};
		\node [cross-green=5pt, rotate=-19.6538, anchor=center] (55) at (10.75*\scalereliiahor,0*\scalereliiaver) {};
		\node [cross-blue=5pt, rotate=19.6538, anchor=center] (56) at (13.6*\scalereliiahor,0.125*\scalereliiaver) {};
		
		\node [cross-purple-green=5pt, rotate=-19.6538, anchor=center] (57) at (10.05*\scalereliiahor,-2.875*\scalereliiaver) {};
		\node [cross-blue=5pt, rotate=19.6538, anchor=center] (58) at (13.6*\scalereliiahor,-3*\scalereliiaver) {};
		\node [cross-yellow=5pt, rotate=19.6538, anchor=center] (59) at (14.65*\scalereliiahor,-2.625*\scalereliiaver) {};
		
		\node [cross-purple=5pt, rotate=-19.6538, anchor=center] (60) at (3.98*\scalereliiahor,-5.225*\scalereliiaver) {};
		\node [cross-green=5pt, rotate=-19.6538, anchor=center] (61) at (3.525*\scalereliiahor,-5.0625*\scalereliiaver) {};
		\node [cross-blue=5pt, rotate=19.6538, anchor=center] (62) at (7.6*\scalereliiahor,-5.375*\scalereliiaver) {};
		\node [cross-yellow=5pt, rotate=19.6538, anchor=center] (63) at (8.125*\scalereliiahor,-5.1875*\scalereliiaver) {};
	\end{tikzpicture}
    \caption{We schematically represent the base of the central fiber of a resolved horizontal Type~II.a model constructed over $\hat{B} = \mathbb{F}_{n}$, with $n \geq 1$, on the left, and various vertical slices of it on the right. A reduced number of global 7-branes are depicted: two in the divisor class $\mathcal{H}_{\infty}^{0}$, one in $\mathcal{H}_{\infty}^{1}$, one in $\mathcal{H}_{0}^{1}$ and one in $\mathcal{F}$. The first, fourth and fifth vertical slices are generic and lead to Kulikov Type II.a models. This is not true for the second vertical slice, since it overlaps with the global 7-brane in the class $\mathcal{F}$ in both components, and for the third vertical slice, since it overlaps with the global 7-brane in the class $\mathcal{H}_{\infty}^{1}$ in the $B^{0}$ component.}
    \label{fig:relative-Type-IIa}
\end{figure}
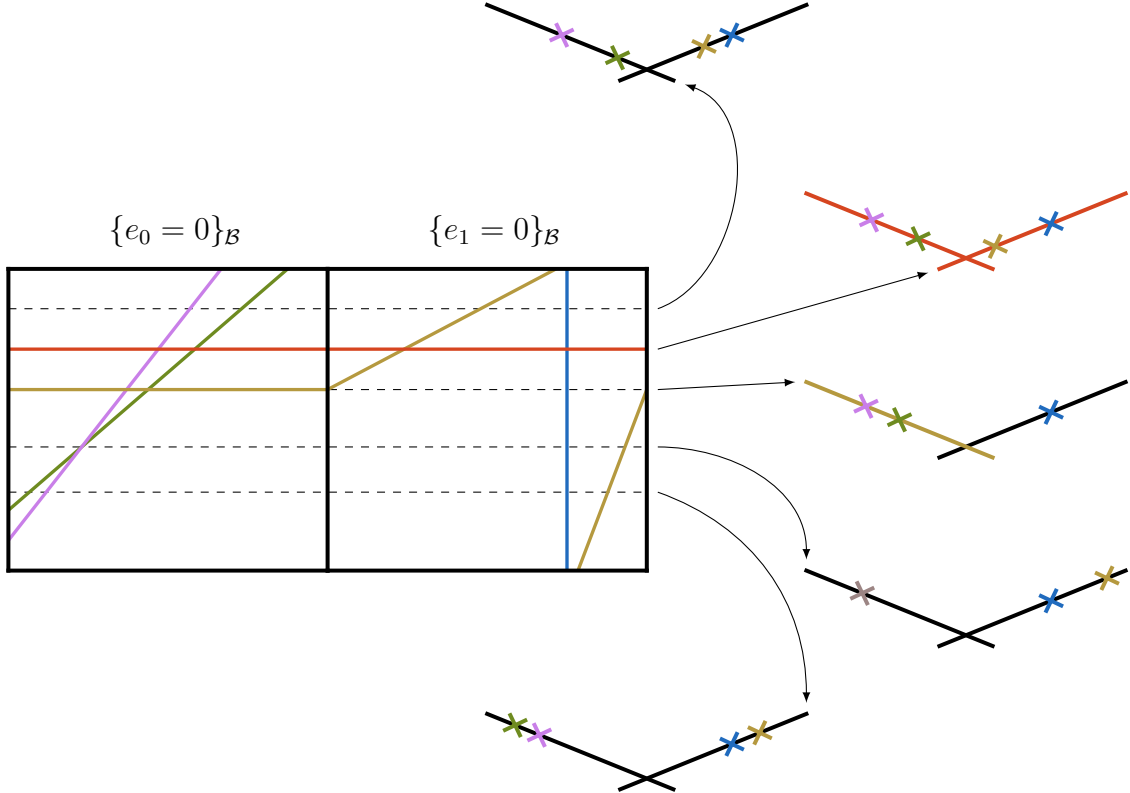

As in a Kulikov Type II.a model, the elliptic fiber of the central fiber $Z_{0}$ of the generic vertical restriction $\sigma: \mathcal{Z} \rightarrow D$ associated with a generic point $p_{b} \in \mathbb{P}^{1}_{b}$ degenerates over 12 points in each of the bases of the $\{Z^{p}\}_{0 \leq p \leq 1}$ components. This  can be read from
\begin{equation}
    \Dphys \cdot \mathcal{F} = 12 + 12\,,
\end{equation}
where we recall the definition of the vertical class $\mathcal{F}$ in \eqref{eq:horizontal-vertical-divisor}.
These degeneration points are given by the restriction of the horizontal (or mixed) global divisors in $B_{0}$ to the chosen generic vertical slice. Regarding $\sigma: \mathcal{Z} \rightarrow D$ as representing an infinite-distance limit in the complex structure moduli space of eight-dimensional F-theory, these are the two sets of 12 7-branes that furnish the two double loop algebras $\hat{\mathrm{E}}_{9}$ sharing a common imaginary root and hence yielding the gauge algebra
\begin{equation}
	G_{\infty} = \left( \hat{\mathrm{E}}_{9} \oplus \hat{\mathrm{E}}_{9} \right)/\sim
\label{eq:enhanced-algebra-Type-II.a-8D}
\end{equation}
as seen from the eight-dimensional standpoint; the algebra $G_{\infty}$ is reinterpreted as
\begin{equation}
	G_{\mathrm{10D}} = \mathrm{E}_{8} \oplus \mathrm{E}_{8}
\end{equation}
in the higher-dimensional theory that results from the limit, as we reviewed in \cref{sec:Kulikov-models-review}, after the $\mathrm{U}(1)_{\mathrm{KK}}^{2}$ are reintegrated into the metric degrees of freedom. The positions of these two sets of 12 7-branes within their components is not relevant for the eight-dimensional asymptotic physics: They appear separated due to the resolution process employed to read off the physical information, but are to be taken as lumped together at the endpoint of the limit. Choosing a different generic point $p_{p}^{\prime} \in \mathbb{P}^{1}_{b}$ to construct the generic vertical slice leads to a Kulikov Type~II.a model whose only difference with the one just discussed is the position of the 7-branes within the components that they belong to. Therefore, the asymptotic physics of the associated eight-dimensional model is the same for all generic vertical slices of the horizontal Type II.a model. This is depicted in \cref{fig:relative-Type-IIa}.

There are 24 representatives of the vertical class $\mathcal{F}$ (counted with multiplicity) that do not lead to a Kulikov Type II.a model presented in the resolved form used in \cite{Lee:2021qkx,Lee:2021usk}, as can be seen from
\begin{equation}
    \Delta_{0}^{\prime} \cdot S_{0} = \Delta_{1}^{\prime} \cdot T_{1} = 24\,.
\label{eq:degeneration-points-heterotic-K3}
\end{equation}
The way in which this can fail is either by one (or both) of the components of the vertical restriction presenting codimension-zero singular fibers, or by the eight-dimensional model exhibiting special fibers at the intersection of the components. We illustrated both situations in \cref{fig:relative-Type-IIa}. Under the adiabaticity assumption explained in \cref{sec:F-theory-heteoric-duality}, the heterotic dual theory is compactified on the elliptic K3 surface $Y^{0} \cap Y^{1}$, with \eqref{eq:degeneration-points-heterotic-K3} corresponding to the 24 points over which the elliptic fiber must degenerate in such a surface. At these points the adiabaticity assumption fails, and we therefore expect purely six-dimensional features to arise, in alignment with the fiberwise  argumentation. Since this only occurs over these isolated loci, the generic vertical slice of the horizontal Type II.a model will inform us about the bulk physics at the endpoint of the infinite-distance limit, while the points of failure of the adiabaticity assumption will relate to localised effects in the asymptotic theory, as we now argue in \cref{sec:horizontal-IIa-F0}.

\subsection{Horizontal Type II.a models over \texorpdfstring{$\hat{B} = \mathbb{F}_{0}$}{B=F0}}
\label{sec:horizontal-IIa-F0}

Horizontal Type II.a models constructed over $\hat{B} = \mathbb{F}_{0}$ constitute the most straightforward relative version of their eight-dimensional counterparts, due to the direct product nature of $\mathbb{F}_{0} = \mathbb{P}^{1}_{f} \times \mathbb{P}^{1}_{b}$. Moreover, let us assume for now the structure
\begin{subequations}
\begin{align}
    f &= p_{8}^{f}([s:t],u) q_{8}^{f}([v:w],u)\,,\\
    g &= p_{12}^{g}([s:t],u) q_{12}^{g}([v:w],u)\,,
\end{align}
\label{eq:F0-horizontal-vertical-separation}%
\end{subequations}
for the defining polynomials of the Weierstrass model of $\hat{\mathcal{Y}}$, which leads to central fibers $Y_{0}$ in which only the global divisor classes $\mathcal{H}_{\infty}^{0}$, $\mathcal{H}_{\infty}^{1}$, $\mathcal{H}_{0}^{1}$ and $\mathcal{F}$ (defined in \cref{sec:horizontal-global-divisors}) can support gauge enhancements. This restricts our attention to the simplest possible horizontal Type II.a models, in which all generic vertical slices lead to, not only equivalent, but identical Kulikov Type II.a models. We now analyse these models in detail, relaxing the assumption \eqref{eq:F0-horizontal-vertical-separation} at the end of the section; the cases in which $\hat{B} = \mathbb{F}_{n}$ with $n \geq 1$ are treated in \cref{sec:horizontal-IIa-Fn}, and only differ slightly.

\subsubsection{Analysis in the adiabatic regime: Decompactification with defects}
\label{sec:horizontal-IIa-heterotic-dual-interpretation}

In order to extract the asymptotic physics associated with horizontal Type II.a models, let us consider the adiabaticity assumption reviewed in \cref{sec:F-theory-heteoric-duality} to hold. On the F-theory side, the hierarchy of volumes $\mathcal{V}_{\mathbb{P}_{b}^{1}} \gg \mathcal{V}_{\mathrm{K3}}$ makes the fiberwise analysis of the geometry accurate, while at the same time granting us explicit access to the heterotic dual models. We can then use the latter, in which the hierarchy of volumes $\mathcal{V}_{\mathbb{P}_{b}^{1}} \gg \mathcal{V}_{T^{2}_{\mathrm{het}}}$ must hold, to confirm the findings on the F-theory side from a different perspective. The difficulties that arise in the interpretation of the F-theory model away from the adiabatic regime are discussed in \cref{sec:horizontal-IIa-away-from-adiabatic-regime}.

Commencing on the F-theory side, we can analyse the horizontal Type II.a model by first taking vertical slices of the central fiber of the degeneration, and then completing the analysis by considering the horizontal slices as well.\footnote{Mixed slices do not provide additional insights, as they essentially are an interpolating case between the previous two, and we therefore do not consider them.}

Consider the vertical slice of a horizontal Type II.a model associated with a generic point $p_{b} := [v_{0}:w_{0}] \in \mathbb{P}^{1}_{b}$. Such slices were already studied in \cref{sec:horizontal-IIa-generic-vertical-slices}: they correspond to Kulikov Type II.a models with two sets of 12 7-branes, one in each of the $\{ Z^{p} \}_{0 \leq p \leq 1}$ components of $Z_{0}$. In a Kulikov Type II.a model, both components $\{ Z^{p} \}_{0 \leq p \leq 1}$ have smooth codimension-zero elliptic fibers. Moreover, the resolution process of the degeneration $\hat{\rho}: \hat{\mathcal{Y}} \rightarrow D$ discussed in \cite{Alvarez-Garcia:2023gdd} ensures that no special fibers are found over the intersection $\left. \left( B^{0} \cap B^{1} \right) \right|_{[v:w]=p_{b}}$. As a consequence, no monodromy acts on the two 1-cycles $\sigma_{i} \in H_{1}(\mathcal{E},\mathbb{Z})$, for $i=1,2$, of the elliptic fiber $\mathcal{E}$ as we move around the base $\left. B_{0} \right|_{[v:w]_{p_{b}}}$ of $Z_{0}$. The geometry of $\left. B_{0} \right|_{[v:w]_{p_{b}}}$ is that of two $\mathbb{P}^{1}$ curves intersecting over a point; a 1-cycle $\Sigma \in H_{1}(\mathbb{P}^{1},\mathbb{Z})$ defined in either of the base components and encircling their intersection point can be deformed to the antipodal point, and is therefore trivial. However, fibering $\{ \sigma_{i} \}_{i=1,2}$ over $\Sigma$ leads to two non-trivial 2-cycles $\{ \gamma_{i} \}_{i=1,2}$ in $H_{2}(Z_{0},\mathbb{Z})$. Taking a string junctions perspective in the generic associated eight-dimensional model, this corresponds to the possibility of defining two non-trivial loop junctions encircling, in one base component, both the intersection point $\left. \left( B^{0} \cap B^{1} \right) \right|_{[v:w] = p_{b}}$ and one of the sets of 12 7-branes.\footnote{The two loop junctions under consideration have no asymptotic charge, which implies that their self-intersection is trivial. They are left invariant by the monodromy of the 12 7-branes that they encircle, and have non-vanishing charge in the junction lattice. This last fact prevents us from trivially contracting them, as this generates additional junctions through the Hanany-Witten effect.} In the generic associated eight-dimensional model, these loop junctions grant the double loop enhancement of the algebras coming from the two sets of 12 7-branes to \eqref{eq:enhanced-algebra-Type-II.a-8D}, see \cite{DeWolfe:1998zf}. The winding strings represented by the loop junctions become asymptotically massless because they are allowed to shrink to the point $\left. \left( B^{0} \cap B^{1} \right) \right|_{[v:w] = p_{b}}$ of the resolved geometry.\footnote{In the unresolved central fiber the loop junctions can also shrink, since all 7-branes of one of the two sets of 12 are located at the same point to produce the non-minimal singularity.} Alternatively, the calibrated volumes of the 2-cycles $\{ \gamma_{i} \}_{i=1,2}$ vanish at the endpoint of the limit and, as a consequence, M2-branes wrapped on them lead to asymptotically massless particles; these are the particle states leading to the aforementioned enhancement \cite{Lee:2021qkx,Lee:2021usk}.

In those patches of $Y_{0}$ that only contain generic vertical slices, the preceding analysis extends locally. Moreover, and due to the adiabaticity assumption, the non-generic vertical slices are well separated and the parameters of the fibral models vary slowly over the base, making the range of validity of such a patch large. Hence, the 2-cycles $\{ \gamma_{i} \}_{i=1,2}$ can be consistently defined over the generic points of $\mathbb{P}^{1}_{b}$. As mentioned in \cref{sec:horizontal-IIa-generic-vertical-slices}, all generic vertical slices lead to associated eight-dimensional models whose asymptotic physics coincides. Altogether, a consistent picture arises over all of $Y_{0}$ but the measure zero set associated to the 24 non-generic points of $\mathbb{P}^{1}_{b}$: M2-branes wrapping the local 2-cycles $\{ \gamma_{i} \}_{i=1,2}$ generate two asymptotically massless towers of particles, signalling a partial decompactification that is to be overlaid with the partial decompactification resulting from enforcing the hierarchy of volumes necessary to situate the model in the adiabatic regime. This results, at least in the adiabatic regime, in a total decompactification to ten dimensions, as we will also argue from the dual heterotic side below. The aforementioned towers also provide the necessary additional states to furnish the enhanced gauge algebras. Said enhancement affects, in each generic vertical slice, the gauge algebra stemming from the restriction of the horizontal 7-branes. Since this applies to almost all points of the internal space, we conclude that the gauge algebra factors obtained from the six-dimensional standpoint from global 7-branes in the divisor classes $\mathcal{H}_{\infty}^{0}$, $\mathcal{H}_{\infty}^{1}$ and $\mathcal{H}_{0}^{1}$, together with the asymptotically massless towers of particles, lead to the double loop algebra
\begin{equation}
	G_{\infty} = \left( \hat{\mathrm{E}}_{9} \oplus \hat{\mathrm{E}}_{9} \right)/\sim\,,
\end{equation}
which is reinterpreted from the point of view of the decompactified theory in ten dimensions as the bulk gauge algebra
\begin{equation}
    G_{\mathrm{10D}} = \mathrm{E}_{8} \times \mathrm{E}_{8}\,.
\end{equation}

The arguments used to construct the local 2-cycles $\{ \gamma_{i} \}_{i=1,2}$ fail over the 24 non-generic points in $\mathbb{P}^{1}_{b}$, and the two asymptotically massless towers associated to M2-branes wrapped on them are, consequently, furnished by non-BPS particles, since the 2-cycles trivialise from the global perspective. To see this, consider first the situation in which $Y_{0}$ presents an enhancement over a representative of $\mathcal{F}$ whose (possibly covering) gauge algebra is in the D or E family. Then, the codimension-zero fibers in the $\{ Z^{p} \}_{0 \leq p \leq P}$ components will be of the corresponding Kodaira type, and so will be the fibers over the intersection point $\left. \left( B^{0} \cap B^{1} \right) \right|_{[v:w]=p_{b}}$. The monodromy action\footnote{See \cite{Lee:2021usk} for the explicit matrix representation of the monodromy action, following the conventions of \cite{DeWolfe:1998zf}.} induced by encircling this point leaves no 1-cycle in $H_{1}(\mathcal{E},\mathbb{Z})$ invariant, and we therefore cannot consistently construct a 2-cycle in $H_{2}(Z_{0},\mathbb{Z})$ by fibering 1-cycles over $\Sigma$. Alternatively, but using the language of string junctions, no monodromy invariant loop junction encircling $\left. \left( B^{0} \cap B^{1} \right) \right|_{[v:w]=p_{b}}$ can be defined. If, instead, all enhancements over representatives of $\mathcal{F}$ are associated with elliptic fibers of Kodaira type $\mathrm{I}_{m}$ and hence in the A family, the monodromy action produced by looping around $\left. \left( B^{0} \cap B^{1} \right) \right|_{[v:w]=p_{b}}$ does leave one 1-cycle in $H_{1}(\mathcal{E},\mathbb{Z})$ invariant, allowing for the consistent construction of a 2-cycle in $H_{2}(Z_{0},\mathbb{Z})$. This was to be expected, since the restriction of a horizontal Type II.a model to a representative of $\mathcal{F}$ supporting $\mathrm{I}_{m}$ fibers leads to a Kulikov Type III.b model. The global picture remains, however, unaltered: While in the restriction to each vertical line of $\mathrm{I}_{m}$ fibers these arguments can be used to claim that one of the local 2-cycles $\{ \gamma_{i} \}_{i=1,2}$ survives, the collection of these vertical enhancements is mutually non-local. The surviving 2-cycle found over one of the vertical enhancements is hence incompatible with the one found over at least one of the others, since the invariant 1-cycle in $H_{1}(\mathcal{E},\mathbb{Z})$ used in its construction is not invariant under the monodromy action induced by encircling the other vertical lines of $\mathrm{I}_{m}$ fibers. We conclude that, indeed, the local 2-cycles $\{ \gamma_{i} \}_{i=1,2}$ cannot be defined globally.

We propose the following interpretation: Away from the non-generic vertical fibers, the theory asymptotically decompactifies to ten dimensions, with two KK towers from wrapped M2-branes along the locally defined vanishing 2-cycles (completed by the supergravity KK tower associated with the large base $\mathbb P^1_b$). The fact that these are not globally defined, and in particular cease to exist over the 24 non-generic vertical fibers means that the degrees of freedom localised there cannot form bound states with the KK towers. They therefore remain as six-dimensional defects in the asymptotically decompactifying bulk. 

This picture is further supported by analysing the horizontal slices of the model. While vertical slices correspond to restrictions of $Y_{0}$ to representatives of $\mathcal{F}$, the number of horizontal curve classes that we can restrict to is greater. In a horizontal Type II.a model the irreducible possibilities are $\mathcal{H}_{\infty}^{0}$, $\mathcal{H}_{0}^{0}$, $\mathcal{H}_{\infty}^{1}$ and $\mathcal{H}_{0}^{1}$, with the horizontal slice obtained by restricting to $\mathcal{H}_{0}^{0}$ corresponding to the dual heterotic internal space in the adiabatic regime. However, since $B^{0} \cong B^{1} \cong \mathbb{F}_{0}$, these curve classes are linearly equivalent; the restricted models obtained from them have the same number
\begin{equation}
    \Delta'_{0} \cdot \mathcal{H}_{\infty}^{0} = \Delta'_{0} \cdot \mathcal{H}_{0}^{0} = \Delta'_{1} \cdot \mathcal{H}_{\infty}^{1} = \Delta'_{1} \cdot \mathcal{H}_{0}^{1} = 24
\end{equation}
of 7-branes, and all lead to the same asymptotic picture. Generic horizontal slices correspond to the restriction of $Y_{0}$ to either a generic point $p_{f} \in \mathbb{P}_{f,0}^{1}$ or a generic point $p_{f} \in \mathbb{P}_{f,1}^{1}$. Their restricted geometry is that of a Kulikov Type I model, with base $\mathbb{P}_{b}^{1}$, for which no asymptotically massless towers arise. Using the string junctions picture, this can be seen from the fact that a non-trivial loop junction encircling 12 of the 7-branes cannot shrink to a point, since there is no intersection point with a second base $\mathbb{P}^{1}$ curve containing the remaining branes, nor are the 12 7-branes lumped together to produce a non-minimal singularity. Hence, the gauge algebra factors arising from the restricted 7-branes do not enhance to gauge contributions that can be reinterpreted as bulk degrees of freedom in the decompactified theory, since the previous argument is common to all generic horizontal slices. Note that the 24 restricted branes stem from the intersections of the vertical classes in the discriminant with the horizontal curve used to slice the model. As a consequence, we can indeed conclude that the vertical gauge algebras supported over representatives of $\mathcal{F}$ remain lower-dimensional, and are localised in the worldvolume of six-dimensional defects.

The collision of horizontal and vertical components of the physical discriminant $\Dphys$ can lead to codimension-two finite-distance non-minimal enhancements\footnote{Codimension-two infinite-distance non-minimal enhancements do not occur, by definition, in single infinite-distance limits \cite{Alvarez-Garcia:2023gdd}.} that can be resolved via a base blow-ups and the appropriate line bundle shifts. Since each exceptional curve arising from such a resolution procedure is completely contained in one of the two base components $\{B^{p}\}_{0 \leq p \leq 1}$, any local gauge enhancement supported on them is automatically a global gauge enhancement. Through the same arguments employed in the analysis of the vertical gauge algebra factors, we conclude that also these contributions remain lower-dimensional in the limit and are localised in the world-volume of six-dimensional defects present in the asymptotic model.

Summarising the discussion, horizontal Type II.a models in the adiabatic regime lead to a decompactification process from six to ten dimensions in which the horizontal 7-branes give rise to the bulk $G_{\mathrm{10D}} = \mathrm{E}_{8} \times \mathrm{E}_{8}$, and the stacks of vertical 7-branes produce six-dimensional defects whose worldvolume is populated by localised gauge degrees of freedom. We hence conclude that the six-dimensional theory does not simply decompactify to the vacuum in ten dimensions, but that we actually need to consider theories containing defects that break higher-dimensional Poincar\'e invariance.

\subsubsection*{Heterotic dual interpretation}

Going now to the heterotic dual model, the complex structure limit corresponding to the stable degeneration limit on the F-theory side translates to the large volume limit $\mathcal{V}_{T^{2}_{\mathrm{het}}} \rightarrow \infty$ of the heterotic elliptic fiber. In order to maintain the hierarchy of volumes $\mathcal{V}_{\mathbb{P}^{1}_{b}} \gg \mathcal{V}_{T^{2}_{\mathrm{het}}}$ demanded by the adiabaticity assumption, we need to superimpose the large volume limit $\mathcal{V}_{\mathbb{P}^{1}_{b}} \rightarrow \infty$ in which the base of the heterotic K3 surface decompactifies faster than the elliptic fiber. Altogether, this implies a decompactification from six to ten dimensions.

The fiberwise application of the duality reviewed in \cref{sec:F-theory-heteoric-duality} is valid away from the non-generic 24 points of $\mathbb{P}^{1}_{b}$ where adiabaticity fails. The generic vertical slice on the F-theory side, from which we extracted the information about the bulk physics at the endpoint of the limit, corresponds to the heterotic model away from the degeneration points of the elliptic fiber of the K3 surface.

Horizontal gauge enhancements on the F-theory side correspond to the perturbative gauge sector on the heterotic side.\footnote{Even though the distinction between horizontal and vertical divisors is arbitrary in a model constructed over $\hat{B} = \mathbb{F}_{0}$, a statement which is dual to heterotic/heterotic duality, a definite choice needs to be made when dealing with concrete scenarios. In this section, we have chosen to call horizontal the direction along which the curve of non-minimal elliptic fibers appears in $\hat{B}_{0}$ and, consequently, the enhancements found over representatives of the global divisor classes $\mathcal{H}_{\infty}^{0}$, $\mathcal{H}_{\infty}^{1}$ and $\mathcal{H}_{0}^{1}$ correspond to the perturbative or bulk gauge algebra factors on the heterotic side of the duality.} The Higgsing of these perturbative contributions in six dimensions arises, in heterotic terms, from the non-trivial background field strength of the $\mathrm{E}_{8}$ bundles defined over the internal space. As the theory undergoes the decompactification process, the bulk physics becomes indistinguishable from that of heterotic string theory on $\mathbb{R}^{1,9}$: the (non-singular) gauge bundle profiles are diluted, and the masses associated to the curvature of the internal space asymptote to zero. Note, however, that the decompactification process is a continuous one, meaning that the topology of spacetime is not changed by it; the loci over which the elliptic fiber of the K3 surface degenerates are still present. The supersymmetry breaking defects can nonetheless be placed infinitely far away from the bulk observer. Altogether, we indeed expect the full perturbative $\mathrm{E}_{8} \times \mathrm{E}_{8}$ perturbative gauge group of the heterotic string to be restored in such a limit.\footnote{Note that, although in six-dimensional models at finite distance we can achieve $\rank(\mathfrak{g}_{\mathrm{hor}}) = 18$ by tuning a line of Kodaira type $\mathrm{II}^{*}$ fibers over the unique representative of $h$, another line of $\mathrm{II}^{*}$ fibers over a representative of $h+nf$, and a third line of IV fibers over a different representative of $h+nf$, such an enhancement pattern is incompatible with tuning the necessary codimension-one non-minimal fibers necessary to achieve a horizontal Type II.a model. Once the latter structure is enforced, we can obtain at most $\rank(\mathfrak{g}_{\mathrm{hor}}) \leq 16$ from the six-dimensional standpoint before any asymptotic enhancements are taken into account. The geometry needs to enforce this in order to be compatible with the perturbative gauge group of the decompactified heterotic side.} The asymptotically massless towers identified in the generic vertical slice of the horizontal Type II.a model correspond to the KK towers associated with the growing 1-cycles in $H_{1}(T^{2}_{\mathrm{het}},\mathbb{Z})$. Their non-BPS nature on the F-theory side was related to the failure of defining the local 2-cycles $\{ \gamma_{i} \}_{i=1,2}$ over the non-generic points in $\mathbb{P}^{1}_{b}$. Likewise, it is the non-mutually local degeneration of the heterotic elliptic fiber over these very same points that makes them 1-chains from a global perspective in the heterotic K3 surface, leading to the same conclusion. The metric $\mathrm{U}(1)_{\mathrm{KK}}$ factors in compactifications of string theory stem from the Killing vectors of the internal space. Since the heterotic K3 surface is a strict Calabi-Yau we do not have any continuous isometries, but as we take the limit, ensuring that the adiabaticity assumption is fulfilled, the local observer in the generic patch far away from the degeneration defects sees the continuous isometries of the torus fiber become symmetries of the internal space in a good approximation. These provide the $\mathrm{U}(1)_{\mathrm{KK}}^{2}$ factors enabling the double loop enhancement of the lower-dimensional algebra to $G_{\infty}$ observed for the generic vertical slice on the F-theory side. The fact that this is only seen asymptotically in the adiabatic regime on the heterotic side corresponds to the fact that we are not demanding any special structure for the Mordell-Weil group of the internal space of the six-dimensional F-theory models.

To complete the heterotic picture of horizontal Type II.a models, we need to also take into account the vertical gauge enhancements appearing on the F-theory side of the duality, which form part of the non-perturbative gauge sector from a heterotic point of view (alongside the gauge algebras supported on the exceptional curves arising from the resolution of codimension-two finite-distance non-minimal singularities). As we discuss in more detail in \cref{sec:defect-algebras-heterotic-dual}, such vertical gauge algebras, supported over representatives of $\mathcal{F}$, correspond to ADE singularities of the heterotic K3 surface probed by point-like instantons. Heterotic ADE singularities do not lead to non-perturbative gauge algebra factors unless they are probed by a singular gauge bundle contribution \cite{Witten:1999fq}. As emphasised earlier, the decompactification process does not change the topology of the internal space, meaning that the ADE singularities are still present at the endpoint of the limit, nor can it dilute the singular gauge bundle contributions, for which the curvature is localised at points. Hence, the non-perturbative gauge factors arising from probed heterotic ADE singularities are unaffected by the infinite-distance limit. Due to their localised nature, they lead to six-dimensional defect gauge algebras in the asymptotic model, as was also observed from the F-theory standpoint. This is particularly clear when on the F-theory side the six-dimensional horizontal gauge algebra is the full unbroken perturbative heterotic gauge algebra $\mathfrak{g}_{\mathrm{hor}} = \mathfrak{e}_{8} \oplus \mathfrak{e}_{8}$: A vertical gauge algebra is then dual to a heterotic ADE singularity probed by point-like instantons with trivial holonomy \cite{Aspinwall:1997ye}, which can be traded for M5-branes via a small instanton transition. The six-dimensional defects surviving in the asymptotic model are then M5-branes located on top of the ADE singularities of the infinitely large heterotic K3 surface, see \cref{sec:defect-algebras-heterotic-dual-unbroken-E8xE8}. More generally, once we allow the horizontal gauge algebra to be broken $\mathfrak{g}_{\mathrm{hor}} \leq \mathfrak{e}_{8} \oplus \mathfrak{e}_{8}$, the heterotic ADE singularities can also be probed by point-like instantons with discrete holonomy. These singular gauge bundle contributions cannot be traded for \mbox{M5-branes} via small instanton transitions, and are associated to one of the Ho\v{r}ava-Witten walls, as we expand on in \cref{sec:defect-algebras-heterotic-dual-broken-E8xE8}. Due to their localised nature, they are also able to survive the infinite-distance limit, again leading to localised algebras living in six-dimensional defects. As explained in \cite{Alvarez-Garcia:2023gdd}, only those local enhancements that patch up into global enhancements yielding a factorisation in $\Dphys$ lead to gauge algebra factors. Vertical gauge algebras are supported over divisors that traverse both base components of the resolved central fiber $Y_{0}$, and such considerations are therefore relevant for them; in \cref{sec:defect-algebras-heterotic-dual} we comment on how this affects the non-perturbative heterotic sector and its relation to the distribution of instanton number between the two heterotic $\mathrm{E}_{8}$ bundles.

\subsubsection{Allowing for mixed enhancements}
\label{sec:horizontal-IIa-mixed-enhancements}

The structure \eqref{eq:F0-horizontal-vertical-separation} assumed for the defining polynomials of the Weierstrass model of $\hat{\mathcal{Y}}$ at the beginning of the section restricts us to horizontal Type II.a models over $\hat{B} = \mathbb{F}_{0}$ with only horizontal and vertical enhancements. We now relax it to allow for gauge algebras supported over mixed global divisors as well.

The maximal gauge rank in the class of models under study is
\begin{equation}
    \rank(\mathfrak{g}_{\mathrm{hor}}) \leq 16\,,\qquad \rank(\mathfrak{g}_{\mathrm{ver}}) \leq 18\,,
\end{equation}
in the horizontal and vertical sectors, respectively.\footnote{These bounds on the gauge algebra rank apply both from the six-dimensional point of view, before the enhancements associated to the asymptotically massless towers have been considered, and from the higher-dimensional perspective. Note that the gauge factors supported over the exceptional curves arising from the resolution of codimension-two finite-distance non-minimal singularities are not accounted for in these bounds. Such factors are ignored in this section, but we comment on them in \cref{sec:defect-algebras-heterotic-dual}.} The most efficient use of the divisor classes available in the discriminant in order to tune gauge algebra factors is to only engineer enhancements of these two types. This can be deduced from the fact that a mixed divisor factoring in the physical discriminant can be split into horizontal and vertical constituents through a finite-distance complex structure deformation. Such a splitting increases the total gauge rank of the model. As mentioned in \cref{sec:horizontal-global-divisors}, a gauge algebra factor supported on a mixed divisor can therefore be seen as the result of Higgsing some horizontal and vertical gauge contributions.

Using the heterotic language,  dropping the assumption \eqref{eq:F0-horizontal-vertical-separation} is hence equivalent to allowing for mixed Higgsings between the perturbative and non-perturbative sectors. Let us analyse this first from a six-dimensional perspective, and then fully consider the asymptotic physics. Given a mixed divisor $\mathcal{D}_{\mathrm{mix}}$ that supports a gauge algebra $\mathfrak{g}_{\mathrm{mix}}$, consider the point
\begin{equation}
    p_{b}^{\mathrm{mix}} := \{ e_{1} = 0 \}_{B^{0}} \cap \left. \mathcal{D}_{\mathrm{mix}} \right|_{B^{0}} = \{ e_{0} = 0 \}_{B^{1}} \cap \left. \mathcal{D}_{\mathrm{mix}} \right|_{B^{1}}
\end{equation}
at which $\mathcal{D}_{\mathrm{mix}}$ intersects $B^{0} \cap B^{1}$. We can then, via a finite-distance tuning, split $\mathcal{D}_{\mathrm{mix}}$ into
\begin{equation}
    \mathcal{D}_{\mathrm{mix}} \longmapsto \sum_{i \in \mathcal{I}} \mathcal{D}_{\mathrm{hor}}^{i} + \mathcal{D}_{\mathrm{ver}}\,,
\end{equation}
where $\mathcal{D}_{\mathrm{ver}}$ is the unique representative of $\mathcal{F}$ passing through $p_{b}^{\mathrm{mix}}$. In this model at finite distance from the original one, the collection of horizontal divisors $\{ \mathcal{D}_{\mathrm{hor}} \}_{i \in \mathcal{I}}$ and the vertical divisor $\mathcal{D}_{\mathrm{ver}}$ support the gauge algebras $\bigoplus_{i \in \mathcal{I}} \mathfrak{g}_{\mathrm{hor}}^{i}$ and $\mathfrak{g}_{\mathrm{ver}}$, respectively. While part of these gauge factors stem from the horizontal and vertical 7-branes that result from the splitting of $\mathcal{D}_{\mathrm{mix}}$, these may not be the only contributions; for example, the original model prior to the finite-distance complex structure deformation may already present a gauge algebra supported on $\mathcal{D}_{\mathrm{ver}}$, which is then further enhanced after the splitting of $\mathcal{D}_{\mathrm{mix}}$.

\begin{figure}[t!]
    \centering
	\begin{tikzpicture}
		\def\scalereliiahor{0.7}
		\def\scalereliiaver{0.8}

		\node (0) at (-6*\scalereliiahor, 2.5*\scalereliiaver) {};
		\node (1) at (-6*\scalereliiahor, -2.5*\scalereliiaver) {};
		\node (2) at (0*\scalereliiahor, 2.5*\scalereliiaver) {};
		\node (3) at (0*\scalereliiahor, -2.5*\scalereliiaver) {};
		\node (4) at (6*\scalereliiahor, 2.5*\scalereliiaver) {};
		\node (5) at (6*\scalereliiahor, -2.5*\scalereliiaver) {};
		\node (6) at (-5*\scalereliiahor, 2.5*\scalereliiaver) {};
		\node (7) at (-1*\scalereliiahor, 2.5*\scalereliiaver) {};
		\node (8) at (-1*\scalereliiahor, -2.5*\scalereliiaver) {};
		\node (9) at (-5*\scalereliiahor, -2.5*\scalereliiaver) {};
		\node (10) at (5*\scalereliiahor, 2.5*\scalereliiaver) {};
		\node (11) at (5*\scalereliiahor, -2.5*\scalereliiaver) {};
		\node (12) at (3*\scalereliiahor, 2.5*\scalereliiaver) {};
		\node (13) at (0*\scalereliiahor, 0*\scalereliiaver) {};
		\node (14) at (-6*\scalereliiahor, 0*\scalereliiaver) {};
		\node (15) at (-6*\scalereliiahor, -1.875*\scalereliiaver) {};
		\node (16) at (6*\scalereliiahor, -1.875*\scalereliiaver) {};
		
		\node (21) at (9*\scalereliiahor, 3.75*\scalereliiaver) {};
		\node (22) at (12.5*\scalereliiahor, 2.5*\scalereliiaver) {};
		\node (23) at (11.5*\scalereliiahor, 2.5*\scalereliiaver) {};
		\node (24) at (15*\scalereliiahor, 3.75*\scalereliiaver) {};
		
		\node (25) at (9*\scalereliiahor, 0.625*\scalereliiaver) {};
		\node (26) at (12.5*\scalereliiahor, -0.625*\scalereliiaver) {};
		\node (27) at (11.5*\scalereliiahor, -0.625*\scalereliiaver) {};
		\node (28) at (15*\scalereliiahor, 0.625*\scalereliiaver) {};
		
		\node (29) at (9*\scalereliiahor, -2.5*\scalereliiaver) {};
		\node (30) at (12.5*\scalereliiahor, -3.75*\scalereliiaver) {};
		\node (31) at (11.5*\scalereliiahor, -3.75*\scalereliiaver) {};
		\node (32) at (15*\scalereliiahor, -2.5*\scalereliiaver) {};
		
		\node (37) at (-6*\scalereliiahor, 1.25*\scalereliiaver) {};
		\node (38) at (6*\scalereliiahor, 1.25*\scalereliiaver) {};
		\node (39) at (6*\scalereliiahor, 0*\scalereliiaver) {};
		\node (42) at (-6*\scalereliiahor, -1.25*\scalereliiaver) {};
		\node (43) at (6*\scalereliiahor, -1.25*\scalereliiaver) {};
		
		\node (56) at (-3*\scalereliiahor, 2.5*\scalereliiaver) {};
		\node (57) at (3*\scalereliiahor, -2.5*\scalereliiaver) {};
		\node (58) at (-3*\scalereliiahor, -2.5*\scalereliiaver) {};
		
		\node [label={[yshift=0cm]$\{e_{0} = 0\}_{\makebox[0pt]{$\scriptstyle \;\;\mathcal{B}$}}$}] (44) at (-3*\scalereliiahor, 2.5*\scalereliiaver) {};
        \node [label={[yshift=0cm]$\{e_{1} = 0\}_{\makebox[0pt]{$\scriptstyle \;\;\mathcal{B}$}}$}] (45) at (3*\scalereliiahor, 2.5*\scalereliiaver) {};
		
		\draw [style=dashed] (37.center) to (38.center);
		\draw [style=dashed] (14.center) to (39.center);
		\draw [style=dashed] (42.center) to (43.center);

		\draw [style=light-green line]  (8.center) to (7.center);
		\draw [style=light-purple line]  (6.center) to (9.center);
		\draw [style=light-blue line] (10.center) to (11.center);
		\draw [style=light-yellow line] (12.center) to (13.center);
		\draw [style=light-yellow line] (13.center) to (56.center);
		\draw [style=light-yellow line] (57.center) to (39.center);
		\draw [style=light-yellow line] (58.center) to (14.center);
		\draw [style=medium-red line] (15.center) to (16.center);
		
		\draw [style=black line] (23.center) to (24.center);
		\draw [style=black line] (22.center) to (21.center);
		\draw [style=black line] (27.center) to (28.center);
		\draw [style=black line] (26.center) to (25.center);
		\draw [style=black line] (31.center) to (32.center);
		\draw [style=black line] (30.center) to (29.center);
		
		\draw [style=black line] (0.center) to (2.center);
		\draw [style=black line] (2.center) to (4.center);
		\draw [style=black line] (4.center) to (5.center);
		\draw [style=black line] (5.center) to (3.center);
		\draw [style=black line] (3.center) to (1.center);
		\draw [style=black line] (1.center) to (0.center);
		\draw [style=black line] (2.center) to (3.center);
		
		\draw [-latex] (38.east) to ([yshift=-0.05cm]21.west);
		\draw [-latex] (39.east) to (25.west);
		\draw [-latex] (43.east) to ([yshift=0.05cm]29.west);

		\node [cross-purple=5pt, rotate=-19.6538, anchor=center] (46) at (9.875*\scalereliiahor,3.4375*\scalereliiaver) {};
		\node [cross-green=5pt, rotate=-19.6538, anchor=center] (47) at (11.275*\scalereliiahor,2.9375*\scalereliiaver) {};
		\node [cross-blue=5pt, rotate=19.6538, anchor=center] (48) at (14.125*\scalereliiahor,3.4375*\scalereliiaver) {};
		\node [cross-yellow=5pt, rotate=-19.6538, anchor=center] (49) at (10.75*\scalereliiahor,3.125*\scalereliiaver) {};
		\node [cross-yellow=5pt, rotate=19.6538, anchor=center] (59) at (13.25*\scalereliiahor,3.125*\scalereliiaver) {};
		
		\node [cross-purple=5pt, rotate=-19.6538, anchor=center] (50) at (9.875*\scalereliiahor,0.3125*\scalereliiaver) {};
		\node [cross-green=5pt, rotate=-19.6538, anchor=center] (51) at (11.275*\scalereliiahor,-0.1875*\scalereliiaver) {};
		\node [cross-blue=5pt, rotate=19.6538, anchor=center] (52) at (14.125*\scalereliiahor,0.3125*\scalereliiaver) {};
		\node [cross-yellow=5pt, rotate=0, anchor=center] (60) at (12*\scalereliiahor,-0.446429*\scalereliiaver) {};
		
		\node [cross-purple=5pt, rotate=-19.6538, anchor=center] (53) at (9.875*\scalereliiahor,-2.8125*\scalereliiaver) {};
		\node [cross-green=5pt, rotate=-19.6538, anchor=center] (61) at (11.275*\scalereliiahor,-3.3125*\scalereliiaver) {};
		\node [cross-blue=5pt, rotate=19.6538, anchor=center] (54) at (14.125*\scalereliiahor,-2.8125*\scalereliiaver) {};
		\node [cross-yellow=5pt, rotate=-19.6538, anchor=center] (55) at (9.35*\scalereliiahor,-2.625*\scalereliiaver) {};
		\node [cross-yellow=5pt, rotate=19.6538, anchor=center] (62) at (14.65*\scalereliiahor,-2.625*\scalereliiaver) {};
	\end{tikzpicture}
    \caption{A schematic representation of the base of the central fiber of a resolved horizontal Type II.a model constructed over $\hat{B} = \mathbb{F}_{0}$ on the left, and various vertical slices of it on the right. We represent two horizontal 7-branes in the classes $\mathcal{H}_{\infty}^{0}$ and $\mathcal{H}_{0}^{1}$, respectively, a vertical 7-brane in the class $\mathcal{F}$, and a mixed 7-brane in the class $\mathcal{D}_{\mathrm{mix}} = \mathcal{H}_{\infty}^{0} + \mathcal{H}_{\infty}^{1} + \mathcal{F}$. The first and the third vertical slices are generic, and lead to the same picture as in the unHiggsed model at finite distance. The second vertical slice has smooth codimension-zero elliptic fibers, but presents special fibers at the intersection of the components, signalling a complete Higgsing of the primordial vertical gauge algebra associated to the representative of $\mathcal{F}$ intervening in the recombination process producing $\mathcal{D}_{\mathrm{mix}}$.}
    \label{fig:relative-Type-IIa-shared-branes}
\end{figure}
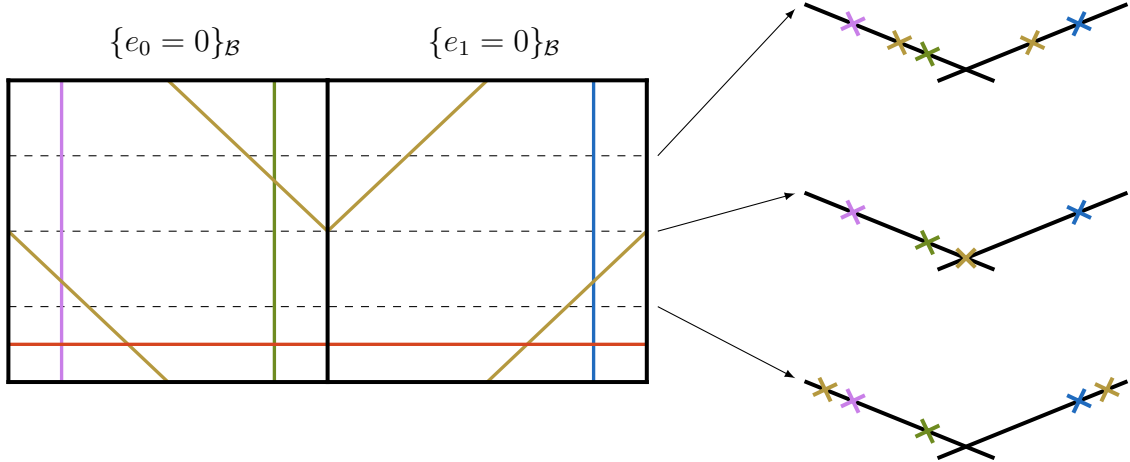

Inverting the logic, we may consider that there is a horizontal Type II.a model constructed over $\hat{B} = \mathbb{F}_{0}$ and satisfying \eqref{eq:F0-horizontal-vertical-separation} that exhibits an original gauge algebra $\bigoplus_{i \in \mathcal{I}} \mathfrak{g}_{\mathrm{hor}}^{i} \oplus \mathfrak{g}_{\mathrm{ver}}$ supported over $\{ \mathcal{D}_{\mathrm{hor}}^{i} \}_{i \in \mathcal{I}}$ and $\mathcal{D}_{\mathrm{ver}}$, and whose asymptotics physics we have already understood in \cref{sec:horizontal-IIa-away-from-adiabatic-regime}. A fine-distance tuning, which does not interfere with the infinite-distance limit and that could also be implemented after the fact, recombines (at least part of) these 7-branes into $\mathcal{D}_{\mathrm{mix}}$. This produces a Higgsing
\begin{equation}
    \bigoplus_{i \in \mathcal{I}} \mathfrak{g}_{\mathrm{hor}}^{i} \oplus \mathfrak{g}_{\mathrm{ver}} \longmapsto \mathfrak{g}_{\mathrm{mix}} \oplus \mathfrak{g}_{\mathrm{hor}}^{\mathrm{res}} \oplus \mathfrak{g}_{\mathrm{ver}}^{\mathrm{res}}\,,
\end{equation}
where $\mathfrak{g}_{\mathrm{hor}}^{\mathrm{res}} \oplus \mathfrak{g}_{\mathrm{ver}}^{\mathrm{res}}$ accounts for the fact that not all 7-branes in the listed divisor classes need to recombine. The remaining ones lead to the residual horizontal and vertical gauge algebras. It may nonetheless occur that no, e.g., vertical gauge algebra remains after the recombination, as is represented in \cref{fig:relative-Type-IIa-shared-branes}.

Turning our attention to the asymptotic physics, we note that such a recombination process into a mixed divisor always requires vertical divisor classes. These are associated with the six-dimensional defects present in the decompactified theory, and therefore the Higgsing should be tied to them. Indeed, the generic vertical slices still lead to the same conclusions for the horizontal contribution to the gauge algebra, since the pertinent part of the discussion in \cref{sec:horizontal-IIa-heterotic-dual-interpretation}  remains unaltered after the finite-distance deformation. After the recombination into a mixed divisor, a pair of branes in different base components of the associated generic eight-dimensional model can belong to the same global mixed divisor in the six-dimensional one, as shown in \cref{fig:relative-Type-IIa-shared-branes} for the yellow 7-brane. Nonetheless, and according to the analysis of the generic vertical slices, they still contribute to a different factor of $G_{\infty}$ each, with the two halves of the global 7-brane only joining at one of the 24 non-generic points in $\mathbb{P}^{1}_{b}$. These union points play a role in the obstruction to globally defining the local 2-cycles $\{ \gamma_{i} \}_{i=1,2}$, since they lead to special fibers at the intersection of the components in the corresponding vertical slice, and hence to the non-BPS nature of the particles furnishing the asymptotically massless towers. We can understand the fact that the two halves of such a mixed 7-brane contribute to different factors of the perturbative heterotic gauge algebra as the dilution of the gauge bundle curvature during the decompactification process. This makes the masses associated with the Higgsing of the bulk gauge degrees of freedom asymptote to zero in the limit. The collection of non-generic vertical slices accounts both for those vertical gauge factors supported over representatives of $\mathcal{F}$ unaffected by the recombination process, and for the residual vertical gauge algebras $\mathfrak{g}_{\mathrm{ver}}^{\mathrm{res}}$ resulting from a localised Higgsing of the primordial defect algebra $\mathfrak{g}_{\mathrm{ver}}$.   Such Higgsings can be complete, as represented in \cref{fig:relative-Type-IIa-shared-branes}. The fact that to produce such mixed enhancements the maximal vertical gauge rank needs to be reduced can be understood from the heterotic dual perspective as the need to deform some of the point-like instantons probing the heterotic ADE singularities into smooth gauge bundle contributions with structure group of positive dimension, as we revisit in \cref{sec:defect-algebras-heterotic-dual}.

\subsubsection{Away from the adiabatic regime}
\label{sec:horizontal-IIa-away-from-adiabatic-regime}

We now comment on the interpretation of the complex structure infinite-distance limit away from the strict adiabatic regime. Consider first the heterotic side of the duality and suppose, for simplicity of the argument, that the K3 surface is described by a smooth Weierstrass model. The K\"ahler volume of the K3 surface then takes the form
\begin{equation}
    \mathcal{V}_{\mathrm{K3}} = \frac{1}{2} \int_{\mathrm{K3}} J^{2} =  \left( \mathcal{V}_{T^{2}_{\mathrm{het}}} \right)^{2} + \mathcal{V}_{T^{2}_{\mathrm{het}}} \mathcal{V}_{\mathbb{P}^{1}_{b}}\,,
\label{eq:vol-K3-het}
\end{equation}
where $\mathcal{V}_{T^{2}_{\mathrm{het}}}$ is the volume of the generic elliptic fiber. In the adiabatic regime, characterised by $\mathcal{V}_{\mathbb{P}^{1}_{b}} \gg \mathcal{V}_{T^{2}_{\mathrm{het}}}$, the term quadratic in $\mathcal{V}_{T^{2}_{\mathrm{het}}}$ is subleading, and the volume asymptotically factorises as it would for a trivial fibration. In this regime, taking both $\mathcal{V}_{\mathbb{P}^{1}_{b}} \rightarrow \infty$ and $\mathcal{V}_{T^{2}_{\mathrm{het}}} \rightarrow \infty$ such that $\mathcal{V}_{\mathbb{P}^{1}_{b}} \gg \mathcal{V}_{T^{2}_{\mathrm{het}}}$ allows us to clearly distinguish between the Kaluza-Klein supergravity towers stemming from decompactification of the base $\mathbb{P}^{1}_{b}$ and the fiber $T^{2}_{\mathrm{het}}$. In fact, to the extent that the geometry of the elliptic fibration asymptotically behaves as it would in a trivial fibration, the fiber contributes two approximate Kaluza-Klein towers, each associated to one of the two local 1-cycles of its generic representative. This picture breaks down near the 24 degeneration loci of the fiber, which, however, are
pushed to infinity in the adiabatic decompactification limit.

This behaviour is contrasted with the limit $\mathcal{V}_{T^{2}_{\mathrm{het}}} \sim \lambda \mathcal{V}_{T^{2}_{\mathrm{het}}}^{0}$ with $\lambda \rightarrow \infty$ and $\mathcal{V}_{T^{2}_{\mathrm{het}}}^{0}$ constant, at finite $\mathcal{V}_{\mathbb{P}^{1}_{b}} = \mathcal{V}_{\mathbb{P}^{1}_{b}}^{0}$. In view of \eqref{eq:vol-K3-het}, the decompactification of the K3 surface can be interpreted as a homogenous rescaling of the K3 volume superimposed with a shrinking of the base $\mathbb{P}^{1}_{b}$ in the original, non-rescaled surface, i.e.\
\begin{equation}
    \mathcal{V}_{\mathrm{K3}} \sim \lambda^{2} \left[ \left( \mathcal{V}_{T^{2}_{\mathrm{het}}}^{0} \right)^2 + \mathcal{V}_{T^{2}_{\mathrm{het}}}^{0} \frac{\mathcal{V}_{\mathbb{P}^{1}_{b}}^{0}}{\lambda} \right] =: \lambda^{2}  \mathcal{V}_{\mathrm{K3}}^{0}\,,\qquad \lambda \rightarrow \infty\,.
\end{equation}
From the form of $\mathcal{V}_{\mathrm{K3}}^{0}$, it is clear that the non-rescaled K3 surface approaches an orbifold point, where the $(-2)$-curve given by the zero-section of the elliptic fibration has shrunk; at the same rate as the orbifold point is reached, the total surface expands homogenously. As a result, the non-adiabatic limit is still a total decompactification to ten dimensions, but the nature of the Kaluza-Klein towers differs from its counterpart in the adiabatic regime. In particular, the hierarchical split of the Kaluza-Klein towers into a tower from the expansion of the base and  two asymptotically independent towers from the generic fiber breaks down. Rather, as we interpolate between the adiabatic and the non-adiabatic regimes at large overall volume, the supergravity towers reorganise in a complicated way. The fiber continues to contribute \mbox{Kaluza-Klein} tower(s), which, however, cannot be treated as two independent towers associated with local isometries along its generic representative.

This suggests that, similarly, the microscopic interpretation of the complex structure infinite-distance limit on the F-theory side must be modified away from the adiabatic regime. Note that, strictly speaking, the F-theory/heterotic duality dictionary in its form reviewed in \cref{sec:F-theory-heteoric-duality} is valid only in the adiabatic limit, so that a priori special care must be applied in translating the heterotic interpretation of the limit to the F-theory side. Nonetheless, there are clear parallels. Away from the adiabatic regime, it is no longer possible to isolate the effect of the vertical fiber degenerations by sending them to infinity. Recall from the discussion in \cref{sec:horizontal-IIa-heterotic-dual-interpretation} that it is at these vertical degenerations that the local picture of asymptotically vanishing \mbox{2-cycles} with the topology of a torus breaks down. Rather than a fibration of such local 2-cycles degenerating at isolated points that asymptotically move to infinity, it is  more appropriate to view the complex structure degeneration as leading to vanishing 3-cycles: These arise by fibering the vanishing 2-chains over the 1-chains between pairs of points on $\mathbb{P}^{1}_{b}$ over which the singular vertical fibers are located. The geometric origin of the massless towers, which in the adiabatic regime arise from M2-branes wrapping the locally well-defined vanishing 2-cycles, is obscured by this. The benefit of the adiabatic limit, superimposed with the complex structure infinite-distance limit, is therefore to provide a clear geometric interpretation of the origin of the towers. In the strict large base limit, these are the two local towers from wrapped M2-branes in M-theory, dual to the two heterotic Kaluza-Klein towers from the adiabatically expanding fiber. The heterotic picture suggests that these two towers reorganise away from the adiabatic limit, but continue to signal a decompactification process also in F-theory. What is more challenging is to read off the end point of this decompactification limit, and in particular the dimensionality of the asymptotic theory. It is natural to speculate that, as on the heterotic side, the decompactification is to a ten-dimensional theory with defect gauge sectors, but a quantitative underpinning of this conjecture is beyond the scope of this paper. 

\subsection{Horizontal Type II.a models over \texorpdfstring{$\hat{B} = \mathbb{F}_{n}$}{B=Fn}}
\label{sec:horizontal-IIa-Fn}

Horizontal Type II.a models constructed over $\hat{B} = \mathbb{F}_{n}$, with $n \geq 1$, behave very similarly to the ones constructed over $\hat{B} = \mathbb{F}_{0}$ and just analysed in \cref{sec:horizontal-IIa-F0}. We therefore only focus here on highlighting some of the differences, which mostly stem from the fact that now $\mathcal{H}_{\infty}^{1}$, defined in \eqref{eq:horizontal-Hpinf-n-geq-1}, is a mixed divisor.

Once $n \geq 1$, the two components $\{ Y^{p} \}_{0 \leq p \leq 1}$ of the central fiber $Y_{0}$ no longer behave identi\-cally, as can be seen from the  line bundles defining the component Weierstrass models,
\begin{subequations}
\begin{align}
    \mathcal{L}_{0} &= S_{0} + (2+n)V_{0}\,,\\
    \mathcal{L}_{1} &= S_{1} + 2V_{1}\,.
\end{align}
\end{subequations}
Since horizontal Type II.a models have smooth codimension-zero elliptic fibers, the $\{ \Delta'_{p} \}_{0 \leq p \leq 1}$ are simply in the divisor class $\Delta'_{p} = 12 \mathcal{L}_{p}$, cf.\ \eqref{eq:Delta-hor}. On the heterotic dual side, this asymmetry corresponds to the uneven distribution of instanton numbers between the two $\mathrm{E}_{8}$ bundles.

The $Y^{0}$ component, corresponding on the heterotic dual side to the $\mathrm{E}_{8}$ bundle with instanton number $c_{2}(V_{0}) = 12 + n$, behaves mostly\footnote{The multiples of $\mathcal{H}_{\infty}^{0}$ may now be irreducible, but this does not change the asymptotic physics.} like in the $\hat{B} = \mathbb{F}_{0}$ case. Its contribution to the six-dimensional horizontal gauge algebra fulfils $\rank(\mathfrak{g}_{\mathrm{hor}}^{0}) \leq 8$, and is saturated by tuning $\mathrm{II}^{*}$~fibers over a representative of $\mathcal{H}_{\infty}^{0}$. Such $\mathfrak{g}_{\mathrm{hor}}^{0} = \mathfrak{e}_{8}$ algebra can be Higgsed without affecting the vertical gauge algebra sector, although this still reduces the rank of the non-perturbative heterotic gauge sector, see \cref{sec:defect-algebras-heterotic-dual}.

For the $Y^{1}$ component, which corresponds to the heterotic $\mathrm{E}_{8}$ bundle with instanton number $c_{2}(V_{1}) = 12 - n$, the situation is different. Apart from the presence of non-Higgsable clusters for $n \geq 3$, which lie in $Y^{1}$ rather than $Y^{0}$, another distinction with respect to the $Y^{0}$ component is how horizontal Higgsings affect the maximal vertical gauge rank that is possible, in addition to the effects on the non-perturbative sector already observed for the $Y^{0}$ component. The contribution of the $Y^{1}$ component to the six-dimensional horizontal gauge algebra fulfils $\rank(\mathfrak{g}_{\mathrm{hor}}^{1}) \leq 8$, and is saturated by tuning $\mathrm{II}^{*}$ fibers over the unique representative of $\mathcal{H}_{0}^{1}$. Since the divisors in the class $\mathcal{H}_{\infty}^{1}$ are now mixed, the $\mathfrak{g}_{\mathrm{hor}}^{1} = \mathfrak{e}_{8}$ algebra cannot be Higgsed without recombining $\mathcal{H}_{0}^{1}$ with a suitable number of copies of $\mathcal{F}$. As a consequence, the maximal vertical gauge rank is reduced, and the model can be seen as the Higgsing of one obtained by a finite-distance deformation, and with maximal primordial horizontal\footnote{\label{foot:maximal-horizontal-rank-not-E8}While in the absence of a vertical enhancement the maximal horizontal gauge algebra rank can always be obtained by tuning $g_{\mathrm{hor}} = \mathfrak{e}_{8} \oplus \mathfrak{e}_{8}$, this is not always viable if we also want to realise the maximal vertical gauge rank simultaneously. For example, in horizontal Type II.a models constructed over $\hat{B} = \mathbb{F}_{3}$ we read from \cref{tab:vertical-rank-bounds} that $\rank(\mathfrak{g}_{\mathrm{ver}}) \leq 16$. This can be achieved by tuning a global line of $\mathrm{I}_{12}^{*\, \mathrm{s}}$ fibers over a representative of $\mathcal{F}$, corresponding to a vertical $\mathrm{D}_{16}$ algebra. Such a vertical tuning is incompatible with having $g_{\mathrm{hor}} = \mathfrak{e}_{8} \oplus \mathfrak{e}_{8}$, but the maximal horizontal gauge rank can still be obtained by tuning instead two horizontal lines of $\mathrm{I}_{4}^{*\, \mathrm{s}}$ fibers, one over a representative of $\mathcal{H}_{\infty}^{0}$ and another one over the unique representative of $\mathcal{H}_{0}^{1}$, each corresponding to a horizontal $\mathrm{D}_{8}$ algebra, i.e.\ we have instead $\mathfrak{g}_{\mathfrak{hor}} = \mathfrak{so}(16) \oplus \mathfrak{so}(16)$.} and vertical gauge algebra factors, cf. \cref{sec:horizontal-IIa-mixed-enhancements}.

For example, consider a horizontal Type II.a model constructed over $\hat{B} = \mathbb{F}_{7}$. According to \cref{tab:vertical-rank-bounds}, we have then $\rank(\mathfrak{g}_{\mathrm{ver}}) \leq 8$. The maximal horizontal and vertical gauge ranks can be obtained in a model with the gauge factors
\begin{equation}
    \mathcal{H}_{\infty}^{0}: \mathrm{E}_{8}\,,\qquad \mathcal{H}_{0}^{1}: \mathrm{E}_{8}\,,\qquad \mathcal{F}: \mathrm{E}_{8}\,.
\end{equation}
Due to the non-Higgsable cluster present in models constructed over $\hat{B} = \mathbb{F}_{7}$, see \cref{tab:non-Higgsable-clusters}, the $\mathrm{E}_{8}$ factor supported on $\mathcal{H}_{0}^{1}$ can be broken, at most, to an $\mathrm{E}_{7}$ factor. To determine the maximal number $\alpha$ of representatives of the divisor class $V_1$ that can factorize in $\Delta'_{1}$ consistent with this, we consider the residual discriminant $\Delta'_{1} - 9S_{1} - \alpha V_{1}$, and demand that it does not factor out an additional copy of $S_{1}$ (which would indicate an enhancement back to $\mathrm{E}_8$).
In particular, this implies that 
\begin{equation}
    \left( \Delta'_{1} - 9S_{1} - \alpha V_{1} \right) \cdot S_{1} = 3-\alpha \geq 0 \Leftrightarrow \alpha \leq 3\,,
\end{equation}
and hence $\rank(\mathfrak{g}_{\mathrm{ver}}) \leq 1$ (saturated for Kodaira type III or type $\mathrm{I}_{2}$ fibers along a representative of $\mathcal{F}$, and hence, in particular, along a representative of $V_{1}$) if we do not want the $\mathrm{E}_{8}$ factor over $\mathcal{H}_{0}^{1}$ to be restored.\footnote{It can be shown that tuning a vertical line of $\mathrm{I}_{3}$ type fibers also restores the $\mathrm{E}_{8}$ factor over $\mathcal{H}_{0}^{1}$.} Therefore, Higgsing the horizontal gauge algebra in the $Y^{1}$ component leads to models in which the maximal enhancement, from the six-dimensional point of view, is
\begin{equation}
    \mathcal{H}_{\infty}^{0}: \mathrm{E}_{8}\,,\qquad \mathcal{H}_{0}^{1}: \mathrm{E}_{7}\,,\qquad \mathcal{F}: \mathrm{A}_{1}\,.
\end{equation}

From the perspective of the asymptotic physics, the horizontal Higgsing is not that relevant, since the horizontal 7-branes will combine to produce the bulk $G_{\mathrm{10D}}$ algebra at the endpoint of the limit. The rank of the maximal defect algebra, however, is greatly reduced as a consequence of the preceding considerations.

Going to the adiabatic regime, we can understand this from a heterotic dual perspective. In the unHiggsed model in which the maximal horizontal gauge rank is realised, none of the instanton number budget $c_{2}(\mathrm{K3}) = 24$ needs to be spent in order to break the perturbative gauge group. The instanton number can then, in particular, be accounted for by point-like instantons with trivial holonomy (which, indeed, leave the bulk gauge sector intact). These point-like instantons are hence available to be placed on top of the geometrical singularities of the heterotic K3, allowing for the manifestation of the maximal non-perturbative gauge algebra possible within the class of models. In order to break the perturbative gauge group, part of these point like instantons need to be smoothed out into a gauge bundle profile with structure group of positive dimension; while the geometrical singularity of the heterotic K3 surface is still present, fewer point-like instantons are able to probe it. This reduces  the maximal possible rank of the non-perturbative gauge algebra. In the decompactification limit the gauge bundle curvature is diluted, and the masses associated with the Higgsing in the perturbative sector asymptote to zero, restoring the bulk gauge algebra. Point-like instantons are singular gauge bundle contributions, and hence do not dilute in the limit: The remaining non-perturbative gauge algebra is maintained in the limit, but its gauge rank is reduced with respect to the situation in which no point-like instantons are smoothed out and diluted.

The decompactified theories arising as the endpoints of infinite-distance limits represented by horizontal Type II.a models are qualitatively not that different from each other, but still can preserve some memory of the Hirzebruch surface $\hat{B} = \mathbb{F}_{n}$ employed in their construction. While point-like instantons with trivial holonomy are not associated with a given Ho\v{r}ava-Witten wall thanks to the possibility of trading them for M5-branes via a small instanton transition, the same is not true for point-like instantons with discrete holonomy. These are singular gauge bundle contributions that also survive the infinite-distance limit, but belong to one of the \mbox{Ho\v{r}ava-Witten} walls. Hence, their presence at the endpoint of the limit preserves some information on the distribution of instanton numbers between the two heterotic $\mathrm{E}_{8}$ bundles or, in F-theory terms, the choice of Hirzebruch surface $\hat{B} = \mathbb{F}_{n}$. As a consequence, two distinct horizontal Type II.a models can either lead to the same asymptotic physics, or their endpoints require an additional finite-distance deformation and non-perturbative transition in order to be connected, see \cref{sec:defect-algebras-heterotic-dual}.

\subsection{Non-minimal singularities of the heterotic K3 surface}
\label{sec:non-minimal-points-heterotic-K3-surface}

Minimal singular elliptic fibers in the heterotic K3 surface do not lead to non-perturbative gauge algebra factors unless they are probed by singular gauge bundle contributions, as we review at the beginning of \cref{sec:defect-algebras-heterotic-dual}. This is due to $\alpha'$-corrections having the effect of smoothing out their moduli space \cite{Witten:1999fq}. It would be conceivable that, in the same way that a minimal geometrical singularity is not enough to produce a non-perturbative gauge algebra factor, a non-minimal geometrical singularity could be brought to finite distance by similar effects. We argue that this is not the case, since non-minimal singularities of the heterotic K3 surface are equivalent to a codimension-one infinite-distance degeneration on the F-theory side of the duality.

In a resolved horizontal Type II.a model with central fiber $Y_{0} = Y^{0} \cup_{\mathrm{K3}} Y^{1}$, the dual heterotic K3 surface is identified with $Y^{0} \cap Y^{1}$. In terms of the Weierstrass model describing the family variety $\mathcal{Y}$ of the resolved degeneration, a non-minimal singularity of the heterotic K3 surface arises if at a point $p \in B^{0} \cap B^{1}$ the interface vanishing orders are
\begin{equation}
    \ord{Y^{0} \cap Y^{1}}(\left. f_{b} \right|_{e_{0} = e_{1} = 0}, \left. g_{b} \right|_{e_{0} = e_{1} = 0}, \left. \Delta_{b} \right|_{e_{0} = e_{1} = 0})_{p} \geq (4,6,12)\,.
\label{eq:heterotic-non-minimal-point}
\end{equation}

Earlier in our discussion, we have already considered what naively seem like worse degenerations in codimension-two over $B_{0}$: Point-like instantons with trivial holonomy correspond to codimension-two finite-distance non-minimal points for which, at the very least, the component vanishing orders are non-minimal. In contrast, we are demanding in \eqref{eq:heterotic-non-minimal-point} only the interface vanishing order, i.e.\ the vanishing orders computed in a non-generic slice, to be non-minimal. Nonetheless, the fact that the non-generic slice corresponds to the heterotic K3 surface is quite relevant from the perspective of the F-theory geometry. A point with non-minimal interface vanishing orders on $B^{0} \cap B^{1}$ corresponds to a (possibly obscured) infinite-distance limit, in the language of \cite{Alvarez-Garcia:2023gdd}. In fact, tuning it takes us away from the single infinite-distance limit class of degenerations, see \cite{Alvarez-Garcia:2023gdd} for the precise definition, since we are indeed overlaying an additional infinite-distance limit on top of the already existing one.

Succinctly summarising the discussion in Appendix C of \cite{Alvarez-Garcia:2023gdd}, such an obscured infinite-distance limit can be transformed  into a codimension-one infinite-distance degeneration by performing a base change
\begin{equation}
	\begin{aligned}
		\delta_{k}: D &\longrightarrow D\\
		u &\longmapsto u^{k}
	\end{aligned}
\end{equation}
with high enough branching degree. The equivalent degeneration obtained after the base change has a central fiber comprised of more than two components. The former point $p \in B^{0} \cap B^{1}$ with non-minimal interface vanishing orders extends, after the base change, into a vertical line of non-minimal component vanishing orders traversing the intermediate components. By blowing the model down to one of said intermediate components, and possibly performing a second base change, we obtain a vertical codimension-one infinite-distance degeneration overlaid on top of the horizontal one that we started with.

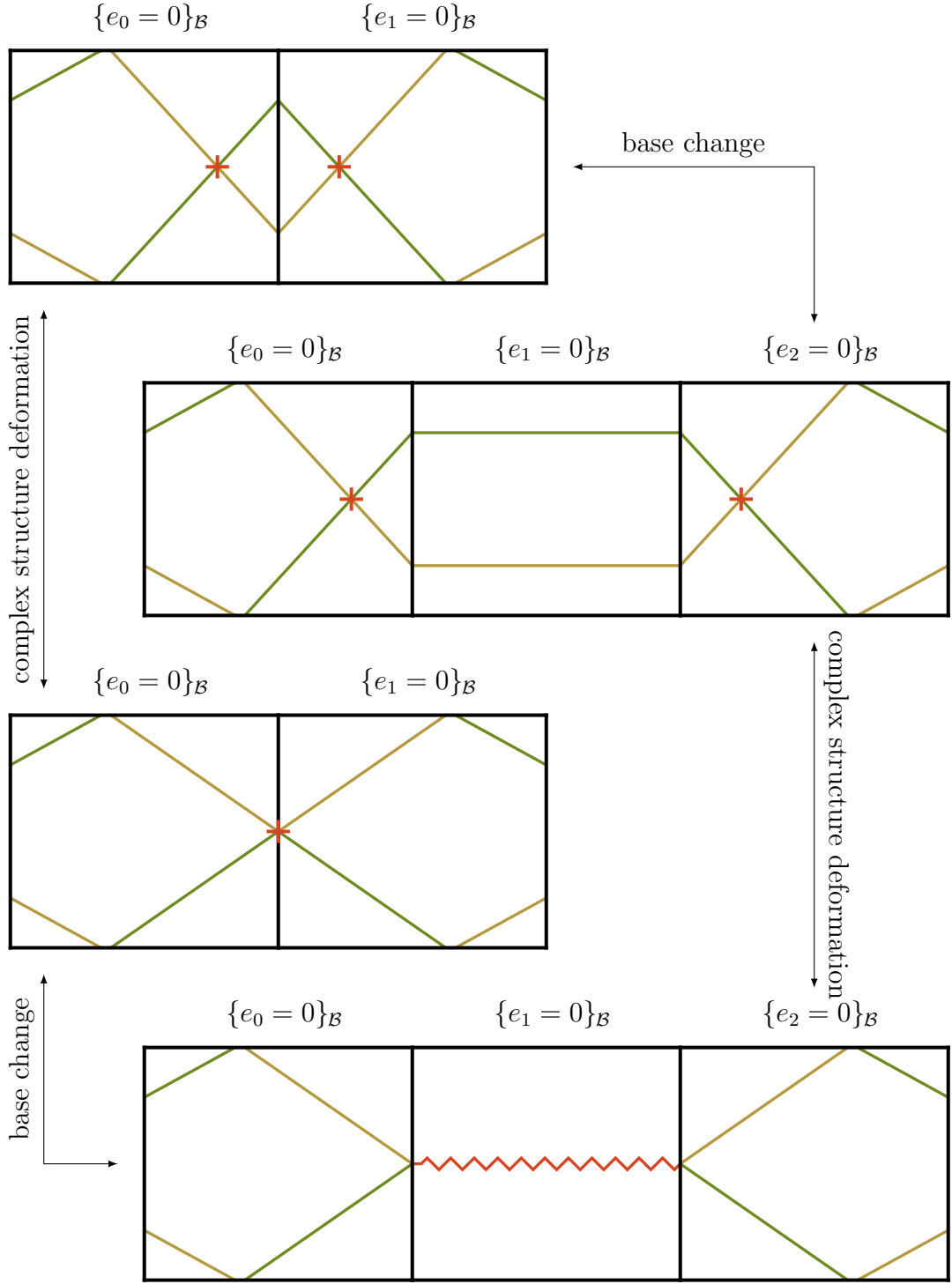
\begin{figure}[tp!]
    \centering
    \begin{tikzpicture}
    		
		\def\diagramoneposxa{-2}
		\def\diagramoneposya{7.5}
    
		\node [] (0) at (0+\diagramoneposxa, 1.75+\diagramoneposya) {};
		\node [] (1) at (0+\diagramoneposxa, -1.75+\diagramoneposya) {};
		\node [] (2) at (4+\diagramoneposxa, 1.75+\diagramoneposya) {};
		\node [] (3) at (4+\diagramoneposxa, -1.75+\diagramoneposya) {};
		\node [] (4) at (-4+\diagramoneposxa, -1.75+\diagramoneposya) {};
		\node [] (5) at (-4+\diagramoneposxa, 1.75+\diagramoneposya) {};
		
		\node [label={[yshift=0cm]$\{e_{0} = 0\}_{\makebox[0pt]{$\scriptstyle \;\;\mathcal{B}$}}$}] (6) at (-2+\diagramoneposxa, 1.75+\diagramoneposya) {};
        \node [label={[yshift=0cm]$\{e_{1} = 0\}_{\makebox[0pt]{$\scriptstyle \;\;\mathcal{B}$}}$}] (7) at (2+\diagramoneposxa, 1.75+\diagramoneposya) {};
        
		\node [] (38) at (2.5+\diagramoneposxa, 1.75+\diagramoneposya) {};
		\node [] (39) at (0+\diagramoneposxa, -1+\diagramoneposya) {};
		\draw [style=light-yellow line] (38.center) to (39.center);
		\node [] (40) at (-2.5+\diagramoneposxa, 1.75+\diagramoneposya) {};
		\node [] (41) at (0+\diagramoneposxa, -1+\diagramoneposya) {};
		\draw [style=light-yellow line] (40.center) to (41.center);
		\node [] (42) at (-2.5+\diagramoneposxa, -1.75+\diagramoneposya) {};
		\node [] (43) at (-4+\diagramoneposxa, -1+\diagramoneposya) {};
		\draw [style=light-yellow line] (42.west) to (43.center);
		\node [] (44) at (2.5+\diagramoneposxa, -1.75+\diagramoneposya) {};
		\node [] (45) at (4+\diagramoneposxa, -1+\diagramoneposya) {};
		\draw [style=light-yellow line] (44.east) to (45.center);
		
		\node [] (46) at (0+\diagramoneposxa, 1+\diagramoneposya) {};
		\node [] (47) at (-2.5+\diagramoneposxa, -1.75+\diagramoneposya) {};
		\draw [style=light-green line] (46.center) to (47.center);
		\node [] (48) at (0+\diagramoneposxa, 1+\diagramoneposya) {};
		\node [] (49) at (2.5+\diagramoneposxa, -1.75+\diagramoneposya) {};
		\draw [style=light-green line] (48.center) to (49.center);
		\node [] (50) at (-4+\diagramoneposxa, 1+\diagramoneposya) {};
		\node [] (51) at (-2.5+\diagramoneposxa, 1.75+\diagramoneposya) {};
		\draw [style=light-green line] (50.center) to (51.west);
		\node [] (52) at (4+\diagramoneposxa, 1+\diagramoneposya) {};
		\node [] (53) at (2.5+\diagramoneposxa, 1.75+\diagramoneposya) {};
		\draw [style=light-green line] (52.center) to (53.east);
		
		\draw [style=black line] (0.center) to (1.center);
		\draw [style=black line] (5.center) to (0.center);
		\draw [style=black line] (0.center) to (2.center);
		\draw [style=black line] (2.center) to (3.center);
		\draw [style=black line] (3.center) to (1.center);
		\draw [style=black line] (1.center) to (4.center);
		\draw [style=black line] (4.center) to (5.center);
		
		\node [cross-red=5pt, rotate=45, anchor=center] (98) at (-0.909091+\diagramoneposxa,0+\diagramoneposya) {};
		\node [cross-red=5pt, rotate=45, anchor=center] (99) at (0.909091+\diagramoneposxa,0+\diagramoneposya) {};
        
    		
		\def\diagramoneposxb{2}
		\def\diagramoneposyb{2.5}
    
		\node [] (8) at (-2+\diagramoneposxb, 1.75+\diagramoneposyb) {};
		\node [] (9) at (-2+\diagramoneposxb, -1.75+\diagramoneposyb) {};
		\node [] (10) at (6+\diagramoneposxb, 1.75+\diagramoneposyb) {};
		\node [] (11) at (6+\diagramoneposxb, -1.75+\diagramoneposyb) {};
		\node [] (12) at (-6+\diagramoneposxb, -1.75+\diagramoneposyb) {};
		\node [] (13) at (-6+\diagramoneposxb, 1.75+\diagramoneposyb) {};
		\node [] (14) at (2+\diagramoneposxb, 1.75+\diagramoneposyb) {};
		\node [] (15) at (2+\diagramoneposxb, -1.75+\diagramoneposyb) {};
		
		\node [label={[yshift=0cm]$\{e_{0} = 0\}_{\makebox[0pt]{$\scriptstyle \;\;\mathcal{B}$}}$}] (16) at (-4+\diagramoneposxb, 1.75+\diagramoneposyb) {};
        \node [label={[yshift=0cm]$\{e_{1} = 0\}_{\makebox[0pt]{$\scriptstyle \;\;\mathcal{B}$}}$}] (17) at (0+\diagramoneposxb, 1.75+\diagramoneposyb) {};
        \node [label={[yshift=0cm]$\{e_{2} = 0\}_{\makebox[0pt]{$\scriptstyle \;\;\mathcal{B}$}}$}] (18) at (4+\diagramoneposxb, 1.75+\diagramoneposyb) {};
		
		\node [] (54) at (4.5+\diagramoneposxb, 1.75+\diagramoneposyb) {};
		\node [] (55) at (2+\diagramoneposxb, -1+\diagramoneposyb) {};
		\draw [style=light-yellow line] (54.center) to (55.center);
		\node [] (56) at (-4.5+\diagramoneposxb, 1.75+\diagramoneposyb) {};
		\node [] (57) at (-2+\diagramoneposxb, -1+\diagramoneposyb) {};
		\draw [style=light-yellow line] (56.center) to (57.center);
		\node [] (58) at (-4.5+\diagramoneposxb, -1.75+\diagramoneposyb) {};
		\node [] (59) at (-6+\diagramoneposxb, -1+\diagramoneposyb) {};
		\draw [style=light-yellow line] (58.west) to (59.center);
		\node [] (60) at (4.5+\diagramoneposxb, -1.75+\diagramoneposyb) {};
		\node [] (61) at (6+\diagramoneposxb, -1+\diagramoneposyb) {};
		\draw [style=light-yellow line] (60.east) to (61.center);
		\draw [style=light-yellow line] (55.center) to (57.center);
		
		\node [] (62) at (-2+\diagramoneposxb, 1+\diagramoneposyb) {};
		\node [] (63) at (-4.5+\diagramoneposxb, -1.75+\diagramoneposyb) {};
		\draw [style=light-green line] (62.center) to (63.center);
		\node [] (64) at (2+\diagramoneposxb, 1+\diagramoneposyb) {};
		\node [] (65) at (4.5+\diagramoneposxb, -1.75+\diagramoneposyb) {};
		\draw [style=light-green line] (64.center) to (65.center);
		\node [] (66) at (-6+\diagramoneposxb, 1+\diagramoneposyb) {};
		\node [] (67) at (-4.5+\diagramoneposxb, 1.75+\diagramoneposyb) {};
		\draw [style=light-green line] (66.center) to (67.west);
		\node [] (68) at (6+\diagramoneposxb, 1+\diagramoneposyb) {};
		\node [] (69) at (4.5+\diagramoneposxb, 1.75+\diagramoneposyb) {};
		\draw [style=light-green line] (68.center) to (69.east);
		\draw [style=light-green line] (62.center) to (64.center);

        \draw [style=black line] (8.center) to (9.center);
        \draw [style=black line] (10.center) to (11.center);
        \draw [style=black line] (12.center) to (13.center);
        \draw [style=black line] (14.center) to (15.center);
        \draw [style=black line] (13.center) to (8.center);
        \draw [style=black line] (8.center) to (14.center);
        \draw [style=black line] (14.center) to (10.center);
        \draw [style=black line] (12.center) to (9.center);
        \draw [style=black line] (9.center) to (15.center);
        \draw [style=black line] (15.center) to (11.center);
        
		\node [cross-red=5pt, rotate=45, anchor=center] (100) at (-2.909091+\diagramoneposxb,0+\diagramoneposyb) {};
		\node [cross-red=5pt, rotate=45, anchor=center] (101) at (2.909091+\diagramoneposxb,0+\diagramoneposyb) {};

    		
		\def\diagramoneposxc{-2}
		\def\diagramoneposyc{-2.5}
    
		\node [] (19) at (0+\diagramoneposxc, 1.75+\diagramoneposyc) {};
		\node [] (20) at (0+\diagramoneposxc, -1.75+\diagramoneposyc) {};
		\node [] (21) at (4+\diagramoneposxc, 1.75+\diagramoneposyc) {};
		\node [] (22) at (4+\diagramoneposxc, -1.75+\diagramoneposyc) {};
		\node [] (23) at (-4+\diagramoneposxc, -1.75+\diagramoneposyc) {};
		\node [] (24) at (-4+\diagramoneposxc, 1.75+\diagramoneposyc) {};
		
		\node [label={[yshift=0cm]$\{e_{0} = 0\}_{\makebox[0pt]{$\scriptstyle \;\;\mathcal{B}$}}$}] (25) at (-2+\diagramoneposxc, 1.75+\diagramoneposyc) {};
        \node [label={[yshift=0cm]$\{e_{1} = 0\}_{\makebox[0pt]{$\scriptstyle \;\;\mathcal{B}$}}$}] (26) at (2+\diagramoneposxc, 1.75+\diagramoneposyc) {};
		
		\node [] (70) at (2.5+\diagramoneposxc, 1.75+\diagramoneposyc) {};
		\node [] (71) at (0+\diagramoneposxc, 0+\diagramoneposyc) {};
		\draw [style=light-yellow line] (70.center) to (71.center);
		\node [] (72) at (-2.5+\diagramoneposxc, 1.75+\diagramoneposyc) {};
		\node [] (73) at (0+\diagramoneposxc, 0+\diagramoneposyc) {};
		\draw [style=light-yellow line] (72.center) to (73.center);
		\node [] (74) at (-2.5+\diagramoneposxc, -1.75+\diagramoneposyc) {};
		\node [] (75) at (-4+\diagramoneposxc, -1+\diagramoneposyc) {};
		\draw [style=light-yellow line] (74.west) to (75.center);
		\node [] (76) at (2.5+\diagramoneposxc, -1.75+\diagramoneposyc) {};
		\node [] (77) at (4+\diagramoneposxc, -1+\diagramoneposyc) {};
		\draw [style=light-yellow line] (76.east) to (77.center);
		
		\node [] (78) at (0+\diagramoneposxc, 0+\diagramoneposyc) {};
		\node [] (79) at (-2.5+\diagramoneposxc, -1.75+\diagramoneposyc) {};
		\draw [style=light-green line] (78.center) to (79.center);
		\node [] (80) at (0+\diagramoneposxc, 0+\diagramoneposyc) {};
		\node [] (81) at (2.5+\diagramoneposxc, -1.75+\diagramoneposyc) {};
		\draw [style=light-green line] (80.center) to (81.center);
		\node [] (82) at (-4+\diagramoneposxc, 1+\diagramoneposyc) {};
		\node [] (83) at (-2.5+\diagramoneposxc, 1.75+\diagramoneposyc) {};
		\draw [style=light-green line] (82.center) to (83.west);
		\node [] (84) at (4+\diagramoneposxc, 1+\diagramoneposyc) {};
		\node [] (85) at (2.5+\diagramoneposxc, 1.75+\diagramoneposyc) {};
		\draw [style=light-green line] (84.center) to (85.east);

        \draw [style=black line] (19.center) to (20.center);
		\draw [style=black line] (24.center) to (19.center);
		\draw [style=black line] (19.center) to (21.center);
		\draw [style=black line] (21.center) to (22.center);
		\draw [style=black line] (22.center) to (20.center);
		\draw [style=black line] (20.center) to (23.center);
		\draw [style=black line] (23.center) to (24.center);
		
		\node [cross-red=5pt, rotate=45, anchor=center] (102) at (0+\diagramoneposxc,0+\diagramoneposyc) {};
        
    		
		\def\diagramoneposxd{2}
		\def\diagramoneposyd{-7.5}
    
		\node [] (27) at (-2+\diagramoneposxd, 1.75+\diagramoneposyd) {};
		\node [] (28) at (-2+\diagramoneposxd, -1.75+\diagramoneposyd) {};
		\node [] (29) at (6+\diagramoneposxd, 1.75+\diagramoneposyd) {};
		\node [] (30) at (6+\diagramoneposxd, -1.75+\diagramoneposyd) {};
		\node [] (31) at (-6+\diagramoneposxd, -1.75+\diagramoneposyd) {};
		\node [] (32) at (-6+\diagramoneposxd, 1.75+\diagramoneposyd) {};
		\node [] (33) at (2+\diagramoneposxd, 1.75+\diagramoneposyd) {};
		\node [] (34) at (2+\diagramoneposxd, -1.75+\diagramoneposyd) {};
		
		\node [label={[yshift=0cm]$\{e_{0} = 0\}_{\makebox[0pt]{$\scriptstyle \;\;\mathcal{B}$}}$}] (35) at (-4+\diagramoneposxd, 1.75+\diagramoneposyd) {};
        \node [label={[yshift=0cm]$\{e_{1} = 0\}_{\makebox[0pt]{$\scriptstyle \;\;\mathcal{B}$}}$}] (36) at (0+\diagramoneposxd, 1.75+\diagramoneposyd) {};
        \node [label={[yshift=0cm]$\{e_{2} = 0\}_{\makebox[0pt]{$\scriptstyle \;\;\mathcal{B}$}}$}] (37) at (4+\diagramoneposxd, 1.75+\diagramoneposyd) {};
		
		\node [] (86) at (4.5+\diagramoneposxd, 1.75+\diagramoneposyd) {};
		\node [] (87) at (2+\diagramoneposxd, 0+\diagramoneposyd) {};
		\draw [style=light-yellow line] (86.center) to (87.center);
		\node [] (88) at (-4.5+\diagramoneposxd, 1.75+\diagramoneposyd) {};
		\node [] (89) at (-2+\diagramoneposxd, 0+\diagramoneposyd) {};
		\draw [style=light-yellow line] (88.center) to (89.center);
		\node [] (90) at (-4.5+\diagramoneposxd, -1.75+\diagramoneposyd) {};
		\node [] (91) at (-6+\diagramoneposxd, -1+\diagramoneposyd) {};
		\draw [style=light-yellow line] (90.west) to (91.center);
		\node [] (92) at (4.5+\diagramoneposxd, -1.75+\diagramoneposyd) {};
		\node [] (93) at (6+\diagramoneposxd, -1+\diagramoneposyd) {};
		\draw [style=light-yellow line] (92.east) to (93.center);
		\draw [style=medium-red zigzag] (87.center) to (89.center);
		
		\node [] (90) at (-2+\diagramoneposxd, 0+\diagramoneposyd) {};
		\node [] (91) at (-4.5+\diagramoneposxd, -1.75+\diagramoneposyd) {};
		\draw [style=light-green line] (90.center) to (91.center);
		\node [] (92) at (2+\diagramoneposxd, 0+\diagramoneposyd) {};
		\node [] (93) at (4.5+\diagramoneposxd, -1.75+\diagramoneposyd) {};
		\draw [style=light-green line] (92.center) to (93.center);
		\node [] (94) at (-6+\diagramoneposxd, 1+\diagramoneposyd) {};
		\node [] (95) at (-4.5+\diagramoneposxd, 1.75+\diagramoneposyd) {};
		\draw [style=light-green line] (94.center) to (95.west);
		\node [] (96) at (6+\diagramoneposxd, 1+\diagramoneposyd) {};
		\node [] (97) at (4.5+\diagramoneposxd, 1.75+\diagramoneposyd) {};
		\draw [style=light-green line] (96.center) to (97.east);

        \draw [style=black line] (27.center) to (28.center);
        \draw [style=black line] (29.center) to (30.center);
        \draw [style=black line] (31.center) to (32.center);
        \draw [style=black line] (33.center) to (34.center);
        \draw [style=black line] (32.center) to (27.center);
        \draw [style=black line] (27.center) to (33.center);
        \draw [style=black line] (33.center) to (29.center);
        \draw [style=black line] (31.center) to (28.center);
        \draw [style=black line] (28.center) to (34.center);
        \draw [style=black line] (34.center) to (30.center);
		
		
		\node [] (105) at (4.25+\diagramoneposxa, 0+\diagramoneposya) {};
		\node [] (106) at (4+\diagramoneposxb, 2.5+\diagramoneposyb) {};
		\draw[latex-latex, rounded corners=0pt] (105.east) -| (106.north)  node[pos=0.25, above, rotate=0] {base change};
		
		\node [] (107) at (-3.5+\diagramoneposxa, -2+\diagramoneposya) {};
		\node [] (108) at (-3.5+\diagramoneposxc, 2+\diagramoneposyc) {};
		\draw[latex-latex] (107.south) -- (108.north) node[midway, above, rotate=90] {complex structure deformation};
		
		\node [] (109) at (4+\diagramoneposxb, -2+\diagramoneposyb) {};
		\node [] (110) at (4+\diagramoneposxd, 2.5+\diagramoneposyd) {};
		\draw[latex-latex] (109.south) -- (110.north) node[midway, above, rotate=270] {complex structure deformation};
		
		\node [] (111) at (-3.5+\diagramoneposxc, -2+\diagramoneposyc) {};
		\node [] (112) at (-6.25+\diagramoneposxd, 0+\diagramoneposyd) {};
		\draw[latex-latex, rounded corners=0pt] (111.south) |- (112.west) node[pos=0.25, above, rotate=90] {base change};
    \end{tikzpicture}
    \caption{Base changing a horizontal Type II.a model has the effect of stretching the 24 points of intersection of the discriminant with $B^{0} \cap B^{1}$ into local vertical branes. As a consequence, complex structure deformations moving these points of intersection and stacking them together are equivalent to local enhancements over the intermediate components. In the figure, we show how this makes moving finite-distance codimension-two non-minimal points to $B^{0} \cap B^{1}$ lead to a local non-minimal codimension-one enhancement in the intermediate components of the equivalent model. Hence, tuning a non-minimal singularity of the heterotic K3 surface corresponds to a complex structure infinite-distance limit.}
    \label{fig:K3-branes-intermediate-components}
\end{figure}

To summarise, although tuning the non-minimal heterotic K3 singularity occurs in codimen\-sion-two in the original degeneration on the F-theory side, the required complex structure deformation  is equivalent to tuning a codimension-one infinite-distance degeneration in a base changed model. We schematically depict this in \cref{fig:K3-branes-intermediate-components}. Hence, we conclude that tuning a non-minimal heterotic K3 singularity corresponds indeed to taking a complex structure infinite-distance limit.

\section{Partial decompactification and weak coupling limits}
\label{sec:horizontal-models-systematics}

After having analysed the asymptotic physics of horizontal Type II.a models in the adiabatic regime, we now move to the study of the remaining types of horizontal models, according to the classification of \cref{sec:classification-horizontal-models}. We briefly comment on horizontal Type II.b models in \cref{sec:horizontal-IIb-models}, which simply are an instance of the Sen limit. Horizontal Type III.a models are analysed in \cref{sec:horizontal-IIIa-models}; they lead, in the adiabatic regime, to a partial decompactification from six to nine dimensions, with the asymptotic theory containing localised defect algebras. We conclude by turning our attention to horizontal Type III.b models in \cref{sec:horizontal-IIIb-models}. They are found to be a total decompactification from six to ten dimensions presenting, once again, localised defect algebras. In \cref{sec:horizontal-IIIb-models-orientifold} we compare horizontal Type III.a and Type III.b models from the point of view of the perturbative Type IIB picture. Some comments on the degenerate K3 double covers that result when insisting on a Sen limit presentation of codimension-one infinite-distance degenerations are relegated to \cref{sec:horizontal-IIIb-models-orientifold-geometry}.

\subsection{Horizontal Type II.b models}
\label{sec:horizontal-IIb-models}

Horizontal Type II.b models correspond to those degenerations $\hat{\rho}: \hat{\mathcal{Y}} \rightarrow D$ of Hirzebruch models in which the central fiber $\hat{Y}_{0}$ of the degeneration develops codimension-zero $\mathrm{I}_{m>0}$ type fibers without presenting any infinite-distance non-minimal vanishing loci.\footnote{For this reason, we can simply refer to this class of degenerations as Type II.b models, since no horizontal, vertical or mixed non-minimal curve appears in $\hat{B}_{0}$. We have nevertheless termed them horizontal Type II.b models because they are a relative version of Kulikov Type II.b models, and hence fit into the classification of \cref{sec:classification-horizontal-models}.} Hence, they are simply an infinite-distance weak coupling limit: the Sen limit \cite{Sen:1997gv} of six-dimensional F-theory models constructed over $\hat{B} = \mathbb{F}_{n}$. Their endpoint is an emergent string limit, the resulting theory corresponding to the perturbative Type IIB orientifold compactification to six-dimensions arising from the F-theory model at weak coupling. The constraints on the 7-brane content discussed in \cref{sec:horizontal-local-global-brane-content,sec:physical-interpretation-of-the-constraints} with the multi-component models in mind also apply here: only gauge enhancements associated with Kodaira type $\mathrm{I}_{m}$ and $\mathrm{I}_{m}^{*}$ fibers, except for $\mathrm{I}_{0}^{*\, \mathrm{ns}}$ fibers, can be realised, in agreement with the perturbative Type IIB picture. The Sen limit has been extensively studied in the F-theory literature, see \cite{Weigand:2018rez} for a review and references, and we do not comment further on it. With applications to the analysis of Type III.b models in mind, we review the concrete geometry relevant for the Sen limit of six-dimensional F-theory models having $B = \mathbb{F}_{n}$ as the base of their internal space in \cref{sec:Sen-limit-six-dimensional-F-theory}.

\subsection{Horizontal Type III.a models}
\label{sec:horizontal-IIIa-models}

Horizontal Type III.a models, as defined in \cref{sec:classification-horizontal-models}, are those whose open-chain resolution $\rho: \mathcal{Y} \rightarrow D$ has generic vertical slices $\sigma: \mathcal{Z} \rightarrow D$ corresponding to Kulikov Type III.a models. Their central fiber is of the form $Y_{0} = \bigcup_{p=0}^{P} Y^{p}$ with $P \geq 1$, and is characterised by having at least one component at weak coupling and at least one component at strong coupling, i.e.\ $n_{p} > 0$ for some $p \in \{0, \dotsc, P\}$ and $n_{q} = 0$ for some $p \neq q \in \{0, \dotsc, P\}$, expressing it in terms of the pattern of codimension-zero singular elliptic fibers. Given the effectiveness bounds \eqref{eq:horizontal-model-effectiveness-constraint-horizontal}, this implies that all intermediate components must be at weak coupling, with at least one of the end-components at strong coupling. This last fact means that they do not correspond to global weak coupling limits and can hence be constructed quite generally without being subject to the types of constraints discussed in \cref{sec:horizontal-restrictions-global-weak-coupling}; comparisons to their geometrically reminiscent global weak coupling limit counterparts are drawn in \cref{sec:horizontal-IIIb-models-orientifold}.

Unlike horizontal Type II.a models, the heterotic dual of horizontal Type III.a models is more obscure. The geometry on the F-theory side is, however, well described by the preceding facts, and we can use it to perform a fiberwise extension of the asymptotic physics of the generic vertical slices in the adiabatic regime, in the same spirit as the discussion of \cref{sec:horizontal-IIa-F0}. When $P =1$ the models can be seen as a further deformation of a horizontal Type II.a model, and we can make some further comments from the heterotic dual perspective.

\subsubsection{Generic vertical slices}
\label{sec:horizontal-IIIa-models-generic-vertical-slices}

Let us start by analysing the asymptotic physics of the would-be eight-dimensional F-theory models associated with the generic vertical slices of horizontal Type III.a models, to see how they paint a consistent picture that will later inform us about the bulk physics at the endpoint of the limit. We conclude the section by commenting on how this generic vertical slice point of view fails over certain loci, at which the purely six-dimensional features of the model will be concentrated, according to our experience from \cref{sec:horizontal-models-heterotic-duals}.

As mentioned earlier, horizontal Type III.a models contain components at weak cou\-pling, which means that their discriminant $\Delta$ and their modified discriminant $\Delta'$ are no longer identical.\footnote{Let us recall that the modified discriminant $\Delta'$ is obtained from the discriminant $\Delta$ of the Weierstrass model of $\mathcal{Y}$ after factoring out the components associated with the codimension-zero singular elliptic fibers in $Y_{0}$, i.e.\ after subtracting the background value of the axio-dilaton. $\Dphys$ is then obtained as the restriction of $\Delta'$ to the base $B_{0}$ of $Y_{0}$. See \cite{Alvarez-Garcia:2023gdd} for definitions and examples.} The information obtained from $\Delta$ concerns the pattern of codimension-zero singular elliptic fibers; for the generic vertical slice, the distribution of weak and strong coupling components is the same as in the full central fiber of the horizontal Type III.a model, as was explained in \cref{sec:classification-horizontal-models}. The restrictions $\{ \Delta'_{p} \}_{0 \leq p \leq P}$ of $\Delta'$ to the components, which are the same as the restrictions of $\Dphys$, provide us with the information about the 7-brane content after the background value of the axio-dilaton has been subtracted. The number of 7-branes in the eight-dimensional model associated with the generic vertical slice, which stem from the restriction of horizontal and mixed branes to the chosen representative of $\mathcal{F}$, are
\begin{subequations}
\begin{align}
	\Delta'_{0} \cdot \left. \mathcal{F} \right|_{E_{0}} &= 12 + n_{0} - n_{1}\,,\\
	\Delta'_{p} \cdot \left. \mathcal{F} \right|_{E_{p}} &= 2n_{p} - n_{p-1} - n_{p+1}\,,\qquad p = 1, \dotsc, P-1,,\\
	\Delta'_{P} \cdot \left. \mathcal{F} \right|_{E_{P}} &= 12 + n_{P} - n_{P-1}\,.
\end{align}
\label{eq:Type-IIIa-horizontal-branes}%
\end{subequations}
Their total amount coincides with the number of 7-branes in a conventional F-theory model in eight dimensions
\begin{equation}
    \Dphys \cdot \mathcal{F} = \sum_{p=0}^{P} \Delta'_{p} \cdot \left. \mathcal{F} \right|_{E_{p}} = 24\,,
\label{eq:Type-IIIa-total-number-horizontal-branes}
\end{equation}
but their distribution is altered by the pattern of codimension-zero singular elliptic fibers. Not only that, but the types of 7-branes that can be found in each component also depend on said pattern, as was analysed from a six-dimensional standpoint in \cref{sec:horizontal-local-global-brane-content}. By restricting the six-dimensional constraints on the (local) 7-brane content to the generic vertical slice, we consistently reproduce the results found in \cite{Lee:2021qkx,Lee:2021usk} for Kulikov Type III.a models: The intermediate components $\{ Y^{p} \}_{1 \leq p \leq P-1}$ can only present singular elliptic fibers of A type supported over representatives of the classes $\{ T_{p} \}_{1 \leq p \leq P-1}$ or multiples of them, which restrict to 7-branes of the corresponding type in the generic vertical slice. Those end-components $\{ Y^{p} \}_{p=0,P}$ at local strong coupling can exhibit singular elliptic fibers of any of the Kodaira types, and this property descends to the 7-branes in the generic vertical slice. If an end-component $\{ Y^{p} \}_{p=0,P}$ is at local weak coupling the defining polynomials of the Weierstrass model of $Y^{p}$ must fulfil the accidental cancellation structure \eqref{eq:codimension-zero-accidental-cancellation}, meaning that $f_{p} = -3h_{p}^{2}$ and $g_{p} = 2h_{p}^{3}$. This leads to D type singular elliptic fibers supported over the (generically irreducible) divisor $\{ h_{p} = 0 \}_{B^{p}}$, which due to
\begin{equation}
    \{ h_{0} = 0\}_{B^{0}} \cdot \left. \mathcal{F} \right|_{E_{0}} = 2\qquad \text{or}\qquad \{ h_{P} = 0\}_{B^{P}} \cdot \left. \mathcal{F} \right|_{E_{P}} = 2
\end{equation}
yield 2 singular elliptic fibers of D type in the generic vertical slice. Additionally, there may be A type singular elliptic fibers supported over representatives of $T_{0}$ or $T_{P}$ and $S_{P}$, depending on the end-component under consideration, which restrict to the corresponding 7-brane type in the generic vertical slice. Altogether, this reproduces Table 3.2 of \cite{Lee:2021qkx}.

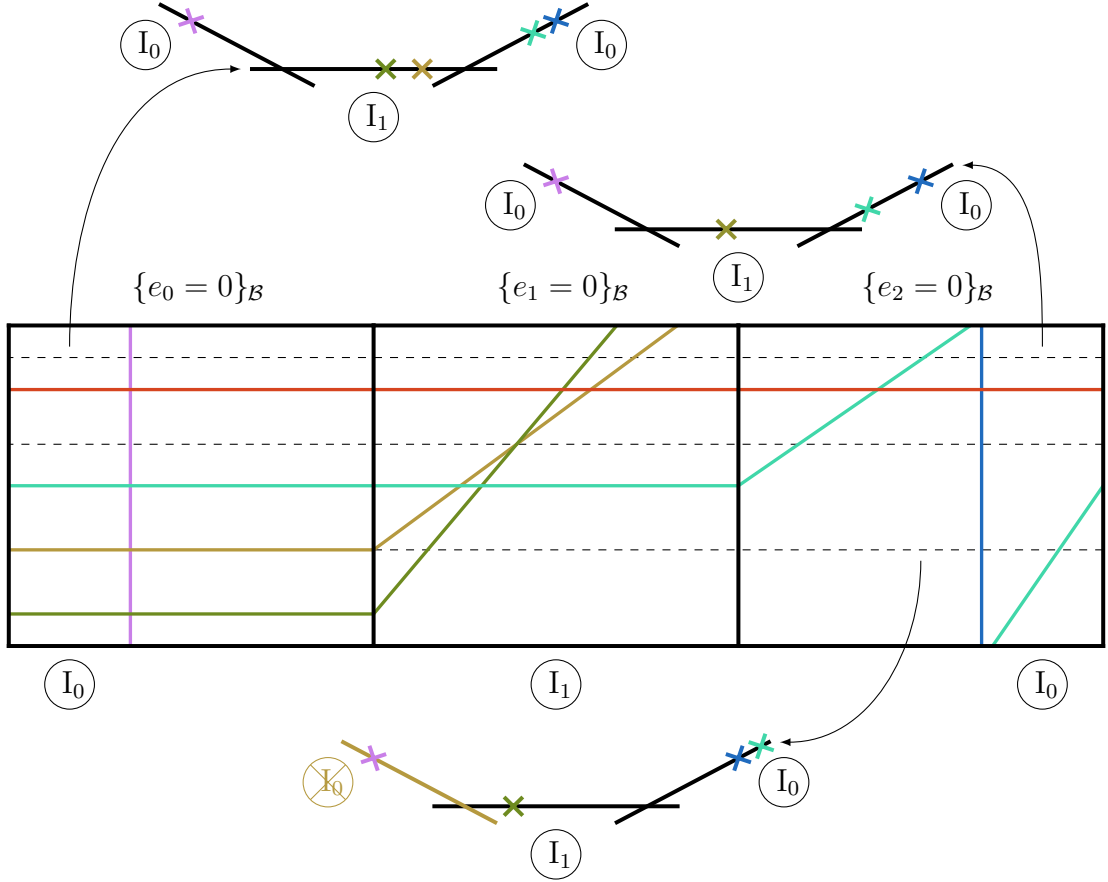
\begin{figure}[t!]
    \centering
	\begin{tikzpicture}
		\def\scalereliiiahor{0.8}
		\def\scalereliiiaver{0.85}
		
		\node (44) at (9*\scalereliiiahor, 2*\scalereliiiaver) {};
		\node (45) at (-9*\scalereliiiahor, 2*\scalereliiiaver) {};
		\draw [style=dashed] (44.center) to (45.center);
		
		\node (46) at (9*\scalereliiiahor, 0.647059*\scalereliiiaver) {};
		\node (47) at (-9*\scalereliiiahor, 0.647059*\scalereliiiaver) {};
		\draw [style=dashed] (46.center) to (47.center);
		
		\node (46) at (9*\scalereliiiahor, -1*\scalereliiiaver) {};
		\node (47) at (-9*\scalereliiiahor, -1*\scalereliiiaver) {};
		\draw [style=dashed] (46.center) to (47.center);
		
		\node (26) at (7*\scalereliiiahor, 2.5*\scalereliiiaver) {};
		\node (27) at (7*\scalereliiiahor, -2.5*\scalereliiiaver) {};
		\node (28) at (-7*\scalereliiiahor, 2.5*\scalereliiiaver) {};
		\node (29) at (-7*\scalereliiiahor, -2.5*\scalereliiiaver) {};
		\node (30) at (2*\scalereliiiahor, 2.5*\scalereliiiaver) {};
		\node (31) at (-3*\scalereliiiahor, -1*\scalereliiiaver) {};
		\node (32) at (-9*\scalereliiiahor, -1*\scalereliiiaver) {};
		\node (33) at (1*\scalereliiiahor, 2.5*\scalereliiiaver) {};
		\node (34) at (-3*\scalereliiiahor, -2*\scalereliiiaver) {};
		\node (35) at (-9*\scalereliiiahor, -2*\scalereliiiaver) {};
		\node (36) at (7*\scalereliiiahor, 2.5*\scalereliiiaver) {};
		\node (37) at (3*\scalereliiiahor, 0*\scalereliiiaver) {};
		\node (38) at (-9*\scalereliiiahor, 0*\scalereliiiaver) {};
		\node (39) at (-9*\scalereliiiahor, 1.5*\scalereliiiaver) {};
		\node (40) at (9*\scalereliiiahor, 1.5*\scalereliiiaver) {};
		\node (48) at (9*\scalereliiiahor, 0*\scalereliiiaver) {};
		
		\draw [style=light-blue line] (26.center) to (27.center);
		\draw [style=light-purple line] (28.center) to (29.center);
		\draw [style=light-yellow line] (30.center) to (31.center);
		\draw [style=light-yellow line] (31.center) to (32.center);
		\draw [style=light-green line] (33.center) to (34.center);
		\draw [style=light-green line] (34.center) to (35.center);
		\draw [style=cyan line] (36.west) to (37.center);
		\draw [style=cyan line] (37.center) to (38.center);
		\draw [style=cyan line] (27.east) to (48.center);
		\draw [style=medium-red line] (39.center) to (40.center);

		\node (0) at (-9*\scalereliiiahor, 2.5*\scalereliiiaver) {};
		\node (1) at (-3*\scalereliiiahor, 2.5*\scalereliiiaver) {};
		\node (2) at (3*\scalereliiiahor, 2.5*\scalereliiiaver) {};
		\node (3) at (9*\scalereliiiahor, 2.5*\scalereliiiaver) {};
		\node (4) at (9*\scalereliiiahor, -2.5*\scalereliiiaver) {};
		\node (5) at (3*\scalereliiiahor, -2.5*\scalereliiiaver) {};
		\node (6) at (-3*\scalereliiiahor, -2.5*\scalereliiiaver) {};
		\node (7) at (-9*\scalereliiiahor, -2.5*\scalereliiiaver) {};
		
		\node (8) at (-6.5*\scalereliiiahor, 7.5*\scalereliiiaver) {};
		\node (9) at (-4*\scalereliiiahor, 6.25*\scalereliiiaver) {};
		\node (10) at (-5*\scalereliiiahor, 6.5*\scalereliiiaver) {};
		\node (11) at (-1*\scalereliiiahor, 6.5*\scalereliiiaver) {};
		\node (12) at (-2*\scalereliiiahor, 6.25*\scalereliiiaver) {};
		\node (13) at (0.5*\scalereliiiahor, 7.5*\scalereliiiaver) {};
		
		\node (14) at (-0.5*\scalereliiiahor, 5.0*\scalereliiiaver) {};
		\node (15) at (2*\scalereliiiahor, 3.75*\scalereliiiaver) {};
		\node (16) at (1*\scalereliiiahor, 4.0*\scalereliiiaver) {};
		\node (17) at (5*\scalereliiiahor, 4.0*\scalereliiiaver) {};
		\node (18) at (4*\scalereliiiahor, 3.75*\scalereliiiaver) {};
		\node (19) at (6.5*\scalereliiiahor, 5.0*\scalereliiiaver) {};
		
		\node (20) at (-3.5*\scalereliiiahor, -4.0*\scalereliiiaver) {};
		\node (21) at (-1*\scalereliiiahor, -5.25*\scalereliiiaver) {};
		\node (22) at (-2*\scalereliiiahor, -5.0*\scalereliiiaver) {};
		\node (23) at (2*\scalereliiiahor, -5.0*\scalereliiiaver) {};
		\node (24) at (1*\scalereliiiahor, -5.25*\scalereliiiaver) {};
		\node (25) at (3.5*\scalereliiiahor, -4.0*\scalereliiiaver) {};

		\draw [style=black line] (0.center) to (1.center);
		\draw [style=black line] (1.center) to (2.center);
		\draw [style=black line] (2.center) to (3.center);
		\draw [style=black line] (7.center) to (6.center);
		\draw [style=black line] (6.center) to (5.center);
		\draw [style=black line] (5.center) to (4.center);
		\draw [style=black line] (0.center) to (7.center);
		\draw [style=black line] (1.center) to (6.center);
		\draw [style=black line] (2.center) to (5.center);
		\draw [style=black line] (3.center) to (4.center);
		
		\draw [style=black line] (10.center) to (11.center);
		\draw [style=black line] (8.center) to (9.center);
		\draw [style=black line] (12.center) to (13.center);
		
		\draw [style=black line] (16.center) to (17.center);
		\draw [style=black line] (14.center) to (15.center);
		\draw [style=black line](18.center) to (19.center);
		
		\draw [style=black line] (22.center) to (23.center);
		\draw [style=light-yellow thick-line] (20.center) to (21.center);
		\draw [style=black line] (24.center) to (25.center);
		
		\node [label={[yshift=0cm]$\{e_{0} = 0\}_{\makebox[0pt]{$\scriptstyle \;\;\mathcal{B}$}}$}] (41) at (-6*\scalereliiiahor, 2.5*\scalereliiiaver) {};
        \node [label={[yshift=0cm]$\{e_{1} = 0\}_{\makebox[0pt]{$\scriptstyle \;\;\mathcal{B}$}}$}] (42) at (0*\scalereliiiahor, 2.5*\scalereliiiaver) {};
        \node [label={[yshift=0cm]$\{e_{2} = 0\}_{\makebox[0pt]{$\scriptstyle \;\;\mathcal{B}$}}$}] (43) at (6*\scalereliiiahor, 2.5*\scalereliiiaver) {};
        
        \node[shape=circle,draw,inner sep=1pt] (73) at (-8*\scalereliiiahor, -3.1*\scalereliiiaver) {$\phantom{.}\mathrm{I}_{0}$};
        \node[shape=circle,draw,inner sep=1pt] (74) at (0*\scalereliiiahor, -3.1*\scalereliiiaver) {$\phantom{.}\mathrm{I}_{1}$};
        \node[shape=circle,draw,inner sep=1pt] (75) at (8*\scalereliiiahor, -3.1*\scalereliiiaver) {$\phantom{.}\mathrm{I}_{0}$};
        
        \node[shape=circle,draw,inner sep=1pt] (76) at (-6.75*\scalereliiiahor, 6.875*\scalereliiiaver) {$\phantom{.}\mathrm{I}_{0}$};
        \node[shape=circle,draw,inner sep=1pt] (77) at (-3*\scalereliiiahor, 5.75*\scalereliiiaver) {$\phantom{.}\mathrm{I}_{1}$};
        \node[shape=circle,draw,inner sep=1pt] (78) at (0.75*\scalereliiiahor, 6.875*\scalereliiiaver) {$\phantom{.}\mathrm{I}_{0}$};
        
        \node[shape=circle,draw,inner sep=1pt] (79) at (-0.75*\scalereliiiahor, 4.375*\scalereliiiaver) {$\phantom{.}\mathrm{I}_{0}$};
        \node[shape=circle,draw,inner sep=1pt] (80) at (3*\scalereliiiahor, 3.25*\scalereliiiaver) {$\phantom{.}\mathrm{I}_{1}$};
        \node[shape=circle,draw,inner sep=1pt] (81) at (6.75*\scalereliiiahor, 4.375*\scalereliiiaver) {$\phantom{.}\mathrm{I}_{0}$};
        
        \node[shape=circle,cross-node,draw,inner sep=1pt,diagLightYellow] (82) at (-3.75*\scalereliiiahor, -4.625*\scalereliiiaver) {\textcolor{diagLightYellow}{$\phantom{.}\mathrm{I}_{0}$}};
        \node[shape=circle,draw,inner sep=1pt] (83) at (0*\scalereliiiahor, -5.75*\scalereliiiaver) {$\phantom{.}\mathrm{I}_{1}$};
        \node[shape=circle,draw,inner sep=1pt] (84) at (3.75*\scalereliiiahor, -4.625*\scalereliiiaver) {$\phantom{.}\mathrm{I}_{0}$};
        
		\node [cross-purple=5pt, rotate=-26.5651, anchor=center] (59) at (-6*\scalereliiiahor,7.25*\scalereliiiaver) {};
		\node [cross-blue=5pt, rotate=26.5651, anchor=center] (60) at (0*\scalereliiiahor,7.25*\scalereliiiaver) {};
		\node [cross-cyan=5pt, rotate=26.5651, anchor=center] (85) at (-0.375*\scalereliiiahor,7.0625*\scalereliiiaver) {};
		\node [cross-yellow=5pt, rotate=0, anchor=center] (61) at (-2.2*\scalereliiiahor,6.5*\scalereliiiaver) {};
		\node [cross-green=5pt, rotate=0, anchor=center] (62) at (-2.8*\scalereliiiahor,6.5*\scalereliiiaver) {};
		
		\node [cross-purple=5pt, rotate=-26.5651, anchor=center] (63) at (0*\scalereliiiahor,4.75*\scalereliiiaver) {};
		\node [cross-blue=5pt, rotate=26.5651, anchor=center] (64) at (6*\scalereliiiahor,4.75*\scalereliiiaver) {};
		\node [cross-cyan=5pt, rotate=26.5651, anchor=center] (86) at (5.125*\scalereliiiahor,4.3125*\scalereliiiaver) {};
		\node [cross-yellow-green=5pt, rotate=0, anchor=center] (65) at (2.8*\scalereliiiahor,4*\scalereliiiaver) {};
		
		\node [cross-purple=5pt, rotate=-26.5651, anchor=center] (66) at (-3*\scalereliiiahor,-4.25*\scalereliiiaver) {};
		\node [cross-blue=5pt, rotate=26.5651, anchor=center] (67) at (3*\scalereliiiahor,-4.25*\scalereliiiaver) {};
		\node [cross-cyan=5pt, rotate=26.5651, anchor=center] (68) at (3.375*\scalereliiiahor,-4.0625*\scalereliiiaver) {};
		\node [cross-green=5pt, rotate=0, anchor=center] (69) at (-0.7*\scalereliiiahor,-5*\scalereliiiaver) {};
		
		\node (70) at (-8*\scalereliiiahor, 2*\scalereliiiaver) {};
		\draw [-latex] (70.north) to [out=90,in=180] ([xshift=0cm, yshift=0cm]10.west);
		
		\node (71) at (8*\scalereliiiahor, 2*\scalereliiiaver) {};
		\draw [-latex] (71.north) to [out=90,in=0] ([xshift=0cm, yshift=0cm]19.east);
		
		\node (72) at (6*\scalereliiiahor, -1*\scalereliiiaver) {};
		\draw [-latex] (72.south) to [out=270,in=0] ([xshift=0cm, yshift=0cm]25.east);
	\end{tikzpicture}
    \caption{We schematically represent a reduced number of global 7-branes in the base of the central fiber of a resolved horizontal Type III.a model. The particular example depicted is constructed over $\hat{B} = \mathbb{F}_{1}$, and we see that the global 7-branes in the representatives of the class $\mathcal{H}_{\infty}^{1}$ account for the difference in the number of intersection points of the discriminant with the curves $B^{0} \cap B^{1}$ and $B^{1} \cap B^{2}$. The two vertical restrictions shown above the central diagram are generic, while the third one, shown below it, is non-generic, since it overlaps with a global 7-brane in a representative of the class $\mathcal{H}_{\infty}^{1}$.}
    \label{fig:relative-Type-IIIa}
\end{figure}

As the point $p_{b} \in \mathbb{P}_{b}^{1}$ over which the vertical slice is defined changes,  the position of the 7-branes of the associated Kulikov Type III.a model move within their components. While their distribution among the various components has a tangible physical effect, their position within them is not relevant for the eight-dimensional asymptotic physics and, as a consequence, all generic vertical slices of a horizontal Type III.a model lead to a consistent picture: As we will argue in \cref{sec:horizontal-IIIa-models-adiabatic-regime}, the bulk physics of the asymptotic model resembles that of Kulikov Type~III.a models, which we reviewed in \cref{sec:K3-degenerations}. We provide a depiction of the central fiber of a horizontal Type III.a model and several global 7-branes within it, as well as a number of generic and non-generic vertical slices, in \cref{fig:relative-Type-IIIa}.

There are non-generic representatives of $\mathcal{F}$ that do not lead to this coherent interpretation: They are the loci over which the adiabaticity assumption must fail and where the six-dimensional features of the asymptotic physics will hence be concentrated at. Each pair of adjacent components $Y^{p}$ and $Y^{p+1}$ of the central fiber $Y_{0}$ intersects over a K3 surface. This can be seen from the fact that all interface curves $\{ B^{p} \cap B^{p+1} \}_{0 \leq p \leq P-1}$ are $\mathbb{P}^{1}$ curves acting as the base of an elliptic fibration in which the elliptic fiber degenerates over a total of
\begin{equation}
	\left. \Delta \right|_{E_{p}} \cdot \left( E_{p} \cap E_{p+1} \right) = \left. \Delta \right|_{E_{p+1}} \cdot \left( E_{p} \cap E_{p+1} \right) = 24\,, \quad p = 0, \dotsc, P-1\,,
\end{equation}
points counted with multiplicity. Since the effectiveness bounds \eqref{eq:horizontal-model-effectiveness-constraint-horizontal} force all intermediate components $\{ Y^{p} \}_{1 \leq p \leq P-1}$ to be at local weak coupling, the interface K3 surfaces present Kodaira type $\mathrm{I}_{m}$ fibers in codimension-zero that are subject to additional A and D type enhancements over points. However, the number of intersections of the physical discriminant $\Dphys$ with the interface curves can differ from one to the other. Using that
\begin{equation}
	\Delta'_{p} = \left. \Delta \right|_{E_{p}} - \left. \sum_{p=0}^{P} n_{p} E_{p} \right|_{E_{p}}\,,
\end{equation}
it can be computed to be
\begin{subequations}
\begin{align}
	\Delta'_{p} \cdot S_{p} &= 24 + n(n_{p+1} - n_{p})\,,\quad p = 0, \dotsc, P-1\,, \label{eq:Type-III.a-intersections-1}\\
	\Delta'_{p} \cdot T_{p} &= 24 + n(n_{p} - n_{p-1})\,,\quad p = 1, \dotsc, P\,, \label{eq:Type-III.a-intersections-2}
\end{align}
\label{eq:Type-III.a-intersections}%
\end{subequations}
where in the first and the second line we compute it from the point of view of the $B^{p}$ and the $B^{p+1}$ component, respectively. Part of the effect of codimension-zero singular elliptic fibers is moving horizontal classes between the restrictions $\{ \Delta'_{p} \}_{0 \leq p \leq P}$ of the modified discriminant; since in a component $B^{p+1}$ a local 7-brane in the class $T_{p+1}$ intersects the interface curve $B^{p} \cap B^{p+1}$, this results in the alteration of the intersection numbers observed above. The representatives of $\mathcal{F}$ passing through these intersection points lead to non-generic vertical slices of the model.

\subsubsection{Asymptotic physics in the adiabatic regime}
\label{sec:horizontal-IIIa-models-adiabatic-regime}

In order to extract the asymptotic physics associated with horizontal Type III.a models we will follow the same strategy employed in the analysis of horizontal Type II.a models: We consider the hierarchy of volumes $\mathcal{V}_{\mathbb{P}_{b}^{1}} \gg \mathcal{V}_{\mathrm{K3}}$ to hold, i.e.\ we study the models in the adiabatic regime. As explained in \cref{sec:horizontal-IIa-away-from-adiabatic-regime}, taking the adiabatic limit makes the geometrical origin of the (dual) KK towers clear; away from this regime the towers that we will identify below still exist and signal a decompactification limit, but they reorganise in a complicated manner, obscuring the unequivocal determination of the endpoint of such a process.

Given the many parallels with the analysis of \cref{sec:horizontal-IIa-F0}, we keep the discussion more concise. In order to start with the simplest relative version of Kulikov Type III.a models, let us assume at first that the horizontal Type III.a models under consideration are constructed over $\hat{B} = \mathbb{F}_{0}$ and that their defining polynomials have the structure \eqref{eq:F0-horizontal-vertical-separation}, i.e.\ only horizontal and vertical enhancements over representatives of the divisor classes $\{ \mathcal{H}_{\infty}^{p} \}_{0 \leq p \leq P}$, $\mathcal{H}_{0}^{P}$ and $\mathcal{F}$ appear.

Consider the vertical slice of a horizontal Type III.a model associated with a generic point $p_{b} := [v_{0}:w_{0}] \in \mathbb{P}_{b}^{1}$. Let us assume without loss of generality that the $Y^{0}$ component has smooth elliptic fibers in codimension-zero. In \cref{sec:horizontal-IIa-F0} the possibility of building the non-trivial shrinking 2-cyles $\{ \gamma_{i} \}_{i=1,2}$ relied on the fact that the monodromy action on the two 1-cycles $\sigma_{i} \in H_{1}(\mathcal{E},\mathbb{Z})$, for $i=1,2$, of the elliptic fiber $\mathcal{E}$ when encircling $\left. (B^{0} \cap B^{1}) \right|_{[v:w]=p_{b}}$ was trivial. This is no longer true for the generic vertical slices of horizontal Type III.a models, but the monodromy action still leaves a 1-cycle $\sigma \in H_{1}(\mathcal{E},\mathbb{Z})$ invariant, allowing for the construction a single non-trivial shrinking 2-cycle $\gamma \in H_{2}(Z_{0},\mathbb{Z})$ on each such slice. Said monodromy action is the same in all generic vertical slices, meaning that the 2-cycle $\gamma$ can be consistently defined in those patches of $Y_{0}$ consisting only of such vertical slices. In this way, we can define a local \mbox{2-cycle} over the generic points of $\mathbb{P}_{b}^{1}$. The M2-branes wrapping the shrinking local 2-cycle $\gamma$ lead to a single tower of asymptotically massless particles, signalling a partial decompactification that is to be superimposed on top of the adiabatic limit. The non-BPS nature of these particles relates to the trivialisation of the local 2-cycle $\gamma$ over the non-generic vertical slices. Note that, if both end-components $\{ Y^{p} \}_{p=0,P}$ have smooth elliptic fibers in codimension-zero, the same construction can be performed on both ends of the open-chain; this still leads to a single local 2-cycle $\gamma$ thanks to the mutual locality of the codimension-zero $\mathrm{I}_{m}$ fibers, which allows us to transport $\gamma$ from one end-component to the other. Compared to the horizontal Type II.a models, we have lost one of the towers of asymptotically massless non-BPS particles associated with the isometries of the heterotic torus fiber. Hence, horizontal Type III.a limits in the adiabatic regime lead to a partial decompactification from six to nine dimensions.

Let us heuristically argue for this picture on the heterotic side of the duality by slightly abusing the dictionary reviewed in \cref{sec:F-theory-heteoric-duality}. To this end, let us focus on those horizontal Type~III.a models for which $P = 1$, which most closely resemble horizontal Type II.a models in their two-component presentation. Starting with a resolved horizontal Type II.a model, the internal space on the heterotic side of the duality corresponds to the K3 surface $Y^{0} \cap Y^{1}$, which undergoes a large volume limit both in the base $\mathcal{V}_{\mathbb{P}_{b}^{1}} \rightarrow \infty$ and in the fiber $\mathcal{V}_{T^{2}_{\mathrm{het}}} \rightarrow \infty$ in a hierarchical way. Deforming the central fiber $Y_{0}$ such that it aligns with the geometry of a horizontal Type III.a model pushes one of its two components to local weak coupling. As a result, the K3 surface $Y^{0} \cap Y^{1}$ develops Kodaira type $\mathrm{I}_{m}$ fibers in codimension-zero. This does not imply that the dual heterotic K3 surface itself develops Kodaira type $\mathrm{I}_{m}$ fibers in codimension-zero, but rather that the generic heterotic torus fiber also undergoes a large complex structure limit $U_{T^{2}_{\mathrm{het}}} \rightarrow \infty$ that competes with the large volume limit $\mathcal{V}_{T^{2}_{\mathrm{het}}} \rightarrow \infty$. In the actual horizontal Type~III.a model the putative generic heterotic torus fiber undergoes both infinite-distance limits at the same time, resulting in a decompactification from six to nine dimensions, instead of to~ten.

Returning to the F-theory discussion, we recall that the bulk physics of the asymptotic model is encapsulated in the patches of $Y_{0}$ containing only generic vertical slices. In the class of models under consideration, the 7-branes in the eight-dimensional model associated to the generic vertical slice stem from the restriction of global 7-branes in the divisor classes $\{ \mathcal{H}_{\infty}^{p} \}_{0 \leq p \leq P}$ and $\mathcal{H}_{0}^{P}$. This set of 7-branes, together with the additional states provided by the tower of asymptotically massless particles, enhance at the endpoint of the limit into a loop algebra, from the six-dimensional point of view. One factor of the asymptotic gauge algebra, which we will denote by $H$, is the gauge algebra associated to those horizontal 7-branes contained in the components at local weak coupling. These must therefore result from the A type branes in the intermediate components and, possibly, the A and D type branes in one of the end-components. In addition to this gauge algebra contribution, we will have one additional factor from each end-component at local weak coupling, consisting of the gauge algebra arising from taking together the set of horizontal 7-branes in the end-component under consideration, since the effect of the resolution process of the degeneration is to artificially separate them. All these factors are subject to a loop enhancement granted by the same imaginary root, as a consequence of the possibility of trivially transporting the local 2-cycle $\gamma$ along the generic vertical slice of the central fiber of the open-chain resolution (without crossing the 7-branes in the end-components).\footnote{Using the string junctions picture, the loop enhancement by a common imaginary root is nicely explained for Kulikov Type III.a models in \cite{Collazuol:2022oey}.} If both end-components $\{ Y^{p} \}_{p=0,P}$ are at local strong coupling, the resulting loop algebra is
\begin{equation}
	G_{\infty} = \left( \hat{\mathrm{E}}_{9-n_{1}} \oplus \hat{H} \oplus \hat{\mathrm{E}}_{9-n_{P-1}} \right)/\sim\,,
\end{equation}
and if only one end-component is at local strong coupling, we have instead
\begin{equation}
    G_{\infty} = \left( \hat{\mathrm{E}}_{9-n_{p}} \oplus \hat{H} \right)/\sim\,,\qquad \text{with}\qquad p = 1, P-1\,.
\end{equation}
From the point of view of the partially decompactified theory these are reinterpreted as the bulk gauge algebras
\begin{equation}
    G_{\mathrm{9D}} = \mathrm{E}_{9-n_{1}} \oplus H \oplus \mathrm{E}_{9-n_{P-1}}\,,\qquad \text{or}\qquad G_{\mathrm{9D}} = \mathrm{E}_{9-n_{p}} \oplus H\,,\quad \text{with}\quad p = 1, P-1\,.
\end{equation}

In the heuristic heterotic dual picture, the asymptotic enhancement to the loop algebra can be understood in terms of the metric $\mathrm{U}(1)_{\mathrm{KK}}$ factor associated with the continuous isometry along the torus radius that is becoming large in the limit. The perturbative heterotic gauge sector is mostly restored as the (non-singular) gauge bundle contributions dilute, but the part of them along the torus direction that remains small still have a tangible effect on the bulk algebra and are responsible for the partial Higgsing observed in the asymptotic model.\footnote{Similarly to how tuning the non-minimal singularities necessary to realise a horizontal Type II.a model tightens the six-dimensional horizontal gauge rank bound $\rank(\mathfrak{g}_{\mathrm{hor}}) \leq 18$ valid at finite-distance to the bound $\rank(\mathfrak{g}_{\mathrm{hor}}) \leq 16$, which matches the dual heterotic interpretation of a decompactification from six to ten dimensions, tuning a horizontal Type III.a models tightens it to $\rank(\mathfrak{g}_{\mathrm{hor}}) \leq 17$ instead, as would correspond for a heterotic dual interpretation as a decompactification from six to nine dimensions.}

We now turn to the non-generic vertical slices of a horizontal Type III.a model. For $\hat{B} = \mathbb{F}_{0}$, and assuming first the structure \eqref{eq:F0-horizontal-vertical-separation}, these are the global vertical 7-branes, which can be brought together in order to realise global vertical enhancements supported over representatives of $\mathcal{F}$. Taking horizontal slices of the model, the same arguments as in \cref{sec:horizontal-IIa-heterotic-dual-interpretation} imply that these gauge algebra contributions remain lower-dimensional at the endpoint of the limit, localised in the worldvolume of six-dimensional defects.

The redistribution of horizontal 7-branes among the components due to the presence of components at local weak coupling reduces the extent to which codimension-two finite-distance non-minimal points can be tuned. As a consequence, the total rank of the gauge algebra factors supported over the exceptional curves arising in the resolution of codimension-two non-minimal points decreases. Heuristically, we can understand this from the heterotic side as a by-product of the residual Higgsing: In order to realise it, part of the instanton number budget $c_{2}(\mathrm{K3}) = 24$ needs to be spent on producing the non-trivial gauge bundle background necessary to realise it, and is therefore not available in order to produce singular gauge bundle contributions capable of probing the geometrical ADE singularities of the internal space. Hence, the bigger the residual Higgsing, the smaller the maximal non-perturbative gauge algebra that can be realised. Similarly, on the F-theory side, the more horizontal 7-branes are moved to the intermediate components, the smaller the vanishing orders that can be attained for the codimension-two non-minimal points and, consequently, the smaller the gauge algebra supported over exceptional curves. In addition, the bounds on the maximal rank of the vertical gauge algebra become stricter in the presence of local weak coupling components, as was explained in \cref{sec:bounds-vertical-gauge-rank}.

Relaxing the condition \eqref{eq:F0-horizontal-vertical-separation} amounts to allowing for mixed enhancements. Their interpretation is as in horizontal Type II.a models discussed in \cref{sec:horizontal-IIa-mixed-enhancements}. Mixed enhancements result from a Higgsing between the perturbative and non-perturbative gauge sectors, using the heterotic language. The Higgsing effect of the perturbative sector is diluted in the limit while only the residual defect algebra survives in the asymptotic model. 

Let us conclude our analysis of horizontal Type III.a models by considering those constructed over $\hat{B} = \mathbb{F}_{n}$, which only require a minor modification of the preceding discussion. The first effect of abandoning the direct product structure for the F-theory base space was already observed for horizontal Type II.a models in \cref{sec:horizontal-IIa-Fn}: While Higgsing the six-dimensional horizontal gauge algebra decreases the maximal rank that can be attained from the gauge sector supported over exceptional curves for all $\hat{B} = \mathbb{F}_{n}$, when $n \geq 1$ the rank of the vertical gauge algebra is also reduced, since $\mathcal{F}$ classes are needed in order to realise the 7-brane recombinations necessary for the horizontal Higgsing. Once we move to horizontal Type III.a models a second, but related, effect occurs: The six-dimensional horizontal gauge algebra is partially Higgsed due to the redistribution of (local) horizontal 7-branes among the components. When $n \geq 1$, this entails part of the horizontal 7-branes that would have formed part of the maximal horizontal gauge algebra in the end-components recombining with a global vertical brane and moving to the intermediate components as mixed branes, which therefore decreases the maximal vertical gauge rank possible in the model (on top of the reduction of the gauge rank of the factors supported over exceptional curves, that was already mentioned for $\hat{B} = \mathbb{F}_{0}$ earlier). This effect was commented on from the geometrical point of view at the end of \cref{sec:bounds-vertical-gauge-rank} using horizontal Type III.a models constructed over $\hat{B} = \mathbb{F}_{7}$ as an example.

\subsection{Horizontal Type III.b models}
\label{sec:horizontal-IIIb-models}

Horizontal Type III.b models are those in which the generic vertical slice $\sigma: \mathcal{Z} \rightarrow D$ of their open-chain resolution $\rho: \mathcal{Y} \rightarrow D$ is a Kulikov Type III.b model. In terms of the pattern of codimension-zero singular elliptic fibers exhibited by their central fiber $Y_{0}$, this class of models is characterised by having all components at local weak coupling, i.e.\ they are global weak coupling limits in which the Type IIB axio-dilaton is driven to infinity $\tau \rightarrow i\infty$.

As was discussed in \cref{sec:horizontal-restrictions-global-weak-coupling}, horizontal Type III.b models can only be constructed over $\hat{B} = \mathbb{F}_{n}$ with $0 \leq n \leq 4$, as can be argued either through their geometrical properties, or by realising that otherwise they would present non-Higgsable clusters supporting exceptional enhancements, which are incompatible with a perturbative Type IIB orientifold picture.

Moreover, the elliptic fibers found in codimension one and higher must be compatible with $j(\tau) \rightarrow \infty$, which imposes strict constraints on the local and global 7-brane content of these models, see \cref{sec:horizontal-local-global-brane-content}. This restricts us to enhancements realising Kodaira type $\mathrm{I}_{m}$ and $\mathrm{I}_{m}^{*}$ fibers. The geometry of horizontal Type III.b models also prevents a $\mathfrak{g}_2$ algebra, associated with $\mathrm{I}_{0}^{*\, \mathrm{ns}}$ fibers, from being realised, by constraining the possible monodromy covers.

\subsubsection{Type IIB orientifold picture}
\label{sec:horizontal-IIIb-models-orientifold}

The central fiber $Y_{0}$ of a horizontal Type III.b model matches the structure of a horizontal Type~III.a model except that both end-components $\{ Y^{p} \}_{p = 0,P}$ have been tuned to be at local weak coupling. Given the similarities between their geometries, it is worth spending a moment comparing them before we completely focus on horizontal Type III.b models.

Horizontal Type III.b models represent global weak coupling limits, and should therefore admit a description in terms of a Sen limit \cite{Sen:1997gv}. A global weak coupling F-theory model obtained by compactification on the elliptic fibration $\pi: Y \rightarrow B$ is interpreted as a Type~IIB orientifold compactification, using the Sen limit language, in the following way: The internal space of the Type IIB theory is given by the Calabi-Yau double cover $\breve{B}$ of the base $B$. The branching locus of the double cover corresponds to the fixed loci of the orientifold involution, which are identified with the O7-planes in the Type IIB internal space. For six-dimensional F-theory models, the resulting double cover $\breve{B}$ is an elliptic K3 surface.

In order to bring a horizontal Type III.b model in this form, we can blow down its central fiber $Y_{0}$ to any of its components $\{ Y^{p} \}_{0 \leq p \leq P}$. The central fiber $\hat{Y}_{0}$ of the resulting degeneration $\hat{\rho}: \hat{\mathcal{Y}} \rightarrow D$ will present, due the global weak coupling nature of horizontal Type III.b models, codimension-zero Kodaira type $\mathrm{I}_{m}$ fibers, and is therefore amenable to the Type IIB orientifold description just discussed above. However, since the blow-down has lead to an unresolved degeneration $\hat{\rho}: \hat{\mathcal{Y}} \rightarrow D$, this central fiber $\hat{Y}_{0}$ supports non-minimal elliptic fibers over some horizontal curves in the base. As a result, the Type IIB internal space $\breve{B}$ is an elliptic K3 surface that has undergone a Kulikov Type II.b degeneration, see \cref{sec:K3-double-cover-horizontal-IIIb-models}. A more detailed account of the process of blowing down the resolved degeneration and constructing the double cover of the base of the central fiber is given in \cref{sec:horizontal-IIIb-models-orientifold-geometry}.

Note that, at least in terms of the geometrical double cover construction, the same Type~IIB orientifold picture can be reached starting from a resolved horizontal Type III.a model. One distinction with the previous case is that horizontal Type III.a models contain at least one component at local strong coupling; the blow down process should be chosen such that the resulting $\hat{Y}_{0}$ corresponds to one of the components $\{ Y^{p} \}_{0 \leq p \leq P}$ at local weak coupling.

In both instances, the end result for the Type IIB internal space is a degenerate elliptic K3 surface $\breve{B}$ of the type appearing as the central fiber of Kulikov Type II.b models. As we see in \cref{sec:K3-double-cover-horizontal-IIIb-models}, the degeneration of this K3 surface is directly tied to the codimension-one non-minimal loci appearing in the blown-down F-theory model; the latter, which correspond to the coalescence of O7-planes in the Type IIB interpretation, force a certain factorization structure for the branching locus of $\breve{B}$. Moving O7-planes on top of each other usually involves strongly coupled dynamics, which competes with the global weak coupling limit. Depending on how they balance against each other, we end up with a horizontal Type III.a or Type III.b model. If the O7-planes are brought together too fast, the strongly coupled nature of the process prevails against the weak coupling limit, and we obtain a horizontal Type III.a model; the central fiber $Y_{0}$ of its open-chain resolution always presents at least one component at local strong coupling as a consequence of this. If the weak coupling limit is instead taken rapidly enough, the model can remain at global weak coupling, and we obtain a horizontal Type III.b model; all components of the central fiber $Y_{0}$ of its open-chain resolution are, for that reason, at local weak coupling. The additional level of tuning necessary to achieve this compared to a horizontal Type III.a model ensures a careful balance between the weak coupling limit and the coalescence process of O7-planes in the Sen limit picture. The same logic underlies the related eight-dimensional F-theory degenerations \cite{Lee:2021usk}.

From a perturbative Type IIB orientifold compactification perspective, the limits under consideration are fairly complicated. Starting from a generic Type IIB orientifold compactification on the K3 double cover, the internal space is subject to a Kulikov Type II.b degeneration at the same time as a limit in the string coupling $g_{\mathrm{IIB}}$ is taken. The shrinking transcendental 2-cycles of the internal space then lead to worldsheet and D-instanton corrections arising from strings wrapped on them. Their importance for the asymptotic physics of the infinite-distance limit depends on the precise way in which $g_{\mathrm{IIB}}$ scales along it. Potentially, the trajectories may even leave the regime of validity of the perturbative description. The analysis in F-theory automatically discriminates between the different possibilities through the resolution process of the degeneration of its internal space.

\subsubsection{Generic vertical slices}
\label{sec:horizontal-IIIb-models-generic-vertical-slices}

As in the analysis of the other horizontal models, let us first study the eight-dimensional \mbox{F-theory} models associated with the generic vertical slices of horizontal Type III.b models. We can be brief because, as emphasised earlier, the geometry corresponds to a further tuning of horizontal Type III.a models analysed in \cref{sec:horizontal-IIIa-models-generic-vertical-slices}. 

The distribution of 7-branes in the eight-dimensional model associated with the generic vertical slice is still described by \eqref{eq:Type-IIIa-horizontal-branes} and \eqref{eq:Type-IIIa-total-number-horizontal-branes}, with the only difference being that now $n_{p} > 0$ for all $p \in \{ 0, \dotsc, P \}$. The restrictions on the 7-brane types that can be found in its components descend again from the constraints on the (local) 7-brane content discussed for six-dimensional models in \cref{sec:horizontal-local-global-brane-content}: All components can contain A type singular elliptic fibers, with the end-components $\{ Y^{p} \}_{p=0,P}$ presenting in addition 2 singular elliptic fibers of D type each.

The non-generic representatives of $\mathcal{F}$ are those passing through the intersection points of $\Dphys$ with the interface curves $\{ B^{p} \cap B^{p+1} \}_{0 \leq p \leq P-1}$ over which the base components of the central fiber meet. The number of these intersection points varies from one pair of adjacent of components to the other due to the redistribution of horizontal classes between the $\{ \Delta'_{p} \}_{0 \leq p \leq P}$, and is given by \eqref{eq:Type-III.a-intersections}.

\subsubsection{Asymptotic physics in the adiabatic regime}
\label{sec:horizontal-IIIb-models-adiabatic-regime}

We now analyse the asymptotic physics of horizontal Type III.b models in the adiabatic regime, i.e.\ by imposing the hierarchy of volumes $\mathcal{V}_{\mathbb{P}_{b}^{1}} \gg \mathcal{V}_{\mathrm{K3}}$. First, we extract the bulk physics at the endpoint of the limit from the generic vertical slices of the model, to then examine the purely six-dimensional features associated with the loci over which the adiabaticity assumption fails.

Consider the vertical slice of a horizontal Type III.b model associated with a generic point $p_{b} := [v_{0},w_{0}] \in \mathbb{P}_{b}^{1}$. All components of its central fiber have codimension-zero $\mathrm{I}_{m}$ fibers that are mutually local to each other, meaning that the 1-cycle $\sigma \in H_{1}(\mathcal{E},\mathbb{Z})$ that collapses to produce the pinched torus fibers is the same in all components. Analogously to the discussion of \cref{sec:horizontal-IIIa-models-adiabatic-regime}, this leads to a shrinking 2-cycle $\gamma \in H_{2}(Z_{0},\mathbb{Z})$ that can be extended to a shrinking local 2-cycle $\gamma$ in the generic patches of $Y_{0}$. The M2-branes wrapping $\gamma$ lead to a single tower of asymptotically massless particles, which are non-BPS due to the trivialisation of $\gamma$ over the non-generic vertical slices.

We can rephrase this in terms of the perturbative Type IIB orientifold interpretation of horizontal Type III.b models. The Type IIB internal space corresponds to a degenerate elliptic K3 surface of the type appearing as the endpoint of Kulikov Type II.b models, as mentioned above. Such a K3 surface has codimension-zero $\mathrm{I}_{m}$ fibers whose singularity type does, in addition, undergo A and D type enhancements over codimension-one loci in the base. Its elliptic fibration is also in the adiabatic regime thanks to the adiabaticity of the internal elliptic fibration of the F-theory model. The M2-branes wrapping the degenerate 1-cycle $\sigma \in H_{1}(\mathcal{E},\mathbb{Z})$ present over the generic points of $p_{b} \in \mathbb{P}_{b}^{1}$ in the F-theory model lead to the non-BPS weakly coupled asymptotically tensionless fundamental string in the Type IIB picture. The M2-branes can not only wrap the local 1-cycle $\sigma$, but also the shrinking local 2-cycle $\gamma$, leading to the tower of asymptotically massless particles discussed earlier. In the Type IIB interpretation, this tower corresponds to the winding modes of the fundamental string wrapping the generic $\mathrm{I}_{m}$ fiber of the degenerate K3 surface. Since the geometrical scaling leading to the string and the particles to be asymptotically massless is the same, it is clear on dimensional grounds that the tower of asymptotically massless particles becomes light at a faster rate, such that horizontal Type III.b models do not represent emergent string limits. In addition to the tower of particles just discussed, there is a second one stemming from the torus direction that becomes large in the large complex structure limit of the generic elliptic fiber of the Type IIB internal K3 surface. Altogether, the adiabatic limit in conjunction with these two towers of asymptotically massless particles result in a total decompactification from six to ten dimensions.

As we have argued in the context of the other horizontal models, the global vertical enhancements remain lower-dimensional at the endpoint of the limit, and are localised in the worldvolume of six-dimensional defects of the decompactified theory. Due to the global weak coupling nature of horizontal Type III.b models, the types of gauge algebras that can appear are restricted to those associated with Kodaira type $\mathrm{I}_{m}$ or $\mathrm{I}_{m}^{*}$ fibers, except for $\mathrm{I}_{0}^{*\, \mathrm{ns}}$ fibers that cannot be realised, see \cref{sec:horizontal-local-global-brane-content,sec:physical-interpretation-of-the-constraints}.

Note that since  the Type IIB internal K3 surface undergoes a Kulikov Type II.b degeneration, this leads to two shrinking 2-cycles $\{ \gamma_{i}^{\mathrm{IIB}} \}_{i=1,2}$ obtained from combining the vanishing 1-cycle of its generic elliptic fiber with the two 1-cycles of the elliptic curve corresponding to the double cover of $\mathbb{P}_{b}^{1}$. By wrapping extended objects on these shrinking 2-cycles, we obtain $\left. \mathrm{D3} \right|_{\gamma_{i}^{\mathrm{IIB}}}$-strings and $\left. \mathrm{F1} \right|_{\gamma_{i}^{\mathrm{IIB}}}$- and $\left. \mathrm{D1} \right|_{\gamma_{i}^{\mathrm{IIB}}}$-instantons. The adiabatic limit makes these contributions irrelevant for the asymptotic physics. It would be interesting to understand if they play a prominent role away from the adiabatic regime, being an avatar of the complicated reorganisation of the towers of asymptotically massless particles observed in the M-/F-theory picture, or if the rapid weak coupling limit is still able to suppress them on its own.

\section{Discussion and future directions}
\label{sec:conclusions}

We have investigated a large class of complex structure degenerations that occur at infinite distance in the moduli space of F-theory compactified to six dimensions. While Part I of our analysis \cite{Alvarez-Garcia:2023gdd} focused on the geometric foundations, our goal here was to interpret the degenerations from the point of view of general quantum gravity expectations. The theories we have explored arise when the internal space develops non-Kodaira singularities in the elliptic fiber over curves on a Hirzebruch surface base. They correspond to certain infinite-distance limits in the non-perturbative open string moduli space, possibly superimposed with a weak coupling limit. As explained in the more general context of \cite{Alvarez-Garcia:2023gdd}, to facilitate the systematics one can first focus on degenerations over non-intersecting curves, so-called single infinite-distance limits. This restricted class of degenerations is possible only over special curves of genus zero, or over an anti-canonical divisor of the base. In this work we have analysed the first type of such models, which come in four classes as listed in \cref{tab:genus-zero-Hirzebruch-summary}. The bulk of this article is devoted to the horizontal models, while the remaining three cases are treated similarly, wherever possible, in \cref{sec:cases-B-C-D-models}.

Horizontal models (as well as the related mixed (bi)sectional models in \cref{tab:genus-zero-Hirzebruch-summary}) arise from non-Kodaira singularities over sections of the Hirzebruch surface. As we have seen, this greatly facilitates the physics interpretation of the degenerations, at least in certain regimes of the moduli space where the base of the Hirzebruch surface is taken to be of asymptotically large volume. In such adiabatic limits, one can convincingly identify the towers of massless states that are expected in view of the Swampland Distance Conjecture \cite{Ooguri:2006in} and its refinement, the Emergent String Conjecture \cite{Lee:2019wij}. These arise, in the language of M-theory, from M2-branes multi-wrapped along local 2-cycles of torus topology which exist away from isolated defects. This picture becomes more and more accurate as the location of the defects is pushed to infinity in the asymptotically adiabatic regime. The result is a relative version of the massless towers appearing in F-theory probing degenerations of K3 surfaces \cite{Lee:2021qkx,Lee:2021usk}. As corroborated by the heterotic dual, the massless towers are interpreted as Kaluza-Klein towers signalling a decompactification limit. Since these are defined only away from isolated points on the base of the Hirzebruch surface, the latter play the role of six-dimensional (gauge) defects within an asymptotically nine or ten-dimensional bulk theory. In addition to these decompactification limits with lower-dimensional defects, some of the infinite distance limits (of Type II.b and III.b) have an interpretation as global weak coupling limits, again as in the eight-dimensional parent theories studied in \cite{Lee:2021qkx,Lee:2021usk}. All this is in perfect agreement with the expectations of the Emergent String Conjecture \cite{Lee:2019wij}. The appearance of lower-dimensional defects in decompactification limits was observed before in a different \mbox{context in \cite{Etheredge:2023odp}}.

The defect theories admit a particularly interesting interpretation from the dual heterotic point of view. They correspond to ADE type singularities on the heterotic K3 surface probed by point-like instantons of trivial or discrete holonomy. In the latter case there arise no tensor branch deformations. This matches the picture on the F-theory side of 7-branes wrapping the fiber of the Hirzebruch surface without being related to finite-distance non-minimal singularities. We have clarified some aspects of these non-perturbative gauge sectors in \cref{sec:defect-algebras-heterotic-dual}.

Numerous open questions lend themselves for  future investigations. Conceptually, the most important, but also most challenging, one is to establish a clear interpretation of the endpoints of the limits away from the adiabatic regime. When the defects no longer move to infinity, the limits are primarily characterised by the appearance of vanishing 3-cycles, while the role of the local 2-cycles becomes more obscure. We expect a reorganisation of the towers of asymptotically massless particles, but in such a way that eventually the limit is still a decompactification limit. As we explained, this picture is suggested by the behaviour on the heterotic dual side, at least if we can qualitatively trust the duality away from the adiabatic limit. Establishing such an interpretation would be important also as a way to tackle geometries in which no adiabatic limit can be taken, for instance the vertical degenerations. A related fruitful direction would be to put our findings in the context of the somewhat complementary approach to studying complex structure infinite-distance limits via asymptotic Hodge theory \cite{Grimm:2018cpv, Grimm:2018ohb, Grimm:2019ixq}. In this framework, the appearance of shrinking 3-cycles at infinite distance is established by studying monodromies around degeneration loci in the moduli space. In compactifications of Type IIB string theory to four dimensions, this makes manifest the appearance of towers of asymptotically massless states from wrapped D3-branes. Irrespective of the way in which the vanishing 3-cycles are deduced, more geometrically or more algebraically as in the Hodge theoretic approach, their role in relation with the towers in six-dimensional F-theory, rather than four-dimensional Type~IIB theory, is rather obscure. Our approach to demystify the situation was to take advantage of the adiabatic regime, where it is manifest that massless states arise from wrapped M2-branes on vanishing 2-chains, though these do not correspond to globally defined 2-cycles. To arrive at a satisfactory understanding of the limits in a more general context, without having to rely on adiabaticity, we will likely have to reconsider also the role of quantum effects as possible obstructions to certain directions in the moduli space. Indeed, both perturbative and non-perturbative quantum effects have the potential to affect infinite-distance limits in non-trivial ways, as studied in various frameworks \cite{Marchesano:2019ifh,Klaewer:2020lfg,Alvarez-Garcia:2021pxo}.

At the level of systematics, we have focused on a subclass of the codimension-one single infinite-distance limits explored more generally in Part I \cite{Alvarez-Garcia:2023gdd}. Natural extensions would be to include also degenerations associated with non-minimal genus-one curves, or degenerations constructed over the complex projective plane, which lacks a fibration structure. One could also allow for codimension-one degenerations outside the single infinite-distance limit class by simultaneously engineering multiple complex structure degenerations, each of which would individually have served as a single infinite-distance limit, still within the framework of \cite{Alvarez-Garcia:2023gdd}. Last but not least, infinite-distance limits can be associated with codimension-two non-Kodaira degenerations, which we have not explored systematically even at the purely geometric level. This state of affairs makes the complex structure degenerations of F-theory a remarkably rich field for future studies. 

\subsection*{\texorpdfstring{\textbf{Acknowledgements}}{Acknowledgements}}

We thank Vicente Cort\'es, Taro Hayashi, Craig Lawrie, Luca Martucci, Fabian R\"uhle, Max \mbox{Wiesner}, Piljin Yi and Michele Del Zotto for useful discussions. R.\,A.-G.\ and T.\,W.\ are supported in part by Deutsche Forschungsgemeinschaft under Germany's Excellence Strategy EXC 2121  Quantum Universe 390833306 and by Deutsche Forschungsgemeinschaft through a German-Israeli Project Cooperation (DIP) grant ``Holography and the Swampland”. The work of S.-J.\,L.\ is supported by IBS under the project code IBS-R018-D1.

\appendix


\section{Discriminant in the weakly coupled components}
\label{sec:discriminant-weakly-coupled-components}

The form that the discriminant takes in a component $Y^{p}$ that is at weak string coupling, i.e.\ one that presents codimension-zero fibers of Kodaira type $\mathrm{I}_{n_{p} > 0}$, is more constrained than in a strongly coupled component. To address this in the explicit global coordinate description that we employ while analysing horizontal and vertical models, let us rewrite the defining polynomials of the Weierstrass model of the family variety $\mathcal{Y}$ as
\begin{subequations}
\begin{align}
    f &= e_{p}^{l_{p}} \check{f}_{p} + f_{p}\,, \label{eq:f-check-definition}\\
    g &= e_{p}^{m_{p}} \check{g}_{p} + g_{p}\,, \label{eq:g-check-definition}\\
    \Delta &= e_{p}^{s_{p}} \check{\Delta}_{p} + \Delta_{p}\,. \label{eq:Delta-check-definition}
\end{align}
\label{eq:polynomials-check-definition}%
\end{subequations}
We have to consider various ways in which the codimension-zero $\mathrm{I}_{n_{p}>0}$ fibers can arise, each demanding more tuning than the previous one.
\begin{enumerate}[label=(\arabic*)]
    \item Single accidental cancellation: In order for the generic elliptic fiber over $B^{p}$ to be of Kodaira type $\mathrm{I}_{n_{p}>0}$, we need $\Delta_{p}$ to vanish. Plugging \eqref{eq:f-check-definition} and \eqref{eq:g-check-definition} into
    \begin{equation}
        \Delta = 4f^{3} + 27g^{2}\,,
    \end{equation}
    we obtain
    \begin{equation}
        \Delta = \left(4f_{p}^{3} + 27g_{p}^{2}\right) + 12f_{p}e_{p}^{l_{p}}\left(\check{f}_{p}f_{p}+e_{p}^{l_{p}}\check{f}_{p}^{2}\right) + 54e_{p}^{m_{p}}g_{p}\check{g}_{p} + \left(4e_{p}^{3l_{p}}\check{f}_{p}^{3} + 27e^{2m_{p}}\check{g}_{p}^{2}\right)\,.
    \end{equation}
    The necessary accidental cancellation for $\Delta_{p}$ to vanish occurs when\footnote{\label{foot:polynomial-accidental-cancellation}In this section, we are considering the coordinate divisors $E_{p} = \{ e_{p} = 0 \}_{\mathcal{B}}$, but we could carry out the discussion for more general divisors $\mathcal{D} = \{ p_{\mathcal{D}} = 0 \}_{\mathcal{B}}$ (working locally, if necessary). Then, the single accidental cancellation in the discriminant can occur not only when the structure \eqref{eq:codimension-zero-accidental-cancellation} is realised, but also when we have $4f_{\mathcal{D}}^{3} + 27g_{\mathcal{D}}^{2} = p_{\mathcal{D}}^{\delta_{\mathcal{D}}} q_{\mathcal{D}}$ with $\delta_{\mathcal{D}} \geq 1$. It is clear that when $p_{\mathcal{D}} = e_{p}$ this second type of accidental cancellation cannot occur, and we therefore neglect it in the remainder of the section. We also use the structure of accidental cancellations in the derivation of the vertical gauge rank bounds for horizontal models printed in \cref{sec:bounds-vertical-gauge-rank} and further discussed in \cref{sec:sec:bounds-vertical-gauge-rank-examples}, where the second possibility needs to and has been taken into account.}
    \begin{equation}
        f_{p} = -3h_{p}^{2}\,,\qquad g_{p} = 2h_{p}^{3}\,,\qquad h_{p} \in H^{0}\left(B^{p},\mathcal{L}_{p}^{\otimes 2}\right)\,,
    \label{eq:codimension-zero-accidental-cancellation}
    \end{equation}
    and assuming this structure leads to
    \begin{equation}
        \Delta = 108 h_{p}^{4}e_{p}^{l_{p}}\check{f}_{p} - 36h_{p}^{2}e_{p}^{2l_{p}}\check{f}_{p}^{2} + 108h_{p}^{3}e_{p}^{m_{p}}\check{g}_{p} + 4e_{p}^{3l_{p}}\check{f}_{p}^{3} + 27 e^{2m_{p}}\check{g}_{p}^{2}\,.
    \label{eq:discriminant-single-accidental-cancellation}
    \end{equation}
    Unless further cancellations take place, we have $n_{p} = s_{p} = \min\left(l_{p},m_{p}\right)$, and
    \begin{equation}
         \left. \check{\Delta}_{p} \right|_{E_{p}} = \left. 108 h_{p}^{3} \left(e_{p}^{l_{p}-n_{p}}\check{f}_{p}h_{p} + e_{p}^{m_{p}-n_{p}}\check{g}_{p}\right) \right|_{e_{p} = 0} = h_{p}^{k} \Delta''_{p}\,,\qquad k \geq 3\,.
    \label{eq:check-component-discriminant-k3}
    \end{equation}

    \item Double accidental cancellation: Additional cancellations can occur when \eqref{eq:codimension-zero-accidental-cancellation} is satisfied and $l_{p} = m_{p} =: r_{s}$, in which case
    \begin{equation}
    	\Delta = 108 h_{p}^{3}e_{p}^{r_{p}} \left( h_{p} \check{f}_{p} + \check{g}_{p} \right) - 36h_{p}^{2}e_{p}^{2r_{p}}\check{f}_{p}^{2} + 4e_{p}^{3r_{p}}\check{f}_{p}^{3} + 27 e_{p}^{2r_{p}}\check{g}_{p}^{2}\,.
    \end{equation}
    We can then have the structure
    \begin{equation}
		h_{p}\check{f}_{p} + \check{g}_{p} = e_{p}^{\rho_{p}} q_{p}\,,\qquad \rho_{p} \geq 1\,,
	\label{eq:double-accidental-cancellation-condition}
	\end{equation}
	leading to
	\begin{equation}
		\Delta = 108 h_{p}^{3}e_{p}^{r_{p} + \rho_{p}} q_{p} - 36h_{p}^{2}e_{p}^{2r_{p}}\check{f}_{p}^{2} + 4e_{p}^{3r_{p}}\check{f}_{p}^{3} + 27 e_{p}^{2r_{p}}\check{g}_{p}^{2}\,.
	\label{eq:discriminant-double-accidental-cancellation}
	\end{equation}
    A particular subcase is the one realised when $q_{p} = 0$, to which we assign $\rho_{p} = \infty$. The resulting component discriminant $\left. \check{\Delta}_{p} \right|_{E_{p}}$ depends on the value of $r_{p} + \rho_{p}$. We assume below that no additional cancellations take place.
	\begin{enumerate}[label=(\arabic{enumi}.\alph*)]
		\item If $r_{p} + \rho_{p} < 2r_{p}$, we have that $n_{p} = r_{p} + \rho_{p} \geq 2$, and the restriction $\left. \check{\Delta}_{p} \right|_{E_{p}}$ is given by
		\begin{equation}
			\check{\Delta}_{p}|_{E_{p}} = \left. 108 h_{p}^{3} q_{p} \right|_{e_{p} = 0}\,.
		\end{equation}
		
		\item If $r_{p} + \rho_{p} = 2r_{p}$, we obtain $n_{p} = 2r_{p} \geq 2$, and the restriction $\left. \check{\Delta}_{p} \right|_{E_{p}}$ is
		\begin{equation}
			\check{\Delta}_{p}|_{E_{p}} = \left. 9h_{p}^{2} \left( 12 h_{p}q_{p} - \check{f}_{p}^{2} \right) \right|_{e_{p} = 0}\,.
		\end{equation}
		
		\item If $r_{p} + \rho_{p} > 2r_{p}$, we still obtain $n_{p} = 2r_{p} \geq 2$, and the restriction $\left. \check{\Delta}_{p} \right|_{E_{p}}$ is now
		\begin{equation}
			\check{\Delta}_{p}|_{E_{p}} = \left. -9h_{p}^{2}  \check{f}_{p}^{2} \right|_{e_{p} = 0}\,.
		\end{equation}
	\end{enumerate}
	The last two cases can give $k = 2$.

    \item Triple accidental cancellation: If the conditions \eqref{eq:codimension-zero-accidental-cancellation} and \eqref{eq:double-accidental-cancellation-condition} are satisfied, and also $r_{p} = \rho_{p} = l_{p} = m_{p} =: t_{p}$, the discriminant can be expressed as
    \begin{equation}
        \Delta = 9h_{p}^{2} e_{p}^{2t_{p}} \left( 12h_{p}q_{p} - \check{f}_{p}^{2} \right) + 2\check{f}_{p}e_{p}^{3t_{p}} \left( 2\check{f}_{p}^{2} - 27h_{p}q_{p} \right) + 27e_{p}^{4t_{p}}q_{p}^{2}\,.
    \end{equation}
    We see that additional cancellations can occur if we have the structure
    \begin{equation}
        12 h_{p} q_{p} - \check{f}_{p}^{2} = e_{p}^{\sigma} \check{q}_{p}\,,\qquad \sigma_{p} \geq 1\,,
    \label{eq:triple-accidental-cancellation-condition}
    \end{equation}
    leading to
    \begin{equation}
        \Delta = 9h_{p}^{2}e_{p}^{2t_{p}+\sigma_{p}}\check{q}_{p} - 6h_{p}e_{p}^{3t_{p}}q_{p}\check{f}_{p} - 4e_{p}^{3t_{p}+\sigma_{p}}\check{f}_{p}\check{q}_{p} + 27e_{p}^{4t_{p}}q_{p}^{2}\,.
    \end{equation}
    Again, to the particular subcase $\check{q}_{p} = 0$ we assign $\sigma_{p} = \infty$. The form of the component discriminant $\left. \check{\Delta}_{p} \right|_{E_{p}}$ depends on the value of $2t_{p} + \sigma_{p}$. Below, we assume that no additional cancellations take place.
    \begin{enumerate}[label=(\arabic{enumi}.\alph*)]
        \item If $2t_{p} + \sigma_{p} < 3t_{p}$, we have that $n_{p} = 2t_{p} + \sigma_{p} \geq 3$, and the restriction $\left. \check{\Delta}_{p} \right|_{E_{p}}$ is
        \begin{equation}
            \left. \check{\Delta}_{p} \right|_{E_{p}} = \left. 9h_{p}^{2}\check{q}_{p} \right|_{e_{p}=0}\,.
        \end{equation}

        \item If $2t_{p} + \sigma_{p} = 3t_{p}$, we obtain $n_{p} = 3t_{p} \geq 3$, and $\left. \check{\Delta}_{p} \right|_{E_{p}}$ is
        \begin{equation}
            \left. \check{\Delta}_{p} \right|_{E_{p}} = \left. h_{p} \left( 9h_{p} \check{q}_{p} - 6q_{p}\check{f}_{p} \right)\right|_{e_{p}=0}\,.
        \end{equation}

        \item If $2t_{p} + \sigma_{p} > 3t_{p}$, then $n_{p} = 2t_{p} + \sigma_{p} \geq 3$, and $\left. \check{\Delta}_{p} \right|_{E_{p}}$ is
        \begin{equation}
            \left. \check{\Delta}_{p} \right|_{E_{p}} = \left. -6h_{p}q_{p}\check{f}_{p} \right|_{e_{p}}=0\,.
        \end{equation}
    \end{enumerate}
    These cases also lead to $k \geq 2$. To see these for the last two of them, note that \eqref{eq:triple-accidental-cancellation-condition} implies that
    \begin{equation}
        \left. \check{f}_{p}^{2} \right|_{e_{p}=0} = \left. 12h_{p}q_{p} \right|_{e_{p}=0} \Rightarrow \left. q_{p} \right|_{e_{p}=0} = \left. h_{p}{q''}_{p}^{2} \right|_{e_{p}=0}\,,
    \end{equation}
    where we have used that the generic $\left. h_{p} = h_{p} \right|_{e_{p}=0}$ polynomial is not a perfect square.

    \item Quadruple accidental cancellation: If the conditions \eqref{eq:codimension-zero-accidental-cancellation}, \eqref{eq:double-accidental-cancellation-condition} and \eqref{eq:triple-accidental-cancellation-condition} are satisfied, and also $t_{p} = r_{p} = \rho_{p} = l_{p} = m_{p}$ and $2t_{p}+\sigma_{p}=3t_{p}$, additional cancellations can occur if we have the structure
    \begin{equation}
        9h_{p}\check{q}_{p} - 6q_{p}\check{f}_{p} = e^{\tau_{p}}\tilde{q}_{p}\,,\qquad \tau_{p} \geq 1\,.
    \end{equation}
    The analysis of this and further accidental cancellation structures would proceed along the same lines as that of the cases previously studied, and we do not perform it explicitly.
\end{enumerate}

To summarise, the component discriminant $\left. \check{\Delta}_{p} \right|_{E_{p}}$ after tuning codimension-zero Kodaira type $\mathrm{I}_{n_{p} > 0}$ fibers in the $Y^{p}$ components takes the form
\begin{equation}
	\check{\Delta}_{p}|_{E_{p}} = h_{p}^{k} \Delta_{p}^{\prime\prime}\,,\qquad k \geq 2\,.
\end{equation}
The minimal value of $k=2$ is associated with the appearance of type $\mathrm{D}_{0}$ singularities, see the discussion in \cref{sec:horizontal-local-global-brane-content}, explaining why we cannot go lower. Note that the object $\left. \check{\Delta}_{p} \right|_{E_{p}}$ considered in this section is different from the restriction $\Delta'_{p}$ of the modified discriminant $\Delta'$ to the component. The two objects are related by
\begin{equation}
    \check{\Delta}_{p} = e_{0}^{n_{0}} e_{1}^{n_{1}} \overset{\overset{p}{\vee}}{\cdots} e_{P-1}^{n_{P-1}} e_{P}^{n_{P}} \Delta' \Rightarrow \check{\Delta}_{p}|_{E_{p}} = e_{0}^{n_{0}} e_{1}^{n_{1}} \overset{\overset{p}{\vee}}{\cdots} e_{P-1}^{n_{P-1}} e_{P}^{n_{P}} \Delta'_{p}\,.
\label{eq:check-prime-discriminants}
\end{equation}

\section{Bounds on \texorpdfstring{$\left| n_{p} - n_{p+1} \right|$}{|np - np+1|}}
\label{sec:bounds-codimension-zero-pattern}

Beyond the effectiveness bounds for the pattern of codimension-zero singular elliptic fibers in the central fiber of horizontal and vertical models derived in \cref{sec:effectiveness-bounds-horizontal,sec:effectiveness-bounds-vertical}, respectively, one can obtain tighter constraints by studying the resolution structure taking us from the original degeneration $\hat{\rho}: \hat{\mathcal{Y}} \rightarrow D$ to its open-chain resolution $\rho: \mathcal{Y} \rightarrow D$. Such bounds were presented in \cref{sec:tighter-np-npplusone-horizontal} for horizontal models; here, we give the technical arguments on which they are sustained and also discuss them for vertical models in \cref{sec:bounds-codimension-zero-pattern-vertical}. They vary slightly depending on how tuned the accidental cancellations giving rise to the weakly coupled components are, see \cref{sec:discriminant-weakly-coupled-components}. We focus on obtaining bounds that apply when the single or double accidental cancellation structure is realised.

\subsection{Horizontal models}
\label{sec:bounds-codimension-zero-pattern-horizontal}

We start by deriving the tighter bounds on $\left| n_{p} - n_{p+1} \right|$ that were printed in \cref{sec:tighter-np-npplusone-horizontal}. It turns out that the inequalities are asymmetric, depending on if we consider $n_{p} - n_{p+1}$, for $p \in \{ 0, \dotsc, P-1 \}$, or $n_{p} - n_{p-1}$, for $p \in \{1, \dotsc, P\}$. We treat each case in turn.

\subsubsection*{Tighter bounds on $n_{p} - n_{p+1}$}

Consider a component $Y^{p}$, with $p \in \{0, \dotsc, P-1\}$, of the central fiber $Y_{0}$ of a resolved horizontal model $\rho: \mathcal{Y} \rightarrow D$ and tune over it codimension-zero $\mathrm{I}_{n_{p}>0}$ fibers. This entails enforcing, at least, the single accidental cancellation structure \eqref{eq:codimension-zero-accidental-cancellation}. To achieve this, certain powers of the exceptional coordinate $e_{p}$ must factorize in some terms of the defining polynomials, which we then write as
\begin{subequations}
\begin{align}
	f-f_{p} &= e_{p}^{l_{p}} \check{f}\,, \label{eq:horizontal-accidental-cancellation-factorization-1}\\
	g-g_{p} &= e_{p}^{m_{p}} \check{g}\,. \label{eq:horizontal-accidental-cancellation-factorization-2}
\end{align}
\label{eq:horizontal-accidental-cancellation-factorization}%
\end{subequations}
But the powers with which the exceptional coordinates $\{e_{p}\}_{0 \leq p \leq P}$ appear in the monomials of $f$ and $g$ are not arbitrary. Leaving a more detailed analysis of this aspect for \cite{ALWClass5}, it suffices for now to note how individual monomials in $f$ and $g$ are affected by the resolution process of a horizontal model.

Consider, without loss of generality, that we are dealing with a horizontal model in which $\hat{\mathscr{C}}_{1} = \{h\}$. We then have that the resolution process acts on the monomials like
\begin{subequations}
\begin{align}
    u^{\mu_{i,0}} s^{i} t^{8-i} v^{j} w^{(8+4n)-(8-i)n-j} \subset f &\longmapsto \prod_{p=0}^{P} e_{p}^{\mu_{i,p}} s^{i} t^{8-i} v^{j} w^{(8+4n)-(8-i)n-j} \subset f\,,\\
    u^{\nu_{i,0}} s^{i} t^{12-i} v^{j} w^{(12+6n)-(12-i)n-j} \subset g &\longmapsto \prod_{p=0}^{P} e_{p}^{\nu_{i,p}} s^{i} t^{12-i} v^{j} w^{(12+6n)-(12-i)n-j} \subset g\,,
\end{align}
\end{subequations}
where
\begin{subequations}
\begin{align}
    \mu_{i,p} &:= \mu_{i,0} + p(i-4)\,,\\
    \nu_{j,p} &:= \nu_{j,0} + p(j-6)\,.
\end{align}
\label{eq:mu-nu-powers}%
\end{subequations}
The slopes of these linear functions are determined by the power with which $s$ appears in the monomial under consideration. Since $f$ and $g$ are global holomorphic sections of the line bundles $F = 4 \overline{K}_{\hat{\mathcal{B}}}$ and $G = 6 \overline{K}_{\hat{\mathcal{B}}}$, respectively, with
\begin{equation}
    \overline{K}_{\hat{\mathcal{B}}} = 2\mathcal{S} + (2+n)\mathcal{V}\,,
\end{equation}
we have that
\begin{subequations}
\begin{align}
	0 &\leq i \leq 8\,,\\
	0 &\leq j \leq 12\,.
\end{align}
\label{eq:i-j-powers-bounds-horizontal}%
\end{subequations}

The factorization \eqref{eq:horizontal-accidental-cancellation-factorization} implies that
\begin{equation}
	\mu_{i,p} \geq l_{p}\,,\qquad \nu_{j,p} \geq m_{p}\,,
\label{eq:horizontal-accidental-cancellation-factorization-condition}
\end{equation}
for all monomials in $f-f_{p}$ and $g-g_{p}$. Defining
\begin{equation}
	(\alpha, \beta) := \mathrm{ord}_{\hat{\mathcal{Y}}}(f,g)_{s=0}\,,
\end{equation}
we get then from \eqref{eq:mu-nu-powers} and \eqref{eq:horizontal-accidental-cancellation-factorization-condition} the bounds
\begin{subequations}
\begin{align}
	\mu_{i,p+1} &\geq \mu_{i,p} - (4 - i) \geq l_{p} - \max_{i' \in \{\alpha,\dotsc,8\}}(4-i') \geq l_{p} - (4 - \alpha)\,,\;\forall  i \in \{\alpha,\dotsc,8\}\,,\label{eq:horizontal-accidental-cancellation-factorization-bound-1}\\
	\nu_{j,p+1} &\geq \nu_{j,p} - (6 - j) \geq m_{p} - \max_{j' \in \{\beta,\dotsc,12\}}(6-j') \geq m_{p} - (6 - \beta)\,,\; \forall  j \in \{\beta,\dotsc,12\}\,,\label{eq:horizontal-accidental-cancellation-factorization-bound-2}
\end{align}
\label{eq:horizontal-accidental-cancellation-factorization-bound}%
\end{subequations}
which apply to the aforementioned monomials. This means that there is a further factorization
\begin{subequations}
\begin{align}
    f &= e_{p}^{l_{p}} e_{p+1}^{\tilde{\mu}_{p+1}} \check{f}'_{p} + f_{p}\,,\\
    g &= e_{p}^{l_{p}} e_{p+1}^{\tilde{\nu}_{p+1}} \check{g}'_{p} + g_{p}\,,
\end{align}
\end{subequations}
where we have introduced
\begin{subequations}
\begin{alignat}{2}
    \tilde{\mu}_{p+1} &:= \min \mu_{i,p+1}\,,\qquad &\tilde{\mu}_{p+1} \geq l_{p} - (4-\alpha)\,,\quad \tilde{\mu}_{p+1} \geq 0\,,\\
    \tilde{\nu}_{p+1} &:= \min \nu_{i,p+1}\,,\qquad &\tilde{\nu}_{p+1} \geq m_{p} - (6-\beta)\,,\quad \tilde{\nu}_{p+1} \geq 0
\end{alignat}
\label{eq:mu-nu-pplusone-tilde}
\end{subequations}
with the minimum is taken over all monomials in $f-f_{p}$ and $g-g_{p}$, respectively.

Assume now that the single accidental cancellation structure \eqref{eq:codimension-zero-accidental-cancellation} is satisfied in the $Y^{p}$ component, without additional cancellations taking place. The form of the discriminant is then \eqref{eq:discriminant-single-accidental-cancellation}, from which one can see that
\begin{equation}
    n_{p+1} \geq \min(\tilde{\mu}_{p+1},\tilde{\nu}_{p+1}) = \min(l_{p}-(4-\alpha),m_{p}-(6-\beta)) = n_{p} - \max(4-\alpha,6-\beta)\,.
\end{equation}
Hence, for a single accidental cancellation structure in $Y^{p}$ we have that
\begin{equation}
    n_{p} - n_{p+1} \leq \max(4-\alpha,6-\beta)\,,\qquad p \in \{0, \dotsc, P-1\}\,.
\label{eq:np-npplusone-single-bound}
\end{equation}

Allow now for a double accidental cancellation to occur in $Y^{p}$, i.e.\ assume that \eqref{eq:codimension-zero-accidental-cancellation} and \eqref{eq:double-accidental-cancellation-condition} are satisfied while $l_{p} = m_{p} =: r_{p}$. Let us rewrite \eqref{eq:double-accidental-cancellation-condition} as
\begin{equation}
    h_{p}\check{f}_{p} + \check{g}_{p} = e_{p}^{\rho_{p}}e_{p}^{\rho_{p+1}} q'_{p}\,,\qquad \rho_{p} \geq 1\,,\quad \rho_{p+1} \geq 0\,.
\label{eq:double-accidental-cancellation-condition-modified}
\end{equation}
By definition, $h_{p}$ contains no powers of $e_{p}$, and the monomials in $\check{f}_{p}$ and $\check{g}_{p}$ are those fulfilling \eqref{eq:horizontal-accidental-cancellation-factorization-condition}, which have been divided by $e_{p}^{-r_{p}}$. Once the necessary cancellations to produce the r.h.s.\ take place, only those monomials whose provenance can be traced back to monomials in $f-f_{p}$ and $g-g_{p}$ satisfying
\begin{equation}
    \mu_{i,p} \geq r_{p}+\rho_{p}\,,\qquad \nu_{j,p} \geq r_{p}+\rho_{p}\,,
\end{equation}
survive, from which we conclude that
\begin{equation}
    \rho_{p+1} \geq \min\left( r_{p}+\rho_{p} - (4-\alpha), r_{p}+\rho_{p} - (6-\beta) \right) = r_{p}+\rho_{p} - \max(4-\alpha,6-\beta)\,.
\end{equation}
Moreover, we see from \eqref{eq:double-accidental-cancellation-condition-modified} and the fact that $h_{p}$ contains no overall powers of $e_{p+1}$ (since these would lead to special fibers at the intersection of the components, see \cite{Alvarez-Garcia:2023gdd}), that for the cancellations on the l.h.s.\ to be possible the lowest power of $e_{p+1}$ appearing in both terms must be the same and smaller than $\rho_{p+1}$, i.e.\ $\tilde{\mu}_{p+1} = \tilde{\nu}_{p+1} < \rho_{p+1}$. Additionally, the double accidental cancellation cases discussed in \cref{sec:discriminant-weakly-coupled-components} showed that
\begin{equation}
	n_{p} =
	\begin{cases}
		r_{p} + \rho_{p}\,,\quad &\textrm{if}\quad r_{p} + \rho_{p} < 2r_{p} \Rightarrow 2r_{p} > r_{p} + \rho_{p} = n_{p}\,,\\
		2r_{p}\,,\quad &\textrm{if}\quad r + \rho_{p} \geq 2r_{p} \Rightarrow r_{p} + \rho_{p} \geq 2r_{p} = n_{p}\,,
	\end{cases}
\end{equation}
and hence the bounds
\begin{equation}
	r_{p} + \rho_{p} \geq n_{p}\,,\qquad 2r_{p} \geq n_{p}\,,
\end{equation}
are always satisfied. From the resulting form \eqref{eq:discriminant-double-accidental-cancellation} of the discriminant, we observe that
\begin{equation}
    n_{p+1} \geq \min\left(\rho_{p+1},2 \min(\tilde{\mu}_{p+1},\tilde{\nu}_{p+1})\right)\,.
\end{equation}
We then distinguish two cases:
\begin{itemize}
    \item If $\min\left(\rho_{p+1},2 \min(\tilde{\mu}_{p+1},\tilde{\nu}_{p+1})\right) = \rho_{p+1}$, the inequalities given above result in
    \begin{equation}
        n_{p+1} \geq \rho_{p+1} \geq n_{p} - \max(4-\alpha,6-\beta) \Leftrightarrow n_{p} - n_{p+1} \leq \max(4-\alpha,6-\beta)\,.
    \end{equation}

    \item If $\min\left(\rho_{p+1},2 \min(\tilde{\mu}_{p+1},\tilde{\nu}_{p+1})\right) = 2 \min(\tilde{\mu}_{p+1},\tilde{\nu}_{p+1})$, we have instead
    \begin{equation}
    \begin{rcases}
        n_{p+1} \geq 2\tilde{\mu}_{p+1} \geq n_{p} - 2(4-\alpha)\\
        n_{p+1} \geq 2\tilde{\nu}_{p+1} \geq n_{p} - 2(6-\beta)
    \end{rcases}
    \Leftrightarrow n_{p} - n_{p+1} \leq 2\min(4-\alpha,6-\beta)\,,
    \end{equation}
    where we have used \eqref{eq:mu-nu-pplusone-tilde}.
\end{itemize}
Altogether, for a double accidental cancellation structure in $Y^{p}$ we have the bound
\begin{equation}
    n_{p} - n_{p+1} \leq \max\left( \max(4-\alpha,6-\beta), 2\min(4-\alpha,6-\beta) \right)\,,\qquad p \in \{0, \dotsc, P-1\}\,,
\label{eq:np-npplusone-double-bound}
\end{equation}
which is laxer than the one found for the single accidental cancellation structure.

For the generic model with a given pattern $\mathrm{I}_{n_{0}} - \cdots - \mathrm{I}_{n_{p}}$ of codimension-zero singular elliptic fibers in $Y_{0}$ obtained through single and double accidental cancellations, we therefore have
\begin{equation}
	n_{p} - n_{p+1} \leq
	\begin{cases}
		8\,, & 0 \leq n \leq 2\,,\\
		4\,, & 3 \leq n \leq 4\,,\\
		2\,, & 5 \leq n \leq 8\,,\\
		1\,, & 9 \leq n \leq 12\,,
	\end{cases}
	\qquad p = 0, \dotsc, P-1\,,
\label{eq:npminusnpplusone-bound-horizontal-appendix}
\end{equation}
where we have used the values of $\alpha$ and $\beta$ associated with the non-Higgsable clusters listed in \cref{tab:non-Higgsable-clusters}. It would be interesting to know if the bounds can be relaxed to match the effectiveness bounds of \cref{sec:effectiveness-bounds-horizontal} by allowing further accidental cancellations to occur as discussed in \cref{sec:discriminant-weakly-coupled-components}.

\subsubsection*{Tighter bounds on $n_{p} - n_{p-1}$}

The same type of arguments can be used to derive bounds in the opposite direction. The starting inequalities that we need to consider now are the analogues of \eqref{eq:horizontal-accidental-cancellation-factorization-bound}, namely
\begin{subequations}
\begin{align}
	\mu_{i,p-1} &\geq \mu_{i,p} + (4 - i) \geq l_{p} + \min_{i' \in \{\alpha,\dotsc,8\}}(4-i') \geq l_{p} - 4\,,\;\forall i \in \{\alpha,\dotsc,8\}\,,\\
	\nu_{j,p-1} &\geq \nu_{j,p} + (6 - j) \geq m_{p} + \min_{j' \in \{\beta,\dotsc,12\}}(6-j') \geq m_{p} - 6\,,\; \forall j \in \{\beta,\dotsc,12\}\,.
\end{align}
\end{subequations}
These have the same form as \eqref{eq:horizontal-accidental-cancellation-factorization-bound} with $(\alpha, \beta) = (0,0)$. Hence, the resulting inequalities for the single and double accidental cancellation structures are \eqref{eq:np-npplusone-single-bound} and \eqref{eq:np-npplusone-double-bound} particularised for this value, which combined result in
\begin{equation}
	n_{p} - n_{p-1} \leq 8\,,\qquad p = 1, \dotsc, P\,.
\label{eq:npminusnpminusone-bound-horizontal-appendix}
\end{equation}
The reason we do not find different bounds depending on the Hirzebruch surface $\hat{B} = \mathbb{F}_{n}$ over which the model is constructed, is that the inequalities in this direction involve the linear functions $\mu_{i,p}$ and $\nu_{j,p}$ with the highest slopes, whose presence in a generic model is unaffected by the existence of non-Higgsable clusters. In the preceding discussion, it was the linear functions with the smallest slopes that set the bounds.

\subsection{Vertical models}
\label{sec:bounds-codimension-zero-pattern-vertical}

The tighter bounds on $|n_{p} - n_{p+1}|$ for vertical models are obtained in an analogous way to the ones applying to horizontal models. Hence, we keep the discussion brief and only point out the differences with respect to the preceding case and the final result.

\subsubsection*{Tighter bounds on $n_{p} - n_{p+1}$ and $n_{p} - n_{p-1}$}

The main difference with respect to the horizontal case is that in the equations \eqref{eq:mu-nu-powers} the variables $i$ and $j$ now refer to the power with which, without loss of generality, $v$ appears in the monomial under consideration. Hence, we have to substitute \eqref{eq:i-j-powers-bounds-horizontal} for
\begin{align}
	0 &\leq i \leq 8+4n\,,\\
	0 &\leq j \leq 12+6n\,,
\label{eq:i-j-powers-bounds-vertical}
\end{align}
where we observe that only the upper bounds are  different. The bounds \eqref{eq:npminusnpplusone-bound-horizontal-appendix} on $n_{p} - n_{p+1}$, with $p \in \{ 0, \dotsc, P-1 \}$, for horizontal models were set by the linear functions $\mu_{i,p}$ and $\nu_{j,p}$ of smallest slope, and therefore remain valid for vertical models, i.e.\ we have again
\begin{equation}
	n_{p} - n_{p+1} \leq
	\begin{cases}
		8\,, & 0 \leq n \leq 2\,,\\
		4\,, & 3 \leq n \leq 4\,,\\
		2\,, & 5 \leq n \leq 8\,,\\
		1\,, & 9 \leq n \leq 12\,,
	\end{cases}
	\qquad p = 0, \dotsc, P-1\,.
\label{eq:npminusnpplusone-bound-vertical}
\end{equation}
The bounds \eqref{eq:npminusnpminusone-bound-horizontal-appendix} on $n_{p} - n_{p+1}$, with $p \in \{ 1, \dotsc, P \}$, for horizontal models were determined by the linear functions $\mu_{i,p}$ and $\nu_{j,p}$ of highest slope, instead. Since in vertical models these functions can be steeper according to \eqref{eq:i-j-powers-bounds-vertical}, we find, \textit{mutatis mutandis}, that
\begin{equation}
	n_{p} - n_{p-1} \leq 2(4+4n)\,.
\label{eq:npminusnpminusone-bound-vertical}
\end{equation}
Both sets of constraints apply to vertical models presenting the single and double accidental cancellation structures.

\section{Bounds on the vertical gauge rank}
\label{sec:sec:bounds-vertical-gauge-rank-examples}

In \cref{sec:bounds-vertical-gauge-rank}, we argued that the rank of the gauge factors supported over representatives of the global divisor $\mathcal{F}$ cannot be arbitrarily high in a horizontal model, with the obvious bound
\begin{equation}
    \rank(\mathfrak{g}_{\mathrm{ver}}) \leq 18
\label{eq:horizontal-gauge-rank-bound-re}
\end{equation}
coming from heterotic/heterotic duality. This bound can be improved by analysing the geometry more carefully, which in horizontal models with no components at weak coupling leads to the bounds given in \cref{tab:vertical-rank-bounds}. To illustrate the process, we derive the bound for horizontal models constructed over $\hat{B} = \mathbb{F}_{8}$ with no components at weak coupling in \cref{sec:B-F8-no-weak-coupling}. In the presence of components at weak coupling, the bounds can become more stringent; we exemplify this in \cref{sec:B-F7-weak-coupling} by deriving a new bound for horizontal models constructed over $\hat{B} = \mathbb{F}_{7}$ with codimension-zero singular elliptic fibers that is stricter than the one printed in \cref{tab:vertical-rank-bounds}.

\subsection{\texorpdfstring{$\hat{B} = \mathbb{F}_{8}$}{B=F8} with no components at weak coupling}
\label{sec:B-F8-no-weak-coupling}

According to \cref{tab:vertical-rank-bounds}, for horizontal models constructed over $\hat{B} = \mathbb{F}_{8}$ with no components at weak coupling the vertical gauge rank is subject to the constraint
\begin{equation}
    \rank(\mathfrak{g}_{\mathrm{ver}}) \leq 4\,.
\end{equation}
Let us derive this result.

The starting point is the rough bound \eqref{eq:horizontal-gauge-rank-bound-re} derived from heterotic/heterotic duality considerations. An improved rough bound can be obtained by estimating how many vertical classes can be factorised before we encounter a curve of non-minimal fibers. In a resolved horizontal model constructed over $\hat{B} = \mathbb{F}_{8}$, the $Y^{P}$ component of the central fiber $Y_{0}$ contains a non-Higgsable cluster with component vanishing orders
\begin{equation}
    \ord{Y^{P}}(f_{P},g_{P},\Delta_{P}^{\prime})_{s=0} = (3,5,9)\,,
\end{equation}
see \cref{tab:non-Higgsable-clusters}. Such a line of singular elliptic fibers can remain of Kodaira type $\mathrm{III}^{*}$, enhance to Kodaira type $\mathrm{II}^{*}$ or become non-minimal. It is then clear that for this class of models\footnote{If the enhancement over $S_{P}$ is of $\mathrm{I}_{m}$ or $\mathrm{I}_{m}^{*}$ type, the inequality no longer holds.}
\begin{equation}
    \ord{Y^{P}}(\Delta_{P}^{\prime})_{s=0} \leq 10\,.
\label{eq:F8-SP-bound}
\end{equation}
Particularizing the discriminant \eqref{eq:Delta-hor-3} to the case in which no component is at weak coupling, we obtain
\begin{equation}
    \Delta_{P}^{\prime} = 12S_{P} + 24V_{P}\,.
\end{equation}
After a given number of tunings involving $S_{P}$ and $V_{P}$ classes, the residual discriminant will not be forced to contain an $S_{P}$ component as long as
\begin{equation}
    \left( \Delta_{P}^{\prime} - \alpha V_{P} - \beta S_{P} \right) \cdot S_{P} \geq 0 \Leftrightarrow n(\beta - 12) + 24 \geq \alpha\,.
\label{eq:F8-rough-VP-bound}
\end{equation}
Using then \eqref{eq:F8-SP-bound} and particularizing to $\hat{B} = \mathbb{F}_{8}$, we obtain that the number $\alpha$ of $V_{P}$ classes that can be factorised in the discriminant must be
\begin{equation}
    \alpha \leq 8\,,
\end{equation}
if we want to avoid non-minimal component vanishing orders over $S_{P}$. With a budget of $8V_{P}$ vertical classes in $\Delta_{P}^{\prime}$ to be distributed over the local vertical gauge enhancements in the $Y^{P}$ component, and hence a maximum of $8\mathcal{F}$ classes in $\Dphys$, we can simply list the a priori possible combinations of vertical enhancements. In \cref{tab:F8-maximal-vertical-tunings} we do this for the subset of them that use all available $8V_{P}$ classes and that therefore are the best candidates to give the highest vertical gauge rank, in principle.
\begin{table}[t!]
    \centering
    \begin{tblr}{columns = {c}, hlines, vlines}
        Vanishing orders & Naive rank & Result of the analysis\\
        $(2,3,8)$ & $\mathrm{D}_{6} \sim 6$ & $(2,3,7) \sim \mathrm{B}_{4} \sim 4$\\
        $(2,3,6) + (0,0,2)$ & $\mathrm{D}_{4} + \mathrm{A}_{1} \sim 5$ & $(2,3,6) \sim \mathrm{G}_{2} \sim 2$\\
        $(1,2,3) + (1,2,3) + (0,0,2)$ & $\mathrm{A}_{1} + \mathrm{A}_{1} + \mathrm{A}_{1} \sim 3$ & $(1,2,3) + (1,2,3) \sim \mathrm{A}_{1} + \mathrm{A}_{1} \sim 2$\\
        $(1,2,3) + (0,0,5)$ & $ \mathrm{A}_{1} + \mathrm{A}_{4} \sim 5$ & $(1,2,3) + (0,0,2) \sim \mathrm{A}_{1} + \mathrm{A}_{1} \sim 2$\\
        $(1,2,3) + (0,0,3) + (0,0,2)$ & $\mathrm{A}_{1} + \mathrm{A}_{2} + \mathrm{A}_{1} \sim 4$ & $(1,2,3) + (0,0,1) + (0,0,1) \sim \mathrm{A}_{1} \sim 1$\\
        $(0,0,8)$ & $\mathrm{A}_{7} \sim 7$ & $(0,0,4) \lesssim \mathrm{A}_{3} \sim 3$\\
        $(0,0,6) + (0,0,2)$ & $\mathrm{A}_{5} + \mathrm{A}_{1} \sim 6$ & $(0,0,3) + (0,0,1) \lesssim \mathrm{A}_{2} \sim 2$\\
        $(0,0,5) + (0,0,3)$ & $\mathrm{A}_{4} + \mathrm{A}_{2} \sim 6$ & $(0,0,3) + (0,0,1) \lesssim \mathrm{A}_{2} \sim 2$\\
        $(0,0,4) + (0,0,4)$ & $\mathrm{A}_{3} + \mathrm{A}_{3} \sim 6$ & $(0,0,2) + (0,0,2) \lesssim \mathrm{A}_{1} + \mathrm{A}_{1} \sim 2$\\
        $(0,0,4) + (0,0,2) + (0,0,2)$ & $\mathrm{A}_{3} + \mathrm{A}_{1} + \mathrm{A}_{1} \sim 5$ & $(0,0,2) + (0,0,1) + (0,0,1) \lesssim \mathrm{A}_{1} \sim 1$\\
        $(0,0,3) + (0,0,3) + (0,0,2)$ & $\mathrm{A}_{2} + \mathrm{A}_{2} + \mathrm{A}_{1} \sim 5$ & $(0,0,2) + (0,0,1) + (0,0,1) \lesssim \mathrm{A}_{1} \sim 1$\\
        $(2,2,4) + (2,2,4)$ & $\mathrm{A}_{2} + \mathrm{A}_{2} \sim 4$ & $(2,2,4) + (2,2,4) \sim \mathrm{A}_{1} + \mathrm{A}_{1} \sim 2$
    \end{tblr}
    \caption{A priori possible maximal vertical gauge ranks over $\mathbb{F}_{8}$.}
    \label{tab:F8-maximal-vertical-tunings}
\end{table}
Going through this list and assigning an optimistic naive rank to each enhancement, i.e.\ assuming that it is possible and that the associated monodromy cover is split, we obtain a new rough bound for the vertical gauge rank in this class of models, namely
\begin{equation}
    \rank(\mathfrak{g}_{\mathrm{ver}}) \leq 7\,,
\end{equation}
that is much tighter than \eqref{eq:horizontal-gauge-rank-bound-re}.

In order to improve on this result and obtain a bound that can actually be saturated, we need to analyse the viability of the possible patterns of vertical enhancements, which we now do case by case.
\begin{itemize}
    \item $\mathrm{I}_{m}$ series: Let us start by considering the case of a single vertical line of $\mathrm{I}_{m}$ fibers in the $Y^{P}$ component obtained through a single accidental cancellation. Let us assume without loss of generality that the tuning occurs over the representative of $V_{P}$ given by $\{v=0\}_{B^{P}}$. We then must have
    \begin{subequations}
    \begin{align}
        f_{P} = v^{l_{v}} \check{f}_{P}^{v} + f_{P}^{v}\,,\\
        g_{P} = v^{m_{v}} \check{g}_{P}^{v} + g_{P}^{v}\,,
    \end{align}
    \end{subequations}
    with
    \begin{equation}
        f_{P}^{v} = -3h_{v}^{2}\,,\qquad g_{P}^{v} = 2h_{v}^{3}\,.
    \end{equation}
    Since
    \begin{equation}
        H_{v} = 2S_{P} + 4V_{P} \Rightarrow h_{v} \propto s^{2}\,,
    \end{equation}
    the obstruction granting minimal component vanishing orders over $S_{P}$ must come from $\check{f}_{P}^{v}$ and $\check{g}_{P}^{v}$. Taking into account the non-Higgsable cluster, these are in the curve classes
    \begin{subequations}
    \begin{align}
        \check{F}_{P}^{v} &= (3S_{P}) + \left[ S_{P} + (8-l_{v})V_{P} \right]\,,\\
        \check{G}_{P}^{v} &= (5S_{P}) + \left[ S_{P} + (12-m_{v})V_{P} \right]\,.
    \end{align}
    \end{subequations}
    As a consequence, $\check{f}_{P}^{v}/s^{3}$ and $\check{g}_{P}^{v}/s^{5}$ are irreducible if $l_{v} \leq 0$ and $m_{v} \leq 4$, respectively. It follows that the best we can obtain through a single accidental cancellation structure is $\mathrm{I}_{4}$~fibers. The associated monodromy cover in the $Y^{P}$ component is always split\footnote{To determine the gauge rank for the global vertical enhancement we would need to analyse the monodromy cover in all components, see \cite{Alvarez-Garcia:2023gdd}. We do not do so because, even if it were globally split, the obtained rank would be surpassed by the one obtained from other series below and therefore not inform us about the bound.}
    \begin{equation}
        \psi + \left.\frac{9g_{P}}{2f_{P}}\right|_{v=0} = \psi - \left.3h_{v}\right|_{v=0} = \psi - s^{2}w^{4} = (\psi + sw^{2})(\psi - sw^{2}) = 0\,,
    \end{equation}
    leading to $\rank(\mathfrak{g}_{\mathrm{ver}}) \leq 3$ from this pattern of enhancements.

    Obtaining a higher cancellation structure demands that $l_{v} = m_{v} \geq 1$, see \cref{sec:discriminant-weakly-coupled-components}, and cannot therefore not be realised in these models, since it would lead to non-minimal component vanishing orders above $S_{P}$.

    Finally, we may consider tuning various lines of vertical $\mathrm{I}_{m}$ fibers, each of them obtained through a single accidental cancellation structure. Considering for example tuning two, the same arguments given above lead, \textit{mutatis mutandis}, to $l_{1} + l_{2} \leq 0$ and $m_{1} + m_{2} \leq 4$. This is less efficient and leads to lower vertical gauge ranks than what can be obtained with a single vertical $\mathrm{I}_{m}$ tuning.

    \item $\mathrm{I}_{m}^{*} + \mathrm{I}_{m'}$ series: This case encompasses both the situation in which the lines of vertical $\mathrm{I}_{m}^{*}$ and $\mathrm{I}_{m'}$ fibers are separated, as well as the one in which the latter are brought on top of the former to produce a higher $\mathrm{I}_{m}^{*}$ enhancement.

    Start by tuning a line of vertical $\mathrm{I}_{0}^{*}$ fibers, which can be located without loss of generality over $\{v=0\}_{B^{P}}$. This enhances the non-Higgsable cluster to
    \begin{equation}
        \ord{Y^{P}}(f_{P},g_{P},\Delta_{P}^{\prime})_{s=0} = (4,5,10)\,.
    \end{equation}
    As a consequence, the only obstruction to having non-minimal component vanishing orders over $S_{P}$ comes from $g_{P}$. Consider tuning an additional $\mathrm{I}_{m}$ type vertical line of fibers over $C = \{p_{C} = 0\}_{B^{P}}$. Writing $g_{P}$ as
    \begin{equation}
        g_{P} = p_{C}^{m_{C}} \check{g}_{P}^{C} + g_{P}^{C}\,,
    \end{equation}
    and noting that
    \begin{equation}
        \check{G}_{P}^{C} = (5S_{P}) + \left[ S_{P} + (12 - 3 - m_{1})V_{P} \right]\,,
    \end{equation}
    we see that $m_{C} \leq 1$ and at most we can obtain $\mathrm{I}_{1}$ fibers through a single accidental cancellation, which by themselves do not lead to an increase in the vertical gauge rank. Hence, we need to analyse the cases in which we have
    \begin{equation}
        \ord{Y^{P}}(f_{P},g_{P},\Delta_{P}^{\prime})_{v=0} = (2,3,6)\quad \text{or}\quad (2,3,7)\,.
    \end{equation}
    
    For the first of these, the monodromy cover in the component is non-split,
    \begin{equation}
        \psi^{3} + \psi \left.\left( \frac{f_{P}}{v^{2}} \right)\right|_{v=0} + \left.\left( \frac{g_{P}}{v^{3}}\right)\right|_{v=0} = \psi^{3} + \psi s^{4}w^{6} + s^{5}(sw^{9}+e_{P-1}w) = 0\,,
    \end{equation}
    leading to $\mathrm{I}_{0}^{*\, \mathrm{ns}}$ fibers. This folding of the algebra affects the global enhancement over $\mathcal{F}$, and we therefore have an associated $\mathfrak{g}_{2}$ algebra.

    For the second case, we have instead the monodromy cover
    \begin{equation}
        \psi^{2} + \left. \frac{1}{4} \left( \frac{\Delta_{P}^{\prime}}{v^{7}} \right) \left( \frac{2 v f_{P}}{9 g_{P}} \right)^{3} \right|_{v=0} = 0\,.
    \end{equation}
    Since $m_{C} = 1$, the structure of $\Delta'_{P}$ must be
    \begin{equation}
        \Delta_{P}^{\prime} = v h_{C}^{k} \Delta_{P}^{\prime\prime}\,,\qquad k \geq 3\,.
    \end{equation}
    The divisor class of $\Delta_{P}^{\prime\prime}$ is then
    \begin{equation}
        \Delta_{P}^{\prime\prime} = \left[(10-2k)S_{P}\right] + \left[2S_{P} + (24 - 1 - 4k)V_{P}\right]\,.
    \end{equation}
    The first term in the preceding expression corresponds to the part of the enhanced non-Higgsable cluster that is not accounted for by the $h_{C}^{k}$ factor. If $k \geq 4$, the second term is generically reducible, factorising $2S_{P}$, and since $h_{C} \propto s^{2}$ we obtain $\Delta_{P}^{\prime} \propto s^{2k}s^{10-2k+2} = s^{12}$, i.e.\ to non-minimal component vanishing orders over $S_{P}$. The monodromy cover for the remaining case $k=3$ takes the form
    \begin{equation}
        \left. \frac{1}{4} \left( \frac{\Delta_{P}^{\prime}}{v^{7}} \right) \left( \frac{2 v f_{P}}{9 g_{P}} \right)^{3} \right|_{v=0} = \left.\left( - \frac{1}{108} \frac{1}{v^{3}} \Delta_{P}^{\prime\prime}\right)\right|_{v=0}\,.
    \end{equation}
    Given that
    \begin{equation}
        \Delta_{P}^{\prime\prime} - 3V_{P} \sim 5S_{P} + T_{P}\,,
    \end{equation}
    we therefore generically have
    \begin{equation}
        \psi^{2} - \frac{1}{108}s^{5} p_{1,8}([s:t],[v:w]) = 0\,.
    \end{equation}
    Forcing the monodromy to be split would require tuning the $t$ term in $p_{1,8}([s:t],[v:w])$ to be zero, but this is the term preventing the non-minimal enhancement over $S_{P}$. Therefore, we have a $\mathrm{I}_{1}^{*\, \mathrm{ns}}$ fiber, associated with the gauge algebra $\mathrm{B}_{4}$. We have realised such a model in an explicit example, meaning that $\rank(\mathfrak{g}_{\mathrm{ver}}) = 4$ is possible; hence, those candidate patters of vertical tunings that do not surpass this rank in the following series do not need to be analysed explicitly for the determination of the bound.

    \item $\mathrm{III} + \mathrm{I}_{m}$ series: The only candidate in this series that could surpass the putative bound of $\rank(\mathfrak{g}_{\mathrm{ver}}) \leq 4$ is $\mathrm{III} + \mathrm{I}_{5}$, but we have noted earlier that tuning $\mathrm{I}_{5}$ type vertical lines is not even possible in the absence of the $\mathrm{III}$ enhancement. It can be seen, through arguments analogous to the ones employed above, that we can at best have
    \begin{equation}
        (1,2,3) + (0,0,2) \sim \mathrm{A}_{1} + \mathrm{A}_{1}\,,
    \end{equation}
    or
    \begin{equation}
        (1,2,3) + (1,2,3) \sim \mathrm{A}_{1} + \mathrm{A}_{1}\,,
    \end{equation}
    after which no additional vertical tunings are possible. Both of these are well below the putative bound.

    \item $\mathrm{IV}$ series: It can be checked that all vertical enhancement patters involving a line of IV fibers lie below the putative bound of $\rank(\mathfrak{g}_{\mathrm{ver}}) \leq 4$. One of these patterns, namely having $\mathrm{IV} + \mathrm{IV}$, can in principle saturate the bound. However, the vertical tuning enhances the non-Higgsable cluster to
    \begin{equation}
        \ord{Y^{P}}(f_{P},g_{P},\Delta_{P}^{\prime})_{s=0} = (4,5,10)\,,
    \end{equation}
    which means that $g_{P}$ must contain a $p_{1,8}([s:t],[v:w])$ factor whose $t$ term cannot vanish if we want to avoid a non-minimal enhancement over $S_{P}$. This makes the monodromy cover for both lines of IV fibers take the form
    \begin{equation}
        \psi^{2} - \left.\frac{g}{w^{2}}\right|_{w=0} = \psi^{2} - s^{5} p_{1,8}([s:t],[v:w]) \left( q_{0,1}([s:t],[v:w]) \right)^{2} = 0\,,
    \end{equation}
    which is always non-split, meaning that at best we can achieve $\mathrm{A}_{1} + \mathrm{A}_{1}$ with this pattern.
\end{itemize}

Altogether, we have obtained an explicit example that reaches $\rank(\mathfrak{g}_{\mathrm{ver}}) = 4$, and checked that those candidates that could in principle surpass this value either cannot be tuned, or have non-split monodromy cover that reduce their rank. We conclude that, for a horizontal model constructed over $\hat{B} = \mathbb{F}_{8}$ with no components at weak coupling, the bound on the vertical gauge rank is
\begin{equation}
    \rank(\mathfrak{g}_{\mathrm{{ver}}}) \leq 4\,.
\end{equation}

\subsection{\texorpdfstring{$\hat{B} = \mathbb{F}_{7}$}{B=F7} with components at weak coupling}
\label{sec:B-F7-weak-coupling}

Once we allow some components to have codimension-zero Kodaira type $\mathrm{I}_{n_{p}>0}$ elliptic fibers, the bounds on the vertical gauge rank computed above can, and indeed in most cases do, become stricter. The reason for this is the restriction on the types of enhancements that can occur over components with $\mathrm{I}_{n_{p} > 0}$ codimension-zero fibers, which, as we explained in \cref{sec:horizontal-local-global-brane-content}, must be compatible with the local weak coupling. This means that, focusing on those types of singular fibers associated to a non-trivial gauge algebra, the global vertical enhancements can only be of Kodaira type $\mathrm{III}$, $\mathrm{IV}$, $\mathrm{I}_{m}$ or $\mathrm{I^{*}_{m}}$ if at least one component is at weak coupling; in a global weak coupling limit only the latter two can be realised. If the bound given in \cref{tab:vertical-rank-bounds} for the models without codimension-zero singular fibers cannot be saturated using only these types of enhancements, the presence of the components at weak coupling will correspondingly decrease the possible maximal vertical gauge rank. Note that, due to the non-generic nature of the Weierstrass models that support such codimension-zero singular fibers, it may be that even if enhancements of these types can saturate the bound in the generic case, they are no longer viable in models with components at weak coupling.

The bounds given in \cref{tab:vertical-rank-bounds} can still be saturated for horizontal models constructed over $\hat{B} = \mathbb{F}_{n}$, with $8 \leq n \leq 12$, even when there are components at weak coupling. We take instead those horizontal models constructed over $\hat{B} = \mathbb{F}_{7}$ to illustrate the reduction in the possible vertical gauge rank. We will assume in the derivation below that the end component $Y^{P}$ is not at weak coupling, since this is less constraining than the alternative, which would likely lead to a stronger bound. According to \cref{tab:vertical-rank-bounds}, in the absence of codimension-zero $\mathrm{I}_{n_{p}>0}$ fibers we have
\begin{equation}
    \rank(\mathfrak{g}_{\mathrm{ver}}) \leq 8\,.
\end{equation}
This bound can most easily be saturated by tuning a vertical line of $\mathrm{II}^{*}$ fibers; this is one type of enhancement that is no longer possible over $\mathcal{F}$ once we have even a single component at weak coupling.

We keep the discussion brief, since the analysis follows the same lines as the one in the previous section. The non-Higgsable in the $Y^{P}$ component of the central fiber $Y_{0}$ of a resolved horizontal model constructed over $\hat{B} = \mathbb{F}_{8}$ is
\begin{equation}
    \ord{Y^{P}}(f_{P},g_{P},\Delta_{P}^{\prime})_{s=0} = (3,5,9)\,,
\end{equation}
see \cref{tab:non-Higgsable-clusters}. Hence, \eqref{eq:F8-SP-bound} and \eqref{eq:F8-rough-VP-bound} still apply, and we have a budget of at most $10V_{P}$ vertical classes in $\Delta_{P}^{\prime}$ to be distributed over the local vertical enhancements in the $Y^{P}$ component, a fact that is mirrored globally for the $\mathcal{F}$ class in $\Dphys$. The subset of a priori possible vertical enhancement patters that both use these $10V_{P}$ available classes and are compatible with some components being at weak coupling are listed in \cref{tab:F7-maximal-vertical-tunings}.
\begin{table}[t!]
    \centering
    \resizebox{\textwidth}{!}{
    \begin{tblr}{columns = {c}, hlines, vlines}
        Vanishing orders & Naive rank & Result of the analysis\\
        $(0,0,10)$ & $\mathrm{A}_{9} \sim 9$ & $(0,0,5) \lesssim \mathrm{A}_{4} \sim 4$\\
        $(1,2,3) + (0,0,7)$ & $\mathrm{A}_{1} + \mathrm{A}_{6} \sim 7$ & $(1,2,3) + (0,0,3) \lesssim \mathrm{A}_{1} + \mathrm{A}_{2} \sim 3$\\
        $(2,2,4) + (0,0,6)$ & $\mathrm{A}_{2} + \mathrm{A}_{5} \sim 7$ & $(2,2,4) + (0,0,3) \lesssim \mathrm{A}_{2} + \mathrm{A}_{2} \sim 4$\\
        $(1,2,3) + (1,2,3) + (1,2,3)$ & $\mathrm{A}_{1} + \mathrm{A}_{1} + \mathrm{A}_{1} \sim 3$ & $(1,2,3) + (1,2,3) \sim \mathrm{A}_{1} + \mathrm{A}_{1} \sim 2$\\
        $(1,2,3) + (2,2,4) + (0,0,3)$ & $\mathrm{A}_{1} + \mathrm{A}_{2} + \mathrm{A}_{2} \sim 5$ & $(1,2,3) + (2,2,4) + (0,0,1) \lesssim \mathrm{A}_{1} + \mathrm{A}_{2} \sim 3$\\
        $(2,2,4) + (2,2,4) + (0,0,2)$ & $\mathrm{A}_{2} + \mathrm{A}_{2} + \mathrm{A}_{1} \sim 4$ & $(2,2,4) + (2,2,4) + (0,0,1) \lesssim \mathrm{A}_{2} + \mathrm{A}_{2} \sim 4$\\
        $(2,3,10)$ & $\mathrm{D}_{8} \sim 8$ & $(2,3,8) \sim \mathrm{B}_{5} \sim 5$\\
        $(2,3,6) + (0,0,4)$ & $\mathrm{D}_{4} + \mathrm{A}_{3} \sim 7$ & $(2,3,6) + (0,0,2) \lesssim \mathrm{D}_{4} + \mathrm{A}_{1} \sim 5$\\
        $(2,3,6) + (1,2,3)$ & $\mathrm{D}_{4} + \mathrm{A}_{1} \sim 5$ & $(2,3,6) + (1,2,3) \lesssim \mathrm{D}_{4} + \mathrm{A}_{1} \sim 5$\\
        $(2,3,6) + (2,2,4)$ & $\mathrm{D}_{4} + \mathrm{A}_{2} \sim 6$ & $(2,3,6) + (2,2,4) \lesssim \mathrm{D}_{4} + \mathrm{A}_{1} \sim 5$
    \end{tblr}
    }
    \caption{Possible maximal vertical gauge ranks over $\mathbb{F}_{7}$ in the presence of codimension-zero $\mathrm{I}_{n_{p}>0}$ fibers over the intermediate components.}
    \label{tab:F7-maximal-vertical-tunings}
\end{table}

Let us now do a detailed analysis, case by case.
\begin{itemize}
    \item $\mathrm{I}_{m}$ series: Considering first the case of a single vertical line of $\mathrm{I}_{m}$ fibers, the same argument of \cref{sec:B-F8-no-weak-coupling} leads in this case to $l_{v} \leq 1$ and $m_{v} \leq 5$ for $\check{f}_{P}^{v}/s^{3}$ and $\check{g}_{P}^{v}/s^{5}$ to be irreducible, respectively. This means that, if the $\mathrm{I}_{m}$ fibers are tuned through a single accidental cancellation structure, we can obtain up to Kodaira type $\mathrm{I}_{5}$. The tuning is less efficient in terms of obtained rank if we try to obtain various lines of $\mathrm{I}_{m}$ fibers.\footnote{It could occur that a single vertical line of $\mathrm{I}_{m}$ fibers is forced, for high $m$, to be non-split, its rank being then surpassed by various independent tunings that, although naively would lead to a smaller rank, allow for a split monodromy cover. This cannot be the case here, since the vertical $\mathrm{I}_{m}$ singularities turn out to be split for horizontal models constructed over $\hat{B} = \mathbb{F}_{7}$.}

    For a double accidental cancellation to be possible, we need $l_{v} = m_{v} =: r_{v}$, which means that $r_{v} \leq 1$ and, as a consequence, we can at best obtain a vertical line of $\mathrm{I}_{2}$ fibers through this type of tuning.

    Higher accidental cancellation structures cannot be realised in this class of models. These build on top of each other, and therefore the first one to be considered is that of triple accidental cancellations. We use the same notation as in \cref{sec:discriminant-weakly-coupled-components}, slightly adapted. From \eqref{eq:double-accidental-cancellation-condition} we see that, since $h_{P}^{v} \propto s^{2}$, we must have $\check{f}_{P}^{v} \propto s^{3}$ at most; otherwise, the terms in $\check{g}_{P}^{v}$ proportional to $s^{5}$ or smaller would need to be set to zero for the accidental cancellations to be possible, which would lead to a non-minimal enhancement over $S_{P}$. In fact, we see from that the presence of the non-Higgsable cluster means that
    \begin{equation}
        \begin{rcases}
        h_{P}^{v} \propto s^{2}\\
        \check{f}_{P}^{v} \propto s^{3}\\
        \check{g}_{P}^{v} \propto s^{5}
        \end{rcases}
        \Rightarrow q_{P}^{v} \propto s^{5}\,.
    \end{equation}
    In the triple accidental cancellation structure \eqref{eq:triple-accidental-cancellation-condition} this means that the first term is $h_{P}^{v}q_{P}^{v} \propto s^{7}$, while the second is $\check{f}_{P}^{v} \propto s^{6}$ and cannot go higher, meaning that the cancellation cannot occur.

    \item $\mathrm{III}$ series: Tuning a vertical $\mathrm{III} + \mathrm{III}$ enhancement is possible without producing non-minimal fibers over $S_{P}$, while adding a third $\mathrm{III}$ forces $4S_{P}$ and $6S_{P}$ to factorise in $F_{P}$ and $G_{P}$, respectively. The result is below the rank that could be achieved with the vertical line of $\mathrm{I}_{m}$ fibers. After tuning $\mathrm{III} + \mathrm{III}$ we see, arguing as we did above, that we can additionally have a single accidental cancellation producing a line of vertical $\mathrm{I}_{1}$ fibers, which does not increase the obtained rank.

    \item $\mathrm{III} + \mathrm{IV}$ series: Tuning a vertical $\mathrm{III} + \mathrm{IV}$ leads to the residual divisors
    \begin{subequations}
    \begin{align}
        F_{P} - V_{P} - 2V_{P} &= 4S_{P} + 5V_{P}\,,\\
        G_{P} - 2V_{P} - 2V_{P} &= 5S_{P} + (S_{P} + 8V_{P})\,,
    \end{align}
    \end{subequations}
    from which we see that no non-abelian gauge algebra involving a factorisation of $V_{P}$ classes in $F_{P}$ and $G_{P}$ can be tuned without leading to a non-minimal enhancement over $S_{P}$. It is possible to tune an additional vertical line of $\mathrm{I}_{1}$ fibers, which does not, however, increase the rank. This means that, even if the $\mathrm{IV}$ singularities have a split monodromy cover, we have $\rank(\mathfrak{g}_{\mathrm{ver}}) \leq 3$ in this series.

    \item $\mathrm{IV}$ series: Arguing analogously, we observe that we can tune a vertical $\mathrm{IV} + \mathrm{IV}$, with room left only for an accidental cancellation yielding an additional vertical line of $\mathrm{I}_{1}$ fibers. Even if both $\mathrm{IV}$ singularities are split, we obtain at most $\rank(\mathfrak{g}_{\mathrm{ver}}) \leq 4$ in this series, which is below the maximal vertical gauge rank that we will achieve below.

    \item $\mathrm{III} + \mathrm{I}_{m'}$ series: Tuning a single vertical line of $\mathrm{III}$ fibers still leaves room to tune a single accidental cancellation up to a vertical line of Kodaira type $\mathrm{I}_{3}$ fibers which, even if split, would yield at most $\rank(\mathfrak{g}_{\mathrm{ver}}) \leq 3$ in this series.

    \item $\mathrm{IV} + \mathrm{I}_{m}$ series: Similarly to the previous case, we can tune a vertical line of $\mathrm{IV}$ fibers and, additionally, a single accidental cancellation up to a vertical line of Kodaira type $\mathrm{I}_{3}$ fibers. This leads, at best, to $\rank(\mathfrak{g}_{\mathrm{ver}}) \leq 4$ in this series.

    \item $\mathrm{I}_{m}^{*} + \mathrm{I}_{m'}$ series: Tuning a vertical line of $\mathrm{I}_{0}^{*}$ fibers leaves enough room for accidental cancellations to occur, with
    \begin{subequations}
    \begin{align}
        \check{f}_{P}^{v} &= (3S_{P}) + \left[ S_{P} + (8 - 2 - l_{v})V_{P} \right]\,,\\
        \check{g}_{P}^{v} &= (5S_{P}) + \left[ S_{P} + (12 - 3 - m_{v})V_{P} \right]\,.
    \end{align}
    \end{subequations}
    This means that only the single accidental cancellation structure can be realised, with $m_{v} \leq 2$, leaving as possible vertical enhancement patterns
    \begin{align}
        (2,3,8) &\lesssim \mathrm{D}_{6}\,,\\
        (2,3,7) + (0,0,1) &\lesssim \mathrm{B}_{5}\,,\\
        (2,3,6) + (0,0,2) &\lesssim \mathrm{D}_{4} + \mathrm{A}_{1}\,.
    \end{align}
    Out of these tunings $\mathrm{I}_{2}^{*}$ reaches the highest possible vertical rank in the series, even if it is non-split. To determine whether it can be split or not, we first look at the monodromy cover in the $Y^{P}$ component. For $m_{v} = 2$, we see that the only term preventing a non-minimal enhancement over $S_{P}$ is the $t$ term in the factor $p_{1,7}([s:e_{P-1}],[v:w]) = S_{P} + 7V_{P}$ in $g_{P}$. The monodromy cover in the component is
    \begin{equation}
        \psi^{2} + \left.\left( \frac{\Delta'_{P}}{v^{7}} \right)\left( \frac{2 v f_{P}}{9g_{P}} \right)^{2}\right|_{v=0} = 0\,,
    \end{equation}
    which will be non-split due to the presence of that very same term. We have explicitly constructed such an enhancement in a horizontal model constructed over $\hat{B} = \mathbb{F}_{7}$ with a $\mathrm{I}_{0} - \mathrm{I}_{1} - \mathrm{I}_{0}$ pattern of codimension-zero singular elliptic fibers.

    \item $\mathrm{I}_{m}^{*} + \mathrm{III}$ series: It is possible to tune a vertical $\mathrm{I}_{m}^{*} + \mathrm{III}$, with no room left for further non-abelian vertical enhancements. This would give, at best, $\mathrm{D}_{4} + \mathrm{A}_{1}$, which would not improve the vertical gauge rank found above.

    \item $\mathrm{I}_{m}^{*} + \mathrm{IV}$ series: Tuning a vertical $\mathrm{I}_{m}^{*} + \mathrm{IV}$ is possible, with no further vertical lines of singular fibers possible. In principle, this can surpass the rank obtained above if both singularities are split, giving $\mathrm{D}_{4} + \mathrm{A}_{2}$. The tuning leads to
    \begin{align}
        F_{P} &= (2V_{P}) + (2V_{P}) + (4S_{P}) + \left[ 4V_{P} \right]\,,\\
        G_{P} &= (3V_{P}) + (2V_{P}) + (5S_{P}) + \left[ S_{P} + 7V_{P} \right]\,,
    \end{align}
    where we see that having non-minimal fibers over $S_{P}$ is only prevented by the $t$ term in the factor $p_{1,7}([s:t],[v:w]) = S_{P} + 7V_{P}$ in $g_{P}$. This factor cannot be set to zero, and it makes the monodromy cover of the IV enhancement in the $B^{P}$ component, which has the form
    \begin{equation}
        \psi^{2} - \left. \left( \frac{g_{P}}{v^{2}} \right) \right|_{v=0} = 0\,,
    \end{equation}
    non-split. This implies that the enhancement is then at most $\mathrm{D}_{4} + \mathrm{A}_{1}$, and we do not need to analyse the case further.
\end{itemize}

We conclude that, in the presence of components at weak coupling, the vertical gauge rank in horizontal models constructed over $\hat{B} = \mathbb{F}_{7}$ must satisfy
\begin{equation}
    \rank(\mathfrak{g}_{\mathrm{ver}}) \leq 5\,.
\end{equation}
The highest possible vertical gauge rank has decreases with respect to the one that can be achieved in the absence of codimension-zero singular elliptic fibers, shown in \cref{tab:vertical-rank-bounds}.

\section{Defect algebras in the heterotic dual}
\label{sec:defect-algebras-heterotic-dual}

Motivated by the goal to understand the vertical gauge algebras as defects in the decompactifi\-cation limits, we use this appendix to elaborate on their interpretation as non-perturbative gauge algebra factors in the heterotic dual. Among these factors, those not related to a tensor branch transition originate from point-like instantons with discrete holonomy probing geometric singularities on the heterotic K3 surface \cite{Aspinwall:1998xj}. These are also known as fractional point-like instantons. As a novel point, we emphasise in \cref{sec:defect-algebras-heterotic-dual-broken-E8xE8} the relevance of the distribution of such point-like instantons between the two Ho\v{r}ava-Witten walls for realising the non-perturbative gauge algebra. This is to be contrasted with the behaviour of point-like instantons with trivial holonomy, which we discuss in \cref{sec:defect-algebras-heterotic-dual-unbroken-E8xE8}. The natural appearance of heterotic ADE singularities probed by fractional point-like instantons in the analysis of the asymptotic physics of codimension-one infinite-distance degenerations of F-theory models has recently sparked interest in a detailed local analysis of the Higgs branch of the associated LSTs \cite{DelZotto:2022ohj,DelZotto:2022xrh,DelZotto:2023nrb,Ahmed:2023lhj,DelZotto:2023ahf,Lawrie:2023uiu}.

Heterotic string theory compactified on a K3 surface with ADE singularities was analysed by Witten in \cite{Witten:1999fq}. Interestingly, such singularities behave rather differently from how they do in M-theory or Type IIA string theory.

As is well known, an ADE singularity in the internal space of M-theory or Type IIA string theory signals a gauge enhancement of  ADE type.
The exceptional $\mathbb{P}^{1}$ curves shrinking to zero volume at the corresponding point in moduli space have an intersection matrix
reproducing the Cartan matrix of the underlying ADE Lie algebra. The M2-branes wrapping these exceptional curves lead to a series of massless particles that furnish the non-abelian gauge algebra. This is precisely how such algebras arise in the F-theory limit of M-theory.\footnote{To be fully precise, the resulting algebra may not be of ADE type after folding by a monodromy action.}

The conclusion of \cite{Witten:1999fq} is that the heterotic string theory near an ADE singularity does not exhibit a corresponding non-perturbative gauge algebra unless it is probed by small instantons. The analysis proceeds by considering heterotic string theory in the absence of small instantons, which makes it tractable in the framework of conformal field theory. The appearance of gauge algebras like the ones discussed is associated to singularities in the moduli space of the theory, which are a possible signal of non-perturbative physics. Working at string tree level, \cite{Witten:1999fq} analyses the moduli space of the theory, finding that it should be smooth once $\alpha'$-corrections are taken into account,\footnote{More concretely, \cite{Witten:1999fq} proposes that the moduli space for heterotic string theory near an ADE singularity of type $G$ is the space of vacua of a minimal supersymmetric 3D $\mathcal{N}=4$ gauge theory with gauge group $G$.} and therefore concluding that no non-perturbative gauge algebra arises.\footnote{The analysis of \cite{Witten:1999fq} holds both for $\mathrm{E}_{8} \times \mathrm{E}_{8}$ and $\mathrm{Spin}(32)/\mathbb{Z}_{2}$ heterotic string theory, since it is performed at string tree level and the two theories are distinguished by the fluctuations around the $F=0$ background appearing once string loops are included.}

The situation changes if the ADE singularity is probed by a small instanton. This can be heuristically understood by looking at the classical equation of motion for the ten-dimensional dilaton \cite{Witten:1999fq}, which can be schematically written as 
\begin{equation}
    \Delta^{2}\phi = \tr F_{ij}F^{ij} - \tr R_{ij}R^{ij}\,.
\end{equation}
Here $F_{ij}$ is the curvature of the heterotic gauge bundle and $R_{ij}$ is the Ricci tensor of the internal space. A small instanton is a skyscraper sheaf, i.e.\ a singular gauge bundle in which all the curvature has been concentrated at a point. Such a configuration hence locally drives the dilaton to strong coupling. It is, therefore, not surprising that in the presence of small instantons probing the ADE singularity non-perturbative features become manifest. In their absence, the geometric curvature associated with the ADE singularity works in the opposite direction, driving the dilaton towards small coupling, and hence preventing non-perturbative effects from arising.

The F-theory/heterotic duality, reviewed in \cref{sec:F-theory-heteoric-duality}, allows us to establish a connection between horizontal Type II.a models in the adiabatic regime and their controlled heterotic duals. This was done in \cref{sec:horizontal-models-heterotic-duals}: The gauge algebras localised at the six-dimensional defects  within the decompactified theory have a heterotic dual interpretation
in terms of ADE singularities of the internal K3 surface probed by small instantons, which we have just discussed above. Since these defect algebras are an important feature of the asymptotic physics of horizontal Type II.a models, it is worth revisiting them in some more detail from the point of view of F-theory/heterotic duality.

Let us recall that, according to this duality, the heterotic K3 surface can be identified with the intersection $Y^{0} \cap Y^{1}$ of the components of the central fiber $Y_{0}$ of the resolved horizontal Type II.a model. The 24 singular elliptic fibers of the heterotic K3, counted with multiplicity, correspond to the 24 non-generic vertical slices on the F-theory side, i.e.\ to the intersection points
\begin{equation}
    \Delta'_{0} \cdot S_{0} = \Delta'_{1} \cdot T_{1} = 24\,.
\end{equation}
The defining polynomials $f_{b}$, $g_{b}$ and $\Delta_{b}$ of the Weierstrass model of the resolved family variety $\mathcal{Y}$ of the horizontal Type II.a model contain the information about the elliptically fibered heterotic K3 surface. Namely, the defining polynomials of its Weierstrass model are
\begin{equation}
    f_{\mathrm{K3}} := \left. f_{b} \right|_{e_{0}=e_{1}=0}\,,\qquad g_{\mathrm{K3}} := \left. g_{b} \right|_{e_{0}=e_{1}=0}\,,\qquad \Delta_{\mathrm{K3}} := \left. \Delta_{b} \right|_{e_{0}=e_{1}=0}\,.
\end{equation}
As a consequence, the types of ADE singularities present in the heterotic K3 surface can be read off from the Kodaira-N\'eron classification of singular elliptic fibers using the interface vanishing orders on the F-theory side. As pointed out in \cite{Alvarez-Garcia:2023gdd}, the same information can be obtained directly from the unresolved horizontal Type II.a model. Denoting the base blow-up map involved in the open-chain resolution process by $\pi: \mathcal{B} \rightarrow \hat{\mathcal{B}}$, we have that
\begin{subequations}
\begin{align}
        \pi_{*}(F_{0}) &= \left. F \right|_{\mathcal{U}} - 4 \left( \mathcal{S} \cap \mathcal{U} \right)\,,\\
        \pi_{*}(G_{0}) &= \left. G \right|_{\mathcal{U}} - 6 \left( \mathcal{S} \cap \mathcal{U} \right)\,,\\
        \pi_{*}(\Delta_{0}) &= \left. \Delta \right|_{\mathcal{U}} - 12 \left( \mathcal{S} \cap \mathcal{U} \right)\,,
\end{align}
\end{subequations}
and as a consequence
\begin{equation}
    \ord{Y^{0} \cap Y^{1}}(f_{\mathrm{K3}}, g_{\mathrm{K3}}, \Delta_{\mathrm{K3}})_{\mathcal{Z}} = \ord{\pi^{*}(\mathcal{S} \cap\;\! \mathcal{U})} \left( \left. \frac{f}{s^{4}} \right|_{u=s=0}, \left. \frac{g}{s^{6}} \right|_{u=s=0}, \left. \frac{\Delta}{s^{12}} \right|_{u=s=0} \right)_{\mathcal{Z}}\,.
\end{equation}
To put it differently, the complete information on the heterotic K3 surface at the endpoint of the limit is contained in the coefficients in $f$, $g$ and $\Delta$ (the defining polynomials of the Weierstrass model of the unresolved family variety $\hat{\mathcal{Y}}$) of the terms of middle homogeneous degree in the coordinates of $\mathbb{P}^{1}_{f}$ and independent on the coordinate of $D$. The remaining terms independent of the coordinate of $D$ encode information on the heterotic gauge bundles at the endpoint of the limit \cite{Morrison:1996pp}.

In horizontal Type II.a limits in the adiabatic regime, the heterotic K3 surface decompactifies. Since this is a continuous process, the topology of the internal space is not changed by it, and the degeneration loci of the elliptic fiber are still present. A local patch around one such locus is equivalent to heterotic string theory on $\mathbb{C}^{2}/\Gamma_{\mathfrak{g}} \times \mathbb{R}^{1,5}$, where $\Gamma_{\mathfrak{g}} \hookrightarrow \mathrm{SU}(2)$ is a finite subgroup of $\mathrm{SU}(2)$ related to the singularity of the K3 surface via the McKay correspondence. In studies oriented towards understanding 6D SCFTs, this is the local picture usually taken, see \cite{Heckman:2018jxk} for a review.

Heterotic ADE singularities in the presence of small instantons were studied in \cite{Aspinwall:1997ye}, assuming that the perturbative $\mathrm{E}_{8} \times \mathrm{E}_{8}$ heterotic gauge group is unbroken, by using the duality to the \mbox{F-theory} side in the stable degeneration limit. We revisit and refine this discussion in \cref{sec:defect-algebras-heterotic-dual-unbroken-E8xE8}, considering the case in which the heterotic bulk gauge group is Higgsed, from a six-dimensional standpoint, in \cref{sec:defect-algebras-heterotic-dual-broken-E8xE8}. The situation in which enough ADE singularities of the heterotic K3 surface coalesce as to produce a non-minimal point was explored in \cref{sec:non-minimal-points-heterotic-K3-surface}.

\subsection{Unbroken horizontal \texorpdfstring{$\mathrm{E}_{8} \times \mathrm{E}_{8}$}{E8xE8} gauge algebra}
\label{sec:defect-algebras-heterotic-dual-unbroken-E8xE8}

Consider the subclass of horizontal Type II.a models constructed over the Hirzebruch surface $\hat{B} = \mathbb{F}_{n}$ in which, from the six-dimensional standpoint, the horizontal gauge algebra is
\begin{equation}
    \mathfrak{g}_{\mathrm{hor}} = \mathfrak{e}_{8} \oplus \mathfrak{e}_{8}\,,
\end{equation}
obtained by supporting the gauge factors
\begin{equation}
    \mathcal{H}_{\infty}^{0}: \mathrm{E}_{8}\,,\qquad \mathcal{H}_{0}^{1}: \mathrm{E}_{8}\,.
\end{equation}
The divisor classes associated with the defining polynomials of the Weierstrass model of the components $\{Y^{p}\}_{0 \leq p \leq 1}$ are
\begin{subequations}
\begin{alignat}{3}
    F_{0} &= 4T_{0} + F^{\mathrm{res}}_{0}\,,\qquad &F^{\mathrm{res}}_{0} &:= 8V_{0}\,,\\
    G_{0} &= 5T_{0} + G^{\mathrm{res}}_{0}\,,\qquad &G^{\mathrm{res}}_{0} &:= T_{0} + 12V_{0}\,,\\
    \Delta'_{0} &= 10T_{0} + \Delta^{\mathrm{res}}_{0}\,,\qquad &\Delta^{\mathrm{res}}_{0} &:= 2T_{0} + 24V_{0}\,,
\end{alignat}
\end{subequations}
in the $B^{0}$ component, and
\begin{subequations}
\begin{alignat}{3}
    F_{1} &= 4S_{1} + F^{\mathrm{res}}_{1}\,,\qquad &F^{\mathrm{res}}_{1} &:= 8V_{1}\\
    G_{1} &= 5S_{1} + G^{\mathrm{res}}_{1}\,,\qquad &G^{\mathrm{res}}_{1} &:= S_{1} + 12V_{1}\\
    \Delta'_{1} &= 10S_{1} + \Delta^{\mathrm{res}}_{1}\,,\qquad &\Delta^{\mathrm{res}}_{1} &:= 2S_{1} + 24V_{1}
\end{alignat}
\end{subequations}
in the $B^{1}$ component. In such a model we have the linear equivalence $\Delta_{p}^{\mathrm{res}} = 2G_{p}^{\mathrm{res}}$, for $p=0,1$, but also the identity of sets
\begin{equation}
    \Delta_{0}^{\mathrm{res}} \cap T_{0} = G_{0}^{\mathrm{res}} \cap T_{0}\,,\qquad \Delta_{1}^{\mathrm{res}} \cap S_{1} = G_{1}^{\mathrm{res}} \cap S_{1}\,.
\end{equation}
This can be directly seen from the polynomials whose vanishing loci describe the concrete divisor representatives associated with the model. Hence, the
\begin{subequations}
\begin{align}
    G_{0}^{\mathrm{res}} \cdot T_{0} &= 12+n\quad \text{intersections in $B^{0}$}\\
    \text{and}\quad G_{1}^{\mathrm{res}} \cdot S_{1} &= 12-n\quad \text{intersections in $B^{1}$}
\end{align}
\label{eq:point-like-instanton-intersections}%
\end{subequations}
of $G_{\mathrm{phys}}^{\mathrm{res}}$ with the two horizontal lines of $\mathrm{II}^{*}$ fibers correspond, in a sufficiently generic  model, to that many distinct nodes of $\Delta_{\mathrm{phys}}^{\mathrm{res}}$; their collision with the representatives of $\mathcal{H}_{\infty}^{0}$ and $\mathcal{H}_{0}^{1}$ supporting the horizontal gauge enhancements leads to a series of codimension-two finite-distance non-minimal points.

As explained in \cite{Aspinwall:1997ye}, these codimension-two finite-distance non-minimal points correspond on the heterotic side to point-like instantons with trivial holonomy. Such singular gauge bundle contributions do indeed preserve the unbroken perturbative heterotic gauge group, as expected from the F-theory side. Since they each contribute one unit of instanton charge to the integrated Bianchi identity, their distribution among the two heterotic $\mathrm{E}_{8}$ bundles must agree with the instanton number splitting $c_{2}(V_{0,1}) = 12 \pm n$. This matches the assignment found in \eqref{eq:point-like-instanton-intersections}. Point-like instantons with trivial holonomy can be traded, through small instanton transitions, for M5-branes moving in the Ho\v{r}ava-Witten interval. Their position in $S^{1}/\mathbb{Z}_{2}$ corresponds to the volume of the exceptional $\mathbb{P}^{1}$ curves appearing on the F-theory side when the codimension-two finite-distance non-minimal points are resolved, as we already reviewed in \cref{sec:F-theory-heteoric-duality}. Of special relevance to our current discussion is the 1 hypermultiplet parametrising the position of a point-like instanton on the heterotic K3 surface. It is unaffected by small instanton transitions, and therefore plays the same role if we wish to take the M5-brane perspective. Consider the position within the $\mathbb{P}^{1}_{b}$ base of the heterotic K3 surface of a point-like instanton with trivial holonomy, corresponding to a certain node of $\Delta_{\mathrm{phys}}^{\mathrm{res}}$ on the F-theory side; it can be mapped to the position of said node within the representative of $\mathcal{H}_{\infty}^{0}$ or $\mathcal{H}_{0}^{1}$ supporting the $\mathrm{E}_{8}$ gauge factor associated with the heterotic gauge bundle to which the point-like instanton belongs. The freedom to move the position of the nodes of $\Delta_{\mathrm{phys}}^{\mathrm{res}}$ through a finite-distance complex structure deformation on the F-theory side allows us to arrange for the point-like instantons to probe or not probe the ADE singularities of the heterotic K3 surface. This is important for the manifestation of non-perturbative physics, as highlighted at the beginning of the section.

\begin{figure}[t!]
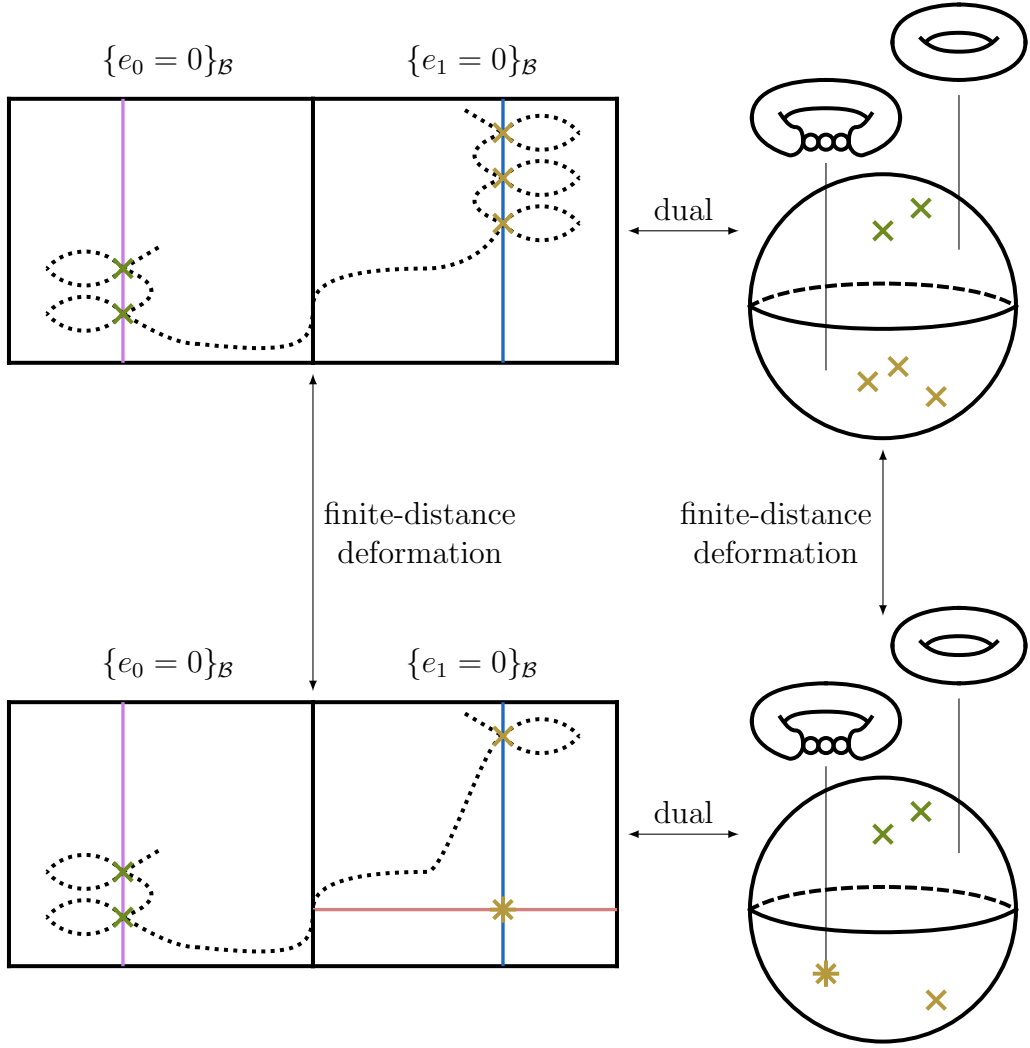

    \centering

    \caption{In the upper-left part of the figure, we schematically represent a horizontal Type~II.a model in which $\Dphys$ tangentially intersects the interface curve $B^{0} \cap B^{1}$. This corresponds on the heterotic side to an ADE singularity not probed by point-like instantons with trivial holonomy, depicted in the upper-right corner. Through a finite-distance deformation, part of the point-like instantons can be moved on top of the ADE singularity, as shown in the lower-right corner. The dual situation in F-theory is depicted in the lower-left corner, where moving the codimension-two finite-distance non-minimal points to align with the interface singularity forces local vertical enhancements in the corresponding components. To produce a global vertical enhancement, it is necessary to align such non-minimal points in the two components or, using the heterotic dual language, the ADE singularity must be probed symmetrically by point-like instantons with trivial holonomy associated with both heterotic $\mathrm{E}_{8}$ bundles.}
    \label{fig:type-IIa-small-instantons}
\end{figure}

Let us start by considering heterotic K3 singularities not probed by singular gauge bundle contributions of any kind. According to the heterotic analysis of \cite{Witten:1999fq}, such singularities should not lead to a non-perturbative gauge algebra contribution. As explained earlier, the Weierstrass model of the heterotic K3 surface is given by the defining polynomials $f_{\mathrm{K3}}$, $g_{\mathrm{K3}}$ and $\Delta_{\mathrm{K3}}$; on the F-theory side they encapsulate information on how the divisors $F_{\mathrm{phys}}$, $G_{\mathrm{phys}}$ and $\Delta_{\mathrm{phys}}$ intersect the interface curve $B^{0} \cap B^{1}$. By making this intersection tangent, it is possible to increase the interface vanishing orders without increasing the component vanishing orders associated with the representative of $f$ passing through the intersection point in $B^{0}$ and $B^{1}$. This means that a heterotic ADE singularity can be tuned over a point of $B^{0} \cap B^{1}$ without producing a \mbox{codimension-one} vertical enhancement on the F-theory side,\footnote{Notice the similarities with the discussion on obscured infinite-distance limits carried out in \cite{Alvarez-Garcia:2023gdd} and revisited in a concrete scenario in \cref{sec:non-minimal-points-heterotic-K3-surface}. The difference between the two situations stems from the fact that the base change revealing the obscured infinite-distance limit at the level of the family vanishing orders now, since the interface vanishing orders are minimal, merely has the effect of producing a minimal local gauge enhancement in the intermediate components of the central fiber of the base-changed model. Since such a local gauge enhancement does not extend into a global gauge enhancement, there is no gauge algebra associated with it. This was to be expected, since the base-changed degeneration is physically equivalent to the original one.} and while maintaining the positions of the codimension-two finite-distance non-minimal points in $\mathbb{P}^{1}_{b}$ away from the position of the heterotic K3 singularity. Such a situation is depicted in the upper half of \cref{fig:type-IIa-small-instantons}.

Through a finite-distance deformation of the model, we can alter the position of the point-like instantons in $\mathbb{P}^{1}_{b}$ to place them on top of the heterotic ADE singularity. On the F-theory side, one moves one of the codimension-two finite-distance non-minimal points occurring over the $\mathrm{E}_8$ branes at $\mathcal{H}_{\infty}^{0}$ and $\mathcal{H}_{0}^{1}$ to align it with the interface ADE singularity. This forces two copies of the representative of the fiber class $f$ passing through that point to factorise in $\Delta'_{0}$ or $\Delta'_{1}$, respectively.\footnote{The divisors $\{ F_{p}^{\mathrm{res}} \}_{0 \leq p \leq P}$ consist purely of vertical classes, which makes their interface vanishing orders at the heterotic K3 singularity agree with the component vanishing orders over the vertical line passing through it. The divisors $\{ G_{p}^{\mathrm{res}} \}_{0 \leq p \leq P}$ have intersection $G_{p}^{\mathrm{res}} \cdot f = 1$, for $p=0,1$, which for the representative of $f$ passing through the heterotic K3 singularity is already accounted for by said point. As explained in \cite{Aspinwall:1997ye}, moving a point-like instanton to lie on the same representative of $f$ would lead to a second point of intersection, a contradiction that the geometry resolves by making $G_{0}^{\mathrm{res}}$ or $G_{1}^{\mathrm{res}}$, depending on the component under consideration, reducible with a copy of the relevant $f$ representative factoring out. Altogether, this means that moving a point-like instanton on top of the heterotic K3 singularity leads to a local vertical $(1,1,2)$ enhancement. This process continues as we probe the ADE singularity with point-like instantons until the interface and component vanishing orders are equal in the component under consideration, at which point the previous arguments no longer apply.} This leads to a local vertical enhancement that closes the gap between the interface vanishing orders corresponding to the heterotic ADE singularity and the component vanishing orders associated with the representative of $f$ passing through it, but only in the component to which the codimension-two finite-distance non-minimal point that was aligned with it belongs to. We depict the result of such a finite-distance deformation in the lower half of \cref{fig:type-IIa-small-instantons}. Hence, we learn from the F-theory side that it is not only important that the heterotic K3 singularity is probed by point-like instantons; the heterotic $\mathrm{E}_{8}$ bundle to which they belong determines which half of the F-theory dual model realises a local vertical enhancement. This raises the question of what happens when the singularity is probed asymmetrically.

From the heterotic point of view, it is clear that the non-perturbative gauge algebra should only depend on the number of point-like instantons with trivial holonomy probing the heterotic ADE singularity. This is because, through a small instanton transition, we can trade them for M5-branes located in the Ho\v{r}ava-Witten interval, at which point they no longer belong to one of the two heterotic $\mathrm{E}_{8}$ bundles and the notion of asymmetric probing looses its meaning.

The same conclusion holds on the F-theory side: As we now explain, the notion of asymmetrically probing a heterotic ADE singularity by point-like instantons with trivial holonomy is an artefact of the resolved degeneration. Even if the stable degeneration limit is necessary to have control over the heterotic dual model, let us for a moment entertain the possibility of studying the codimension-two finite-distance non-minimal singularities in a conventional six-dimensional F-theory model in which the base is an irreducible surface $B = \mathbb{F}_{n}$. After performing a base blow-up leading to $\hat{B} = \mathrm{Bl}_{1}(\mathbb{F}_{n})$, the base geometry is agnostic to its origin as the blow-up of $B = \mathbb{F}_{n}$ or, say, $B' = \mathbb{F}_{n-1}$. This can be easily seen from the toric fans of these three surfaces.\footnote{This is nicely discussed in Section 3 of \cite{Braun:2018ovc}.} This reflects the fact that, once we consider $\hat{B}$ as the base of the F-theory model, we have moved the M5-brane into the Ho\v{r}ava-Witten interval in the putative heterotic dual, and the F-theory geometry no longer knows the initial distribution of instanton number between the two heterotic $\mathrm{E}_{8}$ bundles. The effect of the stable degeneration limit is to separate the information pertaining to the two heterotic $\mathrm{E}_{8}$ bundles, which for the reducible central fiber $Y_{0} = Y^{0} \cup_{\mathrm{K3}} Y^{1}$ is encoded in $\mathrm{Def}(Y^{0})$ and $\mathrm{Def}(Y^{1})$, respectively. The blow-up of the reducible base surface $B_{0}$ with centre a codimension-two finite-distance non-minimal point is now different depending on whether this point is located on $B^{0}$ or on $B^{1}$: Even after moving into the tensor branch, there is a clear notion of the heterotic $\mathrm{E}_{8}$ bundle to which the M5-brane belongs, namely the one associated to the component in which the exceptional curve is located. In this sense, asymmetrically probing heterotic ADE singularities by point-like instantons with trivial holonomy has a definite meaning, but the resulting physics should not depend on this notion.

Indeed, the non-perturbative heterotic gauge sector corresponds on the F-theory side to the collection of gauge algebra factors supported both on the exceptional curves arising from the resolution of codimension-two finite-distance non-minimal points and on global vertical divisors. The former are always contained within a component, while the latter traverse the whole base geometry. This means that a discrepancy between vertical local and global gauge enhancements can occur. This discrepancy corresponds to the heterotic ADE singularity being asymmetrically probed by $(k_{0},k_{1})$ point-like instantons with trivial holonomy. The vertical gauge algebra that manifests is the same one appearing if the heterotic ADE singularity is symmetrically probed by $k:=\min(k_{0},k_{1})$ point-like instantons with trivial holonomy. The additional point-like instantons on one side enhance the non-perturbative gauge factors supported over the exceptional curves. The vertical gauge algebra is singled out among the collection of non-perturbative gauge factors by the resolution process of the degeneration, and the total non-perturbative gauge algebra associated to the heterotic ADE singularity is simply that corresponding to $k_{0}+k_{1}$ point-like instantons with trivial holonomy probing it, irrespectively of their distribution. The bounds on the vertical gauge rank given in \cref{tab:vertical-rank-bounds} affect the vertical gauge algebras,\footnote{One can use the fact that symmetrically probing a heterotic ADE singularity is only possible up to a certain extent in horizontal Type II.a models constructed over $\hat{B} = \mathbb{F}_{n}$, due to the distribution of instanton numbers $c_{2}(V_{0,1}) = 12 \pm n$, to derive rough bounds on the rank of the vertical gauge algebra. To this end, simply equate each unit of instanton number to a local vertical $(1,1,2)$ enhancement in the corresponding component. For most $\hat{B} = \mathbb{F}_{n}$ the bounds naively derived in this way are not as tight as the ones provided in \cref{tab:vertical-rank-bounds}, since they do not take into consideration the possibility of a monodromy action reducing the rank. In some select cases, namely for horizontal Type II.a models constructed over $\hat{B} = \mathbb{F}_{3}$ and $\hat{B} = \mathbb{F}_{1}$, the naive bounds derived in this fashion are, in fact, tighter than the ones printed in \cref{tab:vertical-rank-bounds}. The reason for this is that in this section we are considering the horizontal gauge algebra to be an unbroken $\mathfrak{g}_{\mathrm{hor}} = \mathfrak{e}_{8} \oplus \mathfrak{e}_{8}$, which is incompatible with the maximal vertical gauge rank possible in the models constructed over these Hirzebruch surfaces, cf.\ \cref{foot:maximal-horizontal-rank-not-E8}.} not the complete non-perturbative sector.

To give a concrete example, consider a horizontal Type II.a model constructed over $\hat{B} = \mathbb{F}_{10}$. The maximal vertical gauge algebra that can be realised in such a model is of $\rank(\mathfrak{g}_{\mathrm{ver}}) = 1$, which can be for example obtained from a vertical line of Kodaira type III fibers. We then have, in terms of the physical vanishing orders, a $(1,2,3)$ enhancement over a representative of $\mathcal{F}$, whose collision with the $(4,5,10)$ enhancements supported over a representative of $\mathcal{H}_{\infty}^{0}$ and the unique representative of $\mathcal{H}_{0}^{1}$, respectively, produces two $(5,7,14)$ points. This corresponds to a heterotic $\mathbb{C}/\mathbb{Z}_{2}$ singularity probed by $(k_{0},k_{1}) = (2,2)$ point-like instantons with trivial holonomy. Resolving the codimension-two finite-distance non-minimal points leads to no additional non-perturbative gauge factors. Consider now a horizontal Type II.a model constructed over $\hat{B} = \mathbb{F}_{11}$, where such a vertical gauge algebra is not possible, see \cref{tab:vertical-rank-bounds}. However, we can still tune a local vertical gauge enhancement in $B^{0}$ with $(1,3,3)$ vanishing orders extending as residual discriminant into $B^{1}$. It collides with the $(4,5,10)$ enhancement supported over a representative of $\mathcal{H}_{\infty}^{0}$ to produce a $(5,9,15)$ point. The exceptional curve arising from the resolution of such a codimension-two finite-distance non-minimal point supports Kodaira type III fibers and, because it is fully contained within $B^{0}$, it leads to a global gauge enhancement. This corresponds to a heterotic $\mathbb{C}^{2}/\mathbb{Z}_{2}$ singularity probed by $(k_{0},k_{1}) = (4,0)$ point-like instantons with trivial holonomy. Hence, we conclude that both models lead to the same asymptotic physics, namely that of the HE string in the presence of a $\mathbb{C}^{2}/\mathbb{Z}_{2}$ singularity with two M5-branes on top. The special role played by the vertical gauge algebra is an artefact of the resolution of the degeneration, and only the total number of M5-branes probing a heterotic ADE singularity is relevant in order to determine the non-perturbative gauge algebra associated with it. This agrees with the common lore applied in the context of studying six-dimensional SCFTs.

At any rate, and as explained in \cite{Alvarez-Garcia:2023gdd}, employing the physical vanishing orders in the resolved horizontal Type II.a model is important in order to correctly determine the gauge algebra contributions, and in particular to determine the vertical gauge algebra and hence the non-perturbative heterotic gauge sector. A heterotic ADE singularity can be completely asymmetrically probed by point-like instantons with trivial holonomy, e.g.\ a heterotic $\mathrm{E}_{8}$ singularity probed by $(k_{0},k_{1}) = (0,10)$ such instantons, not leading to a vertical gauge enhancement at all. The vertical line of fibers prior to the degeneration is also not indicative of the vertical gauge algebra found at the endpoint of the limit, since the physical vanishing orders found over a vertical locus may be bigger than the family vanishing orders over it (meaning that the finite-distance enhancement occurs at the same time that the infinite-distance limit is taken).\footnote{These last two facts contrast with some comments made in \cite{Aspinwall:1997ye}, without this discrepancy altering the picture put forward in that work due to the preceding discussion in this section.}

\subsection{Broken horizontal \texorpdfstring{$\mathrm{E}_{8} \times \mathrm{E}_{8}$}{E8xE8} gauge algebra}
\label{sec:defect-algebras-heterotic-dual-broken-E8xE8}

Above, we have discussed the non-perturbative heterotic gauge algebras associated with heterotic ADE singularities probed by point-like instantons with trivial holonomy. These are not the only singular gauge bundle contributions that can probe such singularities: The non-perturbative gauge algebras can also arise if the heterotic ADE singularity is probed by point-like instantons with discrete holonomy, a situation that we now describe.

In models with unbroken horizontal gauge algebra $\mathfrak{g}_{\mathrm{hor}} = \mathfrak{e}_{8} \oplus \mathfrak{e}_{8}$, any collision of the component residual discriminant $\{ \Delta_{p}^{\mathrm{res}} \}_{0 \leq p \leq 1}$ with the representatives of $\mathcal{H}_{\infty}^{0}$ and $\mathcal{H}_{0}^{1}$ supporting it is always non-minimal. This is what allows us to perform a base blow-up, followed by an appropriate line bundle shift, in order to resolve the codimension-two finite-distance non-minimal points. These points are identified with point-like instantons with trivial holonomy, and the resolution procedure moves us to the interior of the tensor branch by separating M5-branes from the Ho\v{r}ava-Witten walls.

If $\mathfrak{g}_{\mathrm{hor}} \neq \mathfrak{e}_{8} \oplus \mathfrak{e}_{8}$, however, not all collisions of the residual discriminant with the horizontal divisors supporting it need to be non-minimal. This means that we can tune vertical gauge algebras without producing codimension-two finite-distance non-minimal points. This is most clear if we only tune a vertical gauge algebra small enough that the model presents no such points whatsoever. Point-like instantons with trivial holonomy are then absent, since they can always be subject to a small instantons transition, which on the F-theory side is tied to the codimension-two finite-distance non-minimal points, as reviewed above. From a heterotic standpoint, this seems reasonable, since point-like instantons with trivial holonomy leave the perturbative gauge group intact and must be deformed into a gauge bundle profile in order to break it. Such deformations are parametrised by the 29 hypermultiplets that must be traded by 1 tensor in the gravitational anomaly cancellation formula in order to realise a small instanton transition, that is hence obstructed after the deformation.

This raises the question of how the vertical gauge algebra associated to the heterotic K3 singularity arises from the heterotic point of view. While on the F-theory side of the duality a model as the one just described clearly shows a vertical gauge algebra factor, we know from the review of \cite{Witten:1999fq} at the beginning of this section that the corresponding ADE singularity of the heterotic K3 surface should be probed a some kind of singular gauge bundle contributions,\footnote{Horizontal Type II.a limits correspond to a (possibly partial) decompactification process, which dilutes the gauge bundle curvature. Since non-perturbative gauge factors associated with a heterotic K3 singularity are of local nature, it seems intuitive that they should not disappear along the decompactification process. Hence, if their manifestation depends on having a gauge configuration probing the singularity, said configuration should indeed be singular, which heuristically aligns with \cite{Witten:1999fq}.} a role played until this moment by the now absent point-like instantons with trivial holonomy.

An alternative source of concentrated gauge bundle curvature are point-like instantons with discrete holonomy. Let us briefly review them on the heterotic side of the duality. In a compactification of $\mathrm{E}_{8} \times \mathrm{E}_{8}$ heterotic string theory on the internal space $X$ we have the two poly-stable $\mathrm{E}_{8}$ bundles
\begin{equation}
    \pi_{i}: V_{i} \longrightarrow X\,,\qquad i = 0,1\,,
\end{equation}
with structure groups $H_{i} \subset \mathrm{E}_{8}$. The gauge group $G_{i}$ associated to each gauge bundle $V_{i}$ in the lower-dimensional theory is given by the centralizer (commutant) $C_{\mathrm{E}_{8}}(H_{i})$ of the structure group $H_{i} \subset \mathrm{E}_{8}$. A point-like instanton corresponds to a skyscraper sheaf, a singular gauge bundle in which all curvature has been concentrated at a point, while having a flat connection at infinity. Hence, they generate a discrete holonomy around their location, associated to those loops that are non-contractible once the support of the point-like instanton has been excised from the space,\footnote{K3 surfaces are simply connected, and hence these are the only non-contractible loops to be considered.} while their curvature contribution to the holonomy vanishes. The discrete holonomy that a point-like instanton can exhibit depends on the fundamental group of a neighbourhood around with the location of the instanton excised \cite{Aspinwall:1996vc}. Over smooth points of the heterotic K3 surface we hence have (after a retraction) $\pi_{1}(S^{3}) = 0$, and the discrete holonomy must be trivial. Hence, over generic points of the heterotic K3 only point-like instantons with trivial holonomy are supported. If the instanton is located at a heterotic $\mathbb{C}^{2}/\Gamma_{\mathfrak{g}}$ singularity we have instead $\pi_{1}(S^{3}/\Gamma_{\mathfrak{g}}) \cong \Gamma_{\mathfrak{g}}$, and the discrete holonomy $H'_{i}$ of the point-like instanton can be any subgroup of $\Gamma_{\mathfrak{g}}$, where $i$ labels from which of the heterotic $\mathrm{E}_{8}$ bundles the point-like instanton stems or, equivalently, to which Ho\v{r}ava-Witten wall it is associated with. The corresponding $\mathrm{E}_{8}$ gauge factor is broken by the presence of a point-like instanton to $C_{\mathrm{E}_{8}}(H'_{i})$. Due to the discrete nature of $H'_{i}$, a configuration with only point-like instantons will either leave $\mathrm{E}_{8} \times \mathrm{E}_{8}$ unbroken or break it into a non-simply connected subgroup, which on the F-theory side corresponds to non-trivial torsion subgroup of the Mordell-Weil group \cite{Aspinwall:1998xj}.

In the absence of codimension-two finite-distance non-minimal points, and hence, using the heterotic language, of point-like instantons with trivial holonomy, the vertical gauge algebras must therefore correspond to heterotic ADE singularities probed by point-like instantons with discrete holonomy. The latter class of point-like instantons cannot be traded for M5-branes via small instanton transitions due to the lack of enough available deformation moduli to be traded for a tensor in the gravitational anomaly cancellation formula. This means that they are stuck at the Ho\v{r}ava-Witten walls, which aligns with the absence of K\"ahler moduli associated with exceptional curves on the F-theory side. Moreover, while the position of point-like instantons with trivial holonomy within the K3 surface could be tuned freely by moving the associated codimension-two finite-distance non-minimal points within $\mathbb{P}_{b}^{1}$, the ones with discrete holonomy are associated to the vertical line. This is in agreement with the fact that they are stuck at the heterotic $\mathbb{C}^{2}/\Gamma_{\mathfrak{g}}$ singularity allowing them to be realised. Such instantons were explicitly studied in \cite{Aspinwall:1998xj}, finding that they are associated to codimension-two minimal enhancement points corresponding to the hypermultiplet matter that populates the representations allowed by the non-simply connected gauge groups; furthermore, they contribute fractionally to the instanton number budget $c_{2}(\mathrm{K3}) = 24$.

By turning on local curvature in a neighbourhood around a heterotic ADE singularity a point-like instanton with discrete holonomy can be continuously deformed into a gauge bundle contribution with structure group of positive dimension, while preserving the point-like instanton at its core.  Such a configuration would be able to probe the heterotic ADE singularities without necessarily enforcing a non-simply connected gauge group to arise, which would explain why on the F-theory side we can tune vertical gauge algebras without enforcing a particular non-trivial torsion subgroup of the Mordell-Weil group. At the endpoint of a horizontal Type~II.a limit the singular gauge bundle part of such a configuration is not diluted away, hence still allowing the non-perturbative gauge algebra to manifest. Away from the six-dimensional defects the bulk gauge group is restored due to the arguments given in \cref{sec:horizontal-IIa-F0}. An observer concerned with the six-dimensional theory living on a defect, and hence placed on top of it, sees the bulk gauge group as a flavour group that is partially broken due to the discrete holonomy of the surviving point-like instanton. This local picture is the one taken in the analyses of the Higgs branch of heterotic LSTs \cite{DelZotto:2022ohj,DelZotto:2022xrh,DelZotto:2023nrb,Ahmed:2023lhj,DelZotto:2023ahf,Lawrie:2023uiu}.

Since a point-like instanton with discrete holonomy is associated to one of the Ho\v{r}ava-Witten walls, it is meaningful to  probe heterotic ADE singularities asymmetrically. In the absence of codimension-two finite-distance non-minimal points no gauge factors can be supported over exceptional curves. The non-perturbative heterotic gauge sector hence entirely corresponds to the vertical gauge algebra. Given a representative of $\mathcal{F}$, tuning a local enhancement over $\left. \mathcal{F} \right|_{B^{i}}$ is associated with $\mathrm{Def}(Y^{i})$, and hence with point-like instantons in the $V_{i}$ heterotic gauge bundle. To produce the vertical gauge enhancement one must symmetrically probe the heterotic ADE singularity, which ensures that the local gauge enhancements on both components lead to a global gauge enhancement factorising in $\Dphys$. This is aspect is nicely captured by the resolved horizontal Type II.a models. Due to these considerations, the asymptotic physics associated with a heterotic ADE singularity does in general not only depend on the total instanton number of the singular gauge bundle contributions probing it, but also on their distribution within the Ho\v{r}ava-Witten walls. In this regard, the point-like instantons with trivial holonomy are special due to their equivalence to M5-branes in the Ho\v{r}ava-Witten interval, as discussed in the previous section. By contrast, moving a point-like instanton with discrete holonomy from one \mbox{Ho\v{r}ava-Witten} wall to the other would entail an additional finite-distance deformation of the model; after bringing it together with other fractional point-like instantons, this turns the factional instantons into a full instanton with trivial holonomy.

Let us consider a concrete example. In the following discussion, the horizontal gauge algebra is kept small enough so that no codimension-two finite-distance non-minimal points arise. Hence, we only have to account for fractional point-like instantons. Consider tuning a global vertical $(1,2,3)$ enhancement, which does lead to an $\mathfrak{su}(2)$ gauge algebra from the F-theory perspective. This is dual to a heterotic $\mathbb{C}^{2}/\mathbb{Z}_{2}$ singularity probed by some singular gauge bundle contribution. Each local vertical $(1,2,3)$ enhancement acting as one half of the global one is independent of the other half. It must therefore concentrate by itself a certain amount of instanton charge, denoted by $c_{\mathrm{III}}$ for concreteness, at the $\mathbb{C}^{2}/\mathbb{Z}_{2}$ point. On the F-theory side, we can enhance the interface vanishing orders of said point to produce a heterotic $\mathbb{C}^{2}/2\mathrm{D}_{4}$ singularity, by making the residual discriminant pass through it tangentially. Since this does not change the global vertical $(1,2,3)$ enhancement, the gauge algebra is unaffected. On the heterotic, side this corresponds to bringing some of the $\mathrm{I}_{1}$ singular elliptic fibers, at which no instanton charge is concentrated, on top of the original orbifold point. This leads to a $\mathbb{C}^{2}/2\mathrm{D}_{4}$ singularity symmetrically probed by $c_{\mathrm{III}} + c_{\mathrm{III}}$ instanton charge and hence, still, to an $\mathfrak{su}(2)$ gauge algebra. Alternatively, we can start by tuning two local vertical $(1,2,3)$ lines in, for example, the $B^{0}$ component that do not extend into global vertical enhancements. In such a case, there arises no vertical gauge group factor. This produces two heterotic $\mathbb{C}^{2}/\mathbb{Z}_{2}$ singularities, each probed by $c_{\mathrm{III}} + 0$ instanton charge. Bringing these two local vertical enhancements together produces a single local vertical $(2,3,6)$ line on the F-theory side, which does not extend into a global vertical enhancement and therefore has no associated gauge algebra. On the heterotic side, this corresponds to bringing together the two $\mathbb{C}^{2}/\mathbb{Z}_{2}$ points alongside the instanton charge concentrated at them, which results in a heterotic $\mathbb{C}^{2}/2\mathrm{D}_{4}$ singularity probed by $2c_{\mathrm{III}} + 0$ instanton charge, but not leading to any gauge algebra, according to the F-theory analysis.

\section{Type IIB orientifold picture of Type III.b models}
\label{sec:horizontal-IIIb-models-orientifold-geometry}

Since Type III.b models represent global weak coupling limits, their endpoints can be described as Type IIB orientifold compactifications in the context of the Sen limit \cite{Sen:1997gv}. In this language, the Type IIB internal space on which we place the O7-planes is the Calabi-Yau double cover $\breve{B}$ of the base $B$ of the F-theory elliptic fibration. The branching points of the double cover correspond to the fixed loci of the orientifold involution, and hence to the position of the \mbox{O7-planes} in the Type IIB compactification. For six-dimensional models with minimal supersymmetry, $\breve{B}$ is a K3 surface.

To interpret the endpoint of an, e.g.\ horizontal, Type III.b degeneration $\rho: \mathcal{Y} \rightarrow D$ as a Sen~limit we can blow down the model such that the new central fiber corresponds to the former component $Y^{p}$. This can be accomplished via the coordinate substitutions
\begin{equation}
	e_{p} \longmapsto u\,,\qquad e_{q \neq p} \longmapsto 1\,,\qquad p,q \in \{0, \dotsc, P\}\,,
\end{equation}
in the defining polynomials $f_{b}$, $g_{b}$ and $\Delta_{b}$ of the family fourfold $\mathcal{Y}$. Since all components of the central fiber  $Y_{0}$ of $\rho: \mathcal{Y} \rightarrow D$ have codimension-zero $\mathrm{I}_{m}$ fibers, the central fiber $\hat{Y}_{0}$ of the blown-down degeneration $\hat{\rho}: \hat{\mathcal{Y}} \rightarrow D$ does as well.\footnote{Alternatively, we can start with a Type III.a model and blow down to one of the components at local weak coupling. This is used in the comparison of horizontal Type III.a and Type III.b models in \cref{sec:horizontal-IIIb-models-orientifold}.} This means that, irrespective of the choice $Y^{p}$ of component that we make, the family variety of the blown-down degeneration has defining polynomials of the form
\begin{subequations}
\begin{align}
    f = e_{p}^{l_{p}} \check{f}_{p} + f_{p} &\longmapsto  f = u^{l_{p}} \check{f} - 3h^{2}\,,\\
    g = e_{p}^{m_{p}} \check{g}_{p} + g_{p} &\longmapsto g = u^{m_{p}} \check{g} + 2h^{3}\,,
\end{align}
\label{eq:blown-down-Type-IIIb-model}%
\end{subequations}
with $h \in H^{0}\left( \mathbb{F}_{n}, \overline{K}_{\mathbb{F}_{n}}^{\otimes 2} \right)$, see \cref{sec:discriminant-weakly-coupled-components}. The divisor defined by the vanishing locus $\{h=0\}_{\hat{B}_{0}}$ corresponds to the branching locus of the double cover of $\hat{B}_{0}$, i.e.\ the fixed locus of the orientifold involution of the Type IIB model. The irreducible components of $\{h=0\}_{\hat{B}_{0}}$ hence correspond to the O7-planes. The parameter $u$ implements the Sen limit, with $u \rightarrow 0$ bringing us (at least in global weak coupling limits) to the perturbative regime.

Before analysing the particular features of the Type IIB orientifolds obtained as the endpoints of Type III.b models, it is convenient to explicitly discuss the geometry related to the Sen limits of generic six-dimensional F-theory models with $B = \mathbb{F}_{n}$ acting as their base.

\subsection{Sen limit in six-dimensional F-theory models}
\label{sec:Sen-limit-six-dimensional-F-theory}

We are interested in understanding the geometric degeneration undergone by the K3 double covers $\breve{B}$ of the bases $B$ of six-dimensional F-theory models explicitly. For concreteness, and since it is the case of interest for what follows, we center our attention on $B = \mathbb{F}_{n}$ with $0 \leq n \leq 4$. We briefly comment on the other possible $B$ geometries at the end.

Let us start by discussing the relevant geometry in eight-dimensional F-theory models. In this context, the base of the internal elliptic fibration must be $B = \mathbb{P}^{1}$ with Calabi-Yau double  $\breve{B} = T^{2}$. The explicit description of this double cover naturally arises when taking the Sen limit of such an F-theory model. Consider the defining polynomials
\begin{align}
    f = u^{l} \check{f} - 3h^{2}\,,\\
    g = u^{m} \check{g} + 2h^{3}
\label{eq:codimension-zero-In-model}
\end{align}
for the Weierstrass model of the internal space $\pi: Y \rightarrow B$, where $h \in H^{0}\left( \mathbb{P}^{1}, \overline{K}_{\mathbb{P}^{1}}^{\otimes 2} \right)$ and $u \rightarrow 0$ implements the weak coupling limit. The Calabi-Yau double cover of $\mathbb{P}^{1}_{[s:t]}$ is obtained by enlarging the set of homogeneous coordinates of $B$ by $\xi$ and considering the hypersurface
\begin{equation}
	\breve{B}:\quad \left\{ P_{\breve{B}} = 0 \right\}_{\mathbb{P}^{2}_{112}}\,,\qquad P_{\breve{B}} := \xi^{2} - h([s:t])\,.
\end{equation}
The ambient space must then be $\mathbb{P}^{2}_{112}$ with homogeneous coordinates $[s:t:\xi]$, such that the defining polynomial $P_{\breve{B}}$ is homogeneous under the $\mathbb{C}^{*}$-action. The new variable $\xi$ appears quadratically in $P_{\breve{B}}$. For every point $[s_{0}:t_{0}]$ in $\mathbb{P}^{1}_{[s:t]}$ we therefore obtain two points, counted with multiplicity, in $\{ P_{\breve{B}} = 0 \}_{\mathbb{P}^{2}_{112}}$. This yields the desired double cover. The ramification points of the cover are given by the roots of $h$. Since $\overline{K}_{\mathbb{P}^{1}} = 2H$, where $H$ is the hyperplane class, $h = h_{4}([s:t])$ is a degree four polynomial, and the double cover has four such ramification points, counted with multiplicity. Since $P_{\breve{B}}$ is of homogeneous degree four, we have a non-generic quadric $\mathbb{P}^{2}_{112}[4]$ in which the linear terms in $\xi$ have been set to zero. Quadrics $\mathbb{P}^{2}_{112}[4]$ are elliptic curves, and therefore we have obtained a double cover of $\mathbb{P}^{1}$ by $T^{2}$ branched at four points. Summarising, the Sen limit of eight-dimensional F-theory models naturally yields a branched cover
\begin{equation}
	\pi: \{ P_{\breve{B}} = 0 \}_{\mathbb{P}^{2}_{112}} \cong T^{2} \longrightarrow \mathbb{P}^{1}
\end{equation}
in which the algebraic extension of the function fields is of degree 2. The cover is in particular an (Abelian) Galois cover in which $G = \mathbb{Z}/2\mathbb{Z} \in \mathrm{Aut}(T^{2})$ with $T^{2}/G \cong \mathbb{P}^{1}$, and the stabilisers are $\mathrm{Stab}_{G}(x) = \mathbb{Z}/2\mathbb{Z}$ for those $x \in \{ P_{\breve{B}} = 0 \}_{\mathbb{P}^{2}_{112}}$ such that $h(x) = 0$, and  $\mathrm{Stab}_{G}(x) = 0$ elsewhere, giving the desired orientifold picture.

Consider now the same construction for a six-dimensional F-theory model whose base is $B = \mathbb{F}_{n}$, with $0 \leq n \leq 4$. This should lead to a K3 double cover $\breve{B}$. The fact that $\breve{B}$ is Calabi-Yau can easily be seen in the smooth case. Namely, consider the dominating morphism of smooth surfaces
\begin{equation}
	\breve{\pi}: \breve{B} \longrightarrow B
\end{equation}
given by the double cover. If we denote the ramification divisor by $R \subset \breve{B}$, the canonical divisors of $\breve{B}$ and $B$ are related by (Theorem 5.5 of \cite{Iitaka1982})
\begin{equation}
	K_{\breve{B}} = \pi^{*} K_{B} + R\,.
\end{equation}
Since the ramification index is 2 for all components of $R$, we have that $2R = \pi^{*}(\{h = 0\}_{B})$. Using then that we are taking $\{ h = 0\}_{B} \sim 2\overline{K}_{B}$, we obtain
\begin{equation}
	2K_{\breve{B}} = 2\pi^{*} K_{B} + 2R = \pi^{*} (2 K_{B} + 2\overline{K}_{B}) = 0\,.
\end{equation}
The fact that it is concretely a K3 surface, as the physics demands, we will see explicitly below without assuming smoothness.

To obtain a global description of $\breve{B}$, enlarge the set of homogeneous coordinates of $B = \mathbb{F}_{n}$ by $\xi$ and consider the hypersurface
\begin{equation}
	\breve{B}:\quad \left\{ P_{\breve{B}} = 0 \right\}_{X}\,,\qquad P_{\breve{B}} := \xi^{2} - h([s:t],[v:w:t])\,.
\end{equation}
In order to make $\mathbb{P}_{\breve{B}}$ homogeneous under the $\mathbb{C}^{*}$-actions, the ambient space must be
\begin{equation}
	\begin{aligned}
		X := \bigslant{\left( \mathbb{C}^{5} \setminus Z \right)}{\mathbb{C}^{*}_{\lambda_{1}} \times \mathbb{C}^{*}_{\lambda_{2}}}\,,\\
		Z := \{ s=t=\xi=0 \} \cup \{ t=v=w=\xi=0 \}
	\end{aligned}
	\quad
	\begin{aligned}
		\mathbb{C}^{*}_{\lambda_{1}}: \left( s,t,v,w,\xi \right) &\longmapsto \left( \lambda_{1}^{1} s,\lambda^{1}_{1} t, v, w, \lambda^{2}_{1} \xi \right)\,,\\
		\mathbb{C}^{*}_{\lambda_{2}}: \left( s,t,v,w,\xi \right) &\longmapsto \left( s,\lambda^{n}_{2} t, \lambda^{1}_{2} v, \lambda^{1}_{2} w, \lambda^{2+n}_{2} \xi \right)\,.
	\end{aligned}
\end{equation}
Since $h \in H^{0}\left( \mathbb{F}_{n}, \overline{K}_{\mathbb{F}_{n}}^{\otimes 2} \right)$, it is a polynomial of homogeneous degrees $4$ and $4+2n$ under the two $\mathbb{C}^{*}$-actions, i.e.\ $h = h_{4,4+2n}([s:t],[v:w:t])$.

The F-theory base $B = \mathbb{F}_{n}$ is a $\mathbb{P}^{1}_{f}$-fibration over the base $\mathbb{P}^{1}_{b} = \mathbb{P}^{1}_{[v:w]}$\,,
\begin{equation}
	\pi_{\mathbb{F}_{n}}: B \longrightarrow \mathbb{P}^{1}_{b}\,,
\end{equation}
with a section $C_{0}$ (and another independent section $C_{\infty}$). Its double cover $\breve{B}$ constructed above is a genus one fibration over the same base $\mathbb{P}^{1}_{b}$. It is given by the flat surjective morphism
\begin{equation}
\begin{aligned}
	\pi_{\breve{B}}: \{ P_{\breve{B}} = 0 \}_{X} = \breve{B} &\longrightarrow \mathbb{P}^{1}_{b}\\
	(s,t,v,w,\xi) &\longmapsto (v,w)
\end{aligned}
\,,
\end{equation}
with $\pi_{\breve{B}} = \pi_{\mathbb{F}_{n}} \circ \breve{\pi}$. The generic fiber of $\pi_{\breve{B}}$ over a generic point $p_{0} = [v_{0}:w_{0}] \in \mathbb{P}_{b}^{1}$ is given by
\begin{equation}
	\pi_{\breve{B}}^{-1} ([v_{0}:w_{0}]) = \{ P_{\breve{B}} = 0 \}_{X} \cap \{ [v:w] = [v_{0}:w_{0}] \}_{X} = \{ \xi^{2} - h_{4,4+2n}([s:t],[v_{0}:w_{0}:t]) = 0 \}_{X_{0}}\,,
\end{equation}
with
\begin{equation}
	X_{0} = X \cap \{ [v:w] = [v_{0}:w_{0}] \} = \mathbb{P}^{2}_{112}\,.
\end{equation}
Therefore,
\begin{equation}
	\pi_{\breve{B}}^{-1} ([v_{0}:w_{0}]) =  \{ \xi^{2} - h_{4}^{0}([s:t]) = 0 \}_{\mathbb{P}^{2}_{112}} = \mathbb{P}^{2}_{112}[4]
\end{equation}
is an elliptic curve, that can degenerate over codimension-one loci in $\mathbb{P}^{1}_{b}$. We use the super- and subscripts $0$ to mark the dependence of the concrete objects on the choice of generic point $p_{0}$. This merely shows the intuitive result that the generic fiber of $\pi_{\breve{B}}$ is the $T^{2}$ double cover of the $\mathbb{P}^{1}_{f}$ fiber of $\pi_{\mathbb{F}_{n}}$. Note that the section of the $\pi_{\mathbb{F}_{n}}$ fibration does not in general lift to a section of $\pi_{\breve{B}}$. The geometry that we have found for $\breve{B}$ is that of a Calabi-Yau $\mathbb{P}^{2}_{112}[4]$-fibration over $\mathbb{P}^{1}_{b}$, which is indeed known to not have a section in general, but a bisection. In our construction, said bisection corresponds to the double cover $\pi_{\breve{B}}^{-1}(\mathbb{P}^{1}_{b})$ of the base of $B$. This type of fibration has been extensively studied in the context of F-theory compactified on genus-one fibrations in relation with the appearance of discrete abelian gauge groups, see \cite{Weigand:2018rez} for a review and references. We now proceed to outline its geometry, referring the reader to the previous sources for more detail.

As we have just determined, $\breve{B}$ is a $\mathbb{P}^{2}_{112}[4]$-fibration over $\mathbb{P}^{1}_{b}$. The most general $\mathbb{P}^{2}_{112}[4]$ curve is described by the vanishing locus of the polynomial
\begin{equation}
	P = b_{\xi} \xi^{2} + b_{0} s^{2} \xi+b_{1} s t \xi + b_{2} t^{2} \xi + c_{0} s^{4} + c_{1} s^{3} t + c_{2} s^{2} t^{2} + c_{3} s t^{3} + c_{4} t^{4}\,.
\label{eq:P112-4-defining-polynomial}
\end{equation}
By taking the $b_{i}$ and $c_{i}$ to be sections of a base $\tilde{B}$ we obtain a $\mathbb{P}^{2}_{112}[4]$-fibration over $\mathbb{P}^{1}_{b}$. Considering the special case in which $b_{\xi} = 1$, the total space of the fibration is Calabi-Yau if
\begin{equation}
\begin{gathered}
	b_{0} \sim \beta\,,\quad b_{1} \sim \overline{K}_{\tilde{B}}\,,\quad b_{2} \sim -\beta + 2\overline{K}_{\tilde{B}}\,,\\ c_{0} \sim 2\beta\,,\quad c_{1} \sim \beta + \overline{K}_{\tilde{B}}\,,\quad c_{2} \sim 2\overline{K}_{\tilde{B}}\,,\quad c_{3} \sim -\beta + 3\overline{K}_{\tilde{B}}\,,\quad c_{4} \sim -2\beta + 4\overline{K}_{\tilde{B}}\,,
\end{gathered}
\label{eq:CY-condition-P112-4}
\end{equation}
where $\beta$ is some divisor class in $\tilde{B}$ such that all the above classes are effective. Such a generic model admits no rational section and contains smooth $\mathrm{I}_{2}$ fibers. Upon tuning $c_{4} = 0$ (or $c_{0} = 0$) the model develops a conifold singularity at
\begin{equation}
	\{ s=0 \}_{\tilde{B}} \cap \{ \xi = 0 \}_{\tilde{B}} \cap \{ b_{2} = 0 \}_{\tilde{B}} \cap \{ c_{3} = 0 \}_{\tilde{B}}
\end{equation}
and admits a holomorphic and a rational section.

Specialising this to the construction of the double cover $\breve{B}$, the hypersurface that we consider in the fiber ambient space $\mathbb{P}^{2}_{112}$ is defined by the polynomial
\begin{equation}
	P_{\breve{B}} = \xi^{2} - h_{4,4+2n}([s:t],[v:w:t])\,.
\end{equation}
We observe that there are no linear terms in $\xi$, meaning that for the double cover
\begin{equation}
	b_{0} = b_{1} = b_{2} = 0\,.
\end{equation}
The section $c_{0}$ is the coefficient of the $s^{4}$ term in $P_{\breve{B}}$, which can be isolated by setting $t=0$. Since $T \cdot \{ h = 0 \}_{\mathbb{F}_{n}} = 4+2n$, we have that
\begin{equation}
	c_{0} = -h_{4,4+2n}([1:0],[v:w:0]) = p^{0}_{4+2n}([v:w]) = (2+n) \overline{K}_{\tilde{B}}\,,
\end{equation}
with $\tilde{B} = \mathbb{P}^{1}_{b}$. Therefore,
\begin{equation}
	\beta = \frac{(2+n)}{2} \overline{K}_{\tilde{B}} = (2+n)H\,.
\end{equation}
For the remaining $c_{i}$ one obtains
\begin{equation}
	c_{0} = (4+2n)H\,,\quad c_{1} = (4+n)H\,,\quad c_{2} = 4H\,,\quad c_{3} = (4-n)H\,,\quad c_{4} = (4-2n)H\,,
\end{equation}
such that the Calabi-Yau condition \eqref{eq:CY-condition-P112-4} is met. Note that for the cases of interest $n =3$ and $n = 4$ the class $c_{4}$ does not cease to be effective, but rather becomes $c_{4} = 0$ due to the non-Higgsable clusters forcing the factorisation $h = s h'$ in those models constructed over these Hirzebruch surfaces. This also means that, for these two geometries, the double cover $\breve{B}$ always is an elliptic fibration. In the models that we work with, and due to the fact that the $b_{i} = 0$, the codimension-two locus
\begin{equation}
	C_{\mathrm{I}}: \{ b_{2} = 0 \}_{\tilde{B}} \cap \{ c_{3} = 0 \}_{\tilde{B}}
\end{equation}
of the conifold singularities that develop upon tuning $c_{4} =  0$ becomes the codimension-one locus
\begin{equation}
	C_{\mathrm{I}}: \{ c_{3} = 0 \}_{\tilde{B}}\,.
\end{equation}

The $\mathbb{P}^{2}_{112}[4]$-fibration has an associated Weierstrass model, which can be obtained by defining the sections
\begin{equation}
\begin{gathered}
	e_{0} = -c_{0} + \frac{1}{4}b_{0}^{2}\,,\quad e_{1} = -c_{1} + \frac{1}{2} b_{0}b_{1}\\
	e_{2} = -c_{2} + \frac{1}{2} b_{0}b_{2} + \frac{1}{4} b_{1}^{2}\,,\quad  e_{3} = -c_{3} + \frac{1}{2} b_{1}b_{2}\,,\quad e_{4} = -c_{4} + \frac{1}{4} b_{2}^{2}\,,
\end{gathered}
\end{equation}
and taking the defining polynomials to be
\begin{align}
	f_{J(\breve{B})} &= e_{1}e_{3} - \frac{1}{3} e_{2}^{2} - 4e_{0}e_{4}\,,\\
	g_{J(\breve{B})} &= -e_{0}e_{3}^{2} + \frac{1}{3} e_{1}e_{2}e_{3} - \frac{2}{27} e_{2}^{3} + \frac{8}{3} e_{0}e_{2}e_{4} - e_{1}^{2}e_{4}\,.
\label{eq:Jacobian-double-cover-Weierstrass-model}
\end{align}
If the $\mathbb{P}^{2}_{112}[4]$-fibration has a section then it is birationally equivalent to this Weierstrass model. Particularizing this to $b_{i} = 0$ and $\tilde{B} = \mathbb{P}^{1}_{b}$, as in the double cover $\breve{B}$, we find that
\begin{equation}
	F \sim 8H = 4\overline{K}_{\mathbb{P}^{1}_{b}}\,,\quad G = 12H = 6\overline{K}_{\mathbb{P}^{1}_{b}} \Rightarrow \Delta = 24H = 12\overline{K}_{\mathbb{P}^{1}_{b}}\,,
\end{equation}
as corresponds to a K3 Weierstrass model. Further tuning $c_{4} = 0$ results in the Weierstrass model supporting singular $\mathrm{I}_{2}$ fibers in codimension-one over $\{ c_{3} = 0\}_{\tilde{B}}$. The total space of a Weierstrass model has ordinary quadratic singularities at the location of $\mathrm{I}_{2}$ singularities, i.e.\ conifold singularities; the aforementioned $\mathrm{I}_{2}$ singularities over $\{ c_{3} = 0\}_{\tilde{B}}$ are the conifold singularities found over the same locus in $\tilde{B}$ for the $\mathbb{P}^{2}_{112}[4]$-fibration.

The Weierstrass model obtained in this way is the Jacobian of the double cover $\breve{B}$. Every Calabi-Yau surface $Y$ with a genus-one fibration over a base $B$
\begin{equation}
	\pi: Y \longrightarrow B
\end{equation}
has an associated Calabi-Yau surface $J(Y)$ that is elliptically fibered over the same base
\begin{equation}
	\pi_{J}: J(Y) \longrightarrow B\,,
\end{equation}
known as the Jacobian of the fibration. The multivalued $\tau(b)$, $b \in B$, functions, $\mathrm{SL}(2,\mathbb{Z})$ representations, and discriminant subvarieties $\Delta \subseteq B$ are identical for the two fibrations $\pi$ and $\pi_{J}$ \cite{Braun:2014oya}. The total spaces $Y$ and $J(Y)$, however, do present some differences. For example, a $\mathbb{P}^{2}_{112}[4]$-fibration like the ones discussed above may have $\mathrm{I}_{2}$ fibers that do not lead to a singularity of the total space of the fibration; this is not possible in a Weierstrass model, and they are therefore contracted into singular $\mathrm{I}_{2}$ fibers in the Jacobian.\footnote{The Jacobian $J(Y)$ of an (even non-singular) genus-one fibered Calabi-Yau threefold $Y$ typically presents $\mathbb{Q}$-factorial terminal singularities \cite{Braun:2014oya}.} The analysis of the relation between F-theory models on $Y$ and $J(Y)$ was started in \cite{Braun:2014oya}. While strictly speaking we are interested in the F-theory model on $Y$, due to the shared properties between both spaces, we can use the Weierstrass model of $J(Y)$ to read off the non-minimal loci of $Y$, which is all the information that we intend to extract.

Before studying the concrete K3 surfaces $\breve{B}$ that arise as the double covers of the endpoints of horizontal Type~III.b limits, let us note a few geometrical facts about the K3 double covers arising for the Sen limit of generic six-dimensional F-theory models with $B = \mathbb{F}_{n}$ as their base. First, recall that if $\breve{B}$ and $B$ are two complex projective surfaces with $B$ smooth, and there is a double cover $\breve{\pi}: \breve{B} \rightarrow B$ with branching locus $\{ h = 0\}_{B} \subset B$, then $\breve{B}$ is smooth if and only if $\{ h = 0\}_{B}$ is smooth, since $\breve{B}$ can only have singularities over the singular points of $\{ h = 0 \}_{B}$. Given that $B = \mathbb{F}_{n}$ is smooth, we are in the conditions in which this applies. Since $h \in H^{0}\left( \mathbb{F}_{n}, \overline{K}_{\mathbb{F}_{n}}^{\otimes 2} \right)$, no factorizations occur for a generic $h$ when $n = 0,1,2$, while
\begin{equation}
	h = s h'\,,\qquad n=3,4\,,\qquad \textrm{with}\quad h' \sim 3S + (4+2n)V\,,
\end{equation}
generically. Thus, we expect the K3 double cover of $\mathbb{F}_{n}$ obtained by taking the Sen limit of a generic F-theory model whose internal space has base $B = \mathbb{F}_{n}$ with $n = 0$, $1$ ,$2$ to be smooth. The same expectation holds for $n=4$, since in spite of the factorization of $h = s h'$, we have that $S \cdot \{ h'=0 \}_{B} = 0$. The same is not true when $n = 3$, since $S \cdot \{ h'=0 \}_{B} = 1$ leading to a transverse intersection of the two components of $\{ h = 0\}_{B}$; we therefore expect a singularity in the K3 double cover of $B = \mathbb{F}_{3}$ obtained from the Sen limit of the generic F-theory model whose internal space has this Hirzebruch surface as its base. Taking into account that the non-Higgsable cluster $(2,2,4)$ over $S \subset B$ enhances to at least $(2,3,5)$ by taking the Sen limit, this is the quadric cone singularity associated to orthogonal groups expected from the analysis of the Donagi-Wijnholt limit of Tate models \cite{Donagi:2009ra} performed in \cite{Esole:2012tf}, and that is avoided by the particularities of the intersection theory of the base in the case of $B = \mathbb{F}_{4}$. The singular K3 double cover $\breve{B}$ of $B = \mathbb{F}_{3}$ admits a crepant resolution, compatible with the orientifold involution, to a smooth K3 surface $\mathrm{Bl}(\breve{B})$ that is the double cover of the blow-up $\mathrm{Bl}(B)$ of the Hirzebruch surface along the self-intersection locus of $\{ h = 0 \}_{B}$.

The physics of the Sen limit of F-theory compactified on $\pi: Y \rightarrow B$ with $B = \mathbb{F}_{n}$ beautifully captures in this way the mathematics of K3 double covers of Hirzebruch surfaces. The smooth quotients of K3 surfaces by finite abelian groups have been analysed in the mathematics literature, see the seminal works of Nikulin \cite{nikulin1979factor,nikulin1979quotient,nikulin1986discrete} and more recent studies on the subject \cite{Hayashi2023nonsymplectic,Hayashi2023finite}. Restricting our attention to $B = \mathbb{P}^{2}$ and $B = \mathbb{F}_{n}$ the following holds.
\begin{theorem}
	Let $X$ be a smooth K3 surface and $G$ be a finite subgroup of $\mathrm{Aut}(X)$ such that $X/G$ is smooth. There exist birational morphisms $f: X/G \rightarrow \mathbb{P}^{2}$ and $f: X/G \rightarrow \mathbb{F}_{n}$ for $n = 0,1,2,3,4,6,8,12$. Moreover, the group $G$ is one in the lists $\mathcal{A}G_{\infty}$ or $\mathcal{A}G_{n}$, respectively.
\end{theorem}
The complete lists $\mathcal{A}G_{\infty}$ and $\mathcal{A}G_{n}$ can be consulted in \cite{Hayashi2023finite}. Since we are interested in double covers of the F-theory base, the group that is relevant for us is $G = \mathbb{Z}/2\mathbb{Z}$. Since
\begin{equation}
	\mathbb{Z}/2\mathbb{Z} \in \mathcal{A}G_{\infty} \cap \mathcal{A}G_{0} \cap \mathcal{A}G_{1} \cap \mathcal{A}G_{2} \cap \mathcal{A}G_{4}\,,
\end{equation}
while
\begin{equation}
	\mathbb{Z}/2\mathbb{Z} \notin \mathcal{A}G_{3} \cup \mathcal{A}G_{6} \cup \mathcal{A}G_{8} \cup \mathcal{A}G_{12}\,,
\end{equation}
we see that smooth K3 double covers of $\mathbb{F}_{n}$ only exist for $n = 0$, $1$, $2$, $4$. The K3 double cover of $\mathbb{F}_{3}$ must always be singular, but can be crepantly resolved into a smooth K3 double cover of the blow-up of $\mathbb{F}_{3}$, as explained above. F-theory captures this as a codimension-two finite-distance singularity, an SCFT point, located at the intersection point $S \cdot \{ h' = 0 \}_{B} = 1$. The tensor branch of the SCFT interpolates between the singular and the smooth K3 double covers.

\subsection{K3 double covers and horizontal Type III.b limits}
\label{sec:K3-double-cover-horizontal-IIIb-models}

Since the Weierstrass model \eqref{eq:blown-down-Type-IIIb-model} is obtained by blowing down a horizontal Type III.b model, in addition to the codimension-zero $\mathrm{I}_{m}$ fibers we must also find some non-minimal curves in the base $\hat{B}_{0}$ of the central fiber $\hat{Y}_{0}$ of the blown-down degeneration. These will correspond to $\mathcal{S} \cap \mathcal{U}$, $\mathcal{T} \cap \mathcal{U}$ and $\mathcal{S} \cap \mathcal{U}$, or $\mathcal{T} \cap \mathcal{U}$ depending on if we have blown down to the $Y^{0}$, $Y^{p} \in \{ Y^{q} \}_{1 \leq q \leq P-1}$, or $Y^{P}$ component. Let us assume, without loss of generality, that we have blown down to the $Y^{0}$ component.

After blowing down to the $Y^{0}$ component, we must have codimension-zero $\mathrm{I}_{m}$ fibers at the endpoint of the limit, implying the structure \eqref{eq:blown-down-Type-IIIb-model} for the Weierstrass model of the blown-down degeneration, and non-minimal fibers over the curve $\mathcal{S} \cap \mathcal{U}$. The latter means that the $h$ polynomial cannot be generic, bur rather must factorize like\begin{equation}
	h = s^{2} h'\,,\qquad h' \sim 2S + (4+2n)V\,.
\end{equation}
Constructing the K3 double cover of $\hat{B}_{0}$ as explained in \cref{sec:Sen-limit-six-dimensional-F-theory}, we that $c_{3} = c_{4} = 0$, in addition to $b_{i} = 0$. Since $c_{4} = 0$, the $\mathbb{P}^{2}_{112}[4]$-fibration does admit a section.\footnote{Blowing down to the $Y^{P}$ component would mean that $c_{0} = 0$ instead, while blowing down to an intermediate component $\{ Y^{p} \}_{1 \leq p \leq P-1}$ means that $c_{0} = c_{4} = 0$. Hence, we have a section in all cases.} Given the fact that also $c_{3} = 0$, the defining polynomials of the Weierstrass model describing the Jacobian $J(\breve{B})$ are
\begin{equation}
	f_{J(\breve{B})} = -3 h_{J(\breve{B})}^{2}\,,\qquad g_{J(\breve{B})} = 2 h_{J(\breve{B})}^{3}\qquad \textrm{with}\qquad  h_{J \left( \breve{B} \right)} = \frac{1}{3} c_{2}\,,
\end{equation}
where $h_{J(\breve{B})}$ is a generic polynomial of homogeneous degree 4 in $[v:w]$. Hence, it has codimension-zero $\mathrm{I}_{m}$ fibers in codimension-zero. Moreover, we find no codimension-one non-minimal singularities due to the genericity of $h_{J(\breve{B})}$. Hence, both $J(\breve{B})$ and its birational equivalent $\breve{B}$ correspond to the central fiber of a Kulikov Type II.b model.

\subsection{K3 double covers and vertical Type III.b limits}
\label{sec:K3-double-cover-vertical-IIIb-models}

We can perform the same analysis for vertical Type III.b models. After blowing down to the $Y^{0}$ component, the curve $\mathcal{V} \cap \mathcal{U}$ will support non-minimal elliptic fibers. This means that the $h$ polynomial must factorize like
\begin{equation}
	h = v^{2} h'\,,\qquad h' \sim 4S + (2+2n)V\,.
\end{equation}
For the K3 double cover of $\hat{B}_{0}$ this implies that, in addition to $b_{i} = 0$, we also have
\begin{equation}
	h = v^{2} h' \Rightarrow c_{i} = v^{2} c'_{i}\,,\qquad i = 0,1,2,3,4\,.
\end{equation}
which leads to
\begin{equation}
	e_{i} \sim c_{i} \propto v^{2}\,,\qquad i = 0,1,2,3,4\,.
\end{equation}
As a consequence of the schematic form of $f_{J(\breve{B})}$ and $g_{J(\breve{B})}$
\begin{equation}
	f \sim e_{i}^{2}\,,\qquad g \sim e_{i}^{3}\,,
\end{equation}
we encounter the non-minimal vanishing orders
\begin{equation}
	\mathrm{ord}_{J(\breve{B})}\left( f_{J(\breve{B})},g_{J(\breve{B})},\Delta_{J(\breve{B})} \right)_{v=0} = (4,6,12)\,.
\end{equation}
The reason for the printed vanishing order for $\Delta_{J(\breve{B})}$ is that, after the factorization of $v^{2}$ has been taken into account in $h$, the remainder polynomial $h'$ is generic in a generic vertical Type~III.b model. Hence, no accidental cancellation increasing this vanishing order should occur generically. The resulting K3 double cover is the endpoint of a codimension-one degeneration of K3 surfaces. For the generic vertical Type III.b model, we expect it to correspond to the endpoint of a Kulikov Type II.a model, since producing codimension-zero $\mathrm{I}_{m>0}$ fibers would entail additional tuning.

\section{Vertical and mixed (bi)section degenerations}
\label{sec:cases-B-C-D-models}

Genus-zero single infinite-distance limit degenerations of Hirzebruch models can be classified into the Cases A to D, see  \cref{tab:genus-zero-Hirzebruch-summary}. In \cref{sec:horizontal-models-general-properties,sec:horizontal-models-heterotic-duals,sec:horizontal-models-systematics} we have analysed the general properties of the horizontal models (Case A), and studied the associated asymptotic physics in the adiabatic regime. We now turn our attention to the remaining types of geometries. We first comment on the vertical models (Case B) in \cref{sec:case-B-models}. After discussing some of their general properties, we point out why the strategy employed in the study of horizontal models is not effective for vertical models. We conclude by analysing mixed section (Case C) and mixed bisection models (Case D) in \cref{sec:case-C-models,sec:case-D-models}, respectively, where we study some of their general properties and extract their asymptotic physics in the adiabatic regime by extrapolating the lessons learnt during the study of horizontal models.

\subsection{Vertical models (Case B)}
\label{sec:case-B-models}

Among the genus-zero single infinite-distance limit degenerations of Hirzebruch models, the vertical models (Case B) are possibly the hardest to analyse in terms of their asymptotic \mbox{physics}. Since they are the degenerations ``orthogonal"  to the horizontal models, the strategy used to study the latter in the adiabatic regime no longer applies, as we elaborate on at the end of this brief section. At the same time, they would be particularly interesting to understand because they represent the F-theory duals to certain infinite-distance limits of the heterotic K3 surface, see \cref{sec:non-minimal-points-heterotic-K3-surface}. For completeness, we therefore collect some of their basic properties, like the constraints on the possible patterns of codimension-zero singular elliptic fibers in \cref{sec:effectiveness-bounds-vertical}, the restrictions on the existence of global weak coupling limits in \cref{sec:vertical-restrictions-global-weak-coupling}, and their generic horizontal slices in \cref{sec:vertical-models-generic-horizontal-slices}.

\subsubsection{Effectiveness bounds}
\label{sec:effectiveness-bounds-vertical}

The most immediate constraint on the pattern $\mathrm{I}_{n_{0}} - \cdots - \mathrm{I}_{n_{P}}$ of codimension-zero singular elliptic fibers of the central fiber $Y_{0}$ of a vertical model is obtained from demanding the divisor classes of the restrictions $\{ \Delta'_{p} \}_{0 \leq p \leq P}$ of the modified discriminant $\Delta'$, see \cref{tab:genus-zero-Hirzebruch-summary}, to be effective. This results in the bounds
\begin{subequations}
\begin{align}
    n_{0} &\geq n_{1} - 12 - 12n\,,\label{eq:vertical-model-effectiveness-constraint-horizontal-1}\\
    n_{p} &\geq \frac{n_{p-1} + n_{p+1}}{2}\,,\qquad p = 1, \dotsc, P-1\,,\label{eq:vertical-model-effectiveness-constraint-horizontal-2}\\
    n_{P} &\geq n_{P-1} - 12\,,\label{eq:vertical-model-effectiveness-constraint-horizontal-3}
\end{align}
\label{eq:vertical-model-effectiveness-constraint-horizontal}%
\end{subequations}
where $n_{p} \in \mathbb{Z}_{\geq 0}$ for all $p \in \{0, \dotsc, P\}$. As an immediate consequence, once at least one component is at local weak coupling all the intermediate components $\{ Y^{p} \}_{1 \leq p \leq P-1}$ must be at local weak coupling as well.

As discussed in \cref{sec:tighter-np-npplusone-horizontal} and \cref{sec:bounds-codimension-zero-pattern-horizontal} for horizontal models, we can obtain tighter bounds for $|n_{p} - n_{p+1}|$, with $p \in \{ 0, \dotsc, P-1 \}$, depending on the type of accidental cancellation structure used to tune the pattern $\mathrm{I}_{n_{0}} - \cdots - \mathrm{I}_{n_{P}}$. We provide the analogous results for vertical models in \cref{sec:bounds-codimension-zero-pattern-vertical}.

\subsubsection{Restrictions on global weak coupling limits}
\label{sec:vertical-restrictions-global-weak-coupling}

In global weak coupling limits all components of $Y_0$ have codimension-zero $\mathrm{I}_{n>0}$ fibers. Such models are highly tuned to fulfil the accidental cancellation structure discussed in \cref{sec:discriminant-weakly-coupled-components}, leading to strong constraints. For example, we found in \cref{sec:horizontal-restrictions-global-weak-coupling} that horizontal global weak coupling limits can only be constructed over the Hirzebruch surfaces $\hat{B} = \mathbb{F}_{n}$ with $0 \leq n \leq 4$. In this section, we carry out the same analysis for vertical models, finding that they are even more constrained, namely, vertical global weak coupling limits can only be constructed over $\hat{B} = \mathbb{F}_{n}$ with $0 \leq n \leq 2$.

In the open-chain resolution of a vertical model all components $B^{p} = \mathbb{F}_{0}$ for $p \in \{1, \dotsc, P\}$, see \cref{tab:genus-zero-Hirzebruch-summary}. Hence, we must focus on the $B^{0} = \mathbb{F}_{n}$ component to analyse forced factorisations of curves with negative self-intersection in the discriminant. In a global weak coupling limit, this component must, at the very least, satisfy the single accidental cancellation structure \eqref{eq:codimension-zero-accidental-cancellation}, with
\begin{equation}
    h_{0} \in H^{0}\left( B^{0}, \mathcal{L}_{0}^{\otimes 2} \right) \Rightarrow H_{0} = 2\mathcal{L}_{0} = 4S_{0} + (2+2n)W_{0}\,.
\end{equation}
Given the intersection numbers
\begin{align}
	H_{0} \cdot S_{0} < 0 &\Leftrightarrow n \geq 2\,,\\
	\left( H_{0} - S_{0} \right) \cdot S_{0} < 0 &\Leftrightarrow n \geq 3\,,
\end{align}
the polynomial $h_{0}$ must factorize like
\begin{equation}
	h_{0} \propto
	\begin{cases}
		s\,, & n = 2\,,\\
		s^{2}\,, & n \geq 3\,.
	\end{cases}
\end{equation}
From the structure of $f_{0}$, $g_{0}$ and $\Delta'_{0}$ we see that the minimal vanishing orders over $S_{0}$ are
\begin{equation}
    \ord{Y^{0}}(f_{0},g_{0},\Delta'_{0})_{s=0} \geq
    \begin{cases}
        (2,3,k+\alpha)\,, &n=2\,,\\
        (4,6,2k+\alpha)\,, &n \geq 3\,,
    \end{cases}
\label{eq:Type-IIIb-restrictions-vanishing-orders}
\end{equation}
where $\alpha$ accounts for additional factorisations forced by the reducibility of
\begin{equation}
    \Delta_{0}^{\prime\prime} := \Delta'_{0} - kH_{0}\,.
\end{equation}
Its value can be computed by considering the intersection numbers of the classes $\Delta_{0}^{\prime\prime}$ and $S_{0}$. The class $\Delta_{0}^{\prime\prime} - \alpha S_{0}$ will still contain $S_{0}$ components as long as
\begin{equation}
	\left( \Delta^{\prime\prime}_{0} - \alpha S_{0} \right) \cdot S_{0} < 0 \Leftrightarrow \alpha < \frac{2k + (n_{1}-n_{0}) - 12}{n} + 12 - 2k\,.
\end{equation}
Therefore, the final value of $\alpha$ is
\begin{equation}
	\alpha = \max \left\{ \left\lceil \frac{2k + (n_{1}-n_{0}) - 12}{n} + 12 - 2k \right\rceil, 0 \right\}\,.
\label{eq:alpha-value-vertical}
\end{equation}
Unlike for horizontal models, we see that $\alpha$ depends not only on $n$ and $k$, but also on $n_{0} - n_{1}$. Let us analyse the possibility of tuning vertical models corresponding to global weak coupling limits depending on the type of Hirzebruch surface $\hat{B} = \mathbb{F}_{n}$ over which they are constructed.

\paragraph{Models with $\mathbf{3 \leq n \leq 12}$}

Vertical models constructed over these Hirzebruch surfaces have components vanishing orders in the $Y^{0}$ component
\begin{equation}
    \ord{Y^{0}}(f_{0},g_{0})_{s=0} = (4,6)\,.
\end{equation}
Hence, tuning this component to be at local weak coupling forces (at the very least) an obscured infinite-distance limit over $S_{0}$. This actually takes us away from the single infinite-distance limit class of degenerations that we are focusing on in this paper. At any rate, the model demands an additional resolution process that will lead to new components at local strong coupling. We therefore conclude that vertical global weak coupling limits are not possible in models constructed over $\hat{B} = \mathbb{F}_{n}$, with $3 \leq n \leq 12$. Note that the non-minimal component vanishing orders over $\{ s=0 \}_{B^{0}}$ may by of the (naively) pathological type, see the comments in \cref{sec:horizontal-restrictions-global-weak-coupling}.

\paragraph{Models with $\mathbf{n = 2}$}

Models constructed over this Hirzebruch surface can indeed be global weak coupling limits. The necessary tuning to achieve this forces an enhancement in the $Y^{0}$ component over the $S_{0}$ curve, with component vanishing orders
\begin{equation}
	\mathrm{ord}_{Y^{0}}(f_{0}, g_{0}, \Delta'_{0})_{s=0} \geq (2,3,k+\alpha)\,,\qquad k \geq 2\,.
\end{equation}
The minimal values of
\begin{equation}
	\mathrm{ord}_{Y^{0}}(f_{0}, g_{0}, \Delta'_{0})_{s=0} \geq (2,3,2)
\end{equation}
can be attained, but this requires at least a double accidental cancellation structure, see \eqref{eq:alpha-value-vertical}, \cref{sec:discriminant-weakly-coupled-components} and the bounds discussed in \cref{sec:bounds-codimension-zero-pattern-horizontal}.

\paragraph{Models with $\mathbf{n = 1}$}

Global weak coupling limits can also be realised in those vertical models constructed over $\hat{B} = \mathbb{F}_{1}$. The necessary tuning leads in the $Y^{0}$ component to the local enhancement
\begin{equation}
	\mathrm{ord}_{Y^{0}}(f_{0}, g_{0}, \Delta'_{0})_{s=0} \geq (0,0,\max (n_{1}-n_{0},0))\,,
\end{equation}
over the $S_{0}$ curve.

\paragraph{Models with $\mathbf{n = 0}$}

Vertical models constructed over $\hat{B} = \mathbb{F}_{0}$ can correspond to global weak coupling limits, without any forced enhancements tied to the necessary tuning, since we generically have
\begin{equation}
	\mathrm{ord}_{Y^{0}}(f_{0}, g_{0}, \Delta'_{0})_{s=0} = (0,0,0)\,.
\end{equation}
This was necessary for consistency with the results of \cref{sec:horizontal-restrictions-global-weak-coupling}, since over $\hat{B} = \mathbb{F}_{0}$ there is no geometrical distinction between horizontal and vertical models.

\subsubsection{Generic horizontal slices}
\label{sec:vertical-models-generic-horizontal-slices}

The generic vertical slices of horizontal models played a prominent role in their study, first for their classification in \cref{sec:classification-horizontal-models} and later in the analysis of their asymptotic physics in \cref{sec:horizontal-models-heterotic-duals,sec:horizontal-models-systematics}. The analogue for vertical models would be to study their generic horizontal slices.

In the base components $B^{p} = \mathbb{F}_{0}$, where $p = 1, \dotsc, P$, there is no distinction between horizontal and vertical directions. Hence, the slicing behaves like  for horizontal models. In the $Y^{0}$ component, however, we do now not have a canonical choice for extending the slice from the other components. Consider first the horizontal slice by the distinguished curve
\begin{equation}
    \mathcal{H}_{0} := \sum_{p=0}^{P} S_{p} = \left. \mathcal{S} \right|_{\Tilde{U}}.
\end{equation}
This leads to an associated eight-dimensional model with
\begin{subequations}
\begin{align}
	\Delta'_{0} \cdot \left. \mathcal{H}_{0} \right|_{E_{0}} &= 12 -12n + n_{0} - n_{1}\,,\\
	\Delta'_{p} \cdot \left. \mathcal{H}_{0} \right|_{E_{p}} &= 2n_{p} - n_{p-1} - n_{p+1}\,,\qquad p = 1, \dotsc, P-1\\
	\Delta'_{P} \cdot \left. \mathcal{H}_{0} \right|_{E_{P}} &= 12 + n_{P} - n_{P-1}\,.
\end{align}
\end{subequations}
7-branes in each component. We could also consider taking horizontal slices restricting to generic representatives of
\begin{equation}
    \mathcal{H}_{\infty} := \sum_{p=0}^{P} T_{p} = \left. \mathcal{T} \right|_{\Tilde{U}}\,,
\end{equation}
which leads to the same results with the exception of
\begin{equation}
    \Delta'_{0} \cdot \left. \mathcal{H}_{\infty} \right|_{E_{0}} = \Delta'_{0} \cdot \left. \mathcal{H}_{0} \right|_{E_{0}} + 24n\,.
\end{equation}
Recombining vertical classes into the divisor used to take the horizontal slice also changes these intersection numbers by multiples of 12. Hence, all generic horizontal slices (where by generic we mean that they do not overlap with components of $\Dphys$ or pass through its intersections with the interface curves between components), lead to the same types of eight-dimensional models that were found in the analysis of horizontal models, with the only difference that they may differ by some sets of 12 7-branes. Because these sets of additional 12 7-branes appearing in the horizontal slices produce a combined trivial monodromy action, a common picture for the generic horizontal slices can still be concocted. Vertical models can therefore still be classified by inheriting the classification of Kulikov models from their generic vertical slices. In that classification, the K3 double covers associated with vertical Type III.b models are the endpoints of, generically, Kulikov Type II.a models, see \cref{sec:K3-double-cover-vertical-IIIb-models}.

However, we can no longer trust an analysis of the asymptotic physics using these slices and then the orthogonal ones, which would simply put the strategy applied to the horizontal models upside-down. The fiberwise analysis of horizontal models relied on the adiabaticity assumption, which was particularly useful thanks to the degeneration forcing the Hirzebruch surface to split along the fibral curve $\mathbb{P}_{f}^{1}$. The adiabatic limit separated, as a consequence, the defects in the orthogonal slices responsible for trivialising the local 2-cycles apart. In vertical models the Hirzebruch surface splits along the base curve $\mathbb{P}_{b}^{1}$, and the same approach is no longer possible. Understanding how the towers of asymptotically massless particles reorganise away from the adiabatic regime already for the simpler horizontal models would shed light also on the vertical models. This would be particularly interesting because, stretching the \mbox{F-theory/heterotic} duality reviewed in \cref{sec:F-theory-heteoric-duality}, they are dual to a non-minimal degeneration along the non-perturbative heterotic sector.

\subsection{Mixed section models (Case C)}
\label{sec:case-C-models}

Mixed section models, Case C in \cref{tab:genus-zero-Hirzebruch-summary}, behave very similarly to horizontal models because the non-minimal curve
\begin{equation}
    C = h + (n+\alpha)f\,,\qquad \text{where}\qquad
    \begin{cases}
        \alpha = 1\,, &\text{with}\quad n \leq 6\,,\\
        \alpha = 2\,, &\text{with}\quad n = 0\,,
    \end{cases}
\end{equation}
is also a section.

\subsubsection{Effectiveness bounds}
\label{sec:effectiveness-bounds-C}

We can constrain the pattern $\mathrm{I}_{n_{0}} - \cdots - \mathrm{I}_{n_{P}}$ of codimension-zero singular elliptic fibers of the central fiber $Y_{0}$ of a mixed section model by demanding that the divisors classes of the restrictions $\{ \Delta'_{p} \}_{0 \leq p \leq P}$ of the modified discriminant $\Delta'$, printed in \cref{tab:genus-zero-Hirzebruch-summary}, are effective. This leads to the bounds
\begin{subequations}
\begin{align}
    n_{0} &\geq n_{1} - 12\,,\label{eq:horizontal-model-effectiveness-constraint-C-1}\\
    n_{p} &\geq \frac{n_{p-1} + n_{p+1}}{2}\,,\qquad p = 1, \dotsc, P-1\,,\label{eq:horizontal-model-effectiveness-constraint-C-2}\\
    n_{P} &\geq n_{P-1} - 12\,,\label{eq:horizontal-model-effectiveness-constraint-C-3}
\end{align}
\label{eq:Case-C-model-effectiveness-constraint-horizontal}%
\end{subequations}
from their horizontal part, and
\begin{equation}
	n_{p-1} - n_{p} \leq \frac{24}{n+2\alpha}\,,\quad p = 1, \dotsc, P-1\,,\qquad n_{P-1} - n_{P} \leq \frac{24-12\alpha}{n+\alpha}\,,
\label{eq:Case-C-model-effectiveness-constraint-vertical}
\end{equation}
from their vertical part, where $n_{p} \in \mathbb{Z}_{\geq 0}$ for all $p \in \{ 0, \dotsc, P \}$.

Comparing with the effectiveness bounds obtained for horizontal models in \cref{sec:effectiveness-bounds-horizontal}, we observe that those obtained from the horizontal part are identical; this owes to the fact that the non-minimal curve $C$ is still a section. Similarly, the bounds derived from the vertical part for the differences $n_{p-1} - n_{p}$, with $p \in \{ 1, \dotsc, P-1 \}$, also coincide with the ones obtained for a horizontal model constructed over $\hat{B} = \mathbb{F}_{n+2\alpha}$, which can be understood from the structure of the central fiber $Y_{0}$ displayed in \cref{tab:genus-zero-Hirzebruch-summary}.

\subsubsection{Restrictions on global weak coupling limits}
\label{sec:C-restrictions-global-weak-coupling}

Tuning a global weak coupling limit in a mixed section model may induce new non-minimal curve,s and hence new components at local strong coupling. 
We now analyse the resulting constraints on the models.

Consider the end-component $B^{P} = \mathbb{F}_{n}$, containing the curve $S_{P}$ with negative self-intersection $S_{P} \cdot S_{P} = -n$; one can check that the other components do not lead to additional constraints. Note that, if components of this curve factorize, we do not only destroy the global weak coupling limit, but also exit the single infinite-distance degeneration class of models. This is due to the non-trivial intersection $C' \cdot S_{P} = \alpha$ between the new non-minimal curve $S_{P}$ and the strict transform $C'$ of the original non-minimal curve $C$. Any local description of the component $Y^{P}$ in terms of local coordinates will necessarily lead to the accidental cancellation structure \eqref{eq:codimension-zero-accidental-cancellation}, and we can therefore exploit the fact that
\begin{equation}
	F_{P} = 2H_{P}\,,\qquad G_{P} = 3H_{P}\,,\qquad H_{P} = 2\mathcal{L}_{P}\,.
\end{equation}

When $n = 0$, the intersection product $S_{P} \cdot S_{P} = 0$ and, as a consequence, no forced factorization of $S_{P}$ will occur. We conclude that mixed section models corresponding to global weak coupling limits are possible when $\alpha = 1$, $2$ and $n=0$.

This leaves us with the cases $\alpha = 1$, with $1 \leq n \leq 6$, to be analysed. From \cref{tab:genus-zero-Hirzebruch-summary} we read that in these cases
\begin{equation}
	H_{P} = 2S_{P} + 2V_{P}\,.
\end{equation}
In view of the intersection products
\begin{align}
	H \cdot S_{P} < 0 &\Leftrightarrow n \geq 2\,,\\
	(H - S_{P}) \cdot S_{P} < 0 &\Leftrightarrow n \geq 3\,,
\end{align}
we find that
\begin{equation}
	\mathrm{ord}_{Y^{P}}(f_{P},g_{P},\Delta'_{P})_{S_{P}} \geq
	\begin{cases}
		(2,3,k+\alpha)\,, &n = 2\,,\\
		(4,6,2k+\alpha)\,, &n \geq 3\,,
	\end{cases}
\label{eq:Type-IIIb-restrictions-vanishing-orders-Case-C}
\end{equation}
where $\alpha$ accounts for additional factors of $S_{P}$ forced by the reducibility
\begin{equation}
    \Delta^{\prime\prime}_{P} := \Delta'_{P} - kH_{P}\,.
\end{equation}
From the intersection product
\begin{equation}
	\left( \Delta^{\prime\prime}_{P} - \alpha S_{P} \right) \cdot S_{P} < 0 \Leftrightarrow \alpha < \frac{2k - (n_{P} - n_{P-1}) - 12}{n} + 12 - 2k\,.
\end{equation}
we conclude that
\begin{equation}
	\alpha = \max \left\{ \left\lceil \frac{2k - (n_{P} - n_{P-1}) - 12}{n} + 12 - 2k \right\rceil, 0 \right\}\,.
\end{equation}
As occurred for vertical models, the final value of $\alpha$ not only depends on $n$ and $k$, but also on the difference $n_{P} - n_{P-1}$.

\paragraph{Models with $\mathbf{3 \leq n \leq 12}$}

A mixed section model constructed over these Hirzebruch surfaces cannot correspond to a global weak coupling limit. Trying to force the geometry to present all components at local weak forces the curve $S_{P}$ in $B^{P}$ to factorise with component vanishing orders
\begin{equation}
    \ord{Y^{P}}(f_{P},g_{P}) \geq (4,6)\,,
\end{equation}
bringing us out of the mixed section model class due to the non-trivial intersection $C' \cdot S_{P} = \alpha$.

\paragraph{Models with $\mathbf{n = 2}$} We can construct mixed section global weak coupling limits in models constructed over $\hat{B} = \mathbb{F}_{2}$. This forces a D type local enhancement over the curve $S_{P}$ in $B^{P}$, as implied by \eqref{eq:Type-IIIb-restrictions-vanishing-orders-Case-C}.

\paragraph{Models with $\mathbf{n = 1}$} Mixed section models constructed over $\hat{B} = \mathbb{F}_{1}$ also allow for global weak coupling limits. The geometry forces a local $\mathrm{I}_{m}$ type enhancement over $S_{P}$ in $B^{P}$, with vanishing orders
\begin{equation}
	\mathrm{ord}_{Y^{P}}(f_{P},g_{P},\Delta'_{P})_{S_{P}} \geq (0, 0, \max (n_{P-1}-n_{P},0))\,,
\end{equation}
where $n_{P-1} - n_{P} \leq 6$ due to \eqref{eq:Case-C-model-effectiveness-constraint-vertical}.

\paragraph{Models with $\mathbf{n = 0}$} Mixed section global weak coupling limits are also allowed in models constructed over $\hat{B} = \mathbb{F}_{0}$, with no local enhancement enforced by the tuning.

\subsubsection{Generic vertical slices}
\label{sec:C-models-generic-horizontal-slices}

In mixed section models, the global vertical divisor in $B_{0}$ is defined in the same way as for horizontal models, namely
\begin{equation}
    \mathcal{F} := \sum_{p=0}^{P} V_{p}\,.
\end{equation}
We see from \cref{tab:genus-zero-Hirzebruch-summary} that the eight-dimensional models obtained by taking generic vertical slices are identical to the ones appearing as the generic vertical slices of horizontal models, presenting the distribution of 7-branes
\begin{subequations}
\begin{align}
	\Delta'_{0} \cdot \left. \mathcal{F} \right|_{E_{0}} &= 12 + n_{0} - n_{1}\,,\\
	\Delta'_{p} \cdot \left. \mathcal{F} \right|_{E_{p}} &= 2n_{p} - n_{p-1} - n_{p+1}\,,\qquad p = 1, \dotsc, P-1\,,\\
	\Delta'_{P} \cdot \left. \mathcal{F} \right|_{E_{P}} &= 12 + n_{P} - n_{P-1}\,.
\end{align}
\end{subequations}
Mixed section models can then be classified according to the Kulikov model type of their generic vertical slice.

Those representatives of $\mathcal{F}$ passing through the intersection points of $\Dphys$ with the interface curves between component do not lead to generic vertical slices. The number of said points of intersection at each interface is
\begin{subequations}
\begin{align}
	\Delta'_{p} \cdot S_{p} &= 24 + (n+2\alpha)(n_{p+1}-n_{p})\,,\qquad p = 0, \dotsc, P-1\,,\\
	\Delta'_{p} \cdot T_{p} &= 24 + (n+2\alpha)(n_{p} - n_{p-1})\,,\qquad p = 1, \dotsc, P-1\,,\\
	\Delta'_{P} \cdot C &= 24 + (n+2\alpha)(n_{P} - n_{P-1})\,.
\end{align}
\end{subequations}

\subsubsection{Fiberwise analysis in the adiabatic regime}
\label{sec:fiberwise-analysis-C}

Mixed section models behave very similarly to horizontal models due to the fact that the non-minimal curve $C$ in $\hat{B}_{0}$ is also a section. Their generic vertical slices  are identical to those found for horizontal models. Moreover, since the local 2-cycles defined over such slices are trivialised over the non-generic vertical slices, and these can be pushed far away from each other by demanding the hierarchy of volumes $\mathcal{V}_{\mathbb{P}_{b}^{1}} \gg \mathcal{V}_{\mathrm{K3}}$ characteristic of the adiabatic limit, we can perform a fiberwise analysis of the asymptotic physics of mixed section models in the adiabatic regime.

The bulk physics of the asymptotic model is extracted from the generic vertical slices, extending their local analysis to the generic patches of $Y_{0}$, as was done in \cref{sec:horizontal-models-heterotic-duals,sec:horizontal-models-systematics}. Since their generic vertical slices are identical, mixed section Type II.a, Type III.a and Type III.b models have the same asymptotic bulk physics as their horizontal counterparts.

The vertical gauge algebras in mixed section models lead to localised algebras living in the worldvolume of six-dimensional defects present in the decompactified theories. It is clear from their geometry that the gauge rank supported over vertical divisors or the exceptional curves in the fiber is smaller than in their horizontal counterparts. This is intuitively clear from the way in which mixed section models are constructed: The would-be non-minimal curves of a horizontal model are recombined with vertical classes and made non-minimal in order to arrive at a mixed section model. Hence, tuning a codimension-one degeneration along the curve $C$ uses up part of the divisor classes that would be associated with the localised defect algebra sector in the horizontal model.

\subsection{Mixed bisection models (Case D)}
\label{sec:case-D-models}

The final class of models correspond to a codimension-one degeneration along the bisection
\begin{equation}
	C = 2h + bf\quad \textrm{with}\quad (n,b) = (0,1), (1,2)\,.
\end{equation}
For this reason, their behaviour shows many parallels to that of horizontal and mixed section models.

\subsubsection{Effectiveness bounds}
\label{sec:effectiveness-bounds-D}

The constraints on the pattern $\mathrm{I}_{n_{0}} - \cdots - \mathrm{I}_{n_{P}}$ of codimension-zero singular elliptic fibers of $Y_{0}$ are given by the bounds
\begin{subequations}
\begin{align}
    n_{0} &\geq n_{1} - 12\,,\label{eq:horizontal-model-effectiveness-constraint-D-1}\\
    n_{p} &\geq \frac{n_{p-1} + n_{p+1}}{2}\,,\qquad p = 1, \dotsc, P-1\,,\label{eq:horizontal-model-effectiveness-constraint-D-2}\\
    n_{P} &\geq n_{P-1}\,,\label{eq:horizontal-model-effectiveness-constraint-D-3}
\end{align}
\label{eq:Case-D-model-effectiveness-constraint-horizontal}%
\end{subequations}
together with
\begin{equation}
	n_{p-1} - n_{p} \leq 6\,,\quad p = 1, \dotsc, P-1\,,\qquad n_{P-1} - n_{P} \leq \frac{12}{n+1}\,,
\label{eq:Case-D-model-effectiveness-constraint-vertical}
\end{equation}
where $n_{p} \in \mathbb{Z}_{\geq 0}$ for all $p \in \{ 0, \dotsc, P \}$.

Note that this time, the bounds \eqref{eq:Case-D-model-effectiveness-constraint-horizontal} imply that, once a single component has codimension-zero $\mathrm{I}_{m}$ fibers, all $\{ Y^{p} \}_{1 \leq p \leq P}$ components must do as well. This is different from the other of models, for which this effect occurs only for the intermediate components $\{ Y^{p} \}_{1 \leq p \leq P-1}$. It will become intuitively clear why the end-component $Y^{P}$ behaves like an intermediate component once we analyse the generic vertical slices of the model in \cref{sec:D-models-generic-horizontal-slices}.

\subsubsection{Restrictions on global weak coupling limits}
\label{sec:D-restrictions-global-weak-coupling}

To constrain the possible global weak coupling limits we analyse the effects of tuning the $Y^{P}$ component to be at local weak coupling; one can check that the other components do not lead to additional constraints. Any description of the component $B^{P}$ in terms of local coordinates will necessarily lead to, at least, the accidental cancellation structure \eqref{eq:codimension-zero-accidental-cancellation}, and therefore
\begin{equation}
	F_{P} = 2H_{P}\,,\quad G_{P} = 3H_{P}\,,\quad H_{P} = 2\mathcal{L}_{P}\,.
\end{equation}
From \cref{tab:genus-zero-Hirzebruch-summary} we see that $H_{P} = 2V_{P}$ and, therefore, we encounter no forced factorization leading to a non-minimal curve in $B^{P}$ as a consequence of tuning the codimension-zero $\mathrm{I}_{n_{P}}$ fibers over the component. This is not too surprising since the bisection degenerations are possible only over Hirzebruch surfaces $\mathbb F_n$ for $n=0,1$.

\subsubsection{Generic vertical slices}
\label{sec:D-models-generic-horizontal-slices}

Let us consider the generic vertical slices of a mixed bisection model. The non-minimal curve $C$ in $\hat{B}_{0}$ is a bisection, as can be seen from $C \cdot f = 2$. In the open-chain resolution of the model, this curve acts as the interface curve $\left. E_{P} \right|_{E_{P-1}}$ between the $B^{P-1}$ and $B^{P}$ base components. This implies that, in the base $B_{0}$ of the central fiber $Y_{0}$ of the resolved degeneration, the global vertical divisors are
\begin{equation}
    \mathcal{F} := \sum_{p=0}^{P-1} 2V_{p} + V_{P}\,.
\end{equation}

Consider now the eight-dimensional model associated with the vertical slice cut by a generic representative of this class. In the $Y^{P}$ component, we are taking a single local vertical cut, leading to one elliptic surface. In the components $\{ Y^{p} \}_{0 \leq p \leq P-1}$, however, we obtain two distinct (due to the generic choice of the representative) local vertical cuts, leading to two elliptic surfaces per component. All these elliptic surfaces intersect their neighbour over an elliptic curve, and their bases intersect in an open chain. Computing the 7-brane content of the eight-dimensional model associated with the generic vertical slice, we find
\begin{subequations}
\begin{align}
	\Delta'_{0} \cdot \left. \mathcal{F} \right|_{E_{0}} &= (12 + n_{0} - n_{1}) + (12 + n_{0} - n_{1})\,,\\
	\Delta'_{p} \cdot \left. \mathcal{F} \right|_{E_{p}} &= (2n_{p} - n_{p-1} - n_{p+1}) + (2n_{p} - n_{p-1} - n_{p+1})\,,\qquad p = 1, \dotsc, P-1\\
	\Delta'_{P} \cdot \left. \mathcal{F} \right|_{E_{P}} &= 2n_{P} - n_{P-1} - n_{P-1}\,.
\end{align}
\end{subequations}
where in the products $\Delta'_{p} \cdot \left. \mathcal{F} \right|_{E_{p}}$ with $p \in \{ 0, \dotsc, P-1 \}$ each term in parentheses corresponds to one of the two local vertical slices. Hence, the generic vertical slice corresponds to the one of a codimension-one degeneration along a section of the Hirzebruch surface, is in a horizontal or a mixed section model, whose resolution leads to a $(2P+1)$-component central fiber, but that has been ``folded" in the $Y^{P}$ component to fit into the $(P+1)$-component central fiber of the Case~D model.

This explains why the effectiveness bounds \eqref{eq:Case-D-model-effectiveness-constraint-horizontal} imply that, in the presence of at least a single component at local weak coupling, all components $\{ Y^{p} \}_{1 \leq p \leq P}$ must be at local weak coupling as well: The vertical slice of the end-component $Y^{P}$ leads to an intermediate component of the generic vertical slice, for which this effect is in place. In order for the pattern \mbox{$\mathrm{I}_{n_{0}} - \cdots - \mathrm{I}_{n_{P}}$} of codimension-zero singular elliptic fiber to match between the central fiber of the six-dimensional model and its generic vertical slice the aforementioned implication must indeed hold. Hence, the end-component $Y^{P}$ behaves in many regards like an intermediate component. Notice as well that the holomorphic line bundle $\mathcal{L}_{P}$ associated with the Weierstrass model of the elliptic fibration $\pi_{P}: Y^{P} \rightarrow B^{P}$ is purely vertical, as occurs for the intermediate components of horizontal or mixed section models.

We can then classify mixed bisectional models in the same way as the other models under consideration, i.e.\ by inheriting the classification of Kulikov models from its generic vertical fiber. Due to the ``folded" nature of such a generic vertical fiber, notice that only those Kulikov Type~III.a models with both components at local strong coupling can arise from the restriction of a mixed bisectional model.

The non-generic vertical slices are those associated with the representatives of $\mathcal{F}$ passing through the intersection points of $\Dphys$ with the interface curve over which the base components intersect. Counting the number of intersection points at each interface, we find
\begin{subequations}
\begin{align}
	\Delta'_{p} \cdot S_{p} &= 24 + 4(n_{p+1}-n_{p})\,,\quad p = 0, \dotsc, P-1\,,\\
	\Delta'_{p} \cdot T_{p} &= 24 + 4(n_{p} - n_{p-1})\,,\quad p = 1, \dotsc, P-1\,,\\
	\Delta'_{P} \cdot \mathcal{C} &= 24 + 4(n_{P} - n_{P-1})\,.
\end{align}
\end{subequations}

\subsubsection{Fiberwise analysis in the adiabatic regime}
\label{sec:fiberwise-analysis-D}

Since the curve $C$ in $\hat{B}_{0}$ is a bisection, the resulting models behave very similarly to horizontal and mixed section models also with respect to the adiabatic limit. The hierarchy of volumes $\mathcal{V}_{\mathbb{P}_{b}^{1}} \gg \mathcal{V}_{\mathrm{K3}}$ separates the non-generic vertical slices, making a fiberwise analysis possible in the adiabatic regime.

The bulk physics at the endpoint of the limit is encoded in the generic vertical slice, as we have seen already for various models. As explained above, the generic vertical slices are identical to the ones obtained for horizontal and mixed section models. Hence, the bulk physics of Case~D Type II.a, Type III.a and Type III.b models will behave as in their horizontal and mixed section model counterparts.

Vertical gauge algebras lead to localised algebras in the worldvolume of six-dimensional defects of the decompactified theory. As for mixed section degenerations, the expenditure of vertical classes in the tuning of a Case D model reduces the maximal gauge rank that can be attained for the defect algebras in the limit with respect to horizontal models. This is most clearly seen by the reduction in the number of local vertical classes in $\Delta'_{P}$.

\bibliography{references}
\bibliographystyle{JHEP}

\end{document}